university of groningen

# Partial-wave analysis of the radiative decay of $J/\psi$ into $p\bar{p}$

**PhD thesis**

to obtain the degree of PhD at the
University of Groningen
on the authority of the
Rector Magnificus Prof. C. Wijmenga
and in accordance with
the decision by the College of Deans.

This thesis will be defended in public on

Monday 29 August 2022 at 11:00 hours

by

**Rosa Kappert**

born on 21 August 1992
in Leeuwarden

## Supervisor
Prof. N. Kalantar-Nayestanaki

## Co-supervisor
Dr. J. G. Messchendorp

## Assessment committee
Prof. D. G. Ireland
Prof. P. Salabura
Prof. R. G. E. Timmermans

*‘Physics is really nothing more than a search for ultimate simplicity, but so far all we have is a kind of elegant messiness.’*

Bill Bryson

# Contents

# 1. Introduction

*How strong the lens, how keen the eyes*
*To see what we hypothesize,*
*To watch so small a thing in motion*
*As what weve christened the "Higgs boson,"*
*A tiny, massive thing that passes*
*For what can best explain the masses*
*Of other things we cannot see*
*But somehow, nonetheless, must be.*
*A thing so small is surely cute,*
*Though weirdly shaped, perhaps hirsute,*
*And just as real as any wraith*
*Imagined with the eyes of faith.*

— Jay Curlin, 2012 [1]

The Higgs mechanism was first postulated by Peter Higgs in 1964 [2] to explain how elementary particles acquire their mass. This mechanism requires that a Higgs boson should exist as an elementary particle in the Standard Model (SM). The SM is a widely accepted theory that describes the properties of the fundamental particles, called quarks and leptons, as well as describing how they interact according to three of the four fundamental forces, namely the electromagnetic, strong, and weak forces. The SM has gradually grown into a very successful theory for particle physics over the last decades of the $20^{\text{th}}$ century. One of its most compelling successes is the correct prediction of the existence of the Higgs boson, which was confirmed at the European Organization for Nuclear Research (CERN) in 2012 [3,4].

But for all its power, there still remain various unanswered questions, including aspects related 'beyond the standard model' and, the focus of this thesis, on the origin of the mass of hadrons, the composite particles in the SM. The Higgs mechanism explains only 1% of the mass of the well-known hadron - and one of the building blocks of atomic nuclei - named the proton [5]. The remaining 99% arises from internal dynamics that are not yet fully understood. Furthermore, certain states that are possible ac-cording to the SM currently have not been observed unambiguously. These non-mesonic and non-baryonic states are called exotic matter, states mainly consisting of gluons (so-called glueballs) being one type of them. To be able to correctly identify possible exotic matter, a full understanding of the spectrum of hadrons is essential. Therefore, it is crucial to study the production of hadrons in different processes, and to cover as many decay modes as possible.

The Beijing Spectrometer (BES) III at the Beijing Electron Positron Collider (BEPC) II is an outstanding setup for experiments which aim to study conventional hadrons, like charmonium, and to search for exotic states, like glueballs.

## 1.1 Particle physics and the Standard Model

Particle physics is the field of study of the fundamental constituents of matter and the forces between them. For more than 40 years these have been described by the SM. The SM is a quantum field theory which contains two types of particles, namely fermions with half integer spin and bosons with integer spin. The SM contains three 'families' of fermions. Each family consists of two leptons, two quarks and for each particle an antiparticle. All the matter we know is composed of (anti)quarks and (anti)leptons. Apart from the fermions, the SM contains gauge bosons which are force carriers of the strong, electromagnetic and weak forces and the Higgs boson, responsible for the masses of SM particles. Figure 1.1 shows an overview of all elementary particles.

The three fundamental forces known in the SM are the strong force, the electromagnetic force and the weak force. Processes due to the strong force generally occur within $10^{-22}$ seconds and processes due to the electromagnetic force take place in $10^{-14}$ to $10^{-20}$ seconds. Processes with the weak interaction are relatively 'slow' and happen typically within $10^{-8}$ to $10^{-13}$ seconds, with extremely faster and slower processes as exceptions [6].

The force carrier of the strong force is the gluon. Gluons come in 8 types and are all massless. The strong interaction is short range even though the force carriers are massless, since gluons are heavily self-interacting and the vacuum screens this interaction at long distances. At long distances the potential manifests itself in so-called flux-tubes and may become large enough

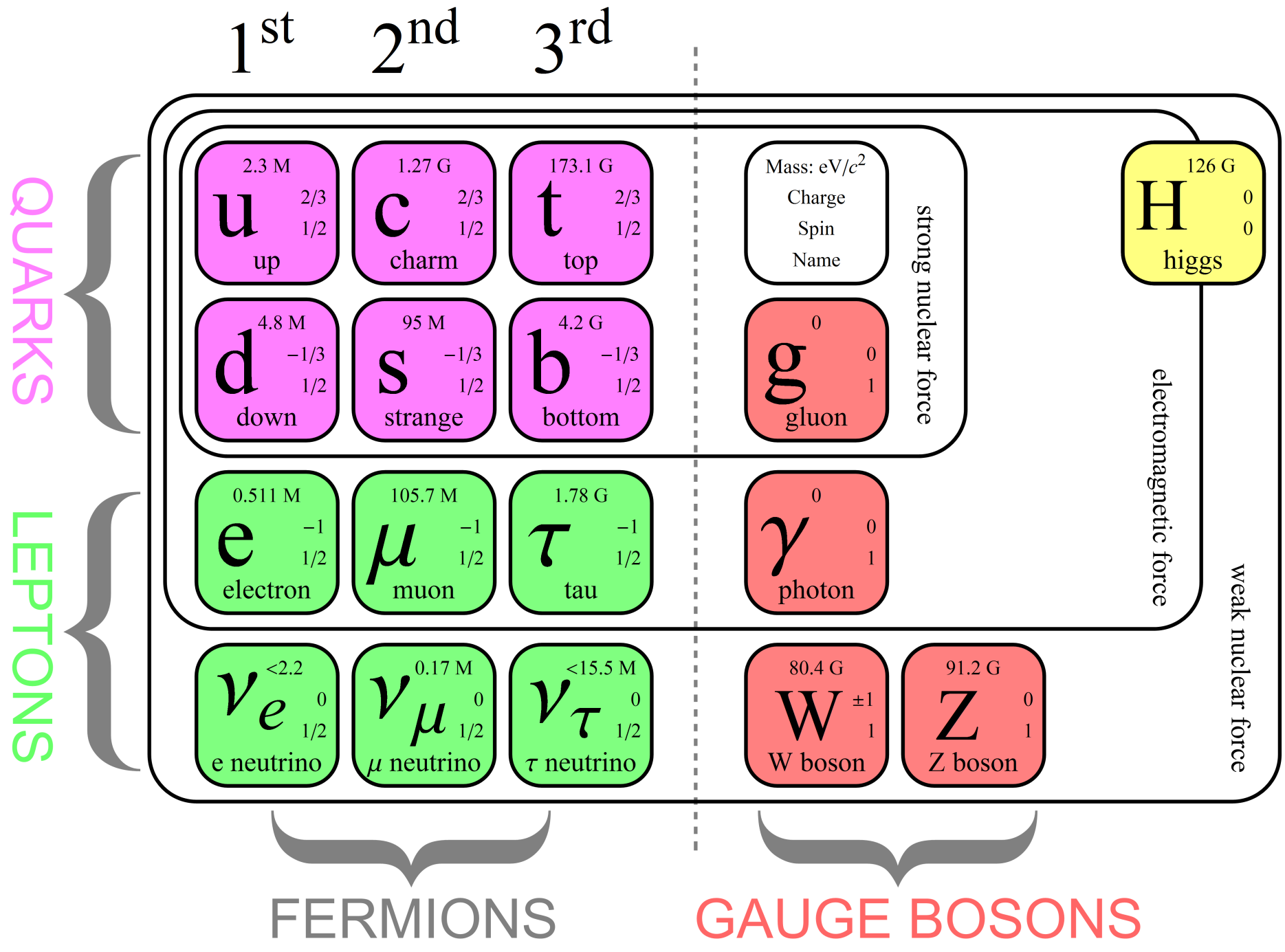

**Figure 1.1:** Overview of the different elementary particles and force carriers[1].

to create new particles, see section 1.2. The carrier of the electromagnetic force is the photon, the carriers of the weak force are the $W^+$, $W^-$ and $Z$ bosons. The weak force has a very short range, and is therefore involved in decay processes that are relatively slow. Naively, one might expect these bosons to be massless to conserve gauge symmetry, however the symmetry breaking induced by the Higgs field gives the $W^+$, $W^-$, and $Z$ bosons their masses and leaves the photon massless. All four of them have spin $S = 1$. For an explanation of the Higgs mechanism we refer to reference [7]. In 2012 the Higgs boson was detected and, therefore, added to the SM. One of the largest shortcomings of the SM is the absence of gravitation. The graviton has been postulated as the carrier of the gravitational force, which would include gravity. On the subatomic scale corresponding to distances of order

[1] Image via Google image search:
http://www.physik.uzh.ch/groups/serra/StandardModel.html.

$10^{-15}$ m, the role of gravitation can be safely ignored since its effect is negligibly small compared to the other forces in the SM.

The lepton family includes the well-known electron $e^-$. Associated with this electron is an elusive particle named the electron neutrino, $\nu_e$. The other members of the lepton family are the muon, $\mu$, with its neutrino, $\nu_\mu$, and the tauon, $\tau^-$, with its neutrino, $\nu_\tau$. All together these leptons are presented in three doublets:

$$\begin{bmatrix} e^- \\ \nu_e \end{bmatrix}, \quad \begin{bmatrix} \mu^- \\ \nu_\mu \end{bmatrix}, \quad \begin{bmatrix} \tau^- \\ \nu_\tau \end{bmatrix}. \tag{1.1}$$

The electron, muon and tauon have an electric charge of $-1e$ and interact with other charged particles via the electromagnetic force and the weak force, whereas the neutral charged neutrinos interact only via the weak force. These properties are replicated for each doublet, called a generation. The only distinctive feature between the generations is the increasing masses of the particles and a difference in lifetime. The electron is stable, the muon has a lifetime of $2.2 \times 10^{-6}$ seconds and the tauon of $2.9 \times 10^{-13}$ seconds. All six leptons have antiparticles with opposite charge. The antiparticle of the electron ($e^+$) is called the positron. The antiparticles of the muon ($\mu^+$), tauon ($\tau^+$) and neutrinos ($\bar{\nu}_e$, $\bar{\nu}_\mu$ and $\bar{\nu}_\tau$) do not have specific names. However, neutrinos could possibly be their own antiparticles, as it is still unclear if neutrinos are Dirac or Majorana[2] fermions [8].

Just like the leptons, quarks come in six types and are divided into three generations. The six types are the up $u$, down $d$, charmed $c$, strange $s$, top $t$, and bottom $b$ quarks, each with their own antiquark, $\bar{u}$, $\bar{d}$, $\bar{c}$, $\bar{s}$, $\bar{t}$ and $\bar{b}$, respectively. The $b$ and $t$ quarks are sometimes called beauty and truth, too. These quarks can be represented by three doublets as well:

$$\begin{bmatrix} u \\ d \end{bmatrix}, \quad \begin{bmatrix} c \\ s \end{bmatrix}, \quad \begin{bmatrix} t \\ b \end{bmatrix}. \tag{1.2}$$

Each doublet contains one up-type and one down-type quark with electric charges equal to $+2/3e$ and $-1/3e$, respectively. Quarks are sensitive to the strong force, and are sensitive to the electromagnetic and weak forces as well. Since the strength of the strong force does not decrease with in-creasing distance, an isolated quark would cause infinite quark self-energy. A quark is thus never found on its own [9]. This principle can be explained by color confinement and will be discussed in more detail in section 1.2.

[2] A Majorana fermion is a fermion that is its own antiparticle, whereas a Dirac fermion has a distinct antiparticle.

The particles that are interacting via the strong interaction and are composed of quarks and/or gluons are known collectively as hadrons. Among the hadrons, the proton and neutron are well-known members of the group of particles called baryons, which are made up of three valence quarks[3]. Another family of strongly interacting particles are the mesons, which are made up of a quark-antiquark pair. In the last few years tetraquark and pentaquark candidates, consisting of at least 4 and 5 quarks respectively, have been observed as well. In 2013, two independent research groups of BESIII and Belle[4] detected the tetraquark candidate $Z_c(3900)$ [10,11] and in 2016, the LHCb experiment[5] has detected four pentaquarks which are between four and five times more massive than a proton [12,13]. An overview of detected tetraquark and pentaquark candidates can be found in reference [14]. Particles mainly consisting of gluons, so-called glueballs, are predicted by the theory describing the strong force within the SM. Glueballs are the most unconventional particles predicted by theory and have not unambiguously been detected thus far [15].

## 1.2 Quantum Chromodynamics (QCD)

In the Standard Model, the strong interaction is described by a theory called Quantum ChromoDynamics (QCD). In contrast to the theory of the electromagnetic force, Quantum ElectroDynamics (QED), that can be and has been tested with a very high accuracy for a large energy range, QCD has only been tested in a similar way at very high energies. In this energy interval, QCD is in agreement with numerous experiments and is not contradicted by any experiment thus far. Here, a conceptual interpretation and some results will be described. For derivations and a more mathematical approach, see for instance reference [16] or [17].

[3] In this description the influence of gluons and sea quarks is neglected. The hadrons are thus described by the minimum-quark-content part of the wave function.

[4] The Belle (II) experiment is located at the SuperKEKB accelerator complex at KEK (High Energy Accelerator Research Organisation) in Tsukuba, Japan.

[5] The Large Hadron Collider beauty (LHCb) experiment is one of the four large detectors at LHC at CERN.

### 1.2.1 Color confinement

QCD describes the strong interaction and thus the interaction between quarks and gluons. The quark theory described in section 1.1 contradicts the Pauli exclusion principle stating that identical quarks with the same quantum numbers are not allowed to occupy the same states. Nevertheless baryons with three identical particles in the same state were found, for instance the $\Omega^-$ baryon which consist of three identical $s$ quarks [15]. To overcome this contradiction the assumption that each quark carries a color charge was introduced in 1964[6]. Color charge is the analogue of electric charge in QED and is thus a conserved quantity in interactions as well. In QCD there are three different kinds of color charge with the names red, green and blue. The quantum parameter color is, except for the naming scheme, unrelated to the ordinary visual interpretation of color. Quarks carry color, antiquarks carry the anticolors anti-red, anti-blue and anti-green. An overview is shown in figure 1.2.

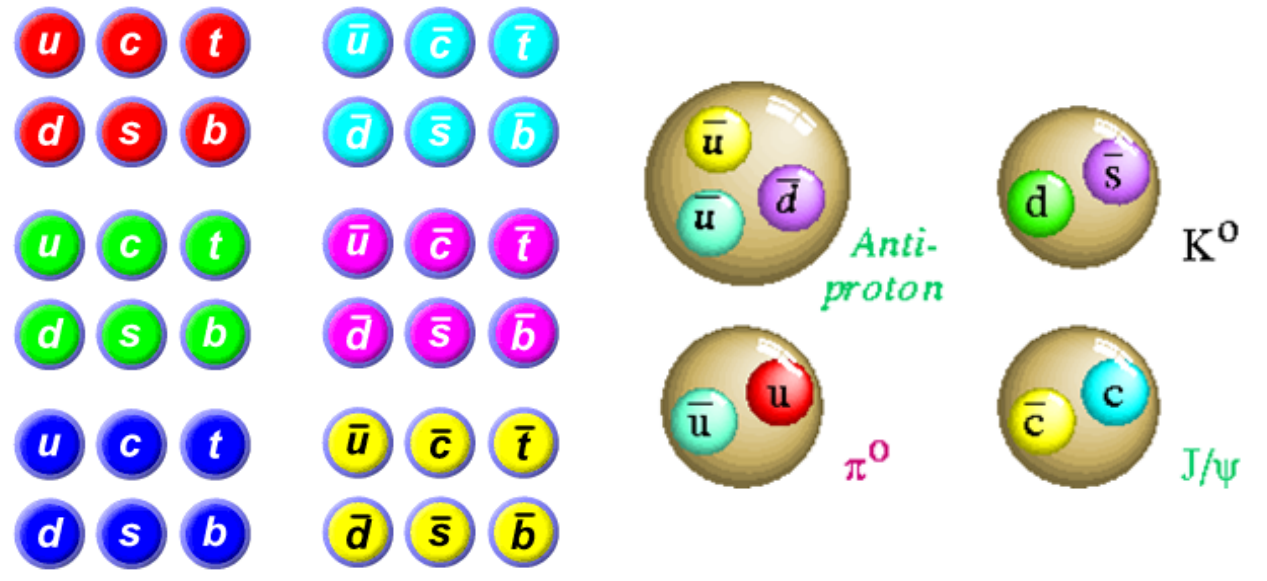

**(a)** Quarks and anti-quarks with (anti)colors.

**(b)** Examples of hadrons with the colored quarks.

**Figure 1.2:** Overview of quarks, antiquarks and hadrons with the three different color charges [7].

For hadrons to occur in nature they need to be uncharged in all color charges, a so-called color singlet [19]. The intuitive, albeit naive, explanation for this is that color-charged particles will attract oppositely color-charged particles, just like oppositely electrically-charged particles attract each other to form neutral composites. Similarly, two quarks with the same color repel, whereas two quarks with different color attract each other. A quark and antiquark will however only attract if the antiquark carries the anticolor corresponding to the color of the quark, otherwise they will repel [18]. The three colors combine to be colorless, similarly the three anticolors combine to be colorless as shown in figure 1.3.

[6] See appendix I in reference [18] for a QCD timeline.

[7] Image via Google image search:
https://www.weltderphysik.de/gebiet/teilchen/bausteine/hadronenaufbau/.

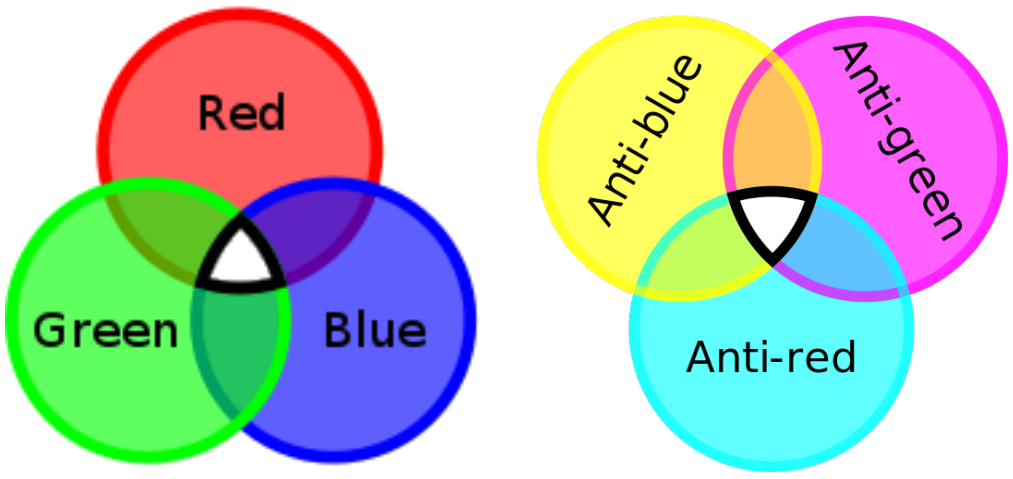

**Figure 1.3:** The three colors combine to be colorless, similarly the three anticolors combine to be colorless. [8]

The property that particles are restricted to colorless states and cannot propagate as free particles with a nonzero color charge is called color confinement. Color confinement is a hypothesis based on the fact that quarks are never observed in isolation while there is plenty of experimental evidence for the existence of quarks [15]. Due to quantum mechanical superpositions there are however not three, but eight (color) charges that must be conserved in QCD, one conserved charge per gluon.

In section 1.1 the eight gluons were introduced as the force carriers of the strong interaction. Gluons have zero electric charge, like photons, but carry and couple to color charge. An example is shown in the Feynman diagram in figure 1.4. Here, and in Feynman diagrams in general, a gluon is by convention represented by a corkscrew line. The diagram shows a quark-quark interaction with a green-antiblue ($g\bar{b}$) gluon exchange. This gluon is one of the eight types of gluon color configurations that exist in nature. In table 1.1 all eight types of gluons are listed. For hadrons the strong force cancels at long distances since forces between the colorless hadrons are given by the residue of the forces between their quark and gluon components [20].

[8] Images taken from Wikipedia: https://en.wikipedia.org/wiki/Color_charge.

| |
|---|
| $r\bar{g}$ |
| $b\bar{g}$ |
| $g\bar{b}$ |
| $r\bar{b}$ |
| $b\bar{r}$ |
| $g\bar{r}$ |
| $\frac{1}{\sqrt{6}}(r\bar{r} + g\bar{g} - 2b\bar{b})$ |
| $\frac{1}{\sqrt{2}}(r\bar{r} - g\bar{g})$ |

**Table 1.1:** The eight color configurations of gluons [18].

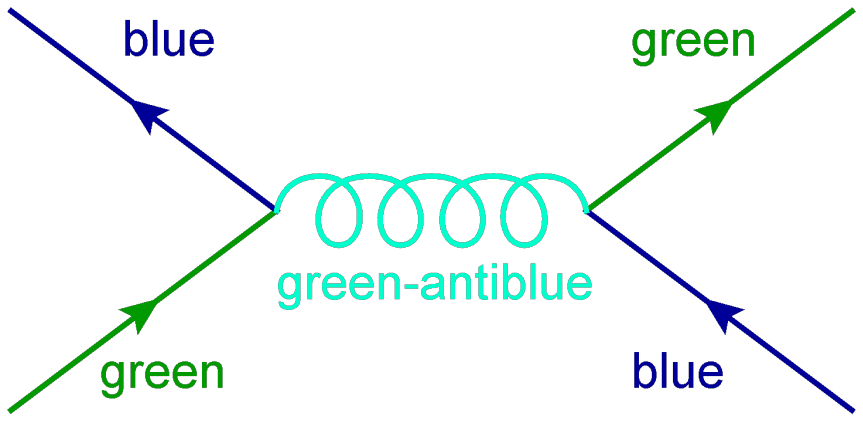

**Figure 1.4:** Quark-quark interaction with a green-antiblue ($g\bar{b}$) gluon exchange.[9]

Due to the self-interaction of gluons, the energy increases with the distance. Therefore, gluons cannot be observed freely, just like quarks. This is captured by the color confinement property, so by stating that only color singlets exist as free particles. Gluon-gluon interactions have no analogue in QED. As a result of the gluon self-interaction bound states of two or more gluons (with a total color charge of zero) could be formed according to theory. Such states are called glueballs and will be discussed in 1.3.2. The principles of color confinement, and how it introduces exactly eight gluons, are a result of symmetry groups, as will be explained in the following section.

[9] Image taken from Wikimedia:
https://commons.wikimedia.org/wiki/File:Gluon_boson.gif

### 1.2.2 Color confinement and the SU(3) symmetry group

The Standard Model is assumed to be invariant under specific local phase transformations and can, therefore, be represented as the symmetry group[10]

$$U(1) \times SU(2) \times SU(3). \tag{1.3}$$

In general, each symmetry is associated with a conserved quantity, such as energy, momentum, or charge. Each interaction has its own symmetry group. $U(1)$ is the symmetry group of the electromagnetic interaction, $SU(2)$ the symmetry group of the weak interaction and $SU(3)$ the symmetry group of the strong interaction. The conserved quantity of a gauge symmetry is the charge (electric charge, isospin, color) [23]. The generator for $U(1)$ is the photon. $SU(2)$ has the $W^0$, $W^+$, and $W^-$ bosons as its three generators, whereas $SU(3)$ has eight generators which are the eight different gluons.

To understand how the $SU(3)$ group represents color symmetry, two additive quantum numbers will be introduced, namely the color isospin $I_3^c$ and color hypercharge $Y^c$. The following derivations are based on the book "Modern Particle Physics" [19]; more background information can be found in this book, too.

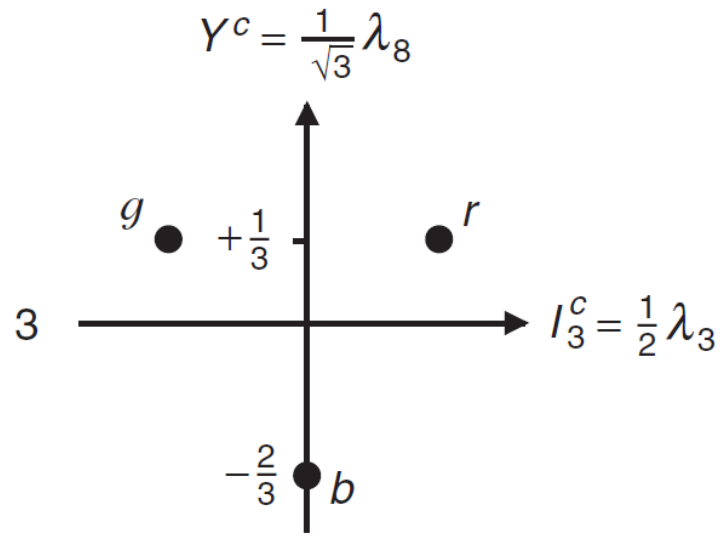

**Figure 1.5:** The representations of the color of quarks and the anticolor of antiquarks.

Color confinement implies that quarks are always observed to be confined to bound colorless, or color singlet, states. For a state to be a color singlet, it is required that the isospin $I_3^c$ and color hypercharge $Y^c$ are both zero, however a state with $I_3^c = Y^c = 0$ is not necessarily a color singlet state. The values of $I_3^c$ and $Y^c$ for the different (anti)colors are shown in figure 1.5. First, consider the possible color wave functions for a bound $q\bar{q}$ state, a meson. Of the three possible states with $I_3^c = Y^c = 0$, only two are linearly independent. Therefore, it can be concluded that one of these states must be in a different $SU(3)$ multiplet, as shown in figure 1.6. The resulting colored multiplets of the $q\bar{q}$ states are thus an octet and a singlet. In terms of the $SU(3)$ group structure this can be expressed as $3 \otimes \bar{3} = 8 \oplus 1$. The color confinement hypothesis implies that all hadrons must be color singlets, so the eight octet states drop out and the only physically-allowed color wave function for mesons is

$$\psi_c(q\bar{q}) = \frac{1}{\sqrt{3}}(r\bar{r} + g\bar{g} + b\bar{b}). \tag{1.4}$$

[10] For an introduction to symmetry groups (in physics) we refer to references [21] and [22].

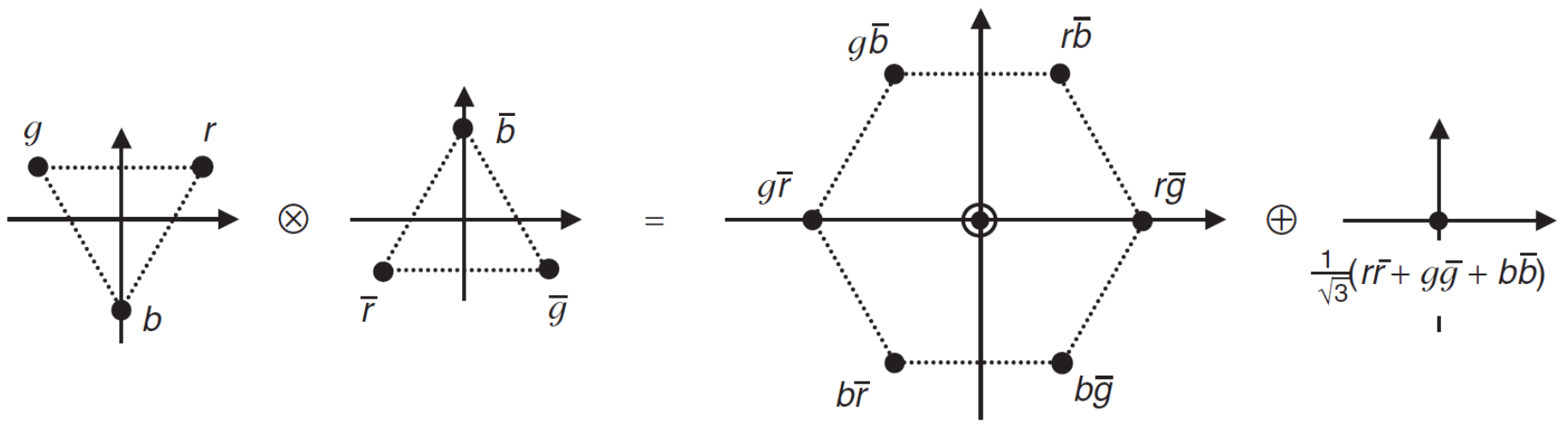

**Figure 1.6:** The representations of the color of quarks and the anticolor of antiquarks.

The addition of another quark (or antiquark) to either the octet or singlet state in figure 1.6 cannot give the required property that both, the color isospin $I_3^c$ and color hypercharge $Y^c$, are equal to zero. Therefore, bound states of $qq\bar{q}$ or $q\bar{q}\bar{q}$ are not allowed by color confinement. With similar reasoning, the combinations of two and three quarks can be checked, as shown in figure 1.7. The combination of two quarks gives a color sextet and a color triplet: $3 \otimes 3 = 6 \oplus 3$. There is no color singlet state and thus bound states of two quarks are always colored objects and are not expected to exist in nature according to color confinement. However, the combination of three colored quarks yields $3 \otimes 3 \otimes 3 = 10 \oplus 8 \oplus 8 \oplus 1$. The singlet state satisfies

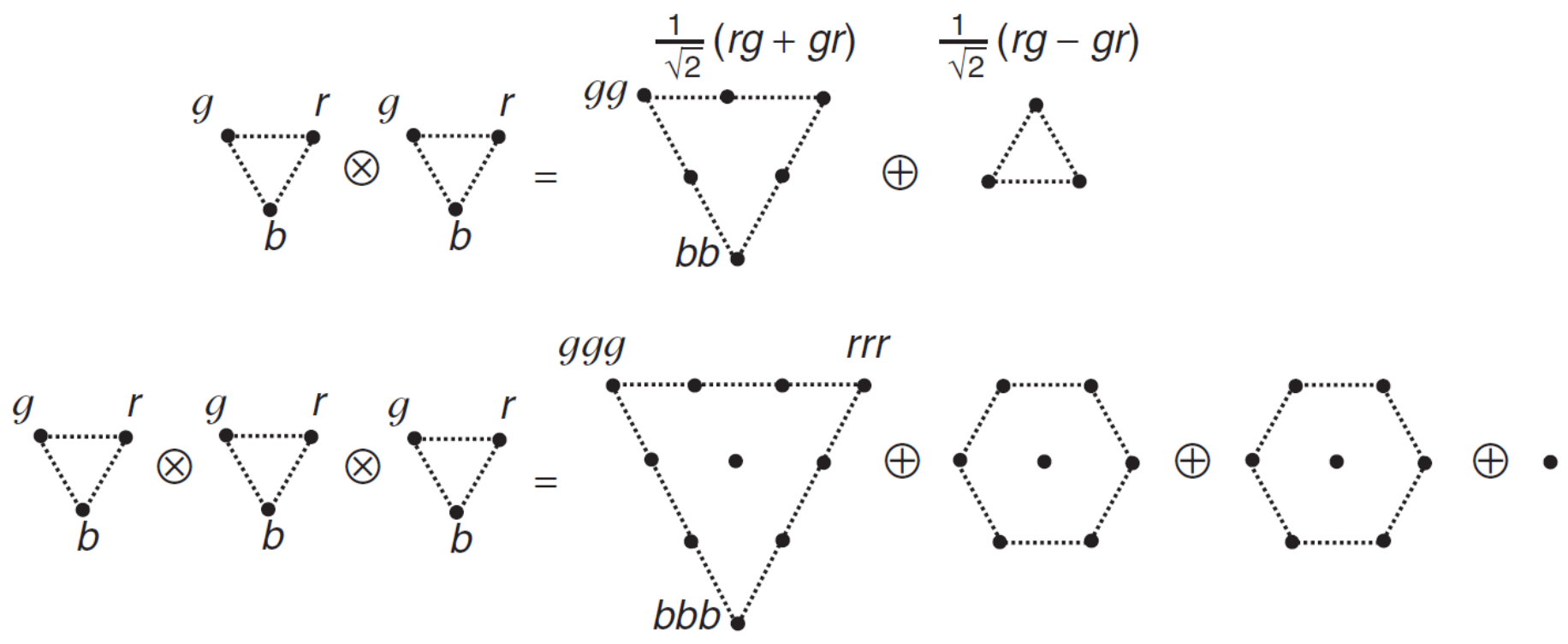

**Figure 1.7:** The multiplets from the color combinations of two and three quarks.

the requirement that $I_3^c = Y^c = 0$ and it has the color wave function

$$\psi_c(qq) = \frac{1}{\sqrt{6}}(rgb - rbg + gbr - grb + brg - bgr). \tag{1.5}$$

These colorless bound states of $qqq$ are the various baryons observed in nature. Color confinement introduces strong restrictions on the possible hadronic states that can be formed of (anti)quarks. All hadronic states confirmed thus far indeed correspond to color singlets in the form of mesons ($q\bar{q}$), baryons ($qqq$) or antibaryons ($\bar{q}\bar{q}\bar{q}$). In principle, combinations of ($q\bar{q}$) and ($qqq$) such as pentaquark states ($qqqq\bar{q}$) or tetraquarks ($q\bar{q}q\bar{q}$) are possible color singlets, either as bound states in their own right or as hadronic molecules like ($q\bar{q}$)-($qqq$). The detected tetra- and pentaquark candidates mentioned in section 1.1 are thus still in agreement with color confinement. However, they are considered to be exotic hadrons. In general, all non-mesonic and non-baryonic states are called exotic matter, including pure glueballs (e.g. $gg$), as well as hybrids (e.g. $q\bar{q}g$).

Another consequence of the color confinement hypothesis is that the colored gluons cannot propagate over large distances, since they are also confined to colorless objects. The eight gluons that were listed in table 1.1 can be derived in a similar way as the meson octet was derived. It is interesting to note that, had nature chosen a $U(3)$ local gauge symmetry, rather than $SU(3)$, there would be a ninth gluon in a state similar to the meson singlet. This gluon would thus be the color singlet, $G_9 = \frac{1}{\sqrt{3}}(r\bar{r} + g\bar{g} + b\bar{b})$.

This colorless gluon state would be unconfined, like a strongly interacting photon, and would result in an infinite range strong force. If such long-range strong interactions between all quarks and nucleons existed, the Universe would be completely different. With the eight colored gluons that do exist we can however create colorless glueballs, since, in terms of the $SU(3)$ group structure, $gg$ can be expressed as [24]

$$8 \otimes 8 = (1 \oplus 8 \oplus 27) \oplus (8 \oplus 10 \oplus \overline{10}). \tag{1.6}$$

Similarly any glueball composed of more than two gluons can be expressed as [24]

$$8 \otimes \cdots \otimes 8 = 1 \oplus 8 \oplus \ldots \tag{1.7}$$

The colored gluons can thus create a color singlet with only gluons! Therefore, glueballs are allowed according to color confinement. To test if glueballs truly exist, a well-defined theoretical framework that can predict their properties is necessary. Then, it is up to experiments to show if glueballs really exist. Note that, in general, experimentally observed 'states' can be mixtures of the afore introduced states. Therefore, it can be a complex quantum-mechanical challenge to extract information from experiments. A partial-wave analysis (see chapter 5) can give more insight in the different constituents of the observed mixture.

Aside from confinement, the gluons and their self-interactions imply another important QCD characteristic: asymptotic freedom.

### 1.2.3 Asymptotic freedom and confinement

It is believed that the origin of color confinement lies in the property of asymptotic freedom. Asymptotic freedom was discovered by Gross, Wilczek [25] and Politzer [26] and is the phenomenon that the strong force between partons[11] becomes weaker when partons are closer together. If the distance between the particles is very small the partons can almost freely move around. On the other hand, when partons are forced to separate, the strong force between them becomes stronger. Partons are thus confined inside a hadron. If the quarks are separated far enough it will be energetically more favorable to create an additional quark-antiquark pair to produce two different hadrons, see figure 1.8.

[11] Quarks and gluons are collectively called partons.

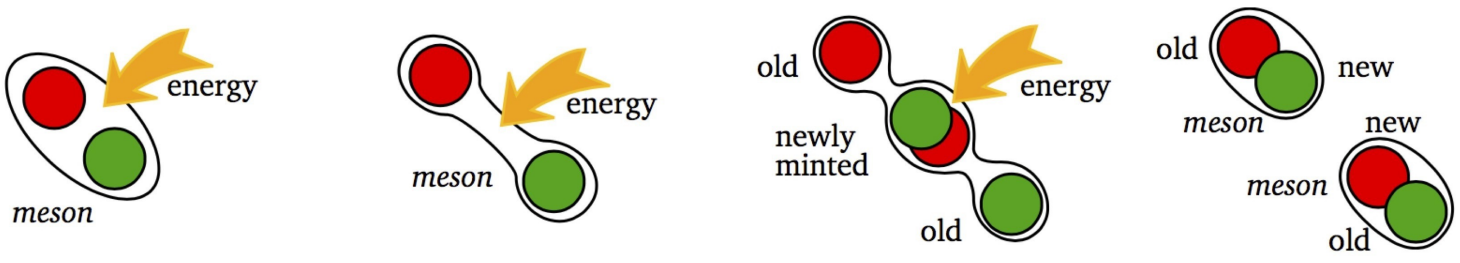

**Figure 1.8:** Quarks cannot be isolated, even when energy is added[12].

Asymptotic freedom has been observed experimentally, as shown in figure 1.9. At distances less than about 0.1 fm the lowest-order diagrams dominate and perturbation theory can be used. For larger distances the coupling becomes stronger, leading to color confinement for distances about 0.7 fm and up. Eventually, for distances larger than one fermi, an effective strong force, called the nuclear force and described by the exchange of colorless mesons, holds the nucleus of an atom together [20, 27]. Since the strong coupling is weak at small distances (high energies), perturbation theory can be reliably used in this regime and calculations based on perturbation theory describe the experimental data. On the contrary, at large distances (low energies), perturbation theory is no longer applicable.

For large distances the interaction becomes very complicated to calculate, limited by computational resources. It has not yet been possible to evaluate all observables of the theory with a high accuracy in this regime. Even if observables can be calculated, there is a lack of understanding. The true principles of strong coupling QCD still need to be unraveled. Several non-perturbative methods are used to find the approximate behavior of the strong force at low energies, each with its own approximations and limitations. Lattice Quantum ChromoDynamics (LQCD), effective field theory (EFT) and the use of potential models are three examples of these non-perturbative methods that will be discussed hereafter. Note that there are more models that aim to describe QCD, such as the Dyson-Schwinger equations (DSEs) [28] and the anti-de Sitter/conformal-field-theory (AdS/CFT) correspondence [29].

[12] Image via Google image search:
https://www.quora.com/What-is-quark-colour-confinement.

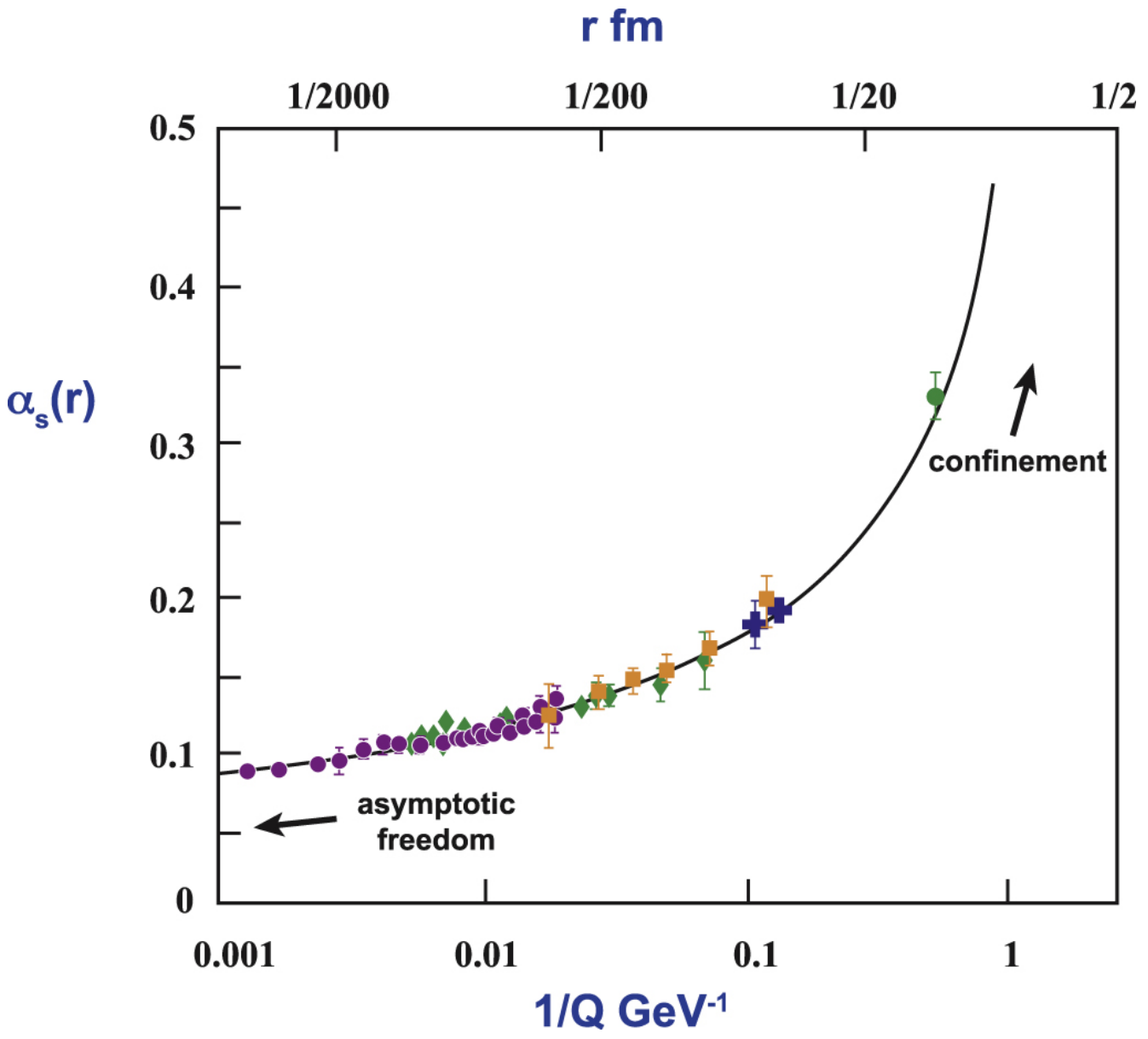

**Figure 1.9:** The strength of the coupling of QCD, $\alpha_s$ varies with distance, $r$ or $1/Q$, where $Q$ is related to the energy scale of the interaction. The line represents calculation from perturbation theory, the dots are data points from several experiments (see reference [15] for references) [27].

### 1.2.4 Potential models

Long before ab-initio non-perturbative QCD evaluations became feasible, potential models turned out to be extremely useful for an understanding of the masses and decay rates of bottomonium ($b\bar{b}$), charmonium ($c\bar{c}$) and $B_c$ ($\bar{b}c/b\bar{c}$) states. The underlying idea of potential models is that for a quark mass $m_q$ that is larger than all other energy scales related to bound states, the time scales that describe the relative motion of a heavy quark are also much larger than time scales associated with the gluon and sea quark dynamics. Feedback effects of the moving heavy quarks onto the surrounding gluons and sea quarks might, therefore, be neglected. If, additionally, the typical relative quark velocity is much smaller than the speed of light, $v = p/m \ll c$, the bound state is obtained by a Schrödinger Hamiltonian,

$$H = T + V(r), \quad H\psi_n = E_n\psi_n, \tag{1.8}$$

where $T$ is the kinetic energy term and $V(r)$ the potential energy term. From constraints given by confinement and asymptotic freedom the potential can be constructed to have the form [30]

$$V(r) = -\frac{4}{3}\frac{a_s(r)\hbar c}{r} + br, \tag{1.9}$$

where $a_s(r)$ is the coupling constant of QCD. The first term is comparable with the Coulomb potential in QED and describes a single-gluon interaction. The distance-dependent coupling constant $a_s(r)$ is given by

$$a_s(r) = \frac{2\pi}{9\ln\frac{1}{r\Lambda_{QCD}}}, \tag{1.10}$$

where $\Lambda_{QCD}$ is the non-perturbative scale of QCD and is found to be of the order of a few hundred MeV [31]. The second term in equation 1.9 represents quark confinement, with $b$ the force constant. The potential given in equation 1.9 can be extended by including three more terms related to the hyperfine, spin-orbit and tensor components. The choice of the full potential is not unique, since QCD only hints to an asymptotic behavior of $V(r)$.

A potential model is a good approximation for charmonium states, like $J/\psi$, since the mass of the charm quark, $m_c = 1.28 \pm 0.03$ GeV$/c^2$, is larger than the other energy scales of bound states and the relative velocity between the quark and antiquark is, with $v^2/c^2 \approx 0.3$, relatively small [31,32]. In reference [33] two potential models for charmonium states are discussed and the results are compared with experimental values. Overall, the two potential models are able to reproduce the experimental charmonium spectrum adequately. In chapter 7, the results obtained from our analysis will be compared to results from potential models.

### 1.2.5 Effective field theories (EFT)

To describe physics with one theory, one has to deal with all scales, from the age of the universe of about $10^{18}$ seconds, to the lifetime of a $W^{\pm}$ or $Z$, a few times $10^{-25}$ seconds. In almost every energy domain there are physical phenomena worth studying. It is thus convenient if physics from a certain domain can be isolated from the rest, so that it can be described without having to understand everything. Setting parameters that are much smaller (larger) than the physical parameters of interest equal to zero (infinity) may result in a simpler approximate description of the physics. The effects of neglected parameters can be included as small perturbations [34]. This is a commonly used trick in physics, Newtonian mechanics is, for example, just the limit of relativistic mechanics at small velocities. If $v \ll c$, relativity can be ignored completely and a full Newtonian approach is more practical. Effective Field Theory (EFT) in QCD is based on a similar idea. The general step-by-step plan is as follows:

- Perform a systematic approximation in a certain energy domain with respect to some scale $\Lambda$;
- Identify the relevant degrees of freedom and symmetries within the domain;
- Construct the most general Lagrangian consistent with the symmetries;
- With standard quantum field theory (QFT), find the relevant equations using this Lagrangian.

In QCD, the dynamics at low energies (large distances) is independent of the dynamics at high energies, so low-energy dynamics can be described using an effective Lagrangian that ignores high-energy dynamics [31, 32]. Instead of the velocity of light in Newtonian physics, this EFT has an intrinsic energy scale $\Lambda_{QCD}$ that defines the limit of validity.

To describe the dynamics of quarkonium, one has to consider a multi-scale problem as well. The non-relativistic nature of quarkonium (like charmonium) imposes the hierarchy $m_Q \gg p \gg E$ and $m_Q \gg \Lambda_{QCD}$, with $m_Q$ the heavy-quark mass, $p$ the relative momentum and $E$ the binding energy [35]. EFT is thus a very useful tool to describe physical processes taking place at lower energy scales. Heavy quarkonium annihilation and production can, for instance, be described by Non-Relativistic QCD (NRQCD) [36] and quarkonium formation by potential NRQCD (pNRQCD) [37, 38]. The results that are obtained from the analysis presented in chapter 7 will be compared to predictions from NRQCD.

### 1.2.6 Lattice quantum chromodynamics (LQCD)

LQCD was first formulated by Kenneth G. Wilson in 1974 [39]. The method starts from first principles, namely the QCD Lagrangian. In LQCD, space-time is discretized to a finite multidimensional lattice with discrete points, as illustrated in figure 1.10. Quarks are located at the sites and gauge fields, representing gluons, are placed on the links between the sites. Generally, the lattice is a hypercube which represents the four spacetime dimensions. On this lattice, the QCD path integral can be estimated stochastically with nonzero lattice spacing [40]. The exact theory can be recovered when the lattice spacing $a$ approaches zero and the number of lattice points becomes infinite. For decreasing lattice spacing and increasing volume the costs for numerical computations become, however, extremely high. Therefore, repeated calculations at different nonzero lattice spacings $a$ are performed to make sure that the results do not depend on the lattice spacing. The quark masses are input variables of LQCD, which creates the interesting opportunity to explore the mass dependence [40].

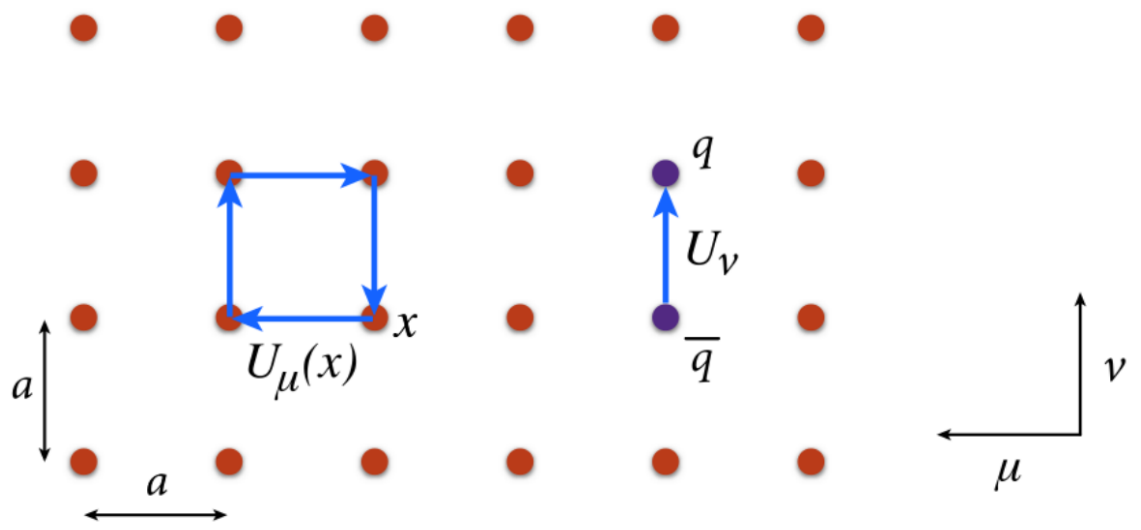

**Figure 1.10:** Sketch of a two-dimensional slice of the $\mu-\nu$ plane of a lattice. Gluon fields lying on links and forming either the product appearing in the gauge action, $U_\mu(x)$, or a component of the covariant derivative connecting quark and antiquark fields[13], $U_\nu$ [32].

Until recently, the computational costs of LQCD calculations at realistically small quark masses were too high, so results needed to be extrapolated to obtain physical quantities. However, recent algorithmic improvements have significantly reduced this problem [15]. Increasingly, LQCD calculations can be performed at, or very close to, the physical quark masses. Still, all LQCD calculations are restricted due to limitations in computational resources and in the efficiency of algorithms. Nonetheless, LQCD calculations yield often the most accurate and straightforward approach to solve some QCD problems and compute the masses of hadrons. In chapter 7,

[13] See chapter "Lattice quantum chromodynamics" in reference [15] for a detailed description.

several LQCD predictions will be consulted to compare with our results.

## 1.3 Hadron spectroscopy

Applying the rules of QCD, numerous different hadrons can be imagined. The subfield of particle physics that studies the masses and decays of the possible hadrons is called hadron spectroscopy. The experimental identification and precise characterization of hadrons provide important tests for the different models that try to describe the confinement regime of QCD. In hadron spectroscopy, the hadrons are commonly described by a spectroscopic notation. This notation represents the most important quantum numbers that characterize a state. One of the represented quantum numbers is the total angular momentum $J$, defined as the the sum of the intrinsic spin $S$ and the angular momentum $L$ between the hadron constituents: $\vec{J} = \vec{L} + \vec{S}$. The projection $J$ is thus restricted to be $|L - S| \leqslant J \leqslant |L + S|$. In the case of conventional mesons, the quark and antiquark can couple to a total intrinsic spin of $S = 0$ or $S = 1$ and the parity can be described by $P = (-1)^{L+1}$. The $C$-parity is only defined for mesons that consist of a quark and an antiquark pair of the same family, like $u\bar{u}$ or $d\bar{d}$. For these cases, the $C$-parity is determined by $C = (-1)^{L+S}$. Here, $P$ and $C$ stand for the parity-symmetry, and charge-symmetry, respectively. A state for which both of these quantum numbers are $+1$ is thus symmetric when the particle is interchanged with its antiparticle (charge change) and also symmetric under an inversion of the spatial coordinates (parity change). For strong and electromagnetic interactions, both $P$ and $C$ are conserved. Additionally, the total angular momentum $J$ is conserved in all interactions, whereas the orbital angular momentum $L$ is not a conserved quantity. The different hadrons can be represented by the spectroscopic spin-parity notation $J^{PC}$, or $J^P$. Another commonly used spectroscopic notation additionally includes the radial quantum excitation number $n_r$, resulting in $(n_r + 1)^{2S+1}L_J$. Note that for ordinary mesons not all combinations of $J^{PC}$ can be achieved using the relations listed above. For instance, the combinations $0^{--}$, $0^{+-}$ and $1^{-+}$ are forbidden for non-exotic quark-antiquark pairs. On the other hand, for ordinary baryons, all half integral $J^P$ quantum numbers are allowed.

When only the three lightest quarks $u$, $d$ and $s$ are considered, in total nine different quark-antiquark states can be constructed. These different states can be presented graphically using their isospin $I_3$ and hypercharge

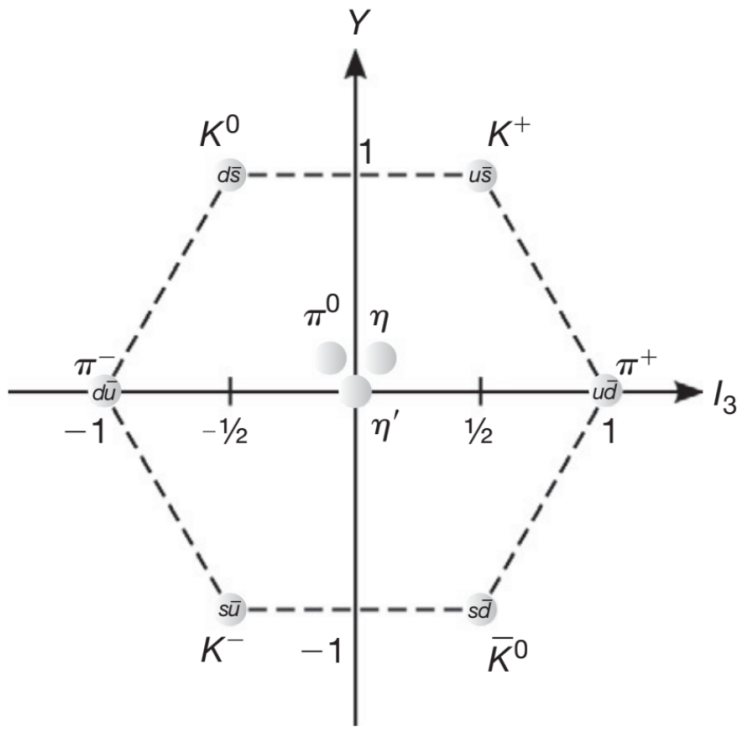

**(a)** $0^-$ meson nonet

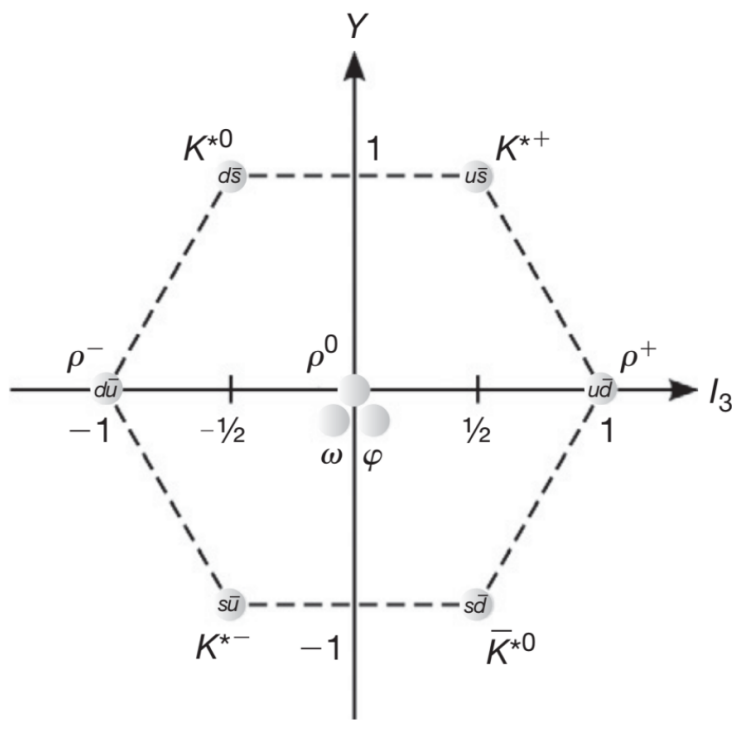

**(b)** $1^-$ meson nonet

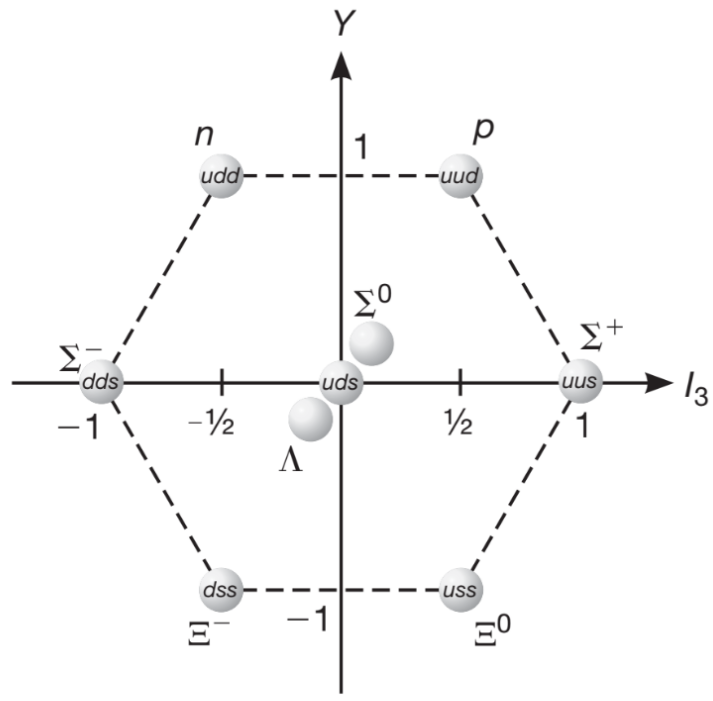

**(c)** $\frac{1}{2}^+$ baryon octet

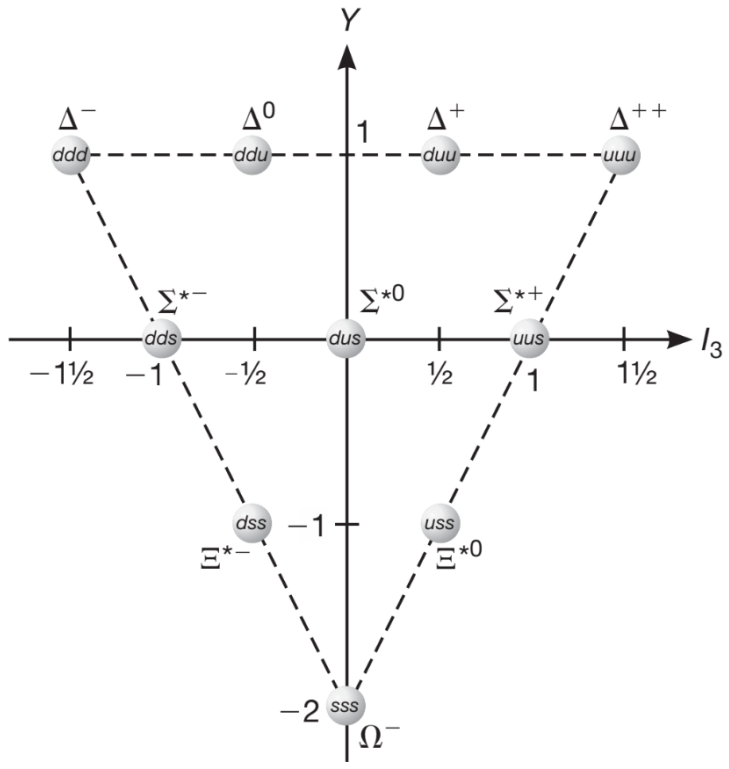

**(d)** $\frac{3}{2}^+$ baryon decuplet

**Figure 1.11:** Light baryons and mesons represented in multiplets [6].

$Y$ properties. The nonets of the lightest mesons with $L = 0$, $S = 0$ and $L = 0$, $S = 1$ are shown in figures 1.11a and 1.11b, respectively. These states are called pseudoscalar ($0^-$) and vector ($1^-$) mesons. Although all the charged mesons shown in figure 1.11 correspond to a specific quark-antiquark pair, the neutral particles correspond to a linear combination of quark states. The $\pi^0$ and $\rho^0$ are a linear combination of $u\bar{u}$ and $d\bar{d}$, and the $\eta$, $\eta'$, $\omega$ and $\phi$ correspond to linear combinations of $u\bar{u}$, $d\bar{d}$ and $s\bar{s}$. In a similar fashion as for the mesons, the lightest baryons can be described by the octet with $J^P = \frac{3}{2}^+$, shown in figure 1.11c, and the decuplet with $J^P = \frac{1}{2}^+$, shown in figure 1.11d. The baryon octet contains the proton, which is an important particle for the analyses presented in this thesis.

When the charm quark is considered as well, the possible multiplets can be further extended, as illustrated in figure 1.12 for the pseudoscalar and vector mesons. In this figure, the different layers represent the number of charm quarks present in the mesons: green for one $c$ present, blue for one $\bar{c}$ present. The mesons in the red layer either consists of the $c\bar{c}$ pair ($J/\psi$ and $\eta_c$), or contain no $c$ or $\bar{c}$ (the rest). Mesons consisting of a $c\bar{c}$ pair are generally known as charmonium. Charmonium mesons are important states for hadron spectroscopy, since they provide good probes to study the confinement regime of QCD, as the properties of charmonium are determined by the strong interaction [30]. Additionally, the charmonium system is considered a very clean environment, as outlined in the next section.

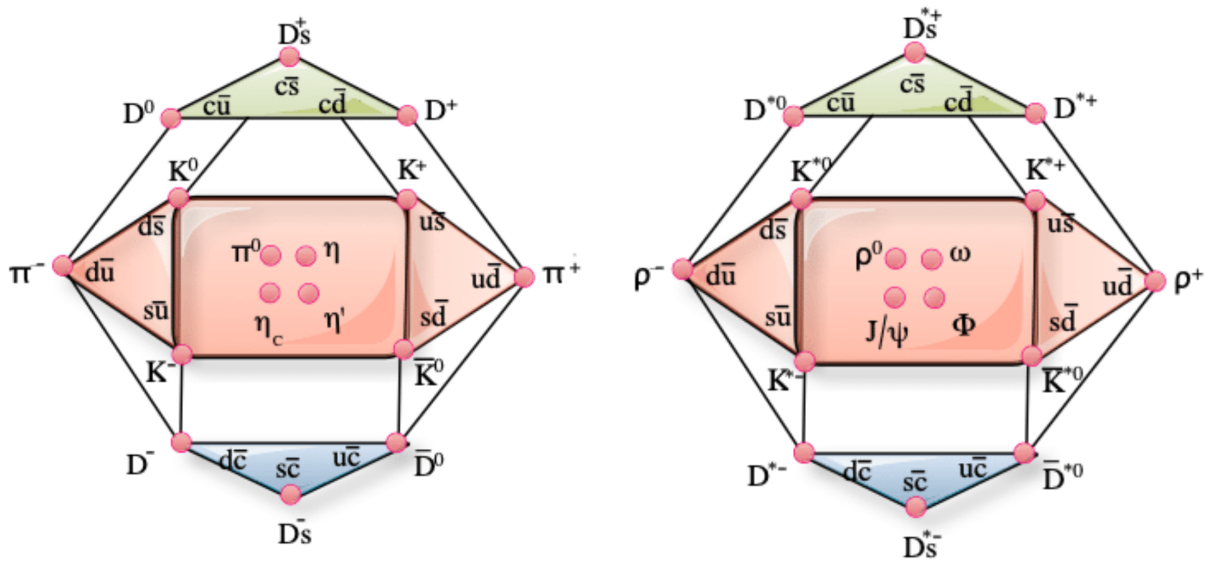

**(a)** Pseudoscalar mesons **(b)** Vector mesons

**Figure 1.12:** The multiplets for mesons made of $u$, $d$, $s$ and $c$ quarks [41].

### 1.3.1 Charmonium

In 1974, the first discovery of charmonium was announced independently by two different groups. The group at the Brookhaven National Laboratory (BNL) in New York called this new, narrow resonance $J$ [42]. The other group at the Standford Linear Accelerator Center (SLAC) in California called the same particle $\psi$ [43]. Hence, the double name for $J/\psi$. The discovery of $J/\psi$ was not just the first observation of charmonium, it was the first particle to be discovered that contained a charm quark at all. With this discovery, the partner for the strange quark was finally found, as predicted by the mechanism of Glashow, Iliopoulos, and Maiani (GIM-mechanism) [44]. Thereby, strengthening the theory of quarks as physical particles instead of mathematical concepts.

Nowadays, charmonium spectroscopy is an active field of physics and there are many charmonium(-like) particles discovered and predicted. An overview is shown in figure 1.13. The dashed lines indicate the open-charm energy thresholds, the lowest line corresponds to twice the mass of the lightest open-charm meson $D$ ($c\bar{u}/c\bar{d}$). Above these thresholds, charmonium can thus decay into a pair of mesons containing both one (anti)charm quark. Therefore, a large number of possible decay channels becomes available, resulting in broader, overlapping charmonium states. Below the thresholds, due to the limited number of available decay channels, the charmonium states are in general long-living, narrow states. Additionally, hadronic decays of charmonia are strongly suppressed by the so-called OZI-rule [45], further increasing the lifetimes and narrowing the widths. Together, this results in a spectrum with an excellent signal-to-background ratio, which is particularly favorable for charmonium, or more general, quarkonium spectroscopy. In figure 1.13, several so-called $XYZ$-states are shown. These states are relatively recently discovered heavy particles whose properties do not fit the standard picture of ordinary charmonium (or bottomonium), even though they definitely contain a charm-anticharm (or bottom-antibottom) pair. Among these new states, those that carry an electric charge are of special interest. As a result of the nonzero charge, these states must contain at least a light quark-antiquark pair besides the heavy $c\bar{c}/b\bar{b}$ pair. Hence, they immediately qualify as exotics. One of these states is the tetraquark candidate $Z_c(3900)$, mentioned in section 1.1. In fact, many of the $XYZ$-states might represent exotic states, like glueballs, hybrids or mesonic molecules. A detailed experimental and theoretical overview of the $XYZ$-states can be found in reference [46].

To gain further insight in this exotic charmonium-like spectrum, it is essential to have a deeper understanding of the charmonium states in the clean region below the open-charm threshold. In this region, there are eight different charmonia established. There are two types of transitions possible between the different states: hadronic transitions and radiative transitions. The following description of both types are based on reference [47].

Hadronic transitions are only possible when the mass difference between two charmonia is large enough to produce one or more pions, and/or an $\eta$ meson. A hadronic transition is for instance not possible between the

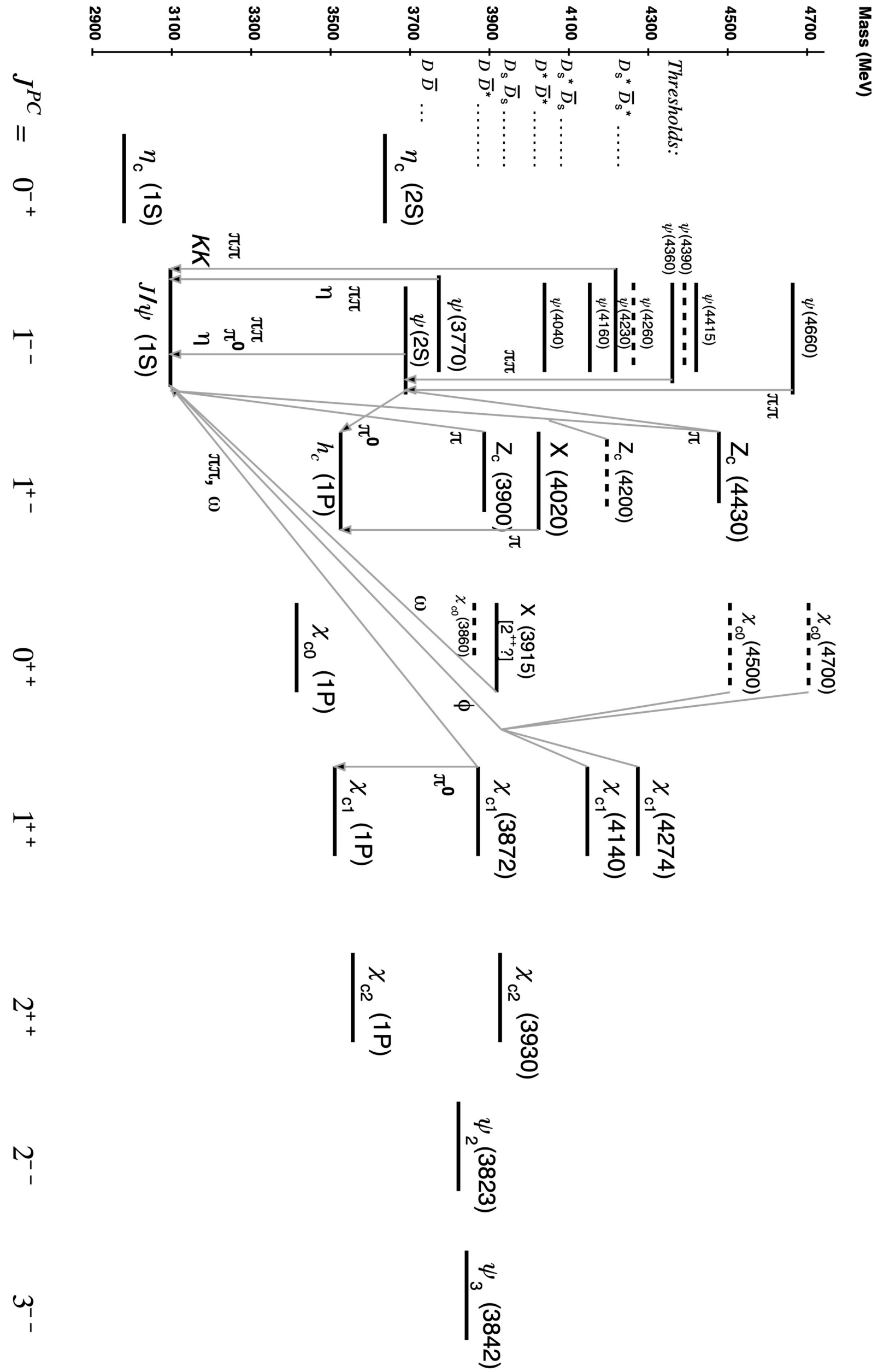

**Figure 1.13:** The spectrum of charmonium and charmonium-like states in 2021 [15].

$J/\psi$ and the charmonium ground-state $\eta_c$. Here, a radiative transition is possible. In general, since the photon has $J^{PC} = 1^{--}$, single photon transitions can only happen between two states with different $C$-parity. These transitions are then either electric- (E) or magnetic-multipole (M) processes, depending on the spins and parities of the initial and final states. If the product of the parities of the initial and final state is equal to $(-1)^{J_\gamma}$, the transition is an E$J_\gamma$ transition, otherwise it is an $MJ_\gamma$ transition. Here, $J_\gamma$ represents the total angular momentum of the photon, which can have the values $|S_i - S_f| \leqslant J_\gamma \leqslant |S_i + S_f|$, with $S_i$ and $S_f$ the initial and final spins, respectively. Generally, when several multipole transitions can occur, the lowest one is dominant. Therefore, the E1 and M1 transitions are the most important. Figure 1.14 demonstrates the possible E1 and M1 transitions below the open-charm threshold. E1 transitions preserve the initial quark spin directions and have relatively large branching fractions up to the order of 10%. On the other hand, M1 transitions require a spin-flip of one of the quarks and are suppressed by a factor $1/m_c$ with respect to E1 transitions. Due to the spin-flip, M1 transitions provide access to spin singlet states that cannot be produced directly. Curiously, the M1 transition rates found experimentally differ significantly from the theoretical rates. Therefore, it is very important to improve the accuracy of M1 transition measurements and to understand the origin of these discrepancies.

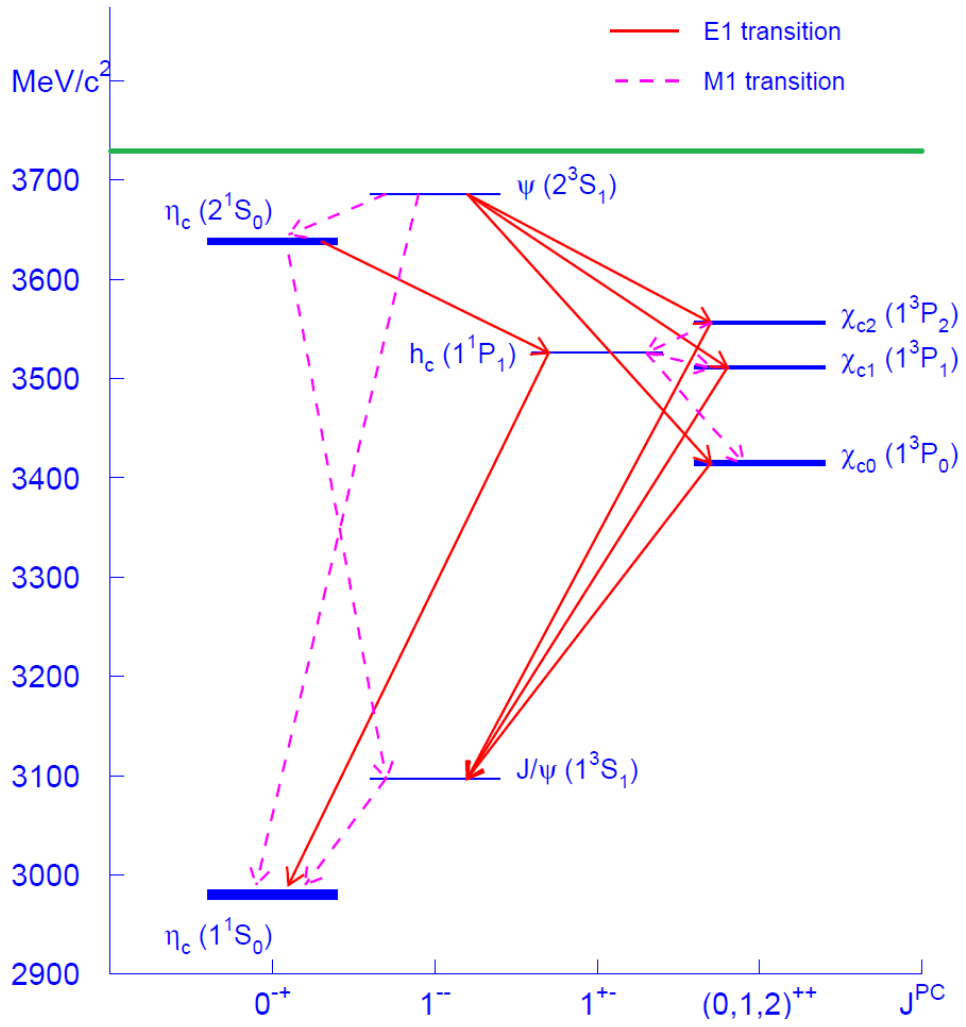

**Figure 1.14:** Radiative transitions between charmonium. The green line represents the open-charm threshold [47].

### 1.3.2 Glueballs

The most basic prediction of LQCD is the level-splitting of the non-exotic hadron spectrum, like the charmonium spectrum discussed in previous section. In figure 1.15, the experimental results for charmonium are shown in black, together with their LQCD predictions in blue and gray. Once the input parameters are fixed, the masses or resonance parameters of all other (exotic) states, including glueballs, can be predicted with LQCD as well [15].

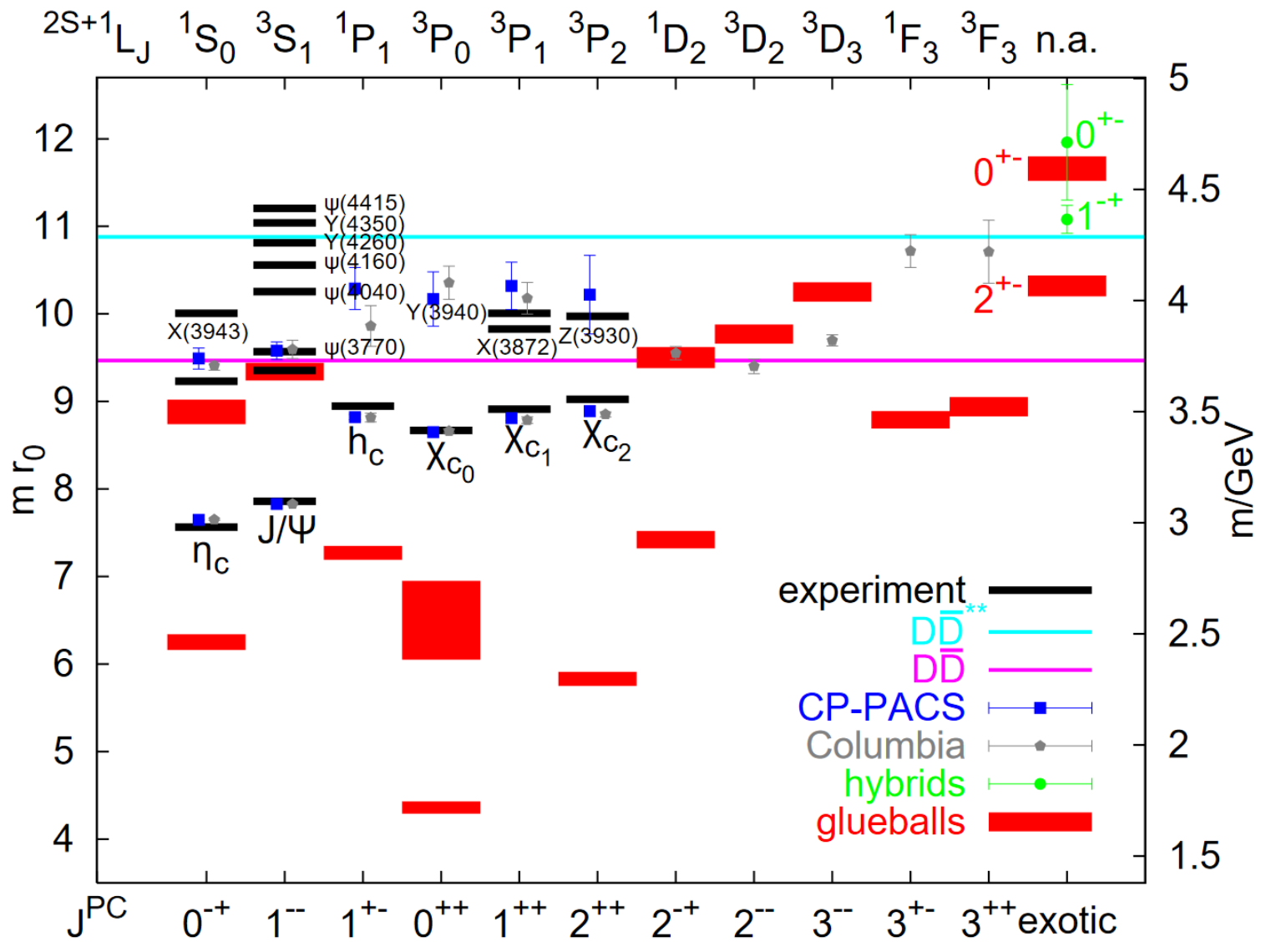

**Figure 1.15:** The charmonium spectrum (Columbia, CP-PACS), glueballs and spin-exotic $c\bar{c}$-glue hybrids as calculated by LQCD, overlaid with the experimental spectrum [48]. The two horizontal lines represent open-charm thresholds. The scale parameter $r_0$ is set with $r_0^{-1} \approx 394$ MeV.

If glueballs exist they will be neutral bosons that interact via the strong interaction, since glueballs are solely composed of gluons. It is, therefore, also expected that their electromagnetic interactions are much weaker than those of non-exotic hadrons with charged (anti)quarks composites. Due to limited theoretical knowledge of confinement, glueball properties cannot yet be calculated accurately. However, LQCD calculations that exclude the existence of quarks indicate that glueballs exist and they predict the lightest glueball to be a scalar meson with quantum numbers $J^{PC} = 0^{++}$ and a mass of around $1.5 - 1.7$ GeV/$c^2$ [20].

As soon as quarks are introduced in the theory, it is likely that glueballs mix with $q\bar{q}$ mesons of similar masses with the same quantum numbers. Instead of pure glueballs, states with both glueball and $q\bar{q}$ components, so-called hybrids, will be possible [49]. Gluon-rich matter will be difficult to distinguish from ordinary mesons, unless they carry exotic quantum numbers. Despite many experimental searches there are currently no observations of pure glueballs, but there is some evidence for mixed states containing both gluon and $q\bar{q}$ components [15]. Additionally, a recent study of combined data of the TOTEM and D0 collaborations provided a strong hint for the existence of a three-gluon compound [50]. One of the reasons that it is hard to unambiguously discover a pure glueball is that a large number of the predicted glueball states carry non-exotic $J^{PC}$ quantum numbers. Therefore, a clear, characteristic identification feature is missing. An overview of the different glueballs as calculated by the quark-free model is shown in figure 1.15 in red, hybrids are shown in green. The search for glueballs and hybrids is still a very active and interesting area of research. Measuring the spectrum of these states will provide valuable information in the confinement regime.

## 1.4 Motivation for $J/\psi \to \gamma p\bar{p}$

This research is based upon an analysis of BESIII data. The channel of interest for the data analysis is $J/\psi \to \gamma p\bar{p}$. Such a radiative $J/\psi$ decay, thus a $J/\psi$ decaying into a photon and something else, is in theory sensitive to an intermediate glueball state. The $J/\psi$ lies below the open-charm-threshold, the $c$ and $\bar{c}$ thus have to annihilate. Since the $J/\psi$ has $J^{PC} = 1^{--}$, the annihilation couples dominantly to three gauge bosons[14]. In a radiative decay, one of these bosons is required to be a photon. The photon can then be used to scan various mass regions of the intermediate particle. In figure 1.16, a schematic diagram of a possible $J/\psi \to \gamma p\bar{p}$ process is shown. In this diagram, an intermediate gluonic state $G$ is present, which then decays into a proton-antiproton pair, so we have $J/\psi \to \gamma G$; $G \to p\bar{p}$. Previous studies revealed some interesting structures in the $p\bar{p}$ invariant-mass spectrum of $J/\psi \to \gamma p\bar{p}$, which will be highlighted in the following sections.

[14] The number of gauge bosons has to be an odd number and one is kinematically not allowed.

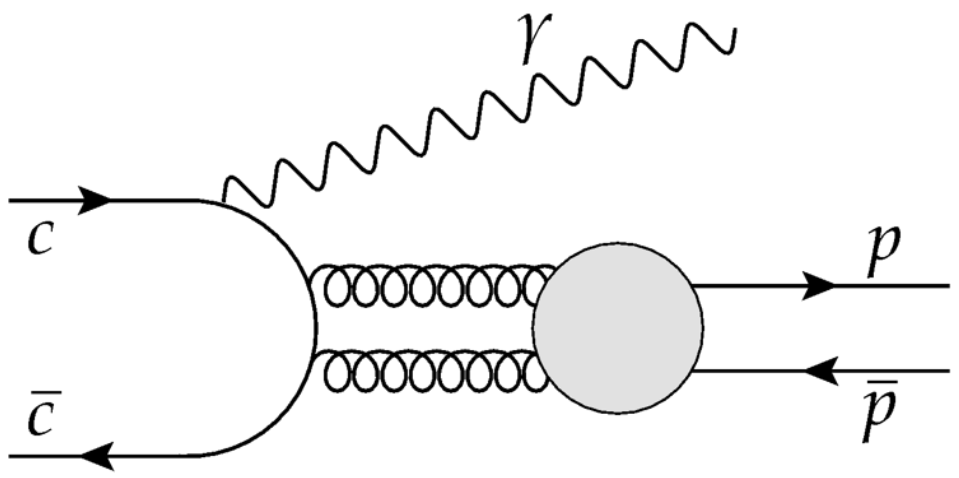

**Figure 1.16:** Schematic representation of $J/\psi \to \gamma G, G \to p\bar{p}$.

### 1.4.1 Near-threshold $X(p\bar{p})$

In previous studies of $J/\psi \to \gamma p\bar{p}$, a clear peak was seen near the $p\bar{p}$-threshold. The peak was first observed by the BESII experiment [51] and has later been confirmed by the BESIII [52] and CLEO-c [53] experiments. In figure 1.17, the $p\bar{p}$ invariant mass spectrum of a previous BESIII $J/\psi \to \gamma p\bar{p}$ study is shown, the spectacular $X(p\bar{p})$ enhancement is clearly visible around $M(p\bar{p}) \approx 1.85$ GeV/$c^2$. Remarkably, a similar strong enhancement does neither show up in related decays like $J/\psi \to xp\bar{p}$ with $x = \omega, \pi, \eta$ or $\psi' \to xp\bar{p}$ with $x = \gamma, \pi, \eta$ [54–62], nor in $B$-meson decays [63] or $\Upsilon \to \gamma p\bar{p}$ [64]. However, in several other radiative $J/\psi$ decays, like $J/\psi \to \gamma\eta'\pi^+\pi^-$ and $J/\psi \to \gamma K_S K_S \eta$, a structure is observed with a similar mass, denoted as $X(1835)$ [65, 66]. To make it even more intriguing, the structure is not present in hadronic decays like $J/\psi \to \omega\eta'\pi^+\pi^-$ [67].

A possible theoretical interpretation of the structure is a (quasi)bound $p\bar{p}$ state [69, 70], which requires a mass of ~1.85 GeV/$c^2$ [68]. Another interpretation of the observed structure is a glueball, or at least that it contains a large glueball component [71]. The glueball interpretation can explain the lack of observation in hadronic decays. Interestingly, although $X(p\bar{p})$ and $X(1835)$ have similar masses and the quantum numbers for both are determined to be $J^{PC} = 0^{-+}$, the widths differ significantly [65, 66, 68]. The width of the $X(1835)$ is determined to be ~190 MeV/$c^2$ [66], whereas the $X(p\bar{p})$ requires a width smaller than 76 MeV/$c^2$ [59]. An overview of recent observations of $X(p\bar{p})$ and $X(1835)$, and approaches to understand the widths and line-shapes, can be found in reference [71]. New high precision measurements can give more insight in the $X(p\bar{p}) - X(1835)$ puzzle.

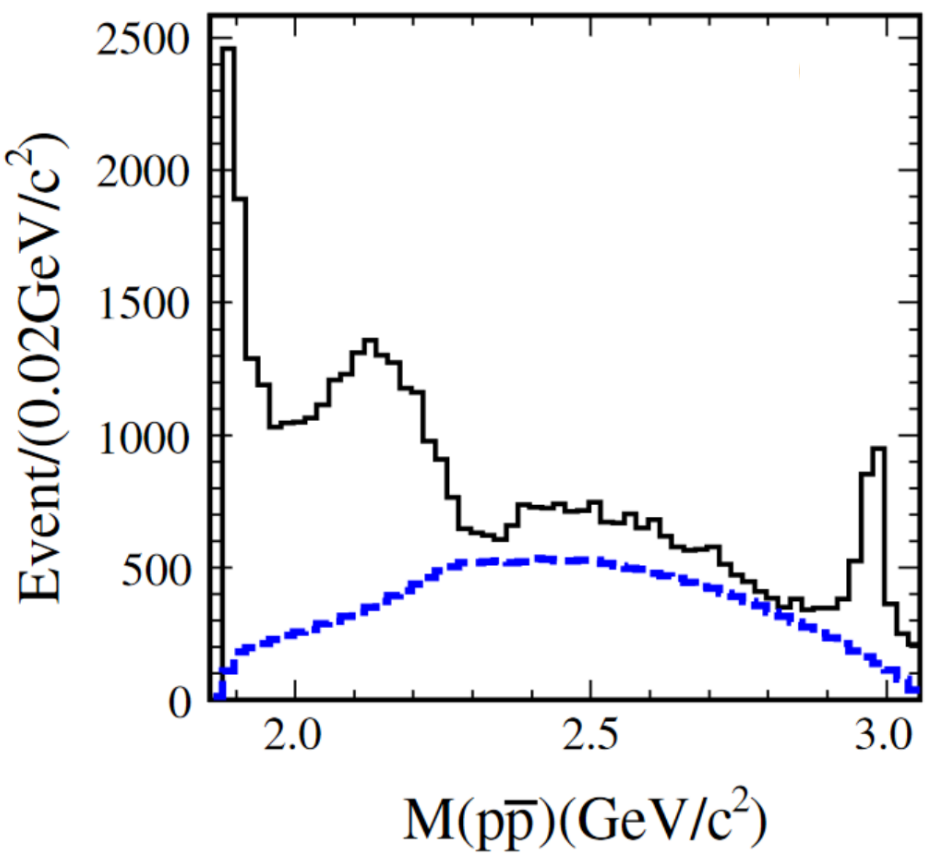

**Figure 1.17:** The $p\bar{p}$ invariant mass spectrum for a BESIII study of $J/\psi \to \gamma p\bar{p}$ preformed in 2012 [68]. The dashed blue line represent the phase-space contribution.

### 1.4.2 The charmonium ground-state $\eta_c$

In the high-mass region of figure 1.17, corresponding to small photon energies, there is a clear $\eta_c$ peak visible around 3 GeV$/c^2$. More than 30 years after the discovery [72] of this lowest lying charmonium state, the knowledge of the $\eta_c$ is still relatively poor. Until now, the summed decay widths of its measured exclusive decays cover less than two-third of its decay width, the $\eta_c$ mass and width have large uncertainties compared to those of other charmonium states and a distortion of its line shape has been observed.

The line shape distortion was first observed by the CLEO-c collaboration in the process $J/\psi \to \gamma\eta_c$ [73]. The resulting photon spectrum is shown in figure 1.18. The dotted black line represents a fit using an unmodified relativistic Breit-Wigner distribution, with the amplitude, mass, and width as free parameters. This fit clearly fails on both the low and high sides of the $\eta_c$ signal. The solid black line describes the fit with the CLEO-c description of the $\eta_c$ line shape. Soon after the observation by CLEO-c, the KEDR collaboration confirmed the distortion and proposed another description of the distortion [74]. In figure 1.19 the same CLEO-c data are fitted with the KEDR description of the line shape. Even though the KEDR description can describe the distorted line shape very well, no theoretical explanation of the line-shape effect was proposed so far.

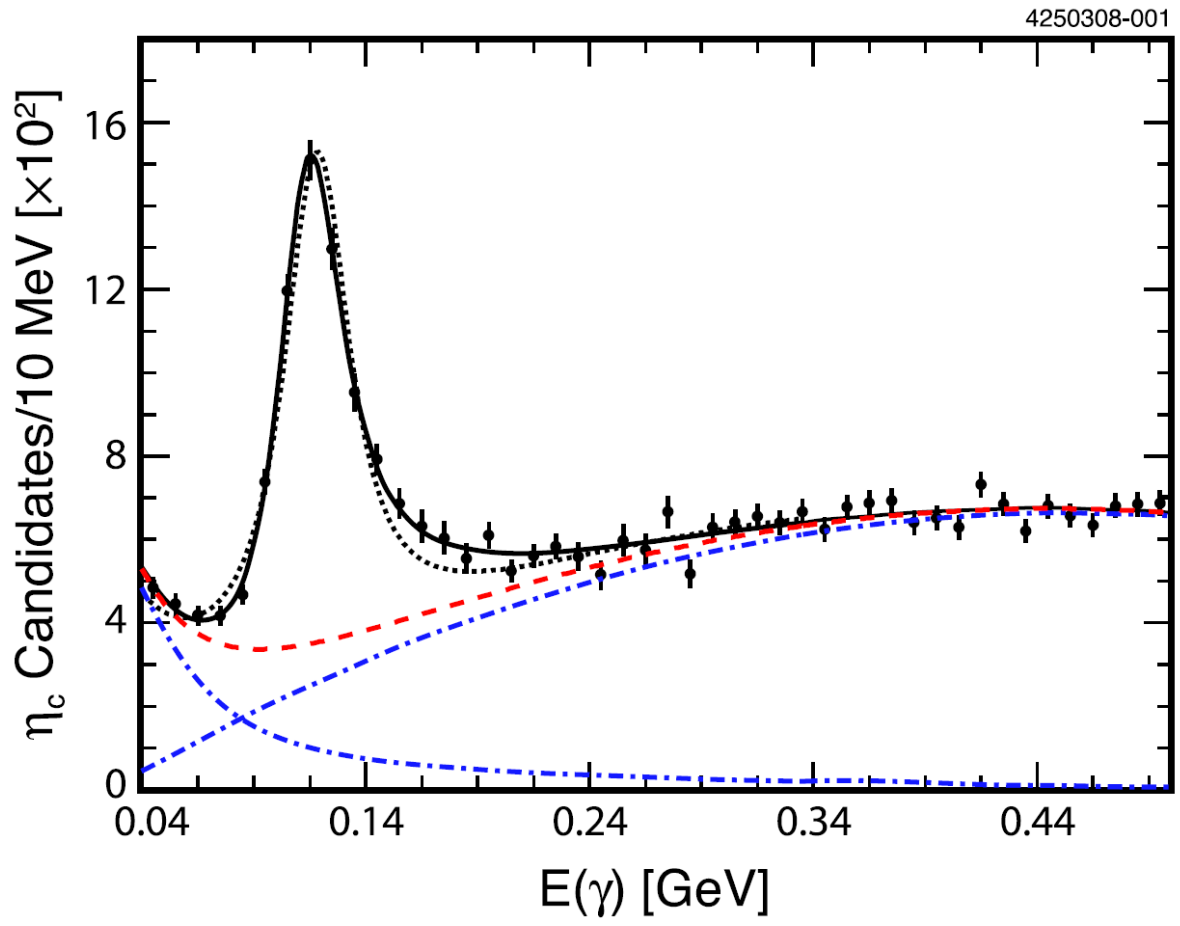

**Figure 1.18:** The CLEO-c data fitted with the $\eta_c$ line shape from CLEO-c [73]. The blue lines represent the two major background components; together they add up to the total background shown in red. The solid line shows the total fit, and the dotted line shows the fit without the CLEO-c corrections included.

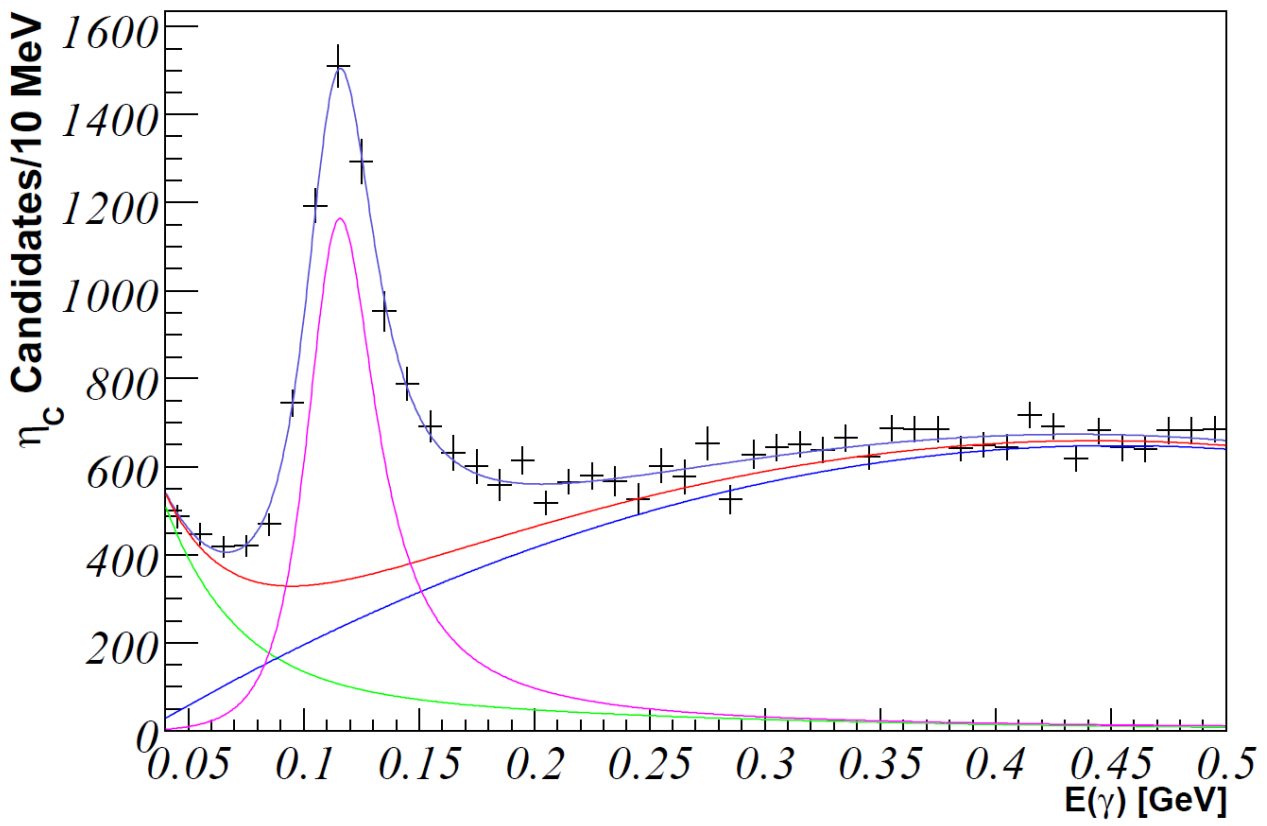

**Figure 1.19:** The CLEO-c data fitted with the $\eta_c$ line shape from KEDR [74]. The two major background components are shown in green and blue, together they add up to the total background shown in red. The magenta line shows the $\eta_c$ signal with the KEDR line shape.

In neither of the analyses from CLEO-c and KEDR, the possibility of interference between the non-resonant background and the $\eta_c$ signal were considered. This was first taken into account by BESIII, where events from the radiative decay $\psi' \to \gamma\eta_c$ were reconstructed using six exclusive decay modes of the $\eta_c$ [75]. The line shape was described by a one-dimensional fit to the $\eta_c$ mass spectrum, using the KEDR line-shape description. Additionally, a free phase parameter was introduced to account for a possible interference between the signal and a background contribution with the same final state without an intermediate $\eta_c$ resonance, referred to as "non-resonant background". Together, the significantly asymmetric $\eta_c$ peak could be described properly. It was found that the significance of the interference was of the order of $15\sigma$, and that the interference affected the $\eta_c$ mass and width significantly. Therefore, the interference can likely explain older deviating measurements. Furthermore, for each decay mode, two solutions were found for the relative phase. One corresponding to constructive interference, and the other to destructive interference. Regardless of which solution, the $\eta_c$ mass and width, and the overall fit quality were unchanged. Nevertheless, a more recent study by BESIII [32] found that the choice between the destructive and constructive interference term has a strong effect on the branching fraction $\mathcal{B}(J/\psi \to \gamma\eta_c)$. In figure 1.20, both solutions of this study are shown, with the interference term in cyan and the $\eta_c$ signal in red. Both fits describe the data equally well, and the $\eta_c$ mass and width are unchanged. However, there is an evident discrepancy between the $\eta_c$ yields, and thus the branching fractions. Both solutions give exactly the same fit quality and, therefore, a one-dimensional fit based on the reconstructed invariant-mass information cannot identify the correct solution.

In figure 1.21, the two solutions for the branching fraction are listed together with theoretical predictions and all experimental values listed by the Particle Data Group (PDG) [15]. Note that the KEDR value listed in the figure is from a later study that did include an interference term. The obtained phase was, however, close to zero and, therefore, affected the measured value only slightly. All other experimental values are well below the predicted values. The M1-transition $J/\psi \to \gamma\eta_c$ is thus still poorly understood. Additionally, $\eta_c$, as the lowest charmonium state, serves as one of the benchmarks for the fine tuning of input parameters for QCD calculations [76]. Therefore, further precision measurements of $\eta_c$ and the M1 transition are necessary to validate the theoretical understanding, and are crucial to normalize $\eta_c$ branching fractions.

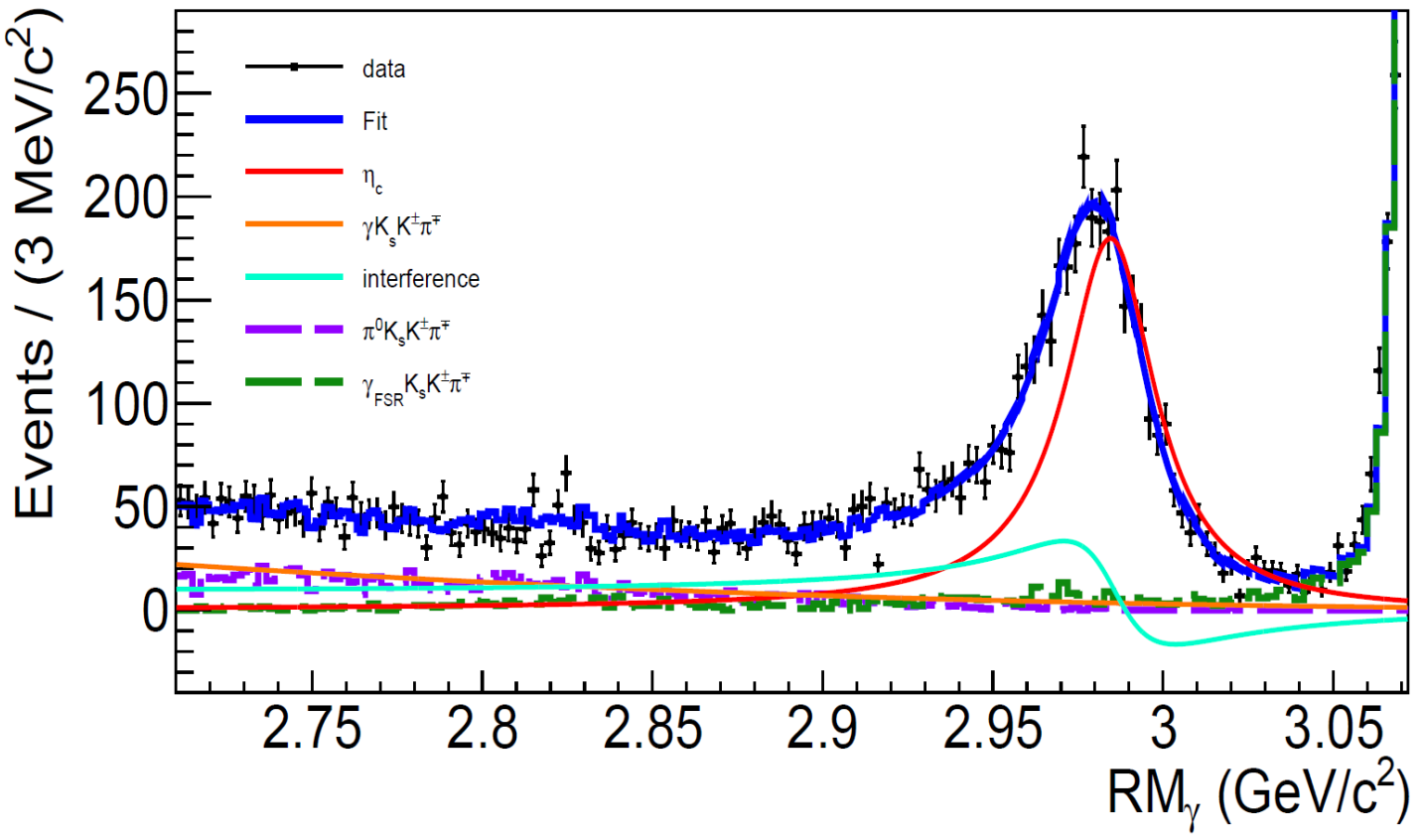

**(a)** Constructive interference

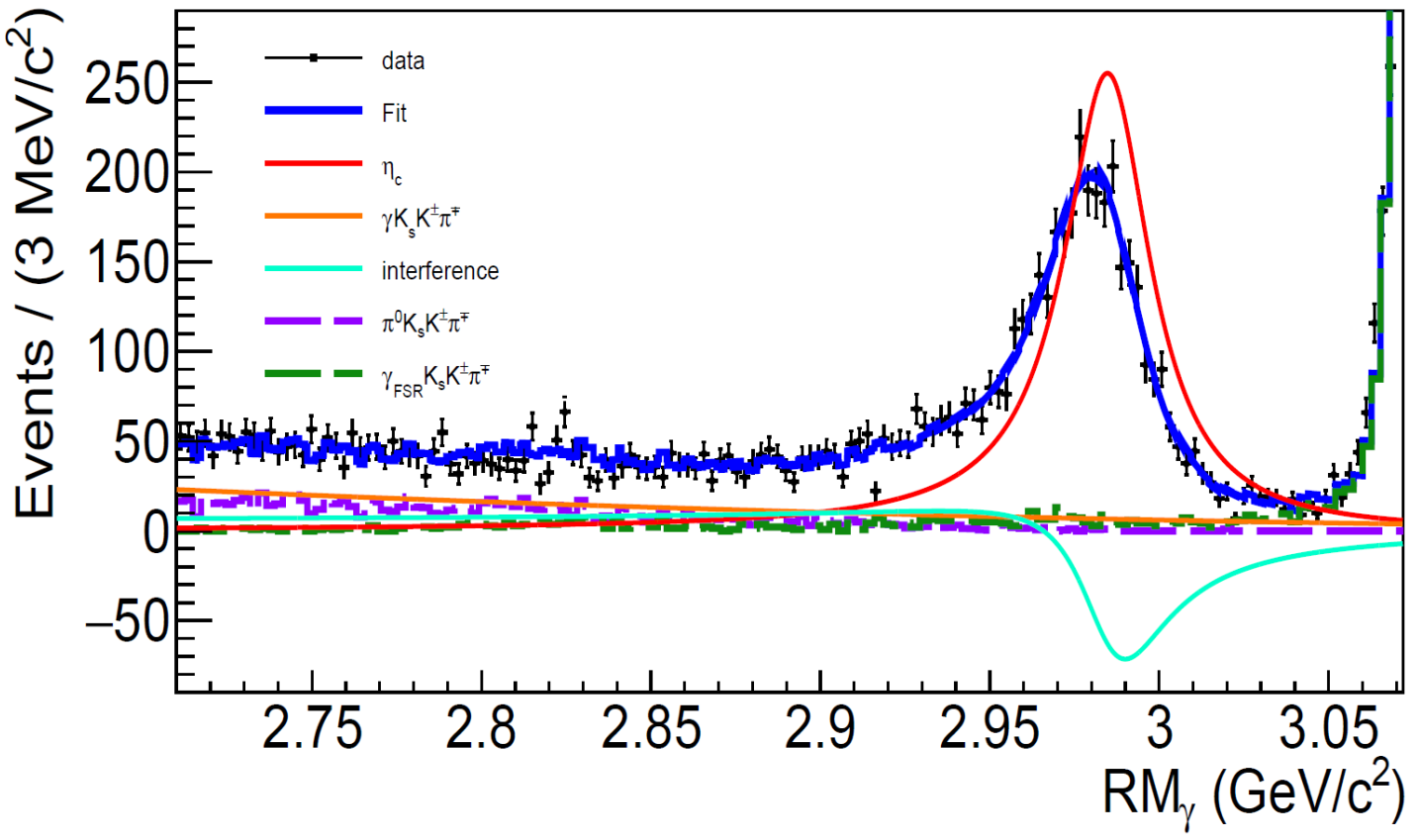

**(b)** Destructive interference

**Figure 1.20:** One dimensional fit to the photon recoil mass spectrum $RM_\gamma$ from reference [32]. The interference between the $\eta_c$ signal (red) and the non-resonant background is shown in cyan.

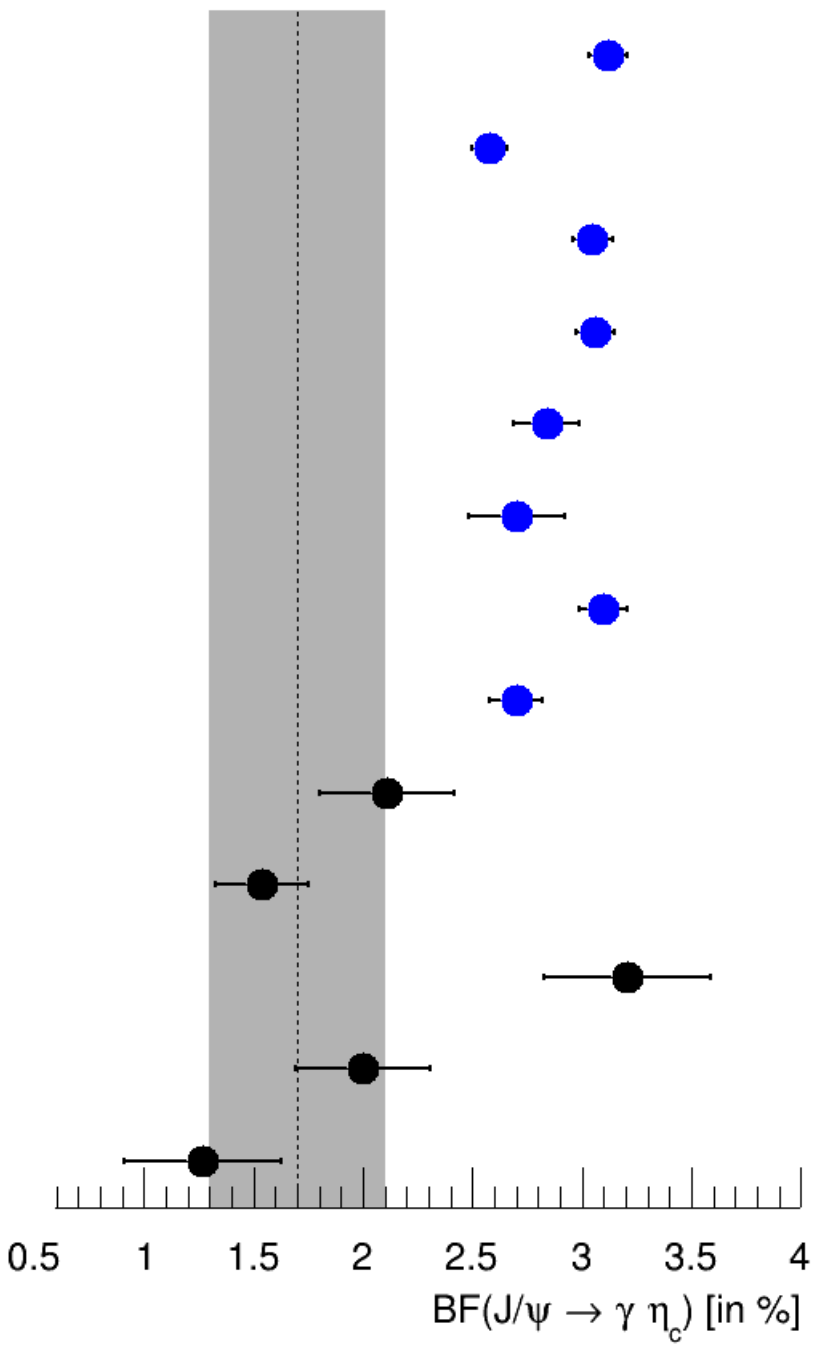

| Source | Year | Ref. |
|---|---|---|
| Non-relativistic potential model | 2008 | [77] |
| Godfrey-Isgur model | 2008 | [77] |
| Non-relativistic EFT | 2008 | [78] |
| Non-relativistic EFT | 2006 | [79] |
| Lattice QCD ($N_f = 2$) | 2013 | [80] |
| Lattice QCD ($N_f = 2 + 1$) | 2012 | [81] |
| Lattice QCD ($N_f = 2$) | 2011 | [82] |
| Lattice QCD ($N_f = 0$) | 2009 | [83] |
| BESIII: $J/\psi \to \gamma K_S^0 K^{\pm}\pi^{\mp}$ (D) | 2017 | [32] |
| BESIII: $J/\psi \to \gamma K_S^0 K^{\pm}\pi^{\mp}$ (C) | 2017 | [32] |
| KEDR: $J/\psi \to \gamma\eta_c$ | 2014 | [84] |
| CLEO-c: $e^+e^- \to \gamma X$ | 2009 | [73] |
| Crystal Ball: $J/\psi \to \gamma X$ | 1986 | [85] |

**Figure 1.21:** Theoretical predictions (blue) and experimental values (black) for the branching fraction of $J/\psi \to \gamma\eta_c$. The gray band represents the PDG value [15]. $N_f$ stands for the number of quark flavors that are considered in the LQCD calculations. The (D) and (C) refer to the destructive or constructive interference term, see figure 1.20.

### 1.4.3 This work

The aim of this analysis is to get a better understanding of the full spectrum of the $p\bar{p}$ invariant mass, and, especially, to get more insight in the puzzles regarding the $\eta_c$ resonance. As discussed before, a one-dimensional fit lacks sufficient information to decide between the destructive or constructive interference term. Therefore, we studied the decay in multiple dimensions, exploiting all internal degrees-of-freedom of the system. This can be achieved by a sophisticated analysis technique called Partial-Wave Analysis (PWA). In a PWA, the angular distributions, as well as all the invariant-mass spec-tra, are taken into account. A PWA may result in unambiguous conclusions about the interference. In previous analyses, a PWA was difficult due to the limited statistics that were available. For example, the KEDR study discussed in previous section was based on a data sample of $6.3 \times 10^6$ $J/\psi$ events [74], and the CLEO-c study on $\sim 8.5 \times 10^6$ $J/\psi$ events [73]. In this study, the PWA of the $\eta_c$ range will be based on a sample of $1.3 \times 10^9$ $J/\psi$ events collected by BESIII in 2009 and 2012 [86].

The same multi-dimensional analysis techniques will be used to perform a mass-independent fit of the full range of the $p\bar{p}$ invariant-mass. In the previous BESIII study of $J/\psi \to \gamma p\bar{p}$, shown in figure 1.17, a mass-dependent PWA was performed on the invariant-mass range $M_{p\bar{p}} < 2.2$ GeV/$c^2$ [68]. This mass-dependent PWA was performed without correction for the Final-State Interaction (FSI), even though it is known that the hadronic FSI in the $p\bar{p}$ system is significant [69, 70, 87–89]. This previous study was based on $225 \times 10^6$ $J/\psi$ events. Nowadays, a total sample of $10^{10}$ $J/\psi$ events is available. The large statistics allow for a mass-independent PWA of the full $p\bar{p}$ invariant-mass range. This PWA can provide additional insights in the near-threshold enhancement $X(p\bar{p})$, and the additional enhancement at an invariant mass of around 2.1 GeV/$c^2$ in figure 1.17. Thus far, the enhancement around 2.1 GeV/$c^2$ has not been studied yet.

Besides the interesting structures in the $p\bar{p}$ invariant mass, this study aims for a significant improvement in the accuracy of the branching fraction $J/\psi \to \gamma p\bar{p}$. The current PDG value of this branching fraction, $(3.8 \pm 1.0) \times 10^{-4}$, is based on one single measurement from 1984 [15]. This branching fraction was extracted from a sample of $1.32 \times 10^6$ $J/\psi$ events [90]. The accuracy could be improved substantially by the $10^{10}$ $J/\psi$ events that are available today, and will result in an important number for further, more detailed studies of this channel.

### 1.4.4 Thesis outline

Chapter 2 gives an overview of the BESIII facilities and the collected datasets, followed by the event selection and background studies presented in chapter 3. The analysis and determination of the full branching fraction of $J/\psi \to \gamma p\bar{p}$ is presented in chapter 4. Subsequently, the concepts of a Partial-Wave Analysis (PWA), and the PWA software package PAWIAN will be introduced in chapter 5. These partial-wave analysis techniques are employed to perform a mass-independent PWA of the full $p\bar{p}$ invariant-mass spectrum, presented in chapter 6. Another PWA is performed to extract the $J/\psi \to \gamma\eta_c$ branching fraction, and the $\eta_c$ mass and width, as will be described in chapter 7. Finally, in chapter 8, an overall summary is given, followed by a discussion on the obtained results and an outlook to future activities.

# 2. Experimental setup

To get a better understanding of QCD at distances comparable to the size of a nucleon, collider experiments are valuable. In the confinement regime, interactions can only be described using non-perturbative methods. To find answers, experiments that are capable to measure observables with high resolution and precision are required. The Beijing Spectrometer (BES) III is designed to study physics in the transition between perturbative and non-perturbative regimes of the strong force [31]. Detailed studies of hadronic-matter formation, via electron-positron annihilations in the charmonium-mass region, will greatly advance our understanding of the internal composition of hadrons. The studies may reveal particles beyond the two and three-quark configuration, some of which are predicted to have exotic properties in the charmonium mass region. BESIII already observed various exotic states, as described in references [91] and [92].

The BESIII facility is located at the Institute of High Energy Physics (IHEP) in West Beijing, China. BESIII is the second upgrade of the original BES detector. The preceding experiments, BES and BESII, operated at the interaction point of the original Beijing Electron Positron Collider (BEPC) from 1989 until 2004. BESIII operates at the interaction point of the BEPCII where it registers the products of the collisions at center-of-mass energies from 2 GeV to 4.6 GeV. The BEPCII is an updated version of the original BEPC and is a double-ring collider built within the existing BEPC tunnel. The construction of BESIII was finished in the summer of 2008 and data taking has officially started in 2009. Nowadays, the BESIII collaboration has about 500 members from 82 institutions in 17 countries.

BESIII has the potential to connect the perturbative and non-perturbative energy domains, interpolating between the limiting scales of QCD. The BESIII collaboration publishes results in topics such as light-hadron and charmonium spectroscopy, tau and charm physics, precision measurements of QCD and Cabibbo-Kobayashi-Maskawa (CKM) parameters and the search for new physics. For the analysis presented in chapter 7, a sample of $1.31 \times 10^9$ $J/\psi$ events recorded by the BESIII experiment in 2009 and 2012 is used, which is larger than any other sample of directly produced $J/\psi$ events recorded ever before. This sample, together with the new $J/\psi$ data collected since 2018, results in a total data sample of 10 billion $J/\psi$-events. The total BESIII sample will be used for the analyses presented in chapters 4 and 6.

Aside from collecting world's largest data sample of $J/\psi$ events, BESIII is an outstanding facility with which one can detect the final-state particles of $J/\psi \to \gamma p\bar{p}$ with high resolution and precision. For this decay channel, one expects photons with energies up to 1 GeV and (anti)protons with momenta up to 1.2 GeV. In this momentum range, the subdetectors of BESIII allow for a clear distinction between (anti)protons and other charged particles, and a (anti)proton-momentum resolution of less than one percent. Additionally, one of the subdetectors is specifically designed to measure the energies and positions of electrons, positrons and photons precisely, resulting in a photon energy resolution of just a few percent and allowing to distinguish the relevant radiative photon from photons originating from $\pi^0$ or $\eta$ decays.

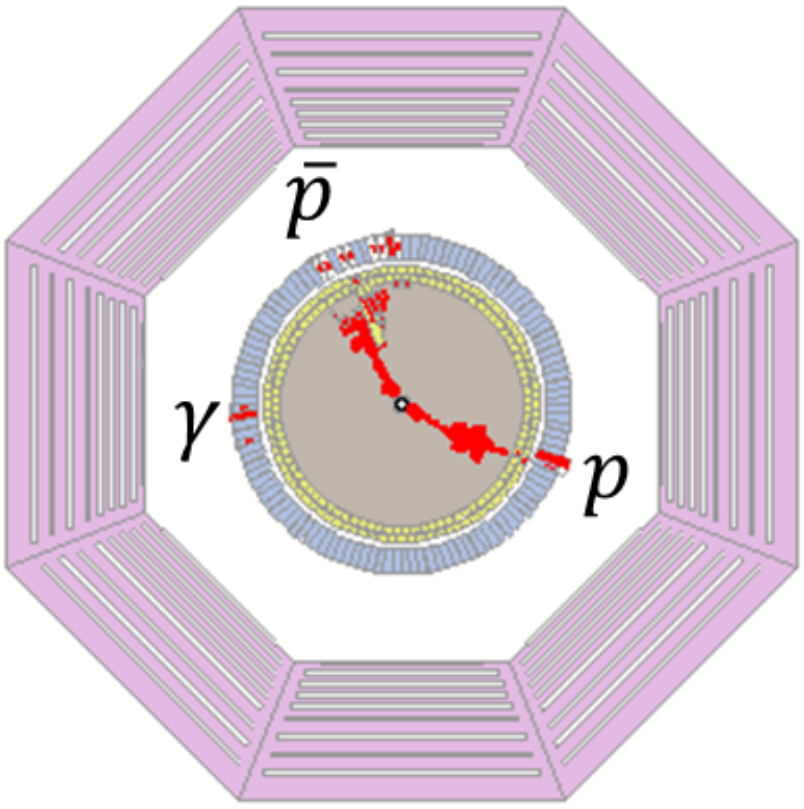

**Figure 2.1:** Simulated BESIII detector response of a Monte Carlo-generated $J/\psi \to \gamma p\bar{p}$ phase-space event, viewed from the plane perpendicular to the beamline.

Figure 2.1 illustrates the response of the different BESIII subdetectors for a certain $J/\psi \to \gamma p\bar{p}$ event. It is clearly visible that the inner subdetectors play an important role in the registration of the $J/\psi \to \gamma p\bar{p}$ final-state particles, whereas the outermost subdetector, the muon counter, has not been used in our analysis since it did not provide any additional valuable information. In the following sections, BEPCII, BESIII, and its subdetectors, will be discussed in more detail.

## 2.1 BEPCII

The double-ring electron-positron collider BEPCII has a circumference of 240 meter. It has a peak luminosity of $10^{33}$ cm$^{-2}$ s$^{-1}$ at a center-of-mass energy of 3.78 GeV, which is just above the open-charm threshold.

The original BEPC was a single-ring electron-positron collider that ran in single bunch mode. BEPCII was built within the existing BEPC tunnel, however, almost all components have been replaced to successfully achieve the BESIII physics program. With the upgrade to BEPCII, the luminosity was improved by two orders of magnitude. The most important upgrade is the installation of two separate beam pipes, one for the electron beam and one for the positron beam. BEPCII runs with 93 bunches in each ring. These electron and positron bunches are injected into the two storage rings by the linear accelerator (LINAC). Every bunch is approximately 1.5 cm long and is spaced by 2.4 m (8 ns) from the neighboring bunches. This results in a single-beam current of 0.91 A in collision mode.

The two beams collide at the interaction point with a horizontal crossing angle of 11 mrad, as illustrated in figure 2.2. A pair of superconducting quadrupoles near the interaction point forces both beams to bend towards the collision point, while reducing the vertical beam width to about 5.7 $\mu$m. The horizontal beam width is about 380 $\mu$m. A detailed description of the design parameters of the BEPCII can be found in references [93] and [94].

After 10 years of successful running, two upgrade plans of BEPCII were proposed and approved. The first increased the maximum beam energy up to 2.45 GeV, thus expanding the range to a center-of-mass energy of 4.90 GeV. The second consisted of a so-called top-up injection to increase the data taking efficiency. The mechanical implementation of these two upgrades began in 2017 and finished by the end of 2019 [95].

In the summer of 2021, a set of further upgrades to optimize BEPCII for higher energies were approved [96]. After these upgrades, a luminosity of $10^{33}$cm$^{-2}$ s$^{-1}$ at a beam energy of 2.35 GeV should be achieved, which is three times higher than the current luminosity for this beam energy. Ad-ditionally, the maximum beam energy will be further expanded to 2.8 GeV. The commissioning is scheduled to start in the beginning of 2025.

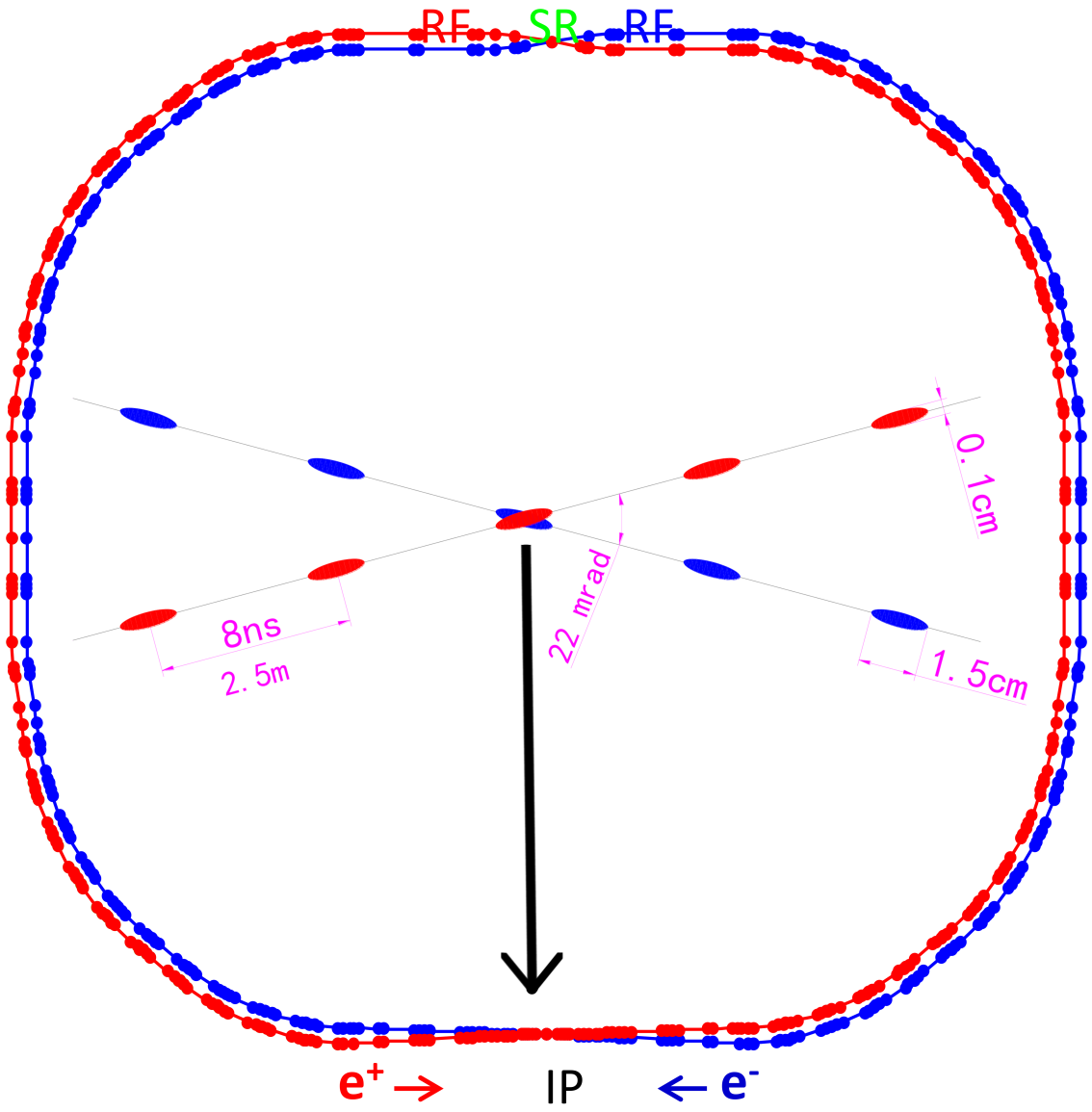

**Figure 2.2:** The BEPCII facility with the interaction point (IP) at the bottom. The BESIII detector is installed at the interaction point.

## 2.2 BESIII

BESIII is a cylindrical detector that covers 93% of the solid angle of $4\pi$ around the interaction point. The detector is developed for the high data rates of the BEPCII and the energy spectra and multiplicity of the expected secondary particles. The detector consists of five main systems:

- The helium-based Multilayer Drift Chamber (MDC);
- The plastic scintillator Time-of-Flight (TOF) system;
- The CsI(Tl) Electromagnetic Calorimeter (EMC);
- The SuperConducting (SC) solenoid magnet with a 1 T magnetic field;
- The Muon Counter (MUC) based on Resistive Plate Chambers (RPCs).

Figure 2.3 shows a schematic overview of the subsystems and their position in BESIII. The most important parameters of each subdetector will be discussed hereafter. The information is based on reference [97], where a more detailed description of BESIII and its subdetectors can be found.

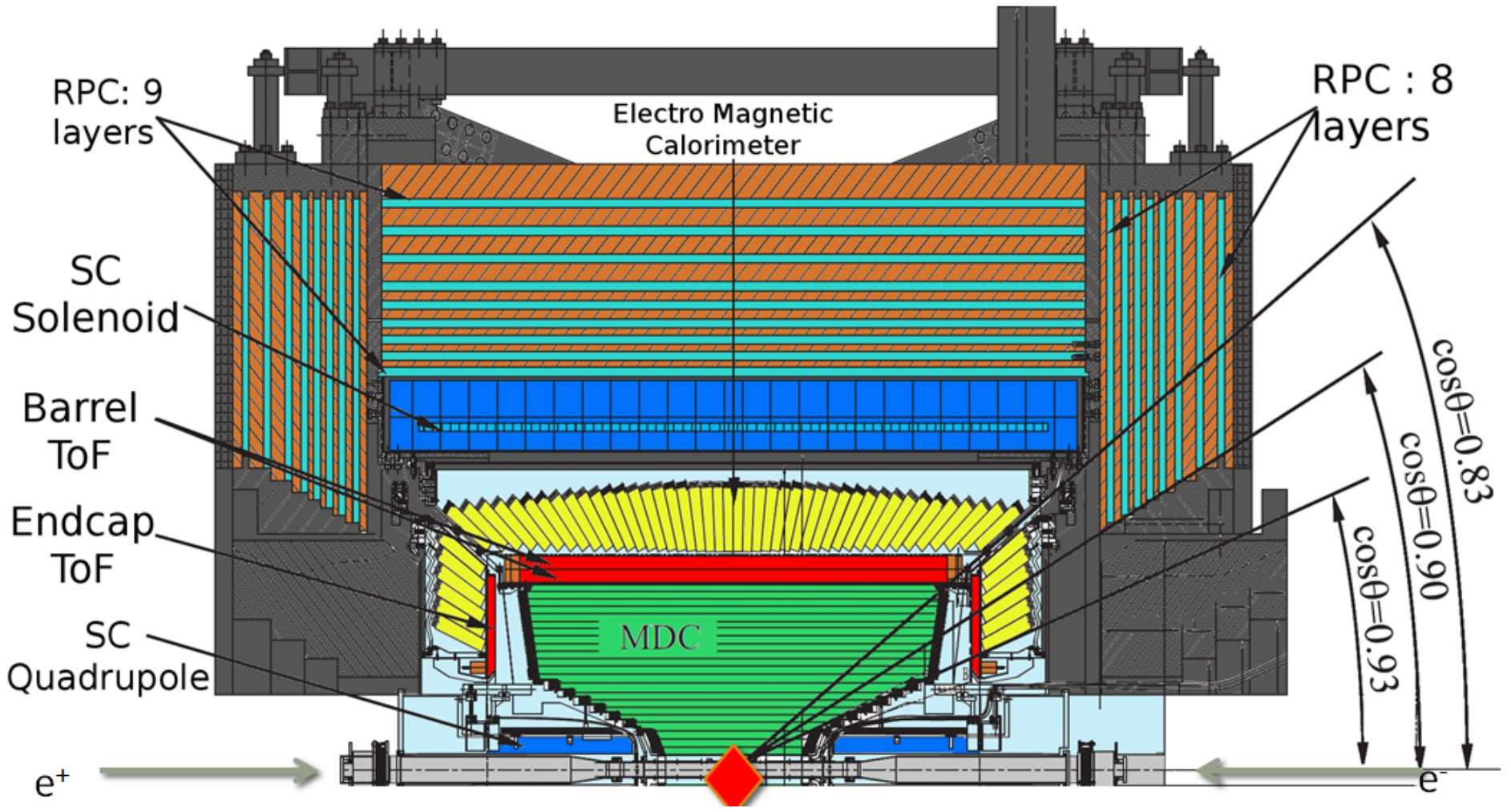

**Figure 2.3:** Schematic representation of the cylindrical BESIII detector. The beam lines run horizontally, as indicated by the gray arrows, with the red diamond representing the interaction point. Note that the detector is symmetric around the beam pipe and that the illustration only shows the upper half of the cross section of the detector.[1]

## 2.2.1 The Multilayer Drift Chamber (MDC)

The Multilayer Drift Chamber (MDC) is the innermost subdetector and is comprised of an inner and outer chamber. The main purposes of the MDC are:

- 3D-reconstruction of charged tracks;
- Momentum measurement of produced charged particles;
- Charged-particle identification by measuring energy deposits ($dE/dx$);
- Reconstruction of the decay-vertex of long-lifetime hadrons that decay inside the MDC;

[1] Image taken from the BESIII homepage.

- Contribution to the L1-trigger to reject background tracks.

The inner and outer drift chambers are filled with gas and consist of arrays of wires at high voltage (anodes) running through a chamber with conducting walls at a ground potential (cathodes). The MDC contains 6796 gold-coated tungsten sense wires, arranged in 43 layers, and additionally 21844 gold-coated aluminum field wires to create a uniform electrical field. The two cylindrical chambers of the MDC are joined at the ends and, therefore, share the same gas volume. Charged particles traversing the gas ionize the gas molecules. These ions and electrons are accelerated due to the high voltage applied to the wires. This produces electrical signals at sense wires, which are digitized and read out.

The MDC has a maximum length of 2582 mm and an outer radius of 810 mm. With an inner radius of 59 mm, the distance to the beryllium beam pipe is only 2 mm. The inner chamber has a conical shape to place the quadrupole beam-focusing magnets as close as possible to the interaction point. Therefore, the wires of the inner layers are shorter than those of the outer layers. This results in a polar angle coverage of $|\cos\theta| \leqslant 0.83$ for the innermost layer. Since the polar angle coverage of the outermost wire layer is $|\cos\theta| \leqslant 0.93$ a solid angle coverage of $\Delta\Omega/4\pi = 93\%$ is achieved.

The complete volume of the MDC, around 4 $\mathrm{m}^3$, is filled with a helium (60%) and propane (40%) gas mixture. This mixture minimizes the effect of multiple Coulomb scattering whilst still providing a $dE/dx$ resolution better than 6% and a spacial resolution of 115 $\mu$m. For charged particles, the momentum resolution at 1 GeV/$c$ is better than 0.5% in the 1 T magnetic field. Furthermore, the tracking efficiency for (anti)protons is better than 95%, even though (anti)protons with a momentum smaller than ~200 MeV are not likely to reach the TOF detector and, therefore, will not be registered.

Figure 2.4 shows a Monte Carlo simulation of the energy loss $dE/dx$ in the MDC versus the incident momentum of particles. For the center-of-mass energy range of BESIII ($2-4.6$ GeV), the momenta of most secondary charged particles generated in the $e^+e^-$-collisions is well below 1 GeV/$c$. It can be seen that the MDC shows a clear (anti)proton separation below 1 GeV/$c$, which already covers the majority of the (anti)proton momenta of $J/\psi \rightarrow \gamma p\bar{p}$. For the few (anti)protons with a momentum higher than 1 GeV/$c$, the additional particle-identification properties of the TOF can be utilized. For the other charged particles, the MDC $e-\pi$ separation works well for momenta of around 0.2 GeV/$c$ and up, while the $K-\pi$ separation only deteriorates beyond 0.6 GeV/$c$. Using only the $dE/dx$ information of the MDC, a $3\sigma$ $K-\pi$ separation can be accomplished for momenta up to 0.77 GeV/$c$.

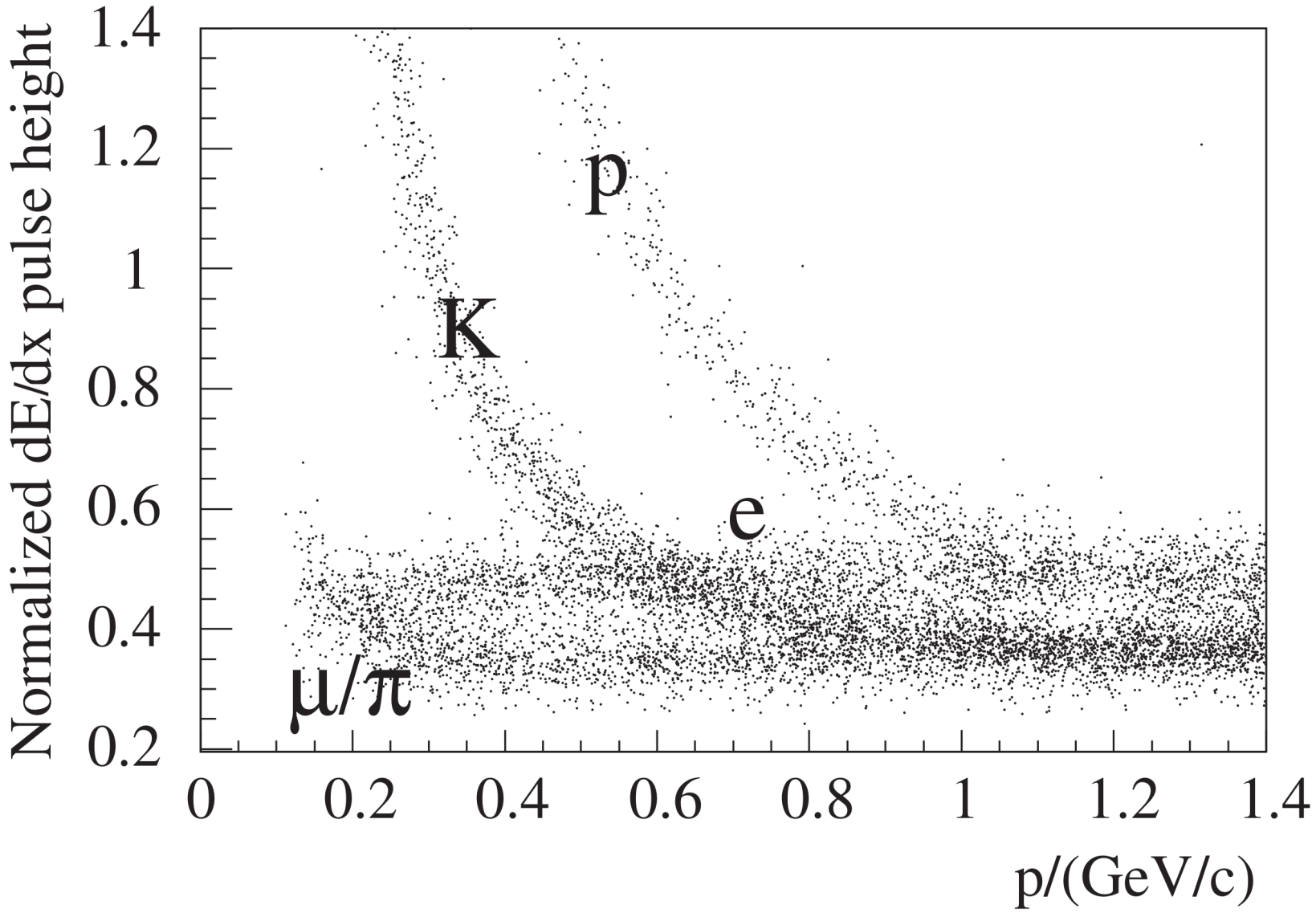

**Figure 2.4:** Specific energy loss ($dE/dx$) versus incident momenta for various particle types [98]. Data are obtained from a detector-simulation study using Monte Carlo generated events.

The modular design of the MDC enables a replacement of the inner chamber in case of radiation damage. Due to aging effects, the inner chamber will be replaced by the three-layer Central Gaseous Electron Multiplier (CGEM) inner tracker [99] in 2022.

### 2.2.2 The Time-of-Flight (TOF) system

The Time-of-Flight (TOF) detector system is installed directly on the outer surface of the MDC (shown in red in figure 2.3). The TOF system measures the flight time of charged particles. The flight time combined with the MDC track length provides the particles' velocity. Together with the momentum derived from the MDC the mass of the particle can be reconstructed, which is essential for particle identification. Furthermore, the fast signals of the TOF system contribute to the L1-trigger to reject background tracks.

The TOF system consists of a double-layered barrel and two single-layered endcaps, as shown in figure 2.3. The first barrel layer has an inner radius of 0.81 m, the second inner radius is 0.86 m. Each layer consists of 88 plastic scintillation bars. The bars are staggered to avoid gaps, resulting in a very high azimuthal acceptance. The bars have a trapezoidal cross-section and are 2300 mm long and 50 mm thick. Fine mesh photomultiplier tubes are attached on both ends of each bar to read out the produced scintillation light at both ends. The two single-layer endcaps are located directly outside the MDC endcaps, at 1.4 m from the interaction point. Each TOF endcap consists of 48 fan-shaped scintillator segments, which together create an almost circular geometry. Each segment is 480 mm long and 50 mm thick. The width varies from 109 mm at the top until 62 mm at the bottom. The scintillator segments of the endcaps are only read out at the outer side.

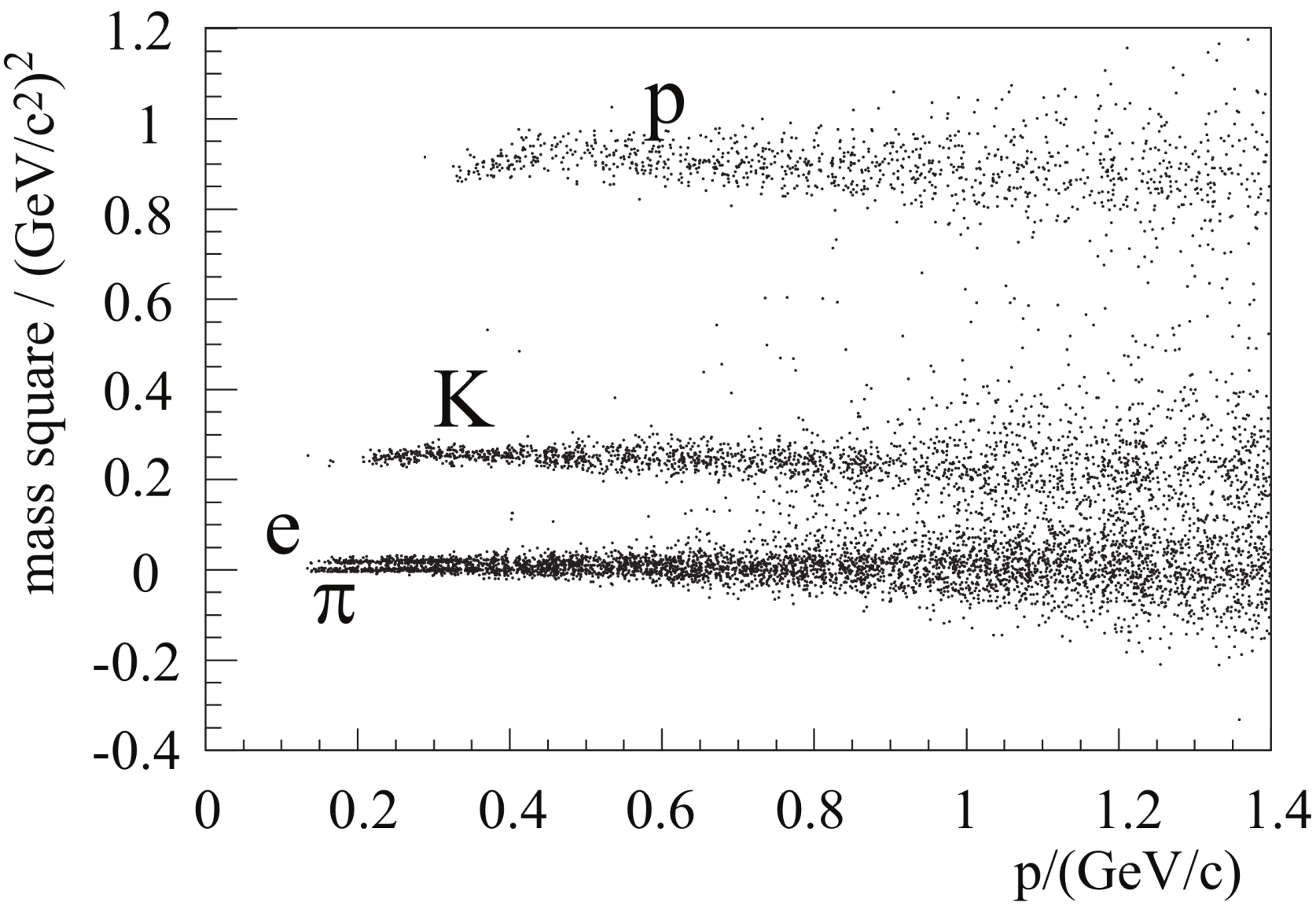

**Figure 2.5:** Extracted particle masses using MDC and TOF information for different particle types versus incident momenta [98]. Data are obtained from a detector-simulation study using Monte Carlo generated events.

The dimensions of the TOF system result in a polar angle coverage of $|\cos\theta| \leqslant 0.83$ for the barrel and $0.85 \leqslant |\cos\theta| \leqslant 0.95$ for the endcaps. The small acceptance gap is a result of space required for service lines and the mechanical support of the MDC. The use of two barrel layers instead of one improves the system reliability and time resolution. The barrel has a time resolution of 68 ps, while the endcaps had a time resolution of 110 ps. In 2015, the endcaps were replaced by a multi-gap resistive plate chamber technology, providing a time resolution of 60 ps [100].

Figure 2.5 shows the result of a MC simulation of the calculated mass square versus the incident particle momentum, obtained from the MDC and TOF information. The TOF resolution provides a $K-\pi$ separation of approximately $3\sigma$ up to a particle momentum of around 0.9 GeV$/c$. The (anti)proton separation is even better for the full (anti)proton-momentum range of $J/\psi \to \gamma p\bar{p}$.

### 2.2.3 The Electromagnetic Calorimeter (EMC)

The Electromagnetic Calorimeter (EMC) is located directly outside the TOF counters and inside the solenoid magnet (the yellow detectors in figure 2.3). The EMC is designed to measure the energies and positions of electrons, positrons and photons precisely. In radiative decay processes, such as $J/\psi \to \gamma p\bar{p}$, directly produced photons must be precisely measured and distinguished from photons originating from $\pi^0$ or $\eta$ decays. The energies of most photons produced in radiative decays are quite low, requiring a low photon threshold for the EMC. On the other hand, to be able to study $e^+e^- \to \gamma\gamma$, the maximum photon energy that has to be measured is the full beam energy. Thus, photons in the energy range of ~20 MeV to 2.3 GeV are expected. The low photon-energy threshold requires a scintillator material with a high light yield. Therefore, the EMC consists of thallium doped cesium iodide (CsI(Tl)) crystals.

In a similar fashion as the TOF, the EMC is comprised of a barrel and two endcaps. The barrel consists of 44 rings, each containing 120 crystals, leading to a total of 5280 crystals in the barrel. All crystals point slightly off the interaction point, with a small tilt of $1° - 3°$ in $\theta$ and $1.5°$ in $\phi$, to prevent leakage of photons originating from the interaction point through the walls of neighboring crystals. The inner radius of the barrel is 94 cm and the inner length 276 cm. The length of each crystal is 28 cm with typical front and rear faces of 5.2 cm $\times$ 5.2 cm and 6.4 cm $\times$ 6.4 cm, respectively.

Both endcaps have an inner radius of 88 cm and an outer radius of 110 cm. Each endcap contains 480 crystals distributed over 6 rings. The endcap crystals have several different shapes, but with dimensions comparable to the barrel crystals. The full EMC thus contains 6240 crystals, resulting in a total weight of 25.6 tons.

The angular coverage of the EMC is $|\cos\theta| < 0.82$ for the barrel and $0.83 < |\cos\theta| < 0.93$ for the two endcaps. The small acceptance gap is again a result of space needed for the mechanical support structures and service lines of the inner detectors. The EMC has a photon energy threshold of 25 MeV in the barrel and 50 MeV in the endcaps.

Figure 2.6 shows the energy resolution, at a particle energy of 1.5 GeV, versus the crystal ring number, corresponding to the polar angle $\theta$. The data are from Monte Carlo (MC) simulations and data runs in 2009 and 2012 at the $J/\psi$ mass center-of-mass energy of 3.097 GeV. The gaps between the endcaps and the barrel are clearly visible around ring numbers 10 and 50. The results are in good agreement with each other and show a resolution of about 4% in the endcaps and about 2.3% in the barrel [101].

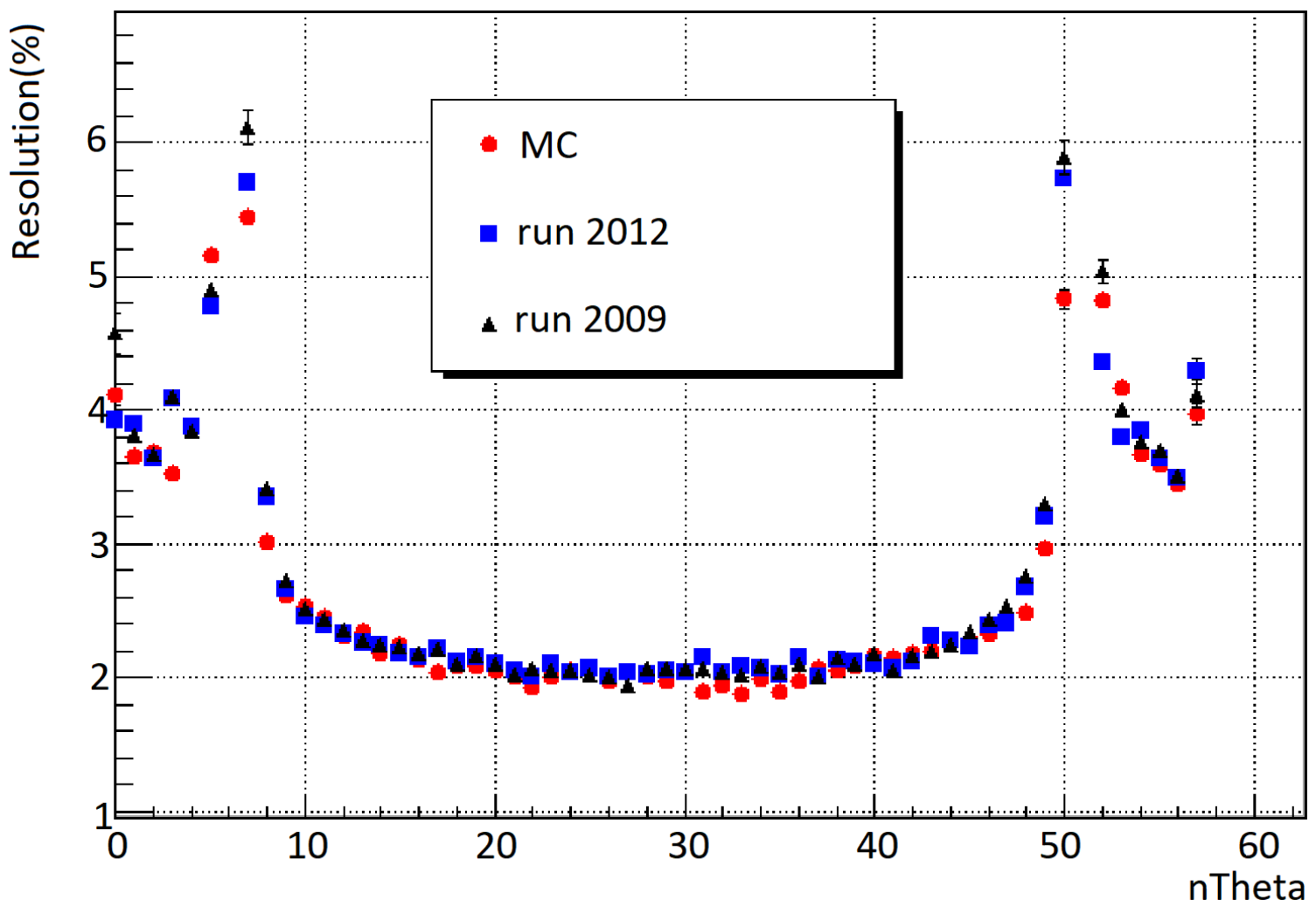

**Figure 2.6:** Energy resolution vs. the crystal-ring number, proportional to the scattering angle. Data correspond to the Bhabha scattering process taken at a center-of-mass energy corresponding to the $J/\psi$ mass [101].

Besides the measurement of electrons, positrons and photons, the EMC contributes to particle identification by improving the electron-hadron separation. They can be distinguished due to a difference in the lateral shape of their showers, however, this feature has not been used and found useful for the study of $J/\psi \to \gamma p\bar{p}$.

### 2.2.4 The superconducting solenoid magnet

The superconducting solenoid magnet surrounds the subdetectors described earlier: the MDC, TOF and EMC. The uniform axial magnetic field of 1.0 T forces charged particles to follow a curved track and consequently enables accurate momentum measurement by its inner subdetectors. Additionally, the steel flux-return yoke provides mechanical support for the inner subdetectors and acts as a hadron absorber, which allows for hadron-muon separation. The total weight of the yoke is around 500 metric tons.

### 2.2.5 The Muon Counter (MUC)

The Muon Counter (MUC) is the outermost subdetector. The main purpose of the MUC is to separate muons from hadrons and other backgrounds. The MUC is especially important to separate muons from charged pions, since their masses are similar. The other subdetectors cannot distinguish muons and pions properly, as for instance can be seen in figure 2.4. Since (anti)protons can already be separated easily by the information provided by the MDC and TOF, the MUC information is irrelevant for the study of $J/\psi \to \gamma p\bar{p}$.

The MUC consists of Resistive Plate Chambers (RCPs) that are triggered by traversing muons. The RPCs are interspersed between the steel plates of the yoke of the solenoid magnet, as shown in schematically in figure 2.3 by the cyan color. The vast majority of particles produced in the $e^+e^-$-collisions are stopped in the EMC or in the coil of the magnet. However, most muons pass trough the full detector and hence create typical hit patterns in the RPCs.

## 2.3 Trigger system and data acquisition

The trigger, data acquisition and online computing systems of BESIII are developed to handle the multi-beam bunches and high data rate of BEPCII. BESIII needs to process a large amount of data in real time, which needs to be reduced rapidly by the trigger system. The BESIII trigger system contains two levels, a hardware trigger (L1) and a software trigger (L3).

The MDC, TOF, and EMC subdetectors provide the input for the L1-trigger[2]. The output of these detectors is continuously stored in a pipelined buffer and processed by the global trigger logic. To prevent any dead time, the data obtained from the different subsystems need to be stored during the trigger latency of 6.4 $\mu$s, as illustrated in figure 2.7. The main purpose of the L1-trigger is to reduce background from cosmic rays and electrons and positrons originating from one of the two beams.

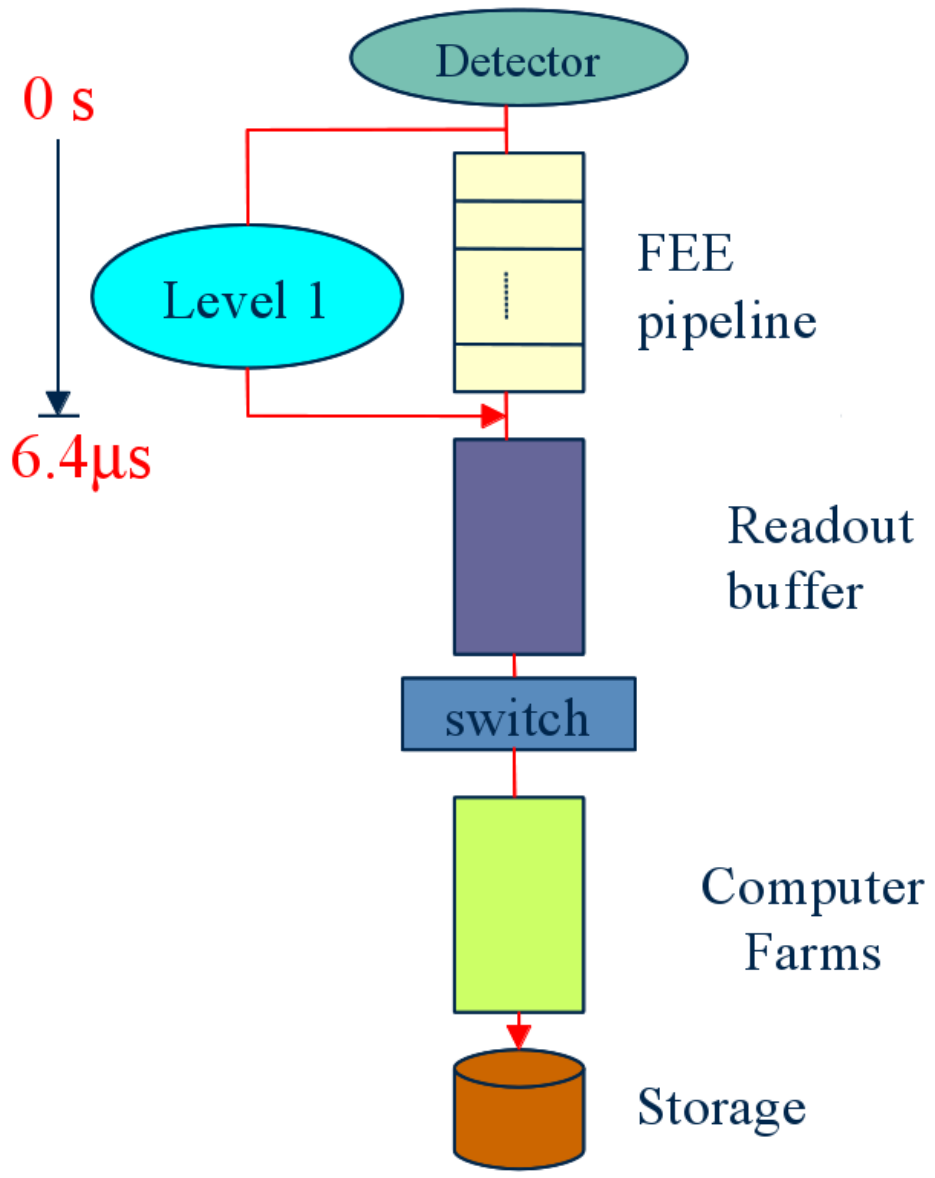

**Figure 2.7:** Schematic overview of the trigger system [32], where FEE stand for front-end electronics.

[2] For more recent runs, the MUC information was included as well. However, this information was not included for the $J/\psi$ trigger set.

Collectively, the subdetectors provide a set of 48 trigger conditions, which are combined by the global trigger logic to create 13 so-called trigger channels. Together with a random trigger and two different prescale factors, a total set of 16 channels is available. For each data run an optimized combination of these channels can be constructed. If any trigger channel in this combination is enabled, the event will be read out. In table 2.1, the combinations of channels for the $J/\psi$ data run of 2009 is shown. For the more recent $J/\psi$ data runs, the combination of trigger channels was only slightly modified. More details about the different trigger conditions and channels can be found in references [102] and [103].

| Channel | Trigger conditions |
|---|---|
| channel 0 | Number of short back-to-back tracks in the MDC $\geqslant 1$<br>Number of TOF endcap hits $\geqslant 1$<br>Number of EMC endcap clusters $\geqslant 1$ |
| channel 1 | Number of long tracks in the MDC $\geqslant 2$<br>Number of TOF barrel hits $\geqslant 2$<br>Number of EMC barrel clusters $\geqslant 1$ |
| channel 2 | Number of long tracks in the MDC $\geqslant 2$<br>Number of TOF barrel hits $\geqslant 2$ |
| channel 4 | Number of long tracks in the MDC $\geqslant 1$<br>Number of TOF barrel hits $\geqslant 1$<br>Total energy deposited in EMC above a lower threshold |
| channel 5 | Number of long tracks in the MDC $\geqslant 2$<br>Number of TOF barrel hits $\geqslant 1$<br>Number of EMC barrel clusters $\geqslant 1$ |
| channel 9 | Random trigger at 60 Hz |
| channel 11 | Number of clusters in the EMC $\geqslant 2$<br>Total energy deposited in EMC above a medium threshold |

**Table 2.1:** Trigger settings for the 2009 $J/\psi$ run. Channel 0 is designed for endcap Bhabha events, channels 1 to 5 for events with charged particles in the barrel region and channel 11 for all-neutral events. For the later $J/\psi$ runs, a slightly altered set was used [102]. Note that only an optimized subset of the total available channels is used.

When BEPCII runs at the $J/\psi$ center-of-mass energy, the physics event rate is about 2 kHz, and the background event rate is estimated to be about 13 MHz. The L1-trigger reduces the background to a lower rate than the physics-event rate, so the L1-trigger rate is below the maximum of 4 kHz such that the data acquisition system can manage the incoming stream of data. The L1-accepted data are transferred to an online computing farm (L3) that runs event-building and filtering software. The L3-trigger further reduces the background event rate to about 1 kHz, resulting in a total storage rate of 40 Mb/s. This rate is acceptable for permanent storage and the additional filtering happens offline.

The applied set of L1 and L3-triggers is thus able to suppress backgrounds by more than three orders of magnitude, while maintaining a signal efficiency of almost 100% for all $J/\psi$ data runs [102,103]. Therefore, the effect of the trigger system on the number of signal events can be neglected in the physics analyses and will not impose any significant systematic uncertainty.

## 2.4 The BES Offline Software System (BOSS)

The offline data analysis and Monte Carlo simulations are carried out with the BES Offline Software System (BOSS). BOSS is developed for the operation system of Scientific Linux CERN (SLC) using C++ language and object-oriented techniques. The software uses several external high-energy physics libraries like CERNLIB, CLHEP, ROOT and Geant4. Furthermore, some codes of Belle, BaBar, ATLAS and GLAST experiments are re-used in the system. The incorporated GAUDI framework provides tools for event simulation, data processing and physics analysis. The full geometry of BESIII is implemented in the Geometry Design Markup Language (GDML). Within BOSS there are three different types of event data available: raw data, reconstructed data and so-called Data Summary Tape (DST) data. In general, further analysis of the data will be performed on the preprocessed and reconstructed DST data. For the study presented in chapter 7, this further analysis was performed with BOSS version 6.6.4.p03, whereas the studies discussed in chapters 4 and 6 are based on BOSS version 7.0.5. For the analyses treated in chapters 6 and 7, another software package was used subsequently. This software package, called PAWIAN, will be introduced in chapter 5.

## 2.5 $J/\psi$ datasets collected by BESIII

The BESIII data taking officially started in 2009. Since then, BESIII has collected different types of datasets, including world's largest samples of several types of $c\bar{c}$ mesons, such as the $J/\psi$. Table 2.2 shows the total number of $J/\psi$ events that BESIII has collected per year. For a comparison, the total number of $J/\psi$ events collected by other $e^+e^-$ collider experiments that produced $J/\psi$ events directly is shown in table 2.3. Note that there are other $e^+e^-$ collider experiments where $J/\psi$ events can be created indirectly, such as Belle and CLEO-c. However, since their total event count is already several orders of magnitude lower, the total number of indirectly populated $J/\psi$ events is by far not comparable to the total number of $J/\psi$ events collected by BESIII.

| **2009** | **2012** | **2017** | **2018** | **Total** |
|---|---|---|---|---|
| $0.2 \times 10^9$ | $1.1 \times 10^9$ | $4.6 \times 10^9$ | $4.2 \times 10^9$ | $1 \times 10^{10}$ |

**Table 2.2:** Number of $J/\psi$ events collected by BESIII over the years.

| **MARKIII** | **DM2** | **Crystal Ball** | **KEDR** |
|---|---|---|---|
| $5.8 \times 10^6$ | $8.6 \times 10^6$ | $2.2 \times 10^6$ | $6.3 \times 10^6$ |

**Table 2.3:** Total number of $J/\psi$ events collected by competitor experiments [84, 104].

# 3. Event Selection

As described in section 1.4, the aim of the study of $J/\psi \to \gamma p\bar{p}$ is to get a better understanding of the full spectrum of the $p\bar{p}$ invariant mass, and, especially, to get more insight in the puzzling $\eta_c$ resonance. In this thesis, the $\eta_c$ range is defined as the range with a $p\bar{p}$ invariant mass of 2.7 GeV or higher. Besides the interesting structures in the $p\bar{p}$ invariant mass, this study aims for a significant improvement in the accuracy of the branching fraction $J/\psi \to \gamma p\bar{p}$, since the experimental value was determined for the first and only time in 1984 to be $(3.8 \pm 1.0) \times 10^{-4}$ [90].

To be able to study the decay $J/\psi \to \gamma p\bar{p}$, the data events collected by BESIII are reconstructed and the selection criteria are optimized. In the following sections, the datasets, both from the experiment and from Monte Carlo (MC) simulations, are briefly presented and the event-selection criteria are described and motivated.

## 3.1 Datasets

The analysis that will be discussed in chapter 7 is based on the $J/\psi$ datasets collected by BESIII in 2009 and 2012. Together, these datasets contain $(1310.6 \pm 7.0) \times 10^6$ events, of which $(223.7 \pm 1.4) \times 10^6$ events are recorded in 2009 and $(1086.9 \pm 6.0) \times 10^6$ in 2012 [86]. For the analyses discussed in chapters 4 and 6, the full available data sample of $N_{J/\psi} = (10086.6 \pm 43.7) \times 10^6$ [105], collected in 2009, 2012, 2017 and 2018, is used for our study. Specifically, the so-called mass-independent study benefits from a higher number of events, since the data will be divided over bins in the invariant $p\bar{p}$ mass. For the analysis part presented in chapter 7, the data collected in 2009 and 2012 provides ample $J/\psi \to \gamma p\bar{p}$ statistics. In the rest of the current chapter, 'data' will always refer to the sample collected in 2009 and 2012.

In addition to experimental data, an inclusive MC sample of $1.225 \times 10^9$ $J/\psi$ is used to enable background-estimation studies and optimization of selection criteria. This inclusive MC sample is generated centrally by the collaboration and includes all known $J/\psi$ decays with the corresponding branching fractions taken from the Particle Data Group (PDG) [15], while the unknown ratios are generated according to the Lundcharm model [106]. Furthermore, an exclusive MC sample of $10^7$ phase-space distributed $J/\psi \rightarrow \gamma p\bar{p}$ events is generated. This exclusive MC sample is used for efficiency studies in chapter 4, and for the fits of the partial-wave analyses described in chapters 6 and 7. For a toy MC study presented in chapter 6, another exclusive MC sample comprising of $10^6$ generated events is used. Here, the $J/\psi \rightarrow \gamma\eta_c$ events are distributed according to the $J/\psi$ to Photon Eta model (JPE). This model is specifically constructed for vector decays into a photon and a pseudoscalar meson, compatible with the reaction of interest $J/\psi \rightarrow \gamma\eta_c$. The subsequent decay, $\eta_c \rightarrow p\bar{p}$, is described by a phase-space model.

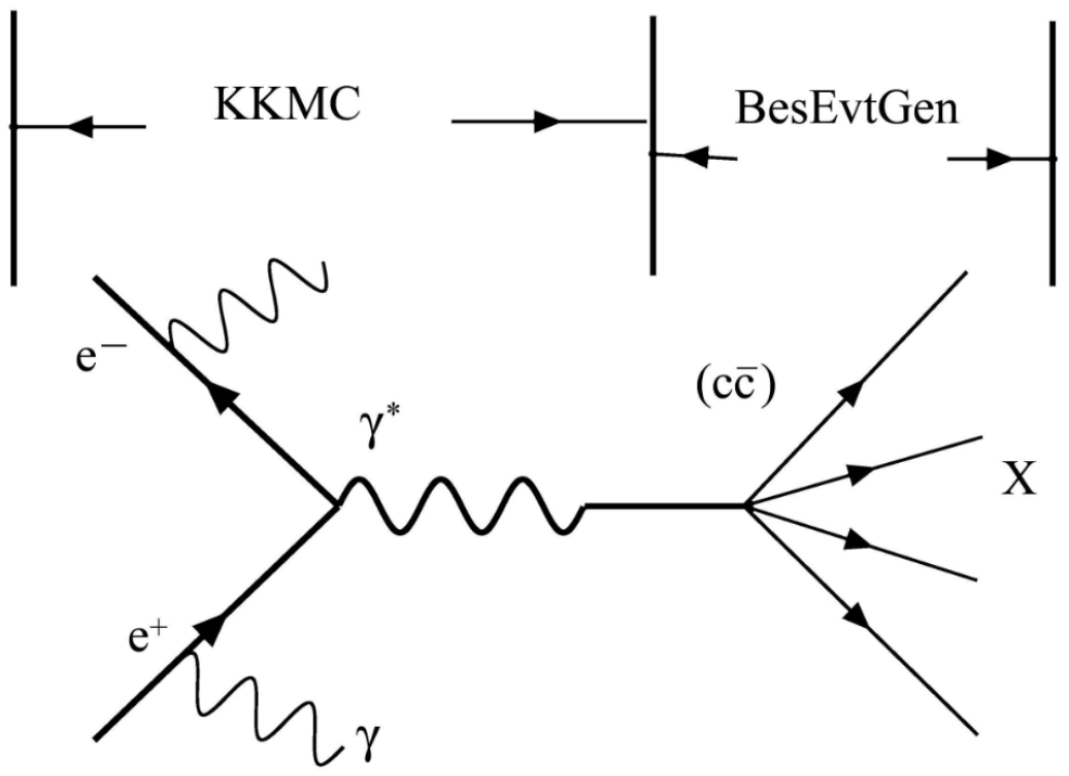

**Figure 3.1:** Illustration of the BESIII generator framework [107].

All mentioned MC samples are based on a combination of software packages for the generation and reconstruction of the events. The production of $J/\psi$ states are simulated by the MC generator KKMC [108], in which the effects of beam-energy spread and initial-state radiation are taken into account. The event generator BesEvtGen [109] is used to generate subsequent $J/\psi$ decays, as illustrated in figure 3.1. Final-state radiation effects are included at the BesEvtGen level by using the PHOTOS package [110]. In the inclusive MC samples, known decays are generated by BesEvtGen using the branching fractions listed by the PDG [15], and the remaining unknown decays are generated with LUNDCHARM [106]. The information on the generated final-state particles and transport through the detector material is subsequently handled by a Geant4 software package [111]. This software package simulates the performance of the BESIII detector and the interaction of the final-state particles with the detector materials. The different MC generators are explained in more detail in reference [107].

The datasets and MC samples are analyzed using BES Offline Software System (BOSS) version 6.6.4.p03. At present, this is the most recent version of BOSS for which the inclusive $J/\psi$ MC sample is available. The analysis of the full available data sample is carried out with BOSS version 7.0.5, as this is required for the data collected in 2017 and 2018.

## 3.2 Selection of $J/\psi \to \gamma p\bar{p}$ candidates

The final-state of $J/\psi \to \gamma p\bar{p}$ consists of two oppositely charged particles and one photon. For a successful analysis of the decay, the correct signal candidates need to be selected with an optimum efficiency, while the amount of background needs to be reduced as much as possible. The first step in achieving this goal is performing a general selection procedure based on the BESIII dimensions and characteristics. Subsequently, the information of several sub-detectors is used to make a further selection based on particle identification (PID) and the reconstructed momenta of the charged particles and photons. Afterwards, a few additional selection criteria are imposed to suppress the surviving background events from $J/\psi \to \pi^0 p\bar{p}$, $\pi^0 \to \gamma\gamma$ and $J/\psi \to p\bar{p}$. Finally, the remaining background events will be modeled and subtracted. In the following parts, the different steps are discussed in more detail.

### 3.2.1 General BESIII selection criteria

The characteristics of the BESIII detector call for a certain set of general selection criteria. These selection criteria are optimized by working groups of the collaboration and are used collaboration-wide.

#### *Charged-particle tracks*

The reconstruction of charged-particle tracks is based on the MDC hit information. For a precise measurement, the track must be completely contained within the MDC volume. Therefore, charged-particle tracks must fulfill the requirement $|\cos\theta| < 0.93$. Since for our channel of interest there are no intermediate resonances with long lifetimes expected, charged-particle tracks should originate from a point close to the electron-positron interaction point. For this reason, the distance between the track origin and the interaction point is required to be smaller than 10 cm in the direction along the beamline, and smaller than 1 cm in the plane perpendicular to the beamline. These numbers are compatible with the expected spread of the interaction point [97].

Figure 3.2 shows the tracking efficiencies and systematic uncertainties that BESIII obtains for protons.

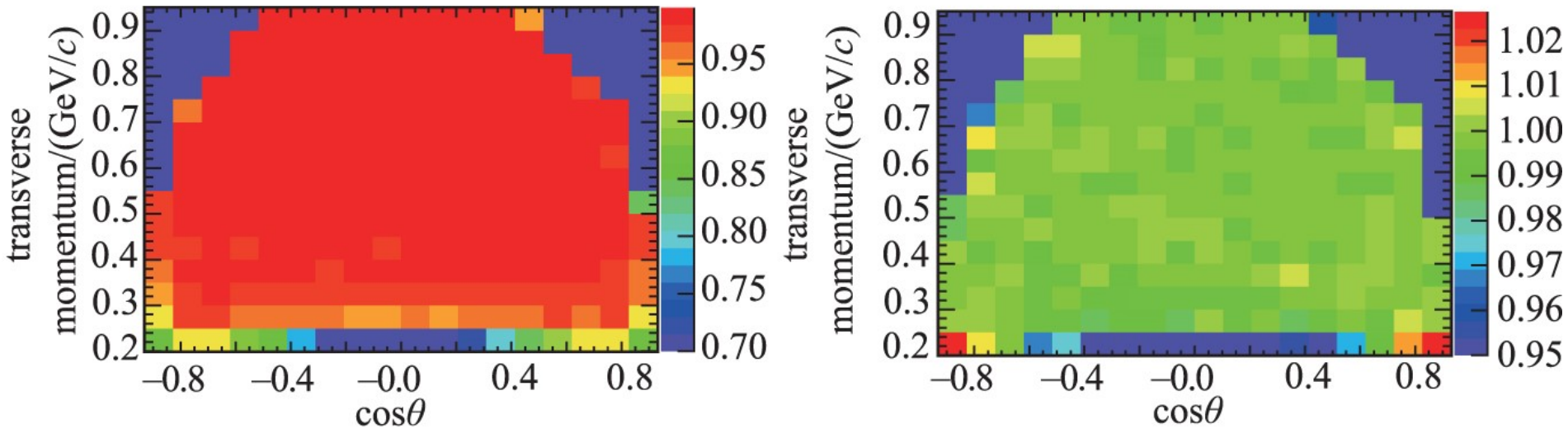

**Figure 3.2:** The two-dimensional tracking efficiencies from data (left), and systematic uncertainties (right) for protons [112]. The colors represent the corresponding efficiency (0-1) or uncertainty (in %).

### *Photons*

The EMC is the only detector of BESIII that is capable to measure the energy and scattering angles of photons with good efficiency and resolution. To suppress background unrelated to the event, the time between the electron-positron collision and the electromagnetic shower is required to be less than 700 ns. Additionally, the angle between a cluster and any charged-particle track should be larger than $10^\circ$ to eliminate showers related to these tracks. To discard background from background radiation or electronic noise, the deposited cluster energy must be higher than 25 MeV for the barrel ($|\cos\theta| < 0.8$) and the endcaps ($0.86 < |\cos\theta| < 0.92$). Note that it is common practice to use a higher minimum energy of 50 MeV for the endcaps. However, for a partial-wave analysis it is beneficial to have a high coverage of the reaction phase space. Therefore, it was decided to use a lower minimum of 25 MeV. We evaluated the possibility to further lower the threshold for photon detection. This turned out not to be opportune, since a smaller value than ∼25 MeV would drastically increase the electronic noise and introduce detector irregularities.

Figure 3.3 shows the tracking efficiencies that BESIII obtains for photons. Additionally, the relative difference in efficiency between data and MC, or systematic uncertainty, is shown.

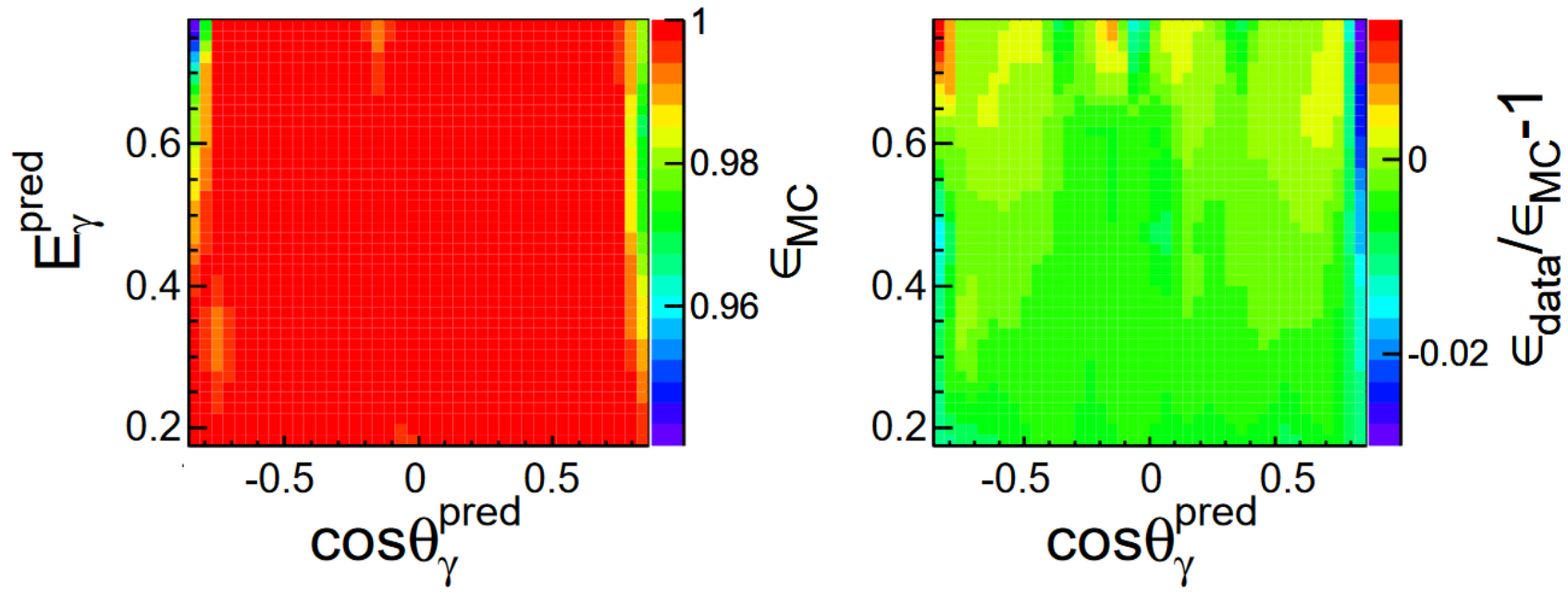

**Figure 3.3:** The two-dimensional tracking efficiencies from MC, $\epsilon_{MC}$, (left), and the relative difference in efficiency between data and MC, $\epsilon_{data}/\epsilon_{MC} - 1$, (right) for photons [113]. Here, $\cos\theta_\gamma^{pred}$ and $E_\gamma^{pred}$ are defined as the polar angle and energy, respectively, of a photon predicted by a kinematic fit.

### 3.2.2 Channel specific selection criteria

For the study of $J/\psi \to \gamma p\bar{p}$ at least two charged-particle tracks and one photon candidate are required in the final state. For each charged-particle track, the specific energy loss, $dE/dx$, obtained by the MDC, and the time-of-flight, obtained by the TOF, will provide a likelihood that the track belongs to an electron ($\mathcal{L}(e)$), a pion ($\mathcal{L}(\pi)$), a kaon ($\mathcal{L}(K)$), a muon ($\mathcal{L}(\mu)$), or a proton ($\mathcal{L}(p)$). Only events where both charged-particle tracks fulfill the requirements $\mathcal{L}(p) > \mathcal{L}(K)$ and $\mathcal{L}(p) > \mathcal{L}(\pi)$ are maintained for further analysis. To summarize, all events are required to have exactly two oppositely-charged-particle tracks with positive proton identification, and at least one photon. The number of photon candidates has no upper limit in our analysis, since additional photons can appear as a result of beam background and electromagnetic split-off. Further selection criteria are applied to select the signal candidates among the remaining set of events.

### *Vertex and kinematic fit*

The two charged-particle tracks, related to the proton and antiproton, are required to originate from a common point. Therefore, a vertex fit is performed. In the fit, the parameters of the two reconstructed tracks are varied within their measured uncertainties. The aim is to minimize the distance between the nominal interaction point and the reconstructed point where the tracks are closest to each other. An event is retained for further analysis if the vertex fit converges.

After the vertex fit, an additional kinematic fit is executed. Where the vertex fit is performed only once per event, the kinematic fit is performed with all possible $\gamma p\bar{p}$ combinations to determine the best photon candidate. Aside from finding the best photon candidate, the kinematic fit improves the resolution of the measured four-vectors, as shown in figure 3.4.

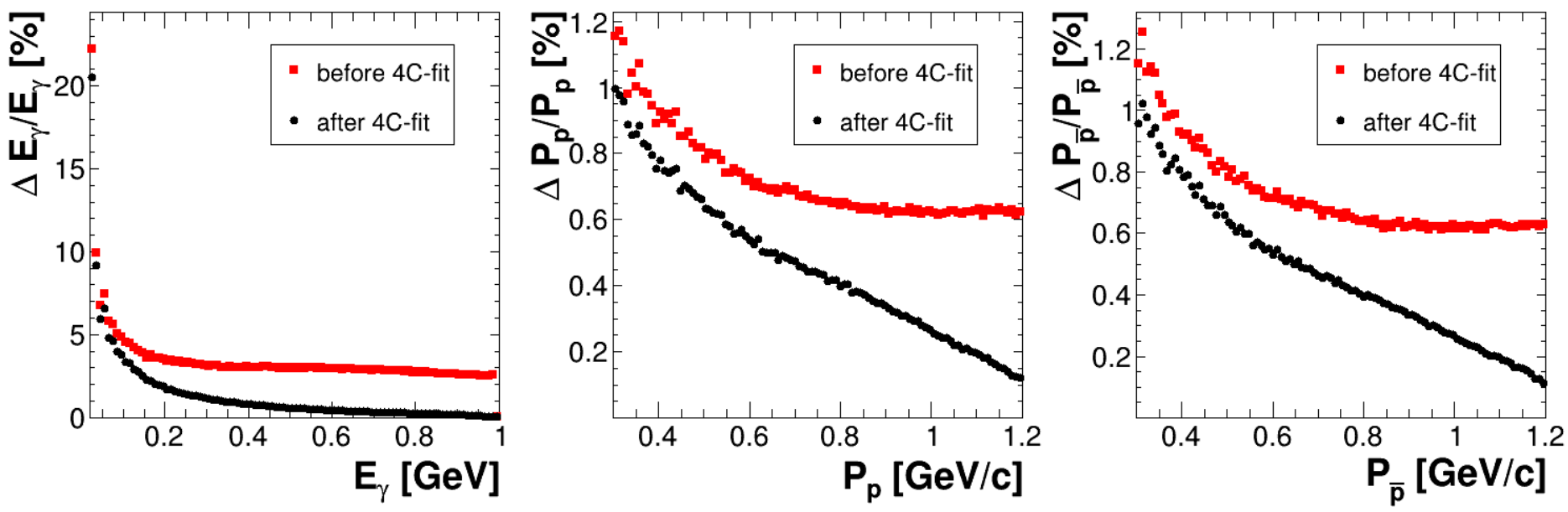

**Figure 3.4:** The relative resolutions of the photon energy $E_\gamma$, proton momentum $P_p$ and antiproton $P_{\bar{p}}$ momentum, respectively, before (red) and after (black) the kinematic 4C-fit. The resolutions are extracted from the exclusive $J/\psi \to \gamma p\bar{p}$ MC sample by calculating the root-mean-square deviation between the measured values versus the MC truth values for each energy or momentum bin.

The kinematic fit is based on the principle of energy and momentum conservation. This conservation law dictates that the sum of the four-vectors of all final-state particles must be equal to the four-vector of the initial state. The energy and the three components of the linear momentum result in four constraints. Therefore, the kinematic fit using just these four constraints is called a 4C-fit. In a 4C-fit, the four-vectors of all final-state particles are varied until the constraints are met. Further constraints could be introduced by requiring that the invariant mass of several final-state particles should correspond to an intermediate (narrow) resonance, as will be discussed in section 3.4 for an intermediate $\pi^0$.

The kinematic-fit method in BOSS is based on the least-squares-method and makes use of Lagrange multipliers. A detailed description of the vertex and kinematic fitting procedures can be found in reference [47].

In this analysis a 4C-fit is performed: the four-vectors of each $\gamma p\bar{p}$ combination are constrained to the four-vector of the initial $e^+e^-$ state. Combinations resulting in a $\chi^2_{4C} > 200$ are discarded, all other combinations are kept for further study. The final $\chi^2_{4C}$ limit that selects the signal candidates is found by optimizing the statistical significance $\frac{S}{\sqrt{S+B}}$, where $S$ and $B$ are the number of signal and background events, respectively. The optimization procedure is done twice: once for the $\eta_c$ range, and a second time for the full $p\bar{p}$ invariant mass range. The limits found are $\chi^2_{4C} < 60$ and $\chi^2_{4C} < 28.5$ for the $\eta_c$ and full range, respectively. More details of the optimization procedure are given in the following section. If an event contains multiple $\gamma p\bar{p}$ combinations after all selection criteria are applied, the combination with the smallest $\chi^2_{4C}$ is selected as the signal candidate.

### *Suppression of specific backgrounds*

The uppermost panel of figure 3.5 shows the $p\bar{p}$ invariant-mass spectrum of the inclusive MC sample for the complete mass range after applying all of the previously discussed selection criteria. It can be seen that the remaining events still contain a non-negligible amount of background. The vast majority of background, in particular at the end of the mass spectrum close to the $\eta_c$, is caused by the channel $J/\psi \to p\bar{p}$. The detected photon candidate can either be caused by a Final-State Radiation (FSR) photon, as shown in green, or by a random photon cluster, as shown in magenta. Another visible background component is caused by the channel $J/\psi \to \pi^0 p\bar{p}, \pi^0 \to \gamma\gamma$, where one of the two photons from the $\pi^0$ decay is not detected. This background contribution is represented by the blue lines in figure 3.5. Note that all channels listed in figure 3.5 include possible intermediate resonances, like $J/\psi \to \gamma\eta_c, \eta_c \to p\bar{p}$ for the signal channel.

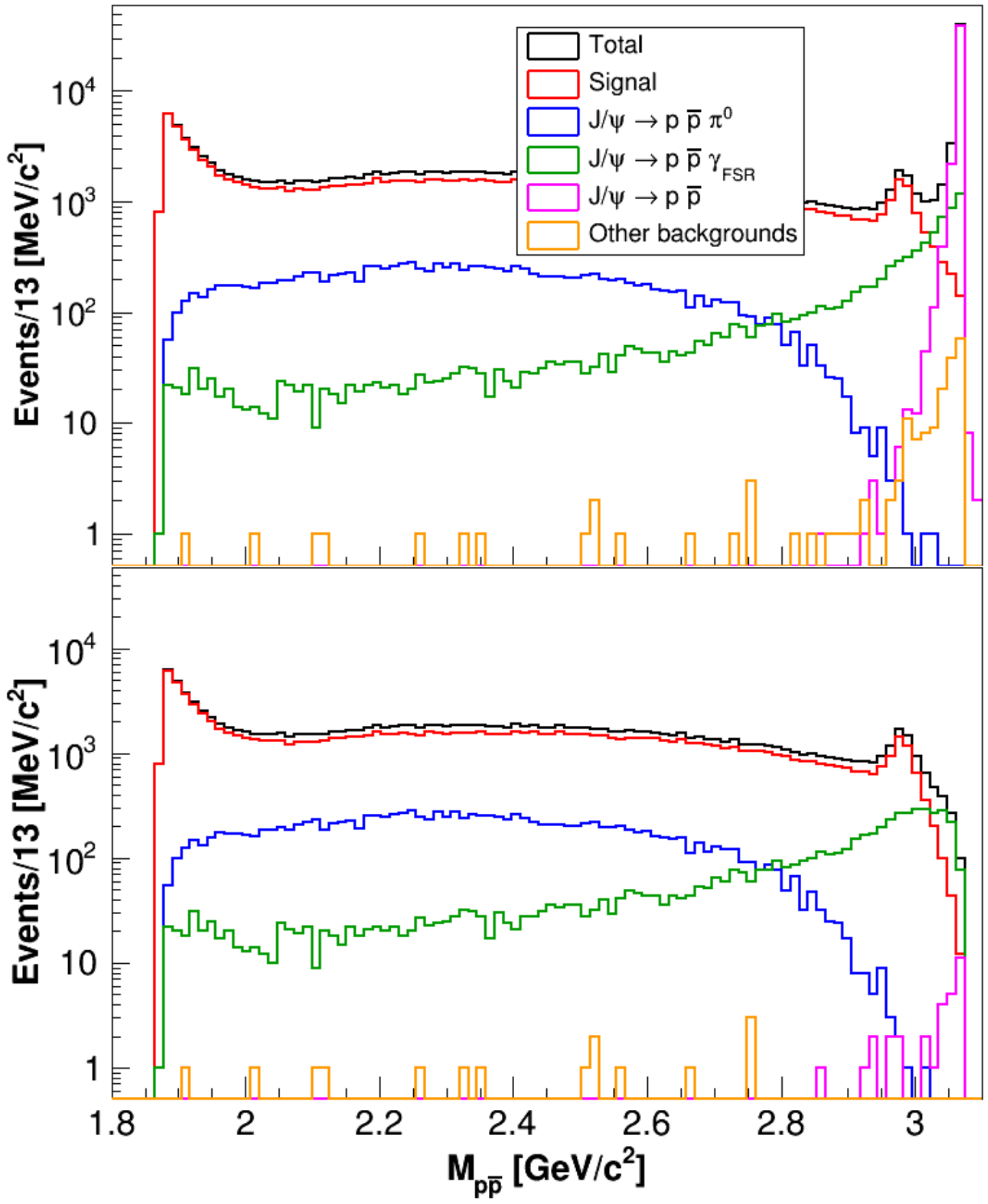

**Figure 3.5:** The $p\bar{p}$ invariant-mass spectrum of the inclusive MC sample before (upper panel) and after (lower panel) the additional $M^2_{miss}$ and $\theta_{p\bar{p}}$ cuts.

In the center-of-mass (CM) frame, a decay of an initial-state into two equal-mass final-states results in a back-to-back emission, so with an angle of 180° between the two final-state particles. To reduce the contribution of the back-to-back $J/\psi \to p\bar{p}$ decay, we have thus set a maximum to the angle between the two charged-particle tracks, $\theta_{p\bar{p}}$. To diminish the $\pi^0$ back-ground, we require the square of the missing mass, $M^2_{miss}$, to be smaller than the mass that corresponds to a $\pi^0$. Here, $M^2_{miss}$ is defined as the squared invariant mass of the four-momentum difference between the initial-state and the combined $p\bar{p}$ state. In the case of $J/\psi \to \pi^0 p\bar{p}, \pi^0 \to \gamma\gamma$, it is expected that $M^2_{miss} \simeq m^2_{\pi^0}$. The $M^2_{miss}$ distribution of the inclusive MC sample is shown in figure 3.6. It can be seen that there is a relatively small number of signal events that have a $M^2_{miss}$ larger than the $\pi^0$ events. Therefore, it is chosen to work with a $M^2_{miss}$ maximum, instead of an exclusion window around the $m^2_{\pi^0}$, hereby minimizing the systematic error.

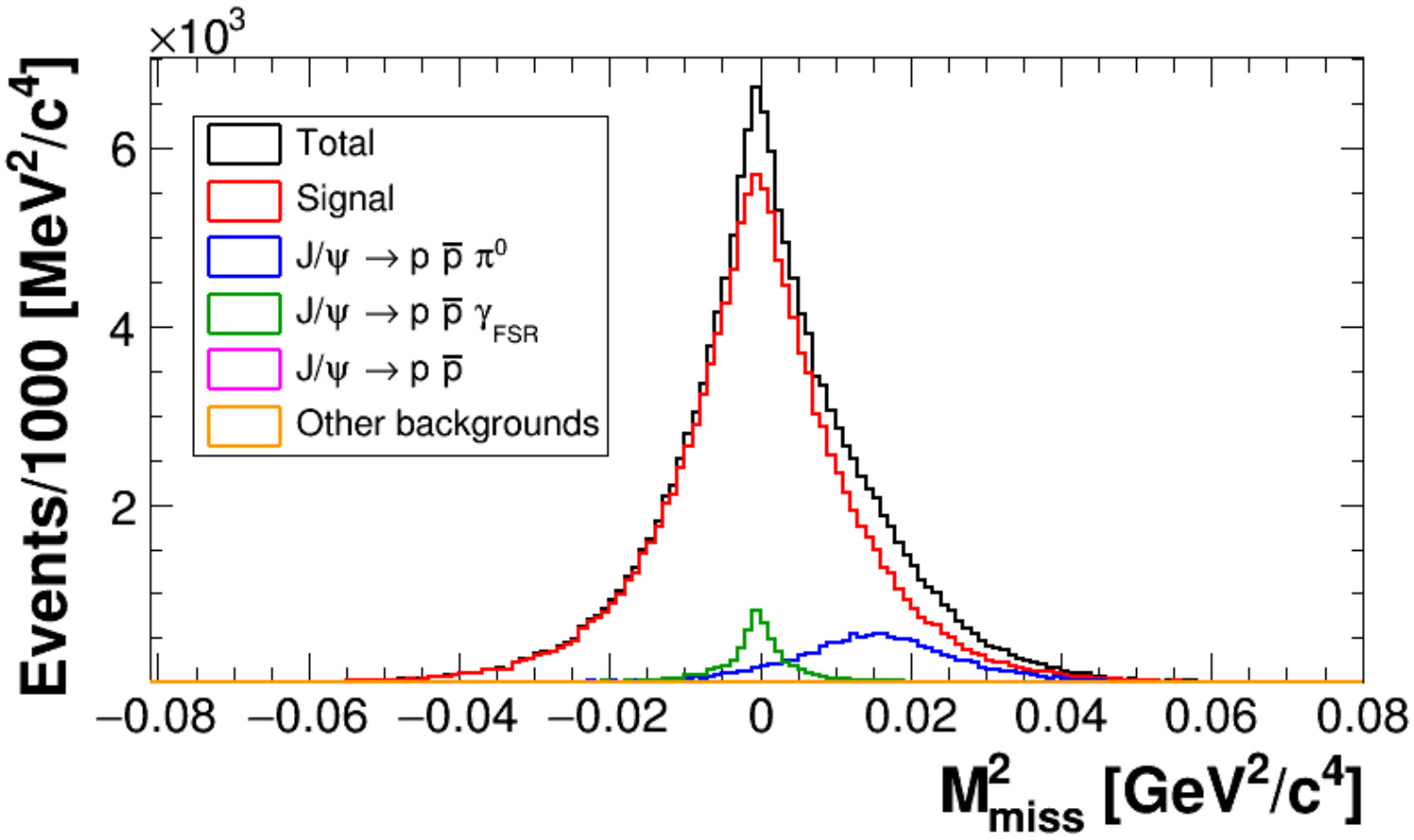

**Figure 3.6:** The $M^2_{miss}$ spectrum of the inclusive MC sample with all the described selection criteria applied, except the cut on $M^2_{miss}$ itself.

The maximum limits used to select the signal candidates are found by optimizing the significance $\frac{S}{\sqrt{S+B}}$. The values of $\chi^2_{4C}$, $M^2_{miss}$ and $\theta_{p\bar{p}}$ are varied simultaneously until the maximum value of the significance is reached. This process is repeated twice. Once to find the optimum cuts for the full $p\bar{p}$ invariant-mass range. The second time is to find the optimum cuts for the study of the $\eta_c$ resonance, so for the $p\bar{p}$ invariant mass range of 2.7 GeV or higher. The results of the optimization, and thus the used limits, can be found in table 3.1. The resulting statistical significances are 332.6 and 136.8 for the full and $\eta_c$ range, respectively. The bottom panel of figure 3.5 shows the resulting inclusive MC sample for the full $p\bar{p}$ invariant-mass range after applying the additional cuts, and figure 3.7 the resulting $p\bar{p}$ mass resolution.

| | $\chi^2_{4C}$ | $\theta_{p\bar{p}}$ | $M^2_{miss}$ [GeV$^2/c^4$] |
|---|---|---|---|
| Maximum value for full range | 28.5 | 177.3° | 0.049 |
| Maximum value for $\eta_c$ range | 60.0 | 177.2° | 0.012 |
| Stepsize | 0.5 | 0.1° | 0.001 |

**Table 3.1:** Significance optimization of the cuts.

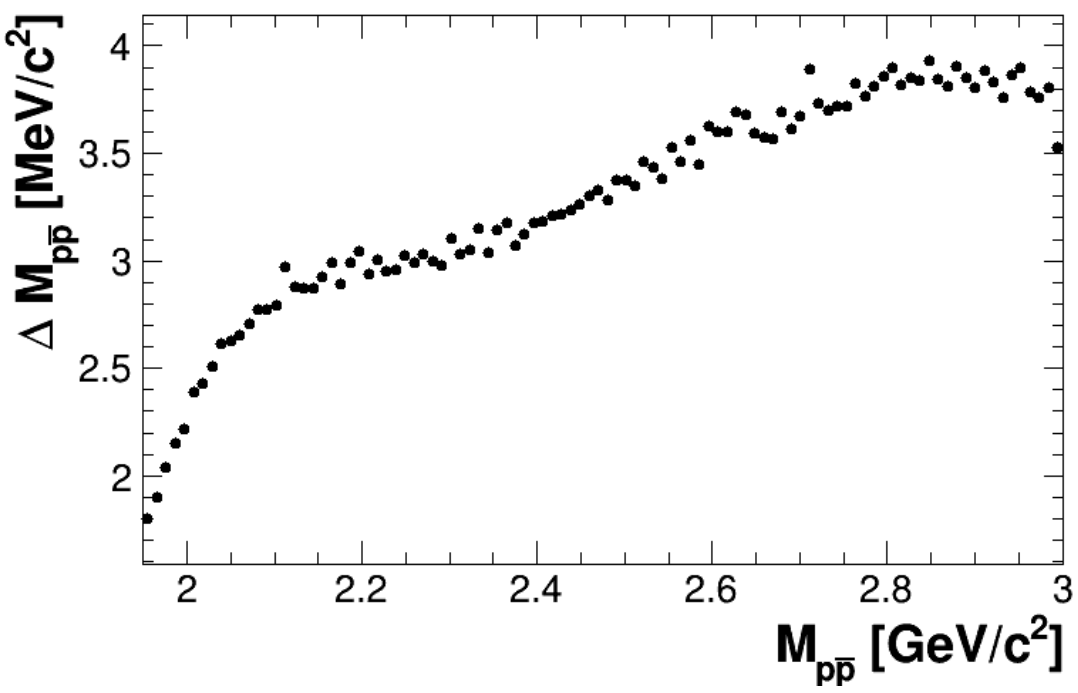

**Figure 3.7:** The resultant $p\bar{p}$ mass resolution after applying all selections. The shown mass resolution distribution was extracted from the exclusive $J/\psi \to \gamma p\bar{p}$ MC sample by calculating the root-mean-square deviation between the measured $p\bar{p}$ values versus the MC truth values for each invariant-mass bin.

In the two following sections it will be discussed how the remaining background contributions of $J/\psi \to p\bar{p}$ and $J/\psi \to \pi^0 p\bar{p}, \pi^0 \to \gamma\gamma$ are handled. According to the inclusive MC sample, all other background channels together compromise only 0.01% and 0.02% of the total selected events in the full and $\eta_c$ range, respectively. Therefore, other background channels are considered negligibly small. Additionally, the background from continuum processes was studied using a data sample recorded at a center-of-mass energy of 3.08 GeV, just below the $J/\psi$ mass. This data sample has an integrated luminosity of $(30.84 \pm 0.04)$ pb$^{-1}$ [86]. The surviving events in the 3.08 GeV dataset are scaled to the luminosity of the $J/\psi$ dataset. In total all expected events from continuum background processes correspond to less than 0.02% of the full and $\eta_c$ selected datasets. The continuum background processes can thus be neglected for the analysis.

## 3.3 Background study $J/\psi \to p\bar{p}$

The main contribution of the remaining $J/\psi \to p\bar{p}$ background is caused by a Final-State-Radiation (FSR) photon due to bremsstrahlung as a secondary vertex process on the production level. In this case, the final-state particles $\gamma_{FSR}p\bar{p}$ are kinematically identical to the final-state particles of the decay of interest. Therefore, the detector information will not help to distinguish the different states. Further restrictions on, for instance, the $\chi^2_{4C}$ value will have no effect: the four-vectors of a $\gamma_{FSR}p\bar{p}$ combination will resemble the initial $J/\psi$ state as well as a signal $\gamma p\bar{p}$ combination.

However, this FSR, or photonic bremsstrahlung, is an effect that can be described by QED. The FSR contribution can be calculated unambiguously by the Feynman rules of QED and can be included at the BesEvtGen level via the PHOTOS algorithm [47,107,110]. PHOTOS is a MC package which corrects simulated events to account for FSR after they have been fully generated. The package has been around for about 30 years and it has a history of use in other experiments like Belle, LHCb, BaBar and DESY. For photonic bremsstrahlung, the PHOTOS package delivers a precision of 0.1% [114–116].

In this study, $10^6$ $J/\psi \to p\bar{p}$ events are generated with the PHOTOS corrections included. The $J/\psi$ decay into the proton-antiproton pair is generated with the model J2BB1 [117]. This model is constructed specifically for $\psi'$ or $J/\psi$ decays into an octet baryon and antibaryon pair, like $p\bar{p}$. The angular distribution of the outgoing proton is generated to take the form $\frac{|M|^2}{d\cos\theta} \propto (1 + \alpha\cos 2\theta)$, where the parameter $\alpha$ can be set manually before generation. For $J/\psi \to p\bar{p}$, the theoretical value described in reference [118] coincides with the experimental value [119]. Therefore, the default theoretical value $\alpha = 0.69$ is used.

The full set of generated $J/\psi \to p\bar{p}$ events will be exposed to exactly the same set of selection criteria as applied to the data[1]. The generated events that survive the selection criteria are weighted and subtracted from the $\gamma p\bar{p}$-selected data. The weight $w$ is chosen such that the number of selected generated events, $N_{gen,sel}$, equals the expected number of selected

[1] All selections as described in section 3.2.

events in the datasets, $N_{dat,exp}$, after weighing:

$$N_{dat,exp} = w \cdot N_{gen,sel}. \tag{3.1}$$

The number of expected data events can be found by correcting the total number of $J/\psi \to p\bar{p}$ events in the data, $N_{p\bar{p}} = N_{J/\psi} \cdot \mathcal{B}(J/\psi \to p\bar{p})$, for the efficiency $\epsilon = N_{gen,sel}/N_{gen,total}$. Therefore, the weight is defined as

$$w = \frac{\epsilon \cdot N_{p\bar{p}}}{N_{gen,sel}} = \frac{N_{J/\psi} \cdot \mathcal{B}(J/\psi \to p\bar{p})}{N_{gen,total}}, \tag{3.2}$$

where the $J/\psi \to p\bar{p}$ branching fraction is $\mathcal{B}(J/\psi \to p\bar{p}) = (2.120 \pm 0.029) \times 10^{-3}$, as stated by the PDG [15].

## 3.4 Background study $J/\psi \to \pi^0 p\bar{p}, \pi^0 \to \gamma\gamma$

Before discussing the subtraction procedure of the remaining $\pi^0$ background, we considered the possibility to further reduce this background by introducing an extra selection criterion. As discussed in section 3.2.2, additional constraints can be included in the kinematic fitting procedure. In this case, the invariant mass of a photon pair was required to correspond to the $\pi^0$ mass, leading to a kinematic fit with five constraints (5C-fit). Events that contained at least one $\gamma\gamma p\bar{p}$ combination for which the 5C-fit converged ($\chi^2_{5C} < 200$) were discarded. Figure 3.8 compares the efficiency distributions with and without this 5C-fit requirement. These distributions are obtained by dividing the Dalitz spectrum of a reconstructed and selected phase-space-distributed MC sample by the Dalitz spectrum of the MC-truth sample. In both scenarios presented in figure 3.8, the maximum limits on $\chi^2_{4C}$, $M^2_{miss}$ and $\theta_{p\bar{p}}$ were optimized and implemented as the final step in the selection procedure.

To limit systematic uncertainties, it is beneficial to have a uniform efficiency distribution without significant structures due to efficiency depletions or enhancements. However, figure 3.8 shows that the introduction of the 5C-fit criterion will lead to a non-uniform efficiency spectrum. A horizontal band corresponding to a lower efficiency can be observed in the $M^2_{\bar{p}\gamma}$ variable. The band is probably caused by interactions of the antiproton with the detector material. Uncertainties in the description of these kind of interactions will lead to a systematic uncertainty in the results. Furthermore,

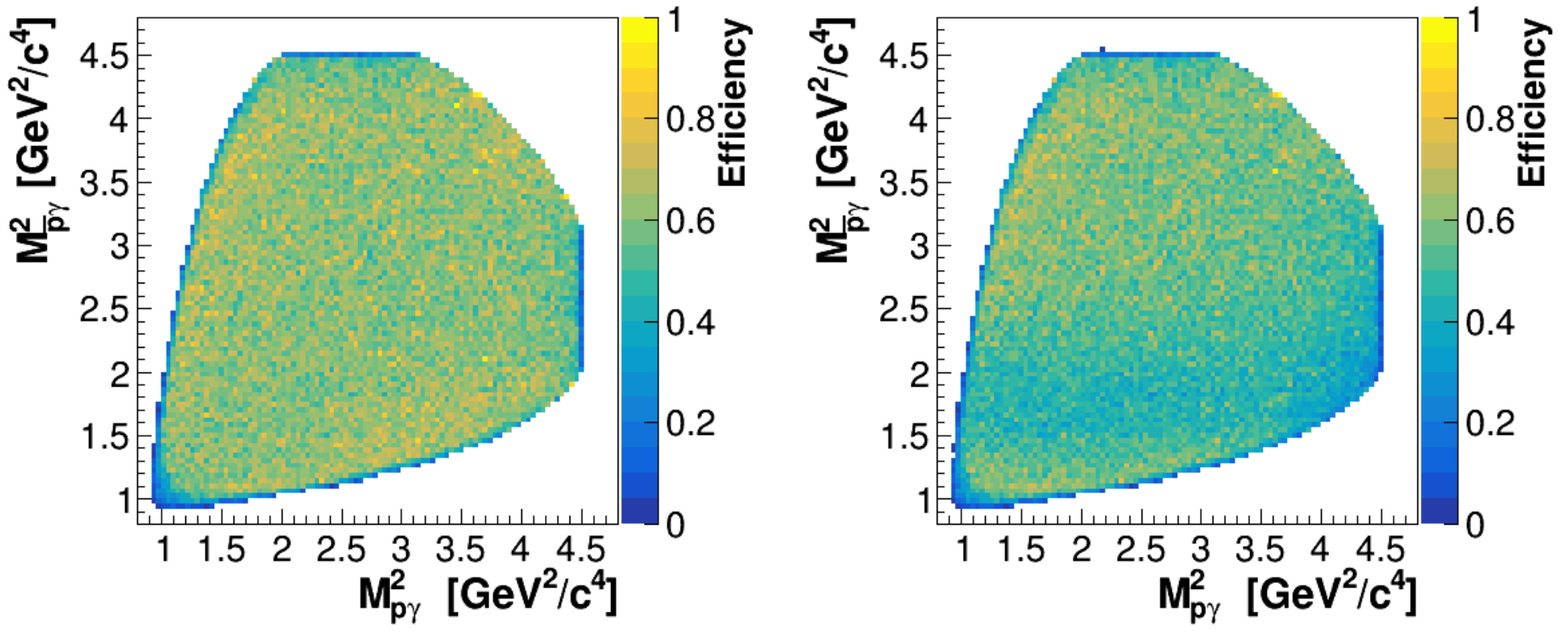

**(a)** Without $\chi^2_{5C}$ requirement **(b)** With $\chi^2_{5C}$ requirement

**Figure 3.8:** The efficiency spectra with and without the rejection of events for which $\chi^2_{5C} < 200$.

introducing the 5C-fit selection did not improve the resulting significance or signal-to-background ratio. Therefore, it was decided to omit such a 5C-fit requirement.

### 3.4.1 Data-driven background model

The resulting background contribution of $J/\psi \to \pi^0 p\bar{p}, \pi^0 \to \gamma\gamma$ cannot be calculated as model independently as the $J/\psi \to p\bar{p}$ background channel. However, $J/\psi \to \pi^0 p\bar{p}, \pi^0 \to \gamma\gamma$ can be measured and used to apply a background subtraction. For the selection of the $\pi^0 p\bar{p}$ candidates, the same standard BESIII criteria as described in section 3.2.1 are imposed. Furthermore, all events are required to have exactly two oppositely-charged-particle tracks with positive proton identification, and at least two photons. The additional selection criteria are partially based on reference [120] and are the following:

- $\chi^2_{4C} < 20$
- $0.015 \text{ GeV}^2/c^4 < M^2_{miss} < 0.021 \text{ GeV}^2/c^4$
- $E_\gamma > 50$ MeV
- $\theta_{\gamma\bar{p}} > 20°$

Here, $\theta_{\gamma\bar{p}}$ represents the angle between a photon cluster and the negatively-charged-particle track, and $M^2_{miss}$ is again defined as the squared invariant mass of the four-momentum difference between the initial-state and the combined $p\bar{p}$ state. For $\pi^0 p\bar{p}, \pi^0 \to \gamma\gamma$, one would expect that $M^2_{miss} \simeq 0.018$ GeV$^2/c^4$. Note that there are now at least two photons present and that for the 4C-fit each $\gamma\gamma p\bar{p}$ combination is thus constrained to the four-vector of the initial $J/\psi$ state. A study of the inclusive MC sample showed that this set of selections leads to a contribution of 0.1% of other channels than $J/\psi \to \pi^0 p\bar{p}, \pi^0 \to \gamma\gamma$. This contribution is considered small enough to proceed without further study of the contamination.

Figure 3.9a shows the Dalitz spectrum of the data after the $\pi^0$ selections. The horizontal and vertical bands present are caused by $N^*$ resonances, like $N(1440, 1520, 1535, 1650, 1710, 2065)$ [120]. To be able to describe the $\pi^0$ contribution in the signal channel $J/\psi \to \gamma p\bar{p}$, the events need to pass the $\gamma p\bar{p}$ selections described in section 3.2. Only a very small fraction of $\pi^0$ events survive both the $\gamma p\bar{p}$ and $\pi^0$ selection. A direct use of the measured $\pi^0$ data would thus lead to a significant statistical uncertainty in the background model. Therefore, the experimental $\pi^0$ Dalitz plot is used as input for an exclusive MC sample that represents the $J/\psi \to \pi^0 p\bar{p}, \pi^0 \to \gamma\gamma$ channel.

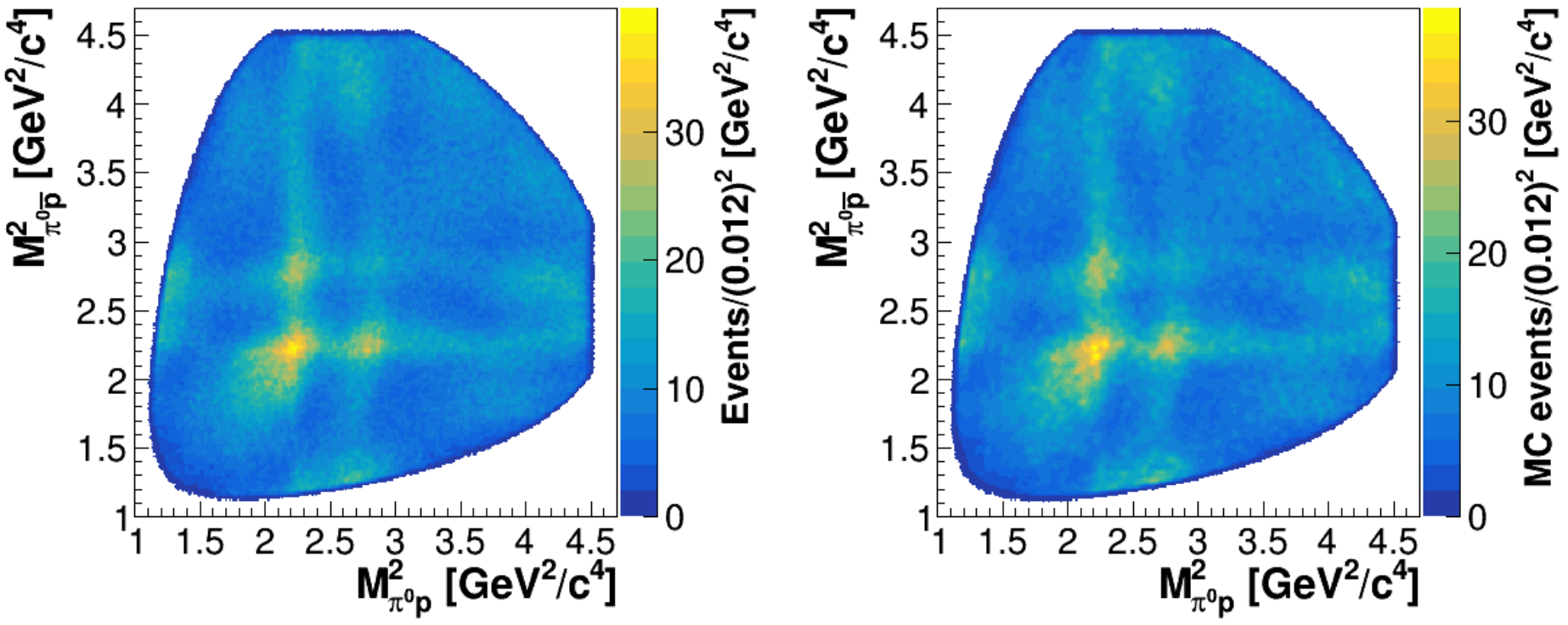

**(a)** Data **(b)** MassH2 MC sample

**Figure 3.9:** Dalitz plots of the $J/\psi \to \pi^0 p\bar{p}, \pi^0 \to \gamma\gamma$ channel with the $\pi^0$ selection criteria applied.

A total of $10^7$ $J/\psi \to \pi^0 p\bar{p}, \pi^0 \to \gamma\gamma$ events are generated. The $J/\psi$ decay into $\pi^0 p\bar{p}$ is generated with the model MassH2 [117], and the subsequent decay $\pi^0 \to \gamma\gamma$ with a generic PHase-SPace model (PHSP). The MassH2 model is constructed for the generation of three-body decays according to a Dalitz plot. The Dalitz plot is based upon the experimentally measured Dalitz spectrum.

Figure 3.9 presents the Dalitz spectra of the data (3.9a) versus the model (3.9b) after the $\pi^0$ selections. As expected, the two spectra are in good agreement with each other. Nonetheless, since the proton and antiproton both carry spin, the three-body decay $J/\psi \to \pi^0 p\bar{p}$ has more degrees of freedom than the two represented by a Dalitz spectrum. Hence, the $\pi^0$ data and model are compared for other parameters as well. The comparison is presented in figures 3.10–3.12, where the model is scaled to match the number of data events. It can be concluded that the MassH2 model gives a proper description of the $J/\psi \to \pi^0 p\bar{p}, \pi^0 \to \gamma\gamma$ data.

With use of the data and the model, the branching fraction $\mathcal{B}(J/\psi \to \pi^0 p\bar{p})$ has been calculated using:

$$\mathcal{B}(J/\psi \to \pi^0 p\bar{p}) = \frac{N_{\pi^0}}{\epsilon \cdot N_{J/\psi} \cdot \mathcal{B}(\pi^0 \to \gamma\gamma)} = (1.23 \pm 0.05) \times 10^{-3}. \quad (3.3)$$

Here, $N_{\pi^0} = 601,954$ represent the total number of $\pi^0$ selected data events, $\epsilon = 3,766,348/10^7$ the efficiency obtained with the model and $\mathcal{B}(\pi^0 \to \gamma\gamma) = (98.823 \pm 0.034)\%$ [15]. The error is estimated based on the uncertainty estimation presented in chapter 4, with the addition of a 1% error for the detection of an extra photon [113], and another 1% for the pion construction [58]. The extracted branching fraction coincides with the PDG value of $(1.19 \pm 0.08) \times 10^{-3}$ [15].

To determine the $\pi^0$ background contribution in $J/\psi \to \gamma p\bar{p}$, the full set of generated $J/\psi \to \pi^0 p\bar{p}, \pi^0 \to \gamma\gamma$ events have been exposed to exactly the same set of $\gamma p\bar{p}$ selection criteria as the data. The generated events that survive the selection criteria are weighted and subtracted in a similar fashion as for $J/\psi \to p\bar{p}$. The weight $w$ is again chosen such that the number of selected generated events, $N_{gen,sel}$, equals the expected number of selected events in the datasets, $N_{dat,exp}$, after weighing:

$$N_{dat,exp} = w \cdot N_{gen,sel}. \quad (3.4)$$

The number of expected data events can be found by correcting the total number of $J/\psi \to \pi^0 p\bar{p}, \pi^0 \to \gamma\gamma$ events in the data, $N_{\pi^0,tot} = N_{J/\psi} \cdot \mathcal{B}(J/\psi \to \pi^0 p\bar{p}) \cdot \mathcal{B}(\pi^0 \to \gamma\gamma)$, for the efficiency $\epsilon = N_{gen,sel}/N_{gen,total}$.

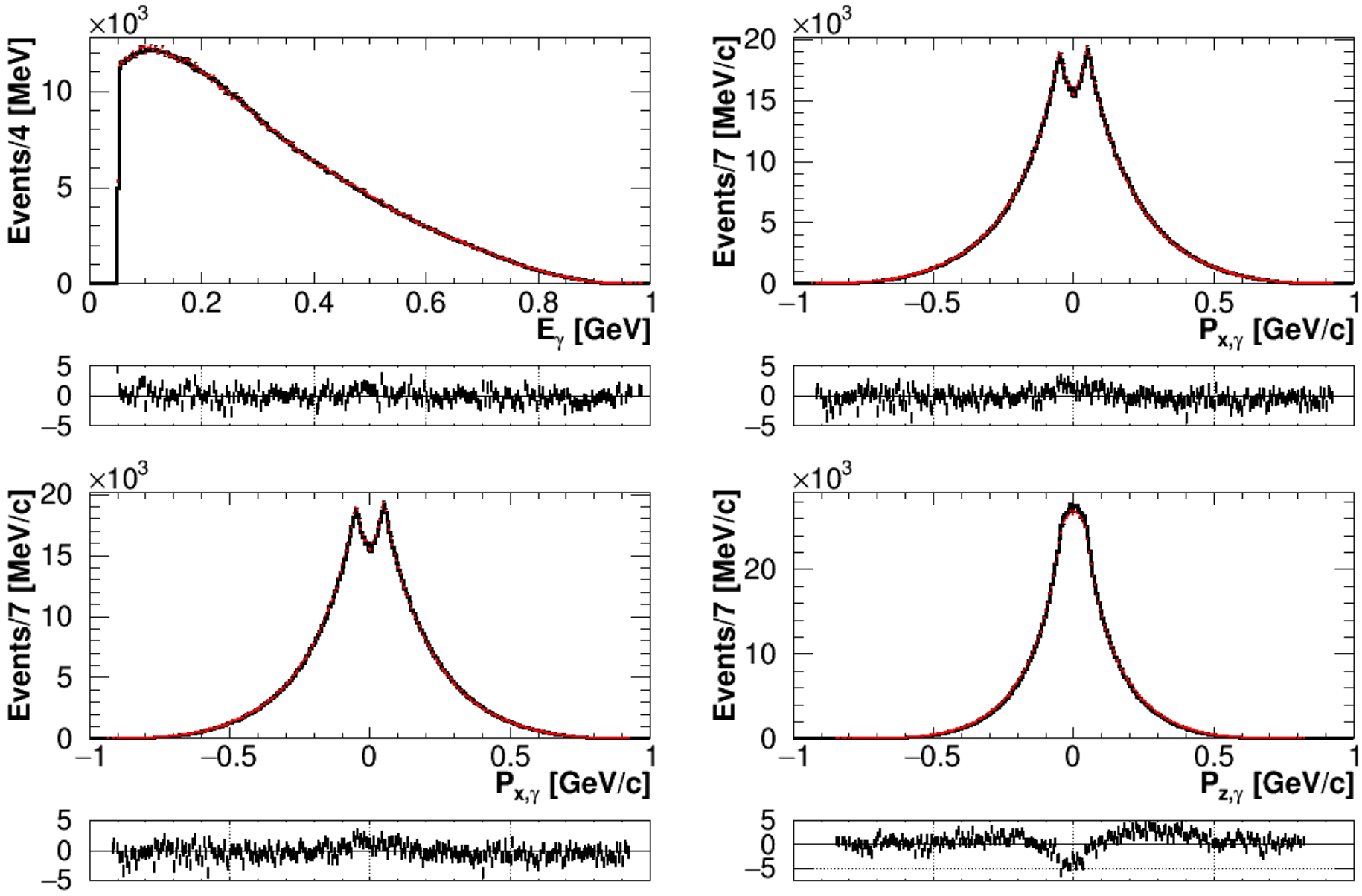

**Figure 3.10:** Data (red) versus MassH2 MC model (black) for the final-state photons of $J/\psi \to \pi^0 p\bar{p}, \pi^0 \to \gamma\gamma$. In the natural reading order, the different panels describe the four components of the photon four-momenta: $E_\gamma$, $P_{x,\gamma}$, $P_{y,\gamma}$ and $P_{z,\gamma}$, respectively. The pull distributions are filled via $(x_{dat} - x_{MC})/\sqrt{\sigma_{dat}^2 + \sigma_{MC}^2}$, where $x_{dat}$ and $x_{MC}$ represent the bin contents of the data and MC histogram, respectively, and $\sigma_{dat}$ and $\sigma_{MC}$ the corresponding bin errors.

Therefore, the weight is defined as

$$w = \frac{\epsilon \cdot N_{\pi^0,tot}}{N_{gen,sel}} = \frac{N_{J/\psi} \cdot \mathcal{B}(J/\psi \to \pi^0 p\bar{p}) \cdot \mathcal{B}(\pi^0 \to \gamma\gamma)}{N_{gen,total}}. \tag{3.5}$$

Here, $\mathcal{B}(J/\psi \to \pi^0 p\bar{p})$ is described by the value found in equation 3.3 and $\mathcal{B}(\pi^0 \to \gamma\gamma)$ as stated by the PDG [15].

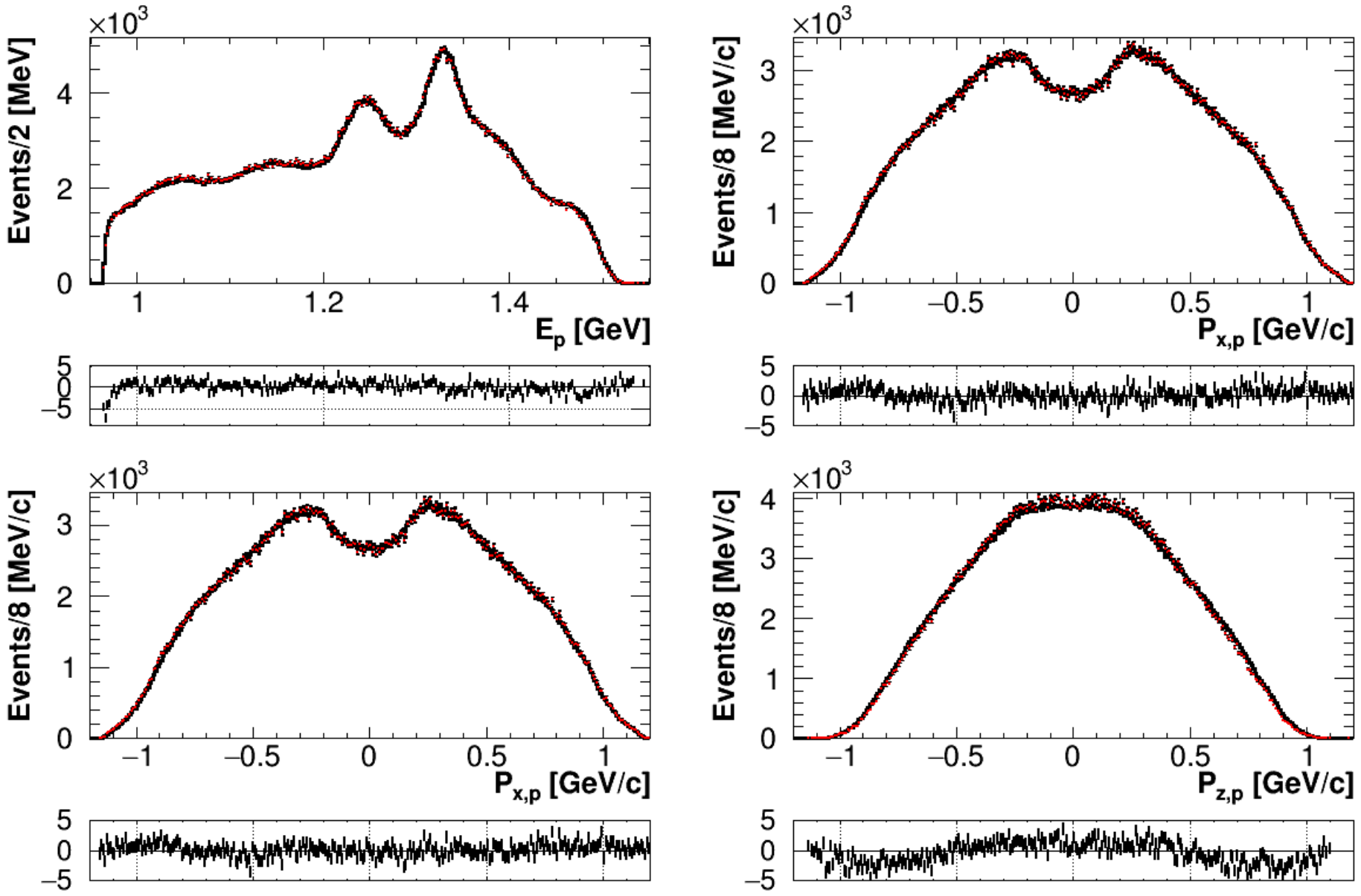

**Figure 3.11:** Same as figure 3.10, except for the final-state protons.

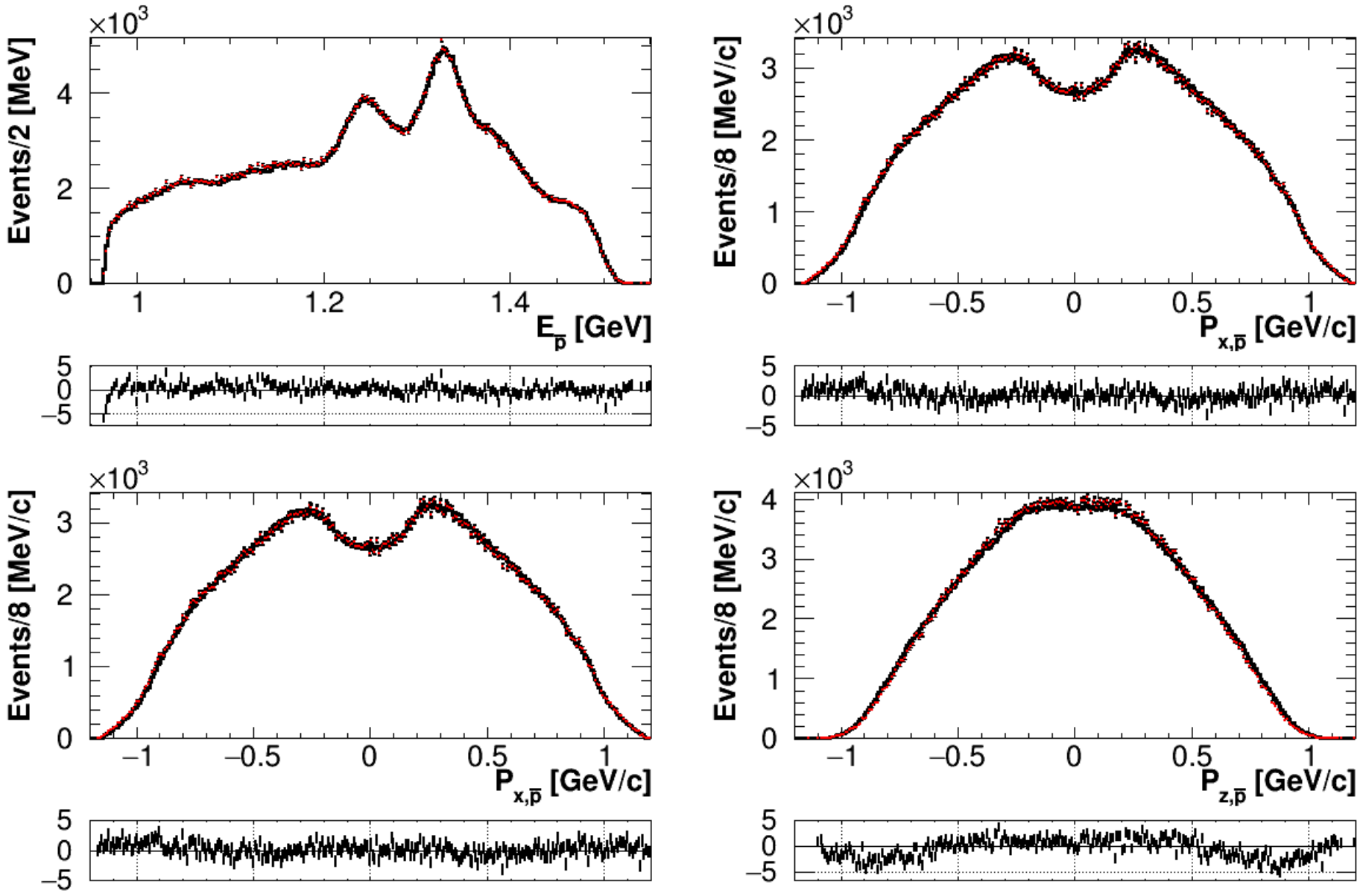

**Figure 3.12:** Same as figure 3.10, except for the final-state antiprotons.

# 4. Results on the branching fraction $J/\psi \to \gamma p\bar{p}$

With the rules of QCD, numerous different exotic and non-exotic hadrons can be imagined. To be able to correctly identify possible exotic matter, a full understanding of the spectrum of conventional hadrons is essential. Therefore, it is crucial to study the production and decay of hadrons in different processes, and cover as many decay modes as possible. The experimentally obtained characterizations of the decays modes provide important tests for the different models that try to describe the confinement regime of QCD. The radiative channel $J/\psi \to \gamma p\bar{p}$ is, for instance, used to study the production and decay of hadrons with baryon number[1] 0, such as $\eta_c$, and their decay into $p\bar{p}$. To get the production rates, one has to normalize with respect to the total production rate. Thus, a proper experimental value for the branching fraction $J/\psi \to \gamma p\bar{p}$ is crucial.

The total branching fraction $J/\psi \to \gamma p\bar{p}$ has only been determined once before in 1984 [15]. The value of $\mathcal{B}(J/\psi \to \gamma p\bar{p}) = (3.8 \pm 1.0) \times 10^{-4}$ is obtained from a data sample of $1.32 \times 10^6$ $J/\psi$ events collected with the MARKII detector [90]. In this determination, the error contains similar contributions related to systematics and statistics. The total dataset that BESIII has collected in 2009, 2012, 2017 and 2018 consists of $10^{10}$ $J/\psi$ events [105]. Therefore, with the available data, the accuracy and precision on the branching fraction can be improved substantially.

In figure 4.1, a Dalitz spectrum of the $J/\psi \to \gamma p\bar{p}$ channel is shown together with its projections. The spectrum is extracted from the full BESIII data sample after applying all the reconstruction, selection and background subtraction steps that are described in chapter 3. Note that the two projections of the spectrum show similar characteristics, reflecting the diagonal bands in the spectrum. In a Dalitz plot, intermediate resonances show up as enhancements in a horizontal, vertical or diagonal band, depending on the decay products of such a resonance and possible interference effects. An intermediate resonance of the type $J/\psi \to \gamma X$, $X \to p\bar{p}$ will show up as a diagonal band, whereas a resonance $J/\psi \to pY$, $Y \to \gamma\bar{p}$ is represented by a horizontal band. Similarly, a resonance $J/\psi \to \bar{p}Z$, $Z \to \gamma p$ would produce a vertical band. Thus, for instance, an intermediate $N^*$ resonance would show up as either an horizontal or vertical band. However, the spectrum only reveals diagonal bands without clear signatures of horizontal or vertical bands. Note that $\Delta$ resonances are not expected due to isospin conservation. The diagonal band in the bottom-left corner corresponds to the $p\bar{p}$ decay of the $\eta_c$ meson, and the narrow, yellow band in the top right corner displays the presence of the $X(p\bar{p})$ resonance. Both these resonances were discussed in section 1.4. Furthermore, a broad diagonal enhancement is visible next to the $X(p\bar{p})$ resonance. Currently, the Particle Data Group (PDG) lists no known decay of a resonance, corresponding to these characteristics, decaying into a $p\bar{p}$ pair. Nevertheless, there are multiple hadrons listed which have masses and quantum numbers that allow for such a decay. Hence, the enhancement could be caused by a complex mixture of various resonances. Chapter 6 presents a study of the possible quantum numbers that contribute to this enhancement.

[1] The baryon number is a conserved quantum number defined as $\frac{1}{3}(n_q - n_{\bar{q}})$, with $n_q$ ($n_{\bar{q}}$) the number of quarks (antiquarks).

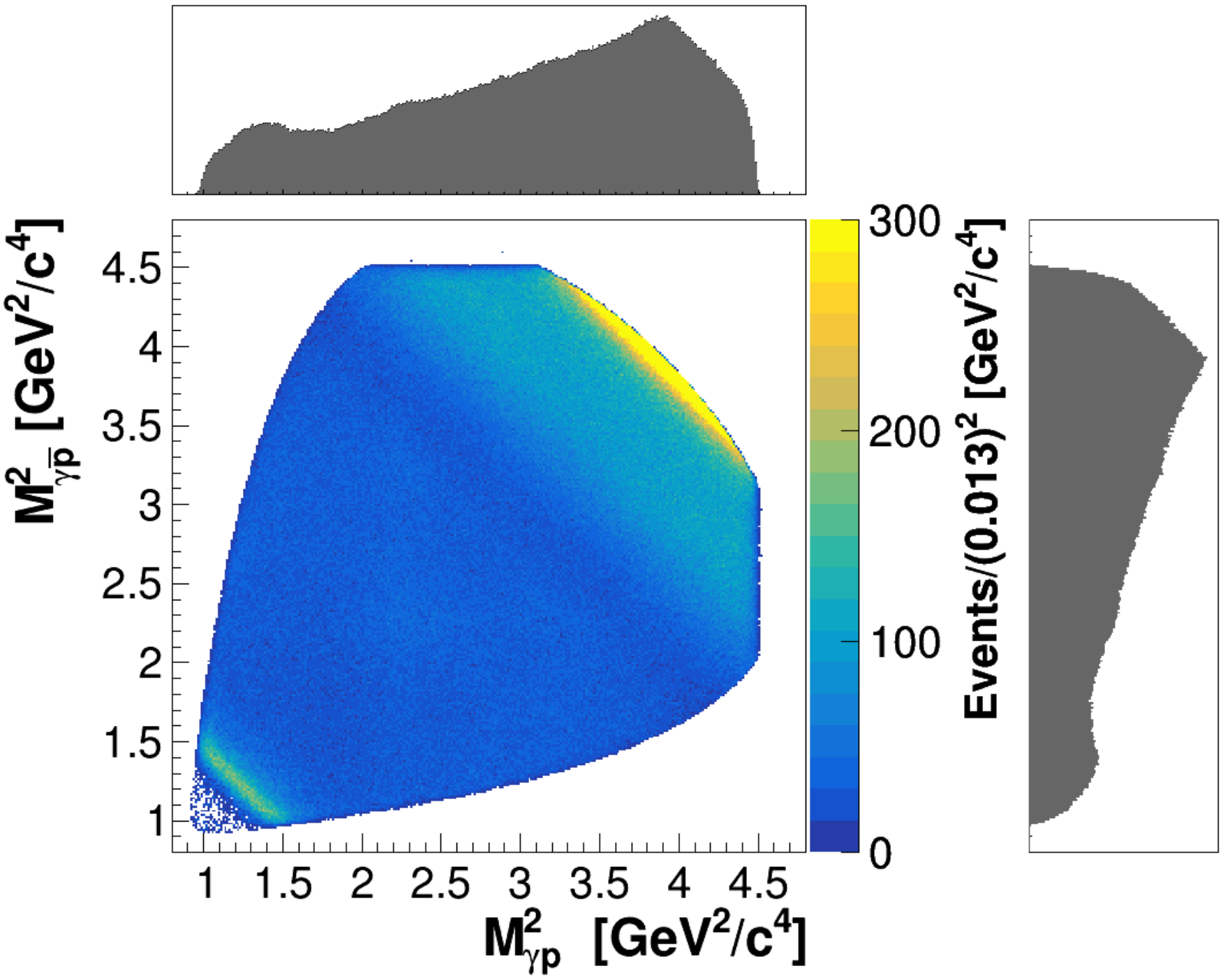

**Figure 4.1:** The resulting Dalitz spectrum extracted from BESIII data taken in 2009, 2012, 2017 and 2018, after reconstruction, selection and background subtraction. The top and right panels show the projections towards $M^2_{\gamma p}$ and $M^2_{\gamma\bar{p}}$.

## 4.1 Determination of the branching fraction

To extract the branching fraction from the data, the obtained yield $N_{\gamma p\bar{p}}$ needs to be corrected for the efficiency $\epsilon$, and divided by the total number of $J/\psi$ events, $N_{J/\psi}$, via

$$\mathcal{B}(J/\psi \to \gamma p\bar{p}) = \frac{N_{\gamma p\bar{p}}}{\epsilon N_{J/\psi}}. \tag{4.1}$$

The efficiency varies as function of kinematic parameters, of which the $M_{\gamma p}$ and $M_{\gamma\bar{p}}$ are the most important. Efficiency variations are primarily induced by the detector acceptance and further selection requirements. Therefore, the efficiency is determined separately for every bin in the Dalitz spectrum. Each bin contains $n_{i,\gamma p\bar{p}}$ events and is corrected by the efficiency $\epsilon_i$ corresponding to that bin, resulting in

$$\mathcal{B}(J/\psi \to \gamma p\bar{p}) = \frac{\sum_i^{N_{bins}} \left(n_{i,\gamma p\bar{p}}/\epsilon_i\right)}{N_{J/\psi}}, \tag{4.2}$$

with $N_{bins}$ the total number of bins. A larger number of bins allows for a more accurate incorporation of the variations in kinematic efficiency. On the other hand, bin widths smaller than the resolution are meaningless. The resolutions in $M^2_{\gamma p}$ and $M^2_{\gamma\bar{p}}$ are studied with a phase-space generated MC sample. In figure 4.2, the resulting resolutions in $M^2_{\gamma p}$ and $M^2_{\gamma\bar{p}}$ are presented. The obtained resolutions are similar for the proton and antiproton. The horizontal, gray, dashed line corresponds to the chosen number of bins $N_{bins} = 300 \times 300$. The other horizontal lines correspond to the number of bins that will be used in the study of the systematic uncertainty, being $N_{bins} = 200 \times 200$ (upper line) and $N_{bins} = 400 \times 400$ (lower line). Note that the resolution gets better towards lower and higher values of $M^2_{\gamma p}$ and $M^2_{\gamma\bar{p}}$, due to the constraints of the kinematic fit.

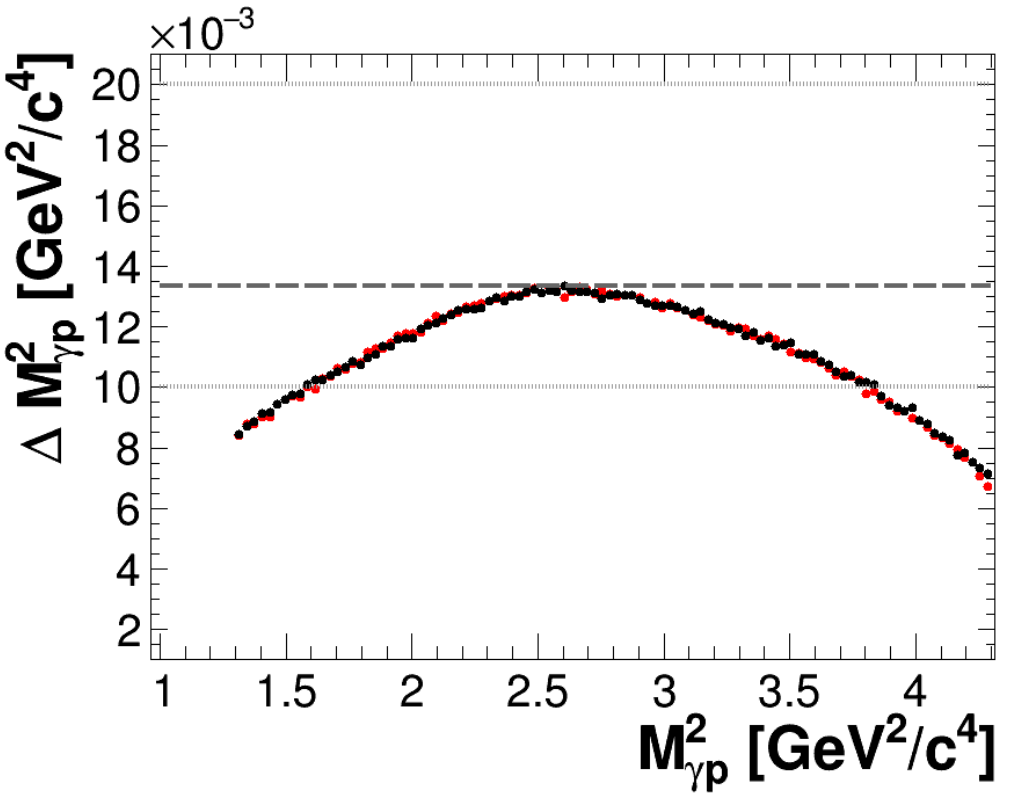

**Figure 4.2:** The detector resolution in $M^2_{\gamma p}$ (black) and $M^2_{\gamma\bar{p}}$ (red). The horizontal, gray, dashed line displays the nominal chosen bin widths, corresponding to $N_{bins} = 300 \times 300$, and the two horizontal, gray, dotted lines show the bin widths used in systematic uncertainty determinations, and correspond to $N_{bins} = 200 \times 200$ (upper line) and $N_{bins} = 400 \times 400$ (lower line).

With the number of bins chosen, the efficiency distribution can be determined with the use of a phase space generated MC sample of $10^7$ signal events. The efficiency in each bin $i$ is calculated via

$$\epsilon_i = \frac{n_{i,sel}}{n_{i,gen}}, \tag{4.3}$$

with $n_{i,sel}$ the number of MC events left in the bin after all reconstruction and selection criteria, and $n_{i,gen}$ the number of total generated events for the bin. The obtained statistical error on the efficiency is of the order $10^{-4}$, and more than a factor two smaller than the statistical error on the data. The final efficiency distribution is shown in figure 4.3a, where the boundary of the total available phase-space is indicated by the red line. The efficiency-corrected Dalitz plot of the data is displayed in figure 4.3b. As discussed in chapter 2, (anti)protons with momenta smaller than about 200 MeV/$c$ move too slowly to be registered by the TOF and EMC subdetectors. This explains the clearly visible efficiency gaps for the two high invariant mass edges, corresponding to small (anti)proton momenta. Since there are no data events registered in these regions, the two gaps are still present after efficiency correction. Another, smaller, efficiency gap is visible for low photon energies in the bottom left corner of the distribution in figure 4.3a. However, in this region, a few data entries are registered. For those few bins that contain a small number of events, but have a bin-efficiency of zero, the finite efficiency of a neighboring bin is taken instead. Hence, after efficiency correction, the few data entries are blown up to a large amount, visible by the yellow region in figure 4.3b. The remaining data entries in this region are probably the result of an imperfect background subtraction related to the decay $J/\psi \rightarrow p\bar{p}$ (see section 3.3), and are thus not relevant for the decay under study.

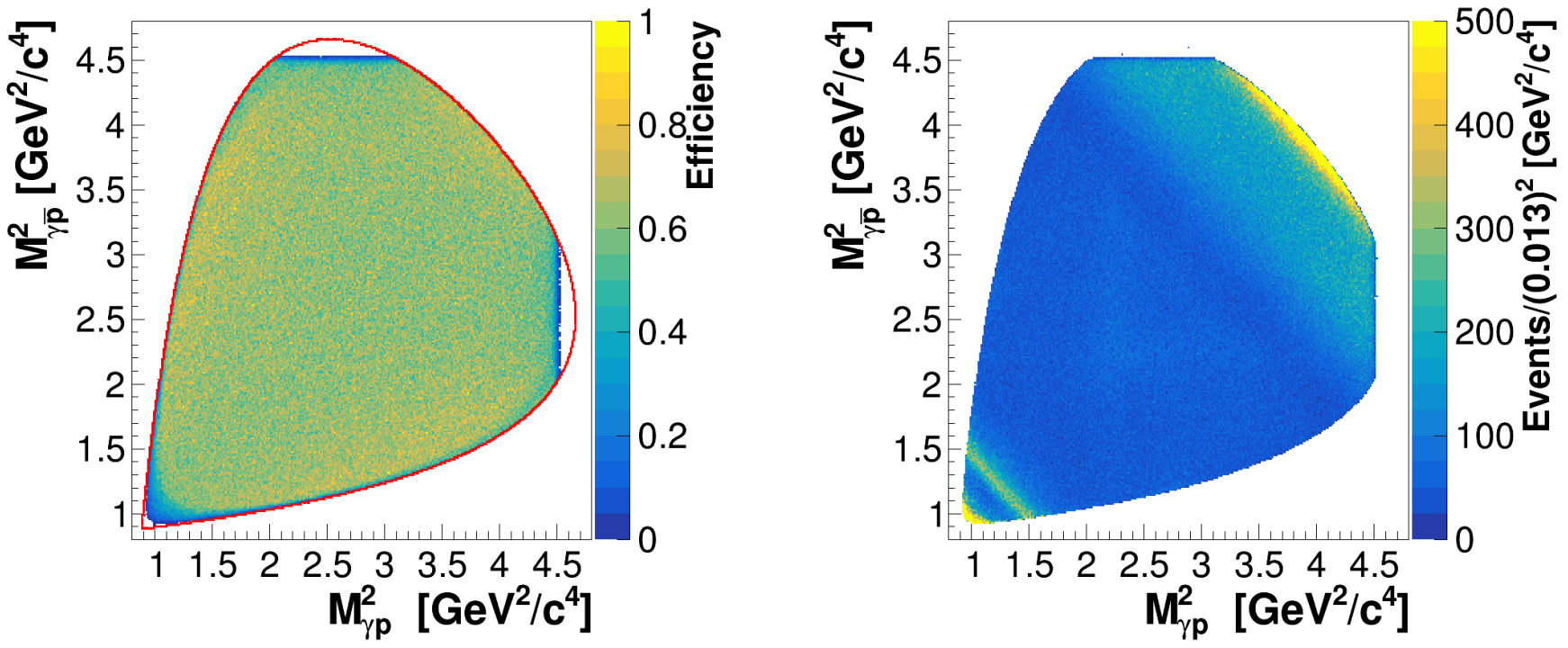

(a) (b)

**Figure 4.3:** The obtained efficiency distribution (a) and the efficiency corrected Dalitz spectrum of the data (b). For those few bins that contain a small number of events, but have a bin-efficiency of zero, the finite efficiency of a neighboring bin is taken instead. The red line in (a) corresponds to the boundary of the total available phase space. The top and right gaps correspond to (anti)protons with momenta smaller than about 200 MeV/$c$, as they move too slowly to be registered by the TOF and EMC subdetectors. Similarly, the smaller, bottom-left gap corresponds to small photon energies.

For a proper estimation of the yields in these three regions, the efficiency-corrected data will be extrapolated with use of MC based fits. The MC sample generated for this estimation is based on all the known intermediate resonances, and their relevant branching fractions, as listed by the PDG [121]. The used fractions and models are listed in table 4.1. The photon energy and (anti)proton momentum distributions from this model are shown in figure 4.4. The lowest values of all the three distributions are fitted with simple polynomial functions, displayed by the red lines.

| Fraction | Decay | Model |
|---|---|---|
| 0.2026 | $\gamma X(1835)$ | PHSP |
| 0.0644 | $\gamma\eta_c$ | JPE |
| 0.7330 | $\gamma p\bar{p}$ | PHSP |
| 1.0 | $X(1835) \to p\bar{p}$ | PHSP |
| 1.0 | $\eta_c \to p\bar{p}$ | PHSP |

**Table 4.1:** The fractions and models applied in the BesEvtGen generated MC sample used for the yield extrapolations.

With all the internal fit-parameters fixed, and an additional free scale parameter, the three extracted functions are scaled to match the efficiency-corrected distributions. The distributions of the efficiency-corrected data, together with the scaled functions, are presented in figure 4.5. Here again, it is visible that the efficiency gap due to low (anti)proton momenta remains empty after efficiency corrections, whereas in the region of low photon-energies the few entries are magnified. In the figure, the vertical gray, dashed lines represent the cut-off values, chosen to be $P_{p/\bar{p}} < 0.25$ GeV/$c$ and $E_\gamma < 0.08$ GeV. For values below the cut-off values, the scaled functions are used to estimate the number of entries in the three regions.

The three yields obtained with the scaled functions, together with the main region inside the cut-offs are summed to get the total efficiency-corrected yield of $\tilde{N}_{\gamma p\bar{p}} = 5,346,190$. A summary of the different yields is listed in table 4.2. As one would expect, the three edge effects are only responsible for a small contribution to the total yield. There is a small difference in the yields related to the small proton versus antiproton momentum. This difference will be used as an estimate for a systematic uncertainty, since there is no evident physical explanation for a difference between these two yields.

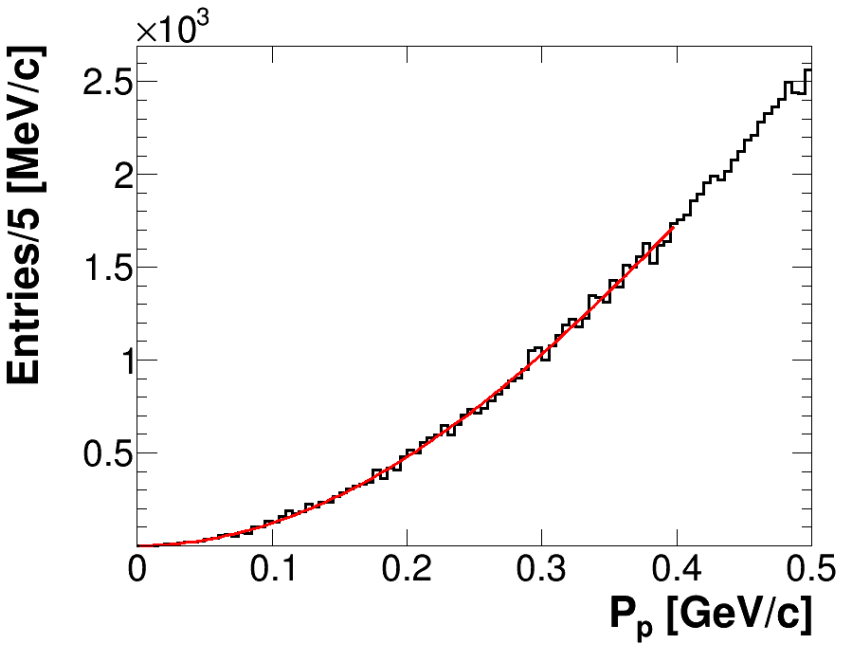

(a)

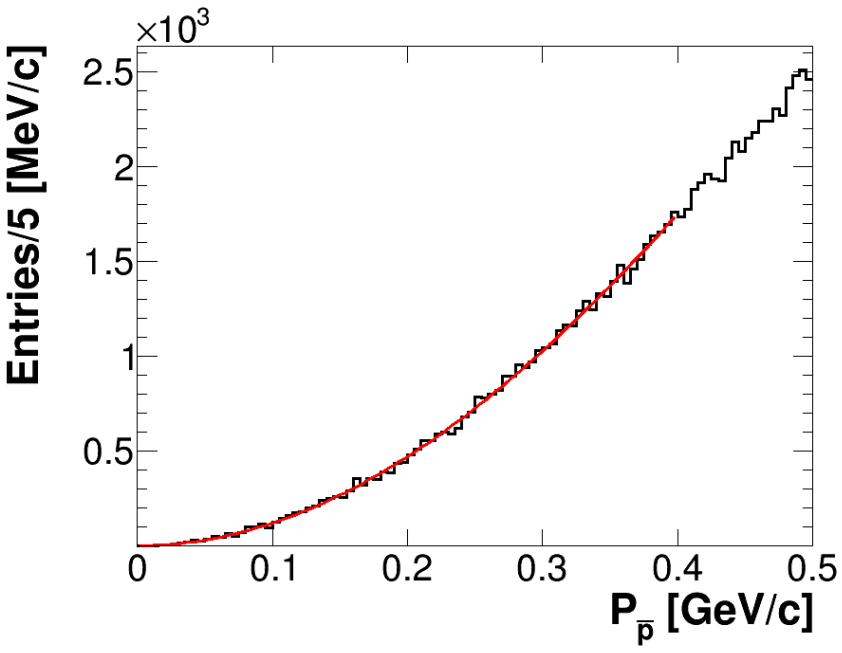

(b)

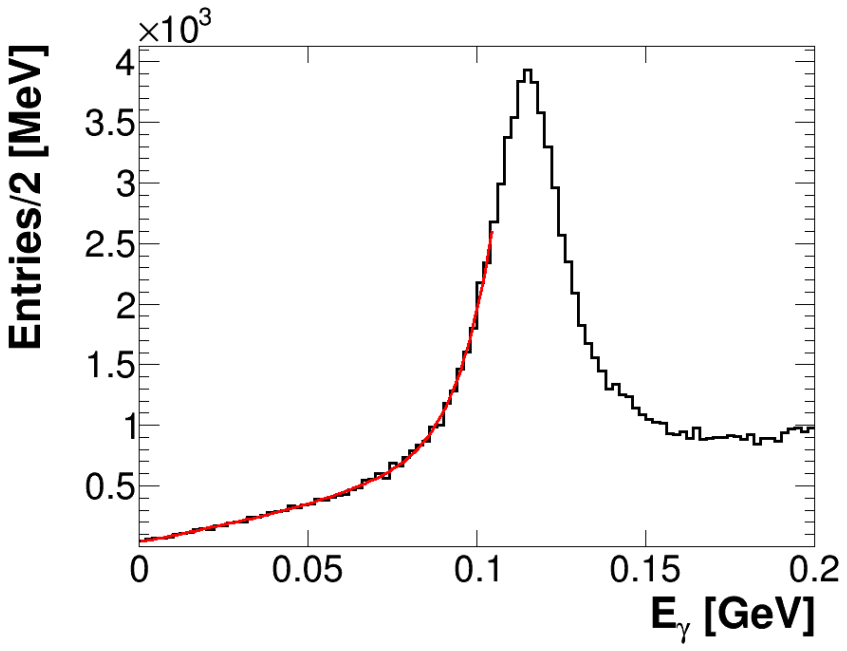

(c)

**Figure 4.4:** The distributions of the lower (anti)proton momentum $(P_{\bar{p}})P_p$, and photon energy $E_\gamma$ for the MC model defined by table 4.1. The red lines represent the fits used for the yield extrapolations.

| | Main | Low $E_\gamma$ | Low $P_p$ | Low $P_{\bar{p}}$ | Total |
|---|---|---|---|---|---|
| $[\tilde{N}]$ | 4,971,760 | 35,045 | 166,526 | 172,859 | 5,346,190 |
| [%] | 93.0 | 0.7 | 3.1 | 3.2 | 100 |

**Table 4.2:** Summary of the obtained yields after efficiency corrections for the cut-off values $P_{p/\bar{p}} < 0.25$ GeV$/c$ and $E_\gamma < 0.08$ GeV.

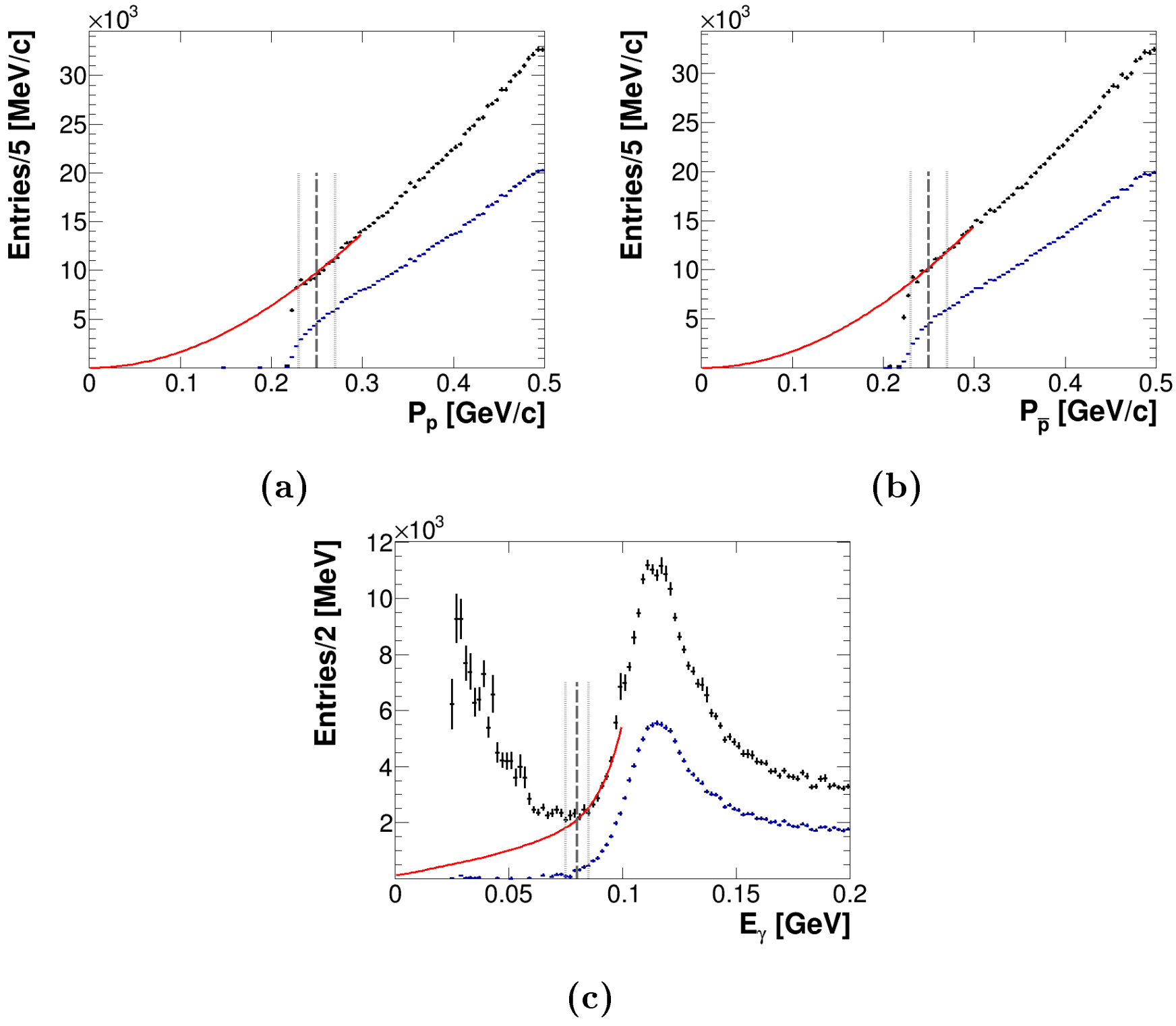

**Figure 4.5:** The distributions of the lower (anti)proton momentum $(P_{\bar{p}})P_p$, and photon energy $E_\gamma$ for the data before (blue) and after (black) efficiency correction. The red lines are the scaled functions extracted from the distributions in figure 4.4. The vertical gray, dashed lines represent the cut-off values: $P_{p/\bar{p}} < 0.25$ GeV/$c$ and $E_\gamma < 0.08$ GeV. The efficiency-corrected data above these cut-off values are used to provide a model-independent measure of the yield. For lower values, the integrals of the functions are used. The vertical light gray, dotted lines represent the cut-off values used for part of the systematic uncertainty estimation.

With the total efficiency-corrected yield $\tilde{N}_{\gamma p\bar{p}}$, and $N_{J/\psi} = 10086.6 \times 10^6$ [105], the resulting branching fraction reads as

$$\mathcal{B}(J/\psi \to \gamma p\bar{p}) = \frac{\tilde{N}_{\gamma p\bar{p}}}{N_{J/\psi}} = 5.30 \times 10^{-4}, \quad (4.4)$$

where the statistical errors of both the data and the efficiency determination are of the order $10^{-7}$.

## 4.2 Systematic error analysis

The error on total number of $J/\psi$ events in the dataset, and the uncertainties related to the different steps in the extraction of the final yield $\tilde{N}_{\gamma p\bar{p}}$, introduce systematic uncertainties. In the following, the determination of the different types of systematic uncertainties will be discussed. Some of the uncertainties arise due to the experimental characteristics of BESIII. These have been evaluated extensively by the collaboration and will be summarized briefly. Additional systematic uncertainties arise due to the specific selection criteria and model choices applied in this analysis. All resulting uncertainty values on the branching fraction are presented in table 4.3 at the end of this section.

### 4.2.1 Photon and charged-track detection

The general procedure in finding the photon and charged-track detection efficiencies starts with choosing an appropriate control sample. The relevant efficiencies are then determined for both the data of the control sample, and data from MC simulations. The differences in the efficiencies between the recorded data and the MC events are then considered as an estimate of the uncertainties in the efficiencies.

The photon detection efficiency of BESIII was studied with a control sample of the initial state radiation process $e^+e^- \to \gamma\mu^+\mu^-$ at center-of-mass energies corresponding to the $J/\psi$ and $\psi(3770)$ resonances [113]. The photon detection efficiency is determined to be 1%. Since there is just one photon to be detected in this analysis, that results in a total uncertainty of 1%.

The BESIII uncertainties of tracking and PID efficiencies for a proton/antiproton were investigated using the almost background-free control sample $J/\psi \to p\bar{p}\pi^+\pi^-$ [112,122]. For both the tracking and PID efficiency, an uncertainty of 1% per track was estimated. Since a proton and an antiproton are required to be detected, an uncertainty of 2% is used for both tracking and PID.

### 4.2.2 Selection criteria

In the selection of the data, three additional channel-specific selection criteria were imposed: a maximum value for $\chi^2_{4C}$, $M^2_{miss}$ and $\theta_{p\bar{p}}$, see section 3.2.2. The values used were found by optimizing the significance $\frac{S}{\sqrt{S+B}}$. To estimate the systematic error, the three step-sizes used in the optimization process are applied to alter the maximum limits of the cuts. The implemented step-sizes are $\chi^2_{4C} = 0.5$, $M^2_{miss} = 0.001$ GeV$^2/c^4$ and $\theta_{p\bar{p}} = 0.1°$. For each of the three cuts, the full analysis is repeated twice: once with the maximum value altered to the nominal value plus the step-size, and a second time with the maximum value set to the nominal value minus the step-size. For all three criteria, the largest deviation from the nominal value is used as a systematic error.

### 4.2.3 Background subtraction

To estimate the errors from the background subtraction of the $J/\psi \to p\bar{p}$ sample, the procedure is repeated twice whereby the weight factor is altered by plus or minus the uncertainty of the $J/\psi \to p\bar{p}$ branching fraction. The largest deviation from the nominal value is included as an uncertainty. Additionally, the model dependent effect is estimated by repeating the procedure with a MC sample generated with a generic phase-space model of $J/\psi \to p\bar{p}$, instead of the J2BB1 model. Furthermore, figure 4.5c shows that the background subtraction related to this decay is not perfect. To estimate the corresponding error, the efficiency-corrected data events on the left-hand side of the photon-energy cut-off are fitted. The fit is extrapolated to the range included in the determination of the branching fraction. The integrated number of events extracted for this range are included as a systematic uncertainty.

For the background subtraction of the data-driven $\pi^0$ model, the model dependence is estimated by repeating the procedure with bins that are a factor 2 narrower, so 0.006 GeV$^2/c^4$ in both directions, instead of 0.012 GeV$^2/c^4$ bins. For the nominal value, the data-driven $\pi^0$ model is based on the $J/\psi$ data collected in 2009 and 2012. To estimate the statistical fluctuations, the data-driven $\pi^0$ model was constructed from a fraction of the 2018 $J/\psi$ data, with roughly the same number of events as the data sample used for the nominal model. Furthermore, the weight factor is changed by plus or minus 1%, to include the uncertainties related to the detection of the extra photon from $J/\psi \to \pi^0 p\bar{p}, \pi^0 \to \gamma\gamma$.

### 4.2.4 Efficiency distribution

To account for kinematic variations, the efficiency is determined for each bin in the Dalitz spectrum separately. For the nominal value, the number of bins in both $M^2_{\gamma p}$ and $M^2_{\gamma\bar{p}}$ direction is $N_{bins} = 300$. To estimate the uncertainty due to choice of the number of bins, the analysis was repeated with $N_{bins} = 200$ and $N_{bins} = 400$. The largest shift compared to the nominal result is added as a systematic uncertainty. Figure 4.2 demonstrates how the different number of bins are related to the detector resolution in terms of $M^2_{\gamma p}$ and $M^2_{\gamma\bar{p}}$.

In this analysis, it is assumed that a Dalitz spectrum provides a good model for the efficiency distribution. However, a Dalitz plot does not include all degrees of freedom of the decay $J/\psi \to \gamma p\bar{p}$. The uncertainty related to this model dependency is estimated with information that will be explained in the following chapters. To account for the model dependency, the results of four different $p\bar{p}$ invariant-mass bins of the analysis in chapter 6 are used to extract the efficiency with the PAWIAN[2] method explained in section 7.2. The four mass bins are chosen to be out of range of the edge effects, and are $2400 < M_{p\bar{p}} < 2405$ MeV/$c^2$, $2500 < M_{p\bar{p}} < 2505$ MeV/$c^2$, $2600 < M_{p\bar{p}} < 2605$ MeV/$c^2$, and $2700 < M_{p\bar{p}} < 2705$ MeV/$c^2$. The resulting PAWIAN efficiency for each mass bin is compared to the average efficiency that one would obtain with the Dalitz model for the given $p\bar{p}$ invariant-mass range. The largest deviation is added as a systematic error.

### 4.2.5 Phase-space edges

In the analysis, three efficiency gaps showed up, related to low photon-energies and (anti)proton momenta. For the nominal analysis, a MC sample based on the PDG information is used to extrapolate the yield beyond the three edges. To estimate the systematic uncertainties due to the chosen model, the analysis was repeated with a phase-space generated MC sample, and the deviation is included as a systematic error. Additionally, the three cut-off values were slightly varied. The nominal values, and the values used for the systematic uncertainty estimations, are shown in figure 4.5. For both, the photon-energy and the (anti)proton-momentum cut-off, the largest deviation is taken as a systematic error. Furthermore, the yields related to the small proton versus antiproton momenta differ slightly. Since there is no evident physical explanation for a difference between these two yields, the difference is taken as an estimate for the systematic uncertainty.

[2] Utilized software package; see section 5.5.

### 4.2.6 Total number of $J/\psi$ and external branching fractions

A study in 2021, described in reference [105], determined the total collected $J/\psi$ events in 2009, 2012, 2017 and 2018 to be $(10086.6 \pm 43.7) \times 10^6$. The given uncertainty of 0.43% will be taken as a source of systematic uncertainty for the branching fraction.

All the systematic errors discussed above are summarized in table 4.3.

| **Source** | **Error [$\times 10^{-3}$]** |
|---|---|
| (anti)proton PID | 20 |
| (anti)proton track | 20 |
| Photon detection | 10 |
| Cuts: $\chi^2_{4C}$ | 1.5 |
| Cuts: $M^2_{miss}$ | 0.5 |
| Cuts: $\theta_{p\bar{p}}$ | 0.3 |
| $p\bar{p}$ background: ± PDG sys. err | 0.3 |
| $p\bar{p}$ background: PHSP model | 0.8 |
| $p\bar{p}$ background: remaining events | 3.0 |
| $\pi^0$ background: ±1% | 0.7 |
| $\pi^0$ background: binning | 8.3 |
| $\pi^0$ background: stat. fluctuations | 8.4 |
| Efficiency: binning | 7.8 |
| Efficiency: model | 6.1 |
| Edges: PHSP fit | 1.9 |
| Edges: cut-off $P_{p/\bar{p}}$ | 0.2 |
| Edges: cut-off $E_\gamma$ | 0.1 |
| Edges: proton vs. antiproton | 1.1 |
| $N_{J/\psi}$ | 4.3 |
| **Total relative error** | **33.9** |
| **No. of efficiency corrected counts** | **5,346,190** |

**Table 4.3:** Summary of all the systematic errors, presented as relative errors with respect to the nominal value. The total error is obtained by the quadratic sum of the individual systematic uncertainties, assuming that the uncertainties are uncorrelated.

## 4.3 Result and discussion

All contributions to the systematic error are listed in table 4.3. Aside from the systematic error, a statistical error is present. However, the statistical error is several orders of magnitude smaller than the systematic error, and can thus be neglected in the final result. Hence, the extracted branching fraction and its error reads as:

$$\mathcal{B}(J/\psi \to \gamma p\bar{p}) = (5.30 \pm 0.18) \times 10^{-4} \tag{4.5}$$

In figure 4.6, the final result of this analysis is compared to the branching fraction extracted in the previous analysis. There is no comparison with results from theory, due to a lack of theoretical predictions for this observable. As expected, there is a significant improvement on the accuracy of the experimental branching fraction. The previous result has a relative error of about 26%, whereas this analysis results in a relative error of roughly 3%, about a factor 9 improvement. In the previous result, the error contained similar contributions related to statistics and systematics, The current result does not only benefit from a dataset with nearly four orders of magnitude more $J/\psi$ events, but also from technical improvements in detector construction. Over the years, the knowledge and technical possibilities in constructing a particle detector, and related trigger-software, have evolved substantially. This results, for instance, in significantly better resolutions in the more sophisticated BESIII detector compared to the MARKII detector. The most relevant improvements are a better signal-to-background ratio due to a photon-energy resolution that is improved by a factor three, and an acceptance of 93% of $4\pi$ for BESIII versus 65% for MARKII [90, 123].

Despite all the improvements, the largest contributions in the error of this result are related to the characteristics of the BESIII detector, as can be seen in table 4.3. Due to little model dependency in the discussed analysis, all other systematic errors are relatively small. Thus, for an even more accurate determination, further improvements in the understanding of the detector response are necessary. For instance, the determination of the standard BESIII errors related to the tracking and PID could be revisited. It could be that the given values overestimate the uncertainty, and that the uncertainties can be reduced by another approach.

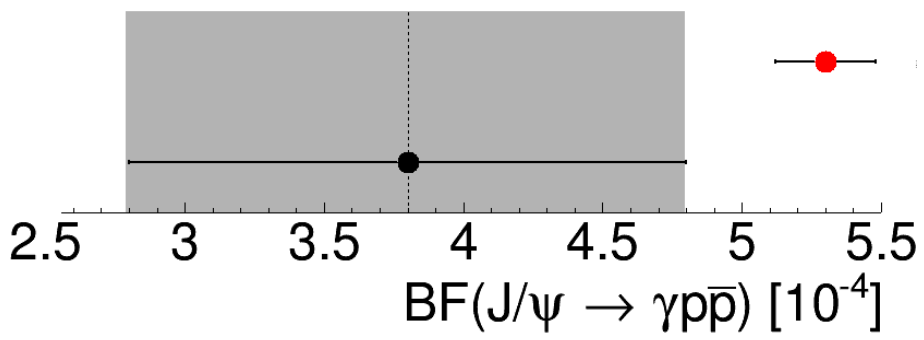

| **Exp.** | **Year** | **Ref.** |
|---|---|---|
| BESIII | 2022 | This work |
| MARKII | 1984 | [90] |

**Figure 4.6:** Final result (red) for the branching fraction, compared to the previous result (black). The gray band represents the PDG value [15], which is solely determined by the previous result.

Other significant contributions are related to the $\pi^0$ background subtraction. In this analysis, the data-driven model is constructed from the Dalitz spectrum of $J/\psi \to \pi^0 p\bar{p}$ data; see section 3.4. For a more accurate description of this background, one could construct a model based on the full decay information, thus including the angular distributions of the data.

Due to the large error in the previous result, the two results on the branching fraction still agree with each other within two standard deviations, even though the new value is almost 1.5 times as high as the previous value. The fact that MARKII has a significantly smaller acceptance and signal-to-background ratio lends more confidence to our data and result, as the high resolution and close to $4\pi$ coverage of BESIII provide a better handle on the systematic effects. A proper, more accurate description of $\mathcal{B}(J/\psi \to \gamma p\bar{p})$ is, for instance, essential for theoretical studies that try to interpret the near-threshold enhancement $X(p\bar{p})$. Additionally, the value of $\mathcal{B}(J/\psi \to \gamma p\bar{p})$ is needed to extract the production rates of meson(-like) hadrons, such as the $\eta_c$ meson. An improved accuracy and precision on the branching fraction thus reduces the uncertainties in these rates.

In the following chapter, the general concepts of a Partial-Wave Analysis (PWA) will be introduced to enable a more thorough study of the dynamics present in $J/\psi \to \gamma p\bar{p}$.

# 5. The Partial-Wave-Analysis toolbox

Partial-wave analyses are widely used in hadron studies in nuclear and high-energy experimental physics. In a Partial-Wave Analysis (PWA), the correlation between the momenta of final state particles is analyzed to determine the spin-parities of the contributing waves, and masses, widths and branching ratios of intermediate resonances. In the BESIII study of $J/\psi \to \gamma p\bar{p}$, the full event topology is obtained by reconstructing the four-momenta of the photon, proton, and antiproton. This allows us to perform a PWA of the complete reaction chain, from the initial $e^+e^-$ state (or $J/\psi$) down to the three final particles of the decay. A full PWA provides the possibility to simultaneously describe all dimensions of the phase space and allows to account for interferences between different components. This is very important in cases whereby one aims to extract properties of resonances that cannot be described by simple Breit-Wigner shapes. This applies to our wish to extract the $\eta_c$ yield of the radiative process $J/\psi \to \gamma\eta_c$, for which we know from previous studies that the mass line shape is heavily distorted due to interference effects. To unambiguously determine the $\eta_c$ yield and interference terms, a PWA is an absolute necessity. Additionally, a PWA can give insights in the proper description of the enhancements in the full $p\bar{p}$-invariant mass range.

The concept of partial-wave analyses originates from quantum scattering theory, where the quantum mechanical state of a particle is interpreted as a plane wave. The scattering of such a plane wave with a given potential, which could for example represent another particle, can be described by a PWA. Each wave is then described by its constituent angular momentum components, the so-called partial waves. The PWA consists of expanding the scattering amplitude into partial-wave amplitudes, as described for instance in reference [124] or [125]. A PWA for the description of decaying resonances consists of a similar procedure. The observed intensity distribution of the phase space, corresponding to the differential cross section of the reaction, can be described by terms of the absolute square of a coherent sum of partial-wave amplitudes. The contributing partial-wave amplitudes are identified by sets of quantum numbers, leading to the spin-parity information of the decay process. Known conservation rules in physics, such as parity and angular-momentum conservation, are implemented as constraints in a PWA.

In section 5.1, the allowed spin-parities for intermediate resonances of $J/\psi \to \gamma p\bar{p}$ are studied. After introducing the framework used to describe the partial-wave amplitudes in section 5.2, the dynamical part of the amplitude will be described in 5.3. This includes the description of the intermediate $\eta_c$ resonance. Finally, the fitting procedure and the used software PArtial Wave Interactive ANalysis (PAWIAN) will be introduced in sections 5.4 and 5.5, respectively.

## 5.1 Fundamental constraints for $J/\psi \to \gamma p\bar{p}$

For the PWA of $J/\psi \to \gamma p\bar{p}$, the amplitude description is based on the isobar model. This empirical model assumes that the full reaction chain can be described as a sequence of two-body decays. The amplitudes are then built by a coherent sum of subsequent two-body decays. The isobar assumption can describe most hadronic reactions in our energy domain reasonably well, although there might be some exceptions [125]. The decay under study has been described by three different combinations of two-body decays:

$$J/\psi \to \gamma X \quad (X \hookrightarrow p\bar{p}) \qquad J/\psi \to Y p \quad (Y \hookrightarrow \gamma\bar{p}) \qquad J/\psi \to Z\bar{p} \quad (Z \hookrightarrow \gamma p)$$

In this analysis, we are interested in the possible intermediate resonances $X$, $\eta_c$ being one of them. The combinations with resonances $Y$ and $Z$ could be achieved with intermediate $N^*$ resonances. However, these combinations are not experimentally observed in $J/\psi \to \gamma p\bar{p}$ data. One might also expect a $\Delta$ baryon as intermediate $Y$ or $Z$ resonance, but these are suppressed due to isospin conservation. Therefore, we will only discuss the scenario $X$. The total decay chain can then be described by the production amplitude of $X$ ($J/\psi \to \gamma X$) and its decay amplitude ($X \to \bar{p}p$).

For a PWA, it is essential to know what quantum numbers are allowed for resonance $X$. The quantum numbers of the initial and final states are

$$\begin{array}{lccccc} & J/\psi & \rightarrow & \gamma & p & \bar{p} \\ J^{P(C)}: & 1^{--} & \rightarrow & 1^{--} & \frac{1}{2}^{+} & \frac{1}{2}^{-} \end{array}$$

Here, $J^{PC}$ represents the quantum numbers as introduced in section 1.3. Note that for the (anti)proton the $C$-parity is omitted, since $C$ is not defined for (anti)baryons. For this purpose the more generalized $G$ parity was introduced, as is done, for instance, in reference [126]. As the decay is an electromagnetic process, both $C$ and $P$ should be conserved. Next to the conservation of angular momentum $J$, the following equations should hold for $a \rightarrow b + c$:

$$P_a = (-1)^L P_b P_c, \tag{5.1}$$

$$C_a = C_b C_c, \tag{5.2}$$

$$|j_b - j_c| \leqslant S_{bc} \leqslant |j_b + j_c|, \tag{5.3}$$

$$|L_{bc} - S_{bc}| \leqslant J_a \leqslant |L_{bc} + S_{bc}|. \tag{5.4}$$

Here, $j$ represents the total angular momentum of particle $b$ or $c$, $S_{bc}$ the combined spin, and $L_{bc}$ the angular momentum between particles $b$ and $c$. Equation 5.2, together with the negative $C$-parity for both $J/\psi$ and the photon, results in a positive $C$-parity for $X$. The symmetry of $X$ decaying into a particle and its antiparticle leads to an additional constraint of $C = (-1)^{L_{p\bar{p}}+S_{p\bar{p}}}$. Therefore, the sum $L_{p\bar{p}} + S_{p\bar{p}}$ should be even, where $S_{p\bar{p}}$ can be either 0 or 1 (eq. 5.3). Finally, equation 5.1 limits the parity to $P_X = (-1)^{L_{p\bar{p}}+1}$. These constraints, together with equation 5.4, allow the combinations

$$J_X^{PC} = (2n)^{-+} \text{ and } J_X^{PC} = (n)^{++},$$

where $n$ can be any non-negative integer. Table 5.1 lists some allowed $J_X^{PC}$ quantum numbers with their conventional naming scheme and allowed $LS$-combinations.

## 5.2 Spin formalisms

The description of the amplitudes of the two-body decays $J/\psi \rightarrow \gamma X$ and $X \rightarrow p\bar{p}$ is based on the helicity formalism. This spin formalism is frequently used for experimentally obtaining the angular distributions of scattering and decay processes. For the processes of interest, the well-known spin-orbit formalism from non-relativistic quantum mechanics is not suitable, since the angular momentum is defined in the center-of-mass frame, whereas the spins of particles are defined in their own rest frames [127]. The helicity formalism, however, is a convenient choice. In the helicity framework, the quantization axis is the direction of movement of a particle. This framework is by definition invariant under rotations and boosts along the momentum-axis of a particle [127] and particles with a zero and nonzero rest mass can be handled equivalently, thereby simplifying calculations [128]. Another physically convenient spin formalism is the Rarita-Schwinger formalism [129]. This formalism is implemented in the used PAWIAN software package [130] as well. However, it introduces a computational disadvantage in PAWIAN, compared to the multipole and helicity formalisms. Therefore, the multipole and helicity formalisms will be used in this work.

| $J^{PC}$ of resonance $X$ | Naming scheme | Allowed $LS$-combinations |
|---|---|---|
| $0^{-+}$ | $\eta$ | $L = 0, S = 0$ |
| $0^{++}$ | $f_0$ | $L = 1, S = 1$ |
| $1^{++}$ | $f_1$ | $L = 1, S = 1$ |
| $2^{-+}$ | $\eta_2$ | $L = 2, S = 0$ |
| $2^{++}$ | $f_2$ | $L = 1, S = 1$<br>$L = 3, S = 1$ |

**Table 5.1:** Allowed quantum numbers for the decay $X \to p\bar{p}$ and the conventional naming scheme.

In the following sections, the formalisms and the relevant amplitudes will be introduced. The information in these sections is based on references [131], [132] and [133], unless stated otherwise. In these references, a more detailed and set-by-step procedure can be found.

### 5.2.1 Quantum mechanical states

The quantum mechanical states of a single, massive particle at rest can be described by $|jm\rangle$, where $j$ is the particle's spin and $m$ the $z$-component of this spin. These $|jm\rangle$ states are the canonical basis vectors in which the angular momentum operators are represented in the way known from non-relativistic quantum mechanics. The canonical basis vectors form an orthogonal set, satisfying the completeness relations

$$\langle j'm'|jm\rangle = \delta_{j'j}\delta_{m'm} \text{ and } \sum_{jm} |jm\rangle\langle jm| = I,$$

with $\delta_{ij}$ the Kronecker delta and $I$ the identity operator. To express such quantum mechanical states in the helicity frame, it is necessary to define an arbitrary rotation of the canonical angular momentum eigenstates. A rotation of a physical system may be written by the unitary operator $R(\alpha, \beta, \gamma)$, with $(\alpha, \beta, \gamma)$ the standard Euler angles. A rotation of the quantum mechanical state $|jm\rangle$ can be described as

$$R(\alpha, \beta, \gamma)|jm\rangle = \sum_{m'} |jm'\rangle D^j_{m'm}(\alpha, \beta, \gamma). \tag{5.5}$$

with $D^j_{m'm}$ the complex unitary Wigner-$D$-matrices, defined as

$$D^j_{m'm}(\alpha, \beta, \gamma) = e^{-im'\alpha} d^j_{m'm}(\beta) e^{-im\gamma}. \tag{5.6}$$

The Wigner-$d$-matrices $d^j_{m'm}$, representing the real part of the complex of the Wigner-$D$-matrices, can be found tabulated for instance in reference [15].

Relativistic one-particle states with momentum $\vec{p}$ may be obtained by applying a Lorentz boost on the states $|jm\rangle$ that takes a particle at rest to a particle of momentum $\vec{p}$. There are two distinct ways of doing this, leading to canonical and helicity descriptions of relativistic free particle states. In both cases the system will be rotated ($R$) such that the new $z$-axis will be aligned with $\vec{p}$, followed by a Lorentz boost $L_z$ along the new $z$-axis. For the canonical description, the system will be rotated back ($R^{-1}$) to its original orientation, whereas this final step is omitted for the helicity system. The resulting systems are shown in figure 5.1. With the previously defined rotation operator, the transformation to the canonical system can be written as

$$|\vec{p}, jm\rangle = R^{-1}(\varphi, \theta, 0) L_z(\vec{p}) R(\varphi, \theta, 0) |\vec{p}, jm\rangle, \tag{5.7}$$

where the angles $\varphi$ and $\theta$ are defined as presented in figure 5.1. Similarly, the description in the helicity frame can be represented as

$$|\vec{p}, j\lambda\rangle = L_z(\vec{p}) R(\varphi, \theta, 0) |\vec{p}, j\lambda\rangle. \tag{5.8}$$

The helicity quantum number $\lambda$ is the projection of a particle's spin along its momentum $\vec{p}$ and is by definition rotationally invariant, since the quantization axis rotates along with the system under rotation. In terms of the original axes $(x, y, z)$, the helicity axes $(x', y', z')$ are now defined as $x' = y' \times z'$, $y' \propto z' \times \vec{p}$ and $z' \propto \vec{p}$.

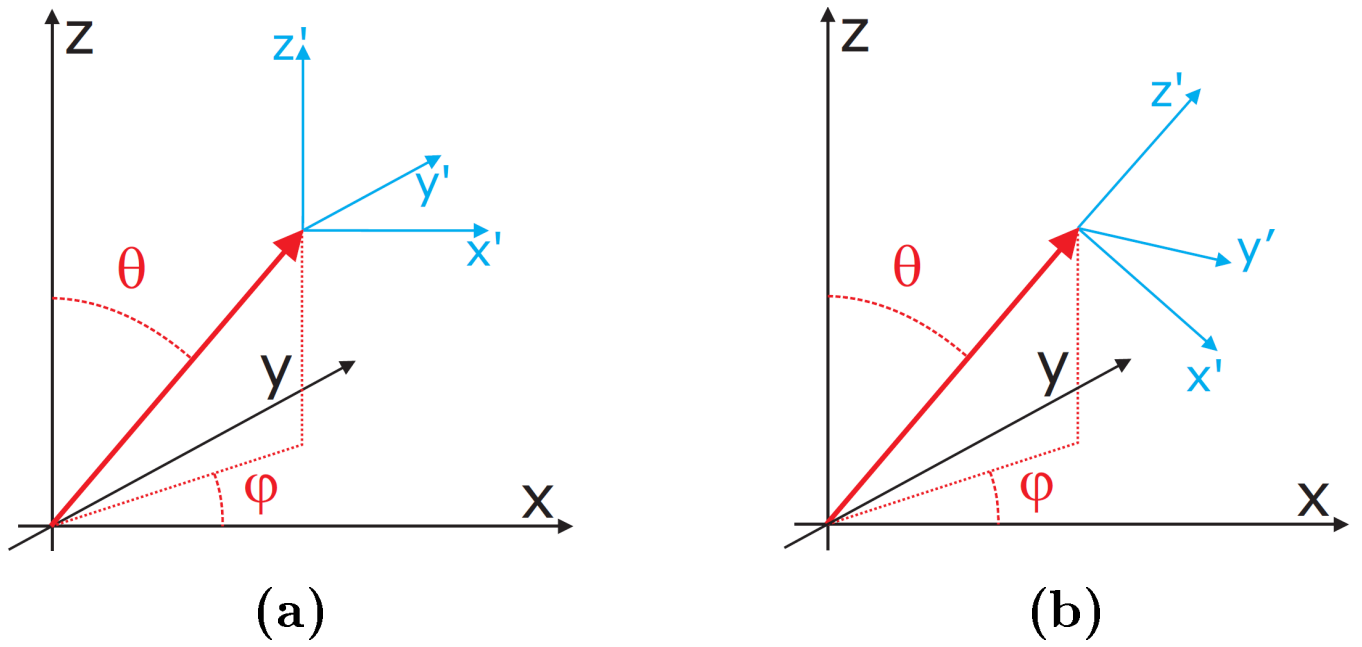

**Figure 5.1:** The canonical (a) versus the helicity (b) coordinate system [125].

### 5.2.2 Helicity amplitudes

In general, the decay amplitude for a decay of particle $a$, with total angular momentum $J_a$ and the projection of the spin $M_a$, into two daughter particles $b$ and $c$ can be written in the helicity basis as

$$A^{J_a,M_a}_{\lambda_b\lambda_c}(a \to b + c) = N_{J_a} D^{J_a*}_{M_a\lambda_{bc}}(\varphi, \theta, 0) F^{J_a}_{\lambda_b\lambda_c}, \tag{5.9}$$

where $\lambda_b$ and $\lambda_c$ are the helicities of the two daughter particles, $\lambda_{bc} = \lambda_b - \lambda_c$ and $N_{J_a} = \sqrt{\frac{2J_a+1}{4\pi}}$ is a normalization factor. The term $F^{J_a}_{\lambda_b\lambda_c}$ is called an helicity amplitude. This complex term is in general chosen as a free parameter, and will be determined by the PWA fit. However, not all of these helicity amplitudes are independent, therefore the number of free parameters in the fit will be reduced. For the channel studied here, only strong and electromagnetic decay processes contribute. This leads to parity conservation, and, hence, to the following relation between the helicity amplitudes:

$$F^{J_a}_{\lambda_b\lambda_c} = P_a \cdot P_b \cdot P_c \cdot (-1)^{-J_a+J_b+J_c} F^{J_a}_{\lambda_b\lambda_c}, \tag{5.10}$$

where $P_i$ and $J_i$ are the intrinsic parity and total angular momentum of particle $i$, respectively. This relation reduces the number of independent helicity amplitudes almost by a factor of two.

The helicity formalism can be easily extended to treat sequential two-body decays, like

$$\begin{aligned} a \rightarrow\ & b + c \\ & \hookrightarrow d + e \end{aligned}$$

If we define the total final state as $d+e+c$ and let $\lambda_{de} = \lambda_d - \lambda_e$, the total amplitude for the sequential decay can be written as

$$\begin{aligned} A^{J_a,M_a}_{\lambda_c\lambda_d\lambda_e}(a \rightarrow d+e+c) = \sum_{\lambda_b} & N_{J_b} N_{J_a} D^{J_b *}_{\lambda_b\lambda_{de}}(\varphi_d, \theta_d, 0) D^{J_a *}_{M_a\lambda_{bc}}(\varphi_b, \theta_b, 0) \\ & \cdot F^{J_b}_{\lambda_d\lambda_e} F^{J_a}_{\lambda_b\lambda_c}. \end{aligned} \tag{5.11}$$

Note the coherent sum over $\lambda_b$, since this value cannot be measured.

From the amplitude describing a decay, the angular distribution can be found by the amplitude squared. So, for the single two-body decay $a \rightarrow b+c$, equation 5.9 can be squared to get

$$\frac{d\Gamma}{d\Omega}_{M_a\lambda_b\lambda_c} = N^2_{J_a} \left| D^{J_a}_{M_a\lambda}(\varphi, \theta, 0) \right|^2 \left| F^{J_a}_{\lambda_b\lambda_c} \right|^2 \tag{5.12}$$

$$= N^2_{J_a} \left| d^{J_a}_{M_a\lambda}(\theta) \right|^2 \left| F^{J_a}_{\lambda_b\lambda_c} \right|^2, \tag{5.13}$$

with $d\Omega = d\theta d\cos\varphi$ an infinitesimally small element of the solid angle, and $\Gamma$ the decay rate. Integration over $\Omega$ gives

$$\Gamma_{M_a\lambda_b\lambda_c} = \left| F^{J_a}_{\lambda_b\lambda_c} \right|^2. \tag{5.14}$$

The magnitude of the helicity amplitude squared thus represents the decay rate into a final state with specified helicities. For radiative transitions, like the production of $X$, the helicity amplitudes can be transformed into the radiative multipole basis, such that the fit parameters of the amplitudes have a physical interpretation. For the decay of $X$, the amplitudes will be expanded in the $LS$-basis. This provides a direct access to the angular-momentum dependent part of the amplitudes.

### 5.2.3 Expansion to the radiative multipole and $LS$ amplitudes

In the helicity basis, the production amplitude depends on the angular momentum and helicity of the $p\bar{p}$ resonance $X$, as well as the angular momentum and spin projection of the $J/\psi$ meson and the radiative photon. Alternatively, in the radiative multipole basis, the dependence on the quantum numbers of the $X$ resonance is converted to a dependence on the angular momentum and helicity of the radiative photon. In the multipole basis, the different terms in the amplitude description are directly related to magnetic and electric multipole transitions treated in section 1.3.1. Recall that a radiative decay is defined as an E$J_\gamma$ transition when the product of the parities of the initial ($P_{J/\psi}$) and the final state ($P_X$) is equal to $(-1)^{J_\gamma}$, while processes for which $P_{J/\psi}P_X = (-1)^{J_\gamma+1}$ are called M$J_\gamma$ transitions. Here, $J_\gamma$ represents the angular momentum carried by the radiative photon, which can have the values $|J_{J/\psi} - J_X| \leqslant J_\gamma \leqslant |J_{J/\psi} + J_X|$. In table 5.2 the allowed multipole amplitudes corresponding to the $J^{PC}$ of the resonance $X$ are listed. Generally, when multiple multipole transitions are allowed, only the lowest one shows a dominant contribution, although exceptions have been observed [15, 134].

| $J^{PC}$ of resonance $X$ | Allowed multipole transitions |
|---|---|
| $0^{-+}$ | M1 |
| $0^{++}$ | E1 |
| $1^{++}$ | E1, M2 |
| $2^{-+}$ | M1, E2, M3 |
| $2^{++}$ | E1, M2, E3 |

**Table 5.2:** Allowed multipole transitions for the production process $J/\psi \to \gamma X$.

In a similar fashion as equation 5.11, with the inclusion of two dynamical functions $V^{a\to b+c}$, the total amplitude of the full decay chain $J/\psi \to X$, $X \to p\bar{p}$ can be constructed as

$$
\begin{aligned}
A^{M,\lambda_\gamma}_{\lambda_p\lambda_{\bar{p}}}(J/\psi \to \gamma p\bar{p}) = \sum_{J_X,J_\gamma,\lambda_X} & N_{J_\gamma} N_{J_X} D^{J_{J/\psi}*}_{M,\lambda_X-\lambda_\gamma}(\varphi^{J/\psi}_\gamma, \theta^{J/\psi}_\gamma, 0) \\
& \cdot D^{J_X*}_{\lambda_X,\lambda_p-\lambda_{\bar{p}}}(\varphi^X_p, \theta^X_p, 0) F^{J_{J/\psi}}_{\lambda_X\lambda_\gamma} F^{J_X}_{\lambda_p\lambda_{\bar{p}}} \\
& \cdot V^{J/\psi\to\gamma X} V^{X\to p\bar{p}}.
\end{aligned} \tag{5.15}
$$

Here, $F^{J_{J/\psi}}_{\lambda_X\lambda_\gamma}$ and $F^{J_X}_{\lambda_p\lambda_{\bar{p}}}$ represent the helicity amplitudes, and the matrices $D^{J_{J/\psi}*}_{M,\lambda_X-\lambda_\gamma}(\varphi^{J/\psi}_\gamma, \theta^{J/\psi}_\gamma, 0)$ and $D^{J_X*}_{\lambda_X,\lambda_p-\lambda_{\bar{p}}}(\varphi^X_p, \theta^X_p, 0)$ are the complex conjugated Wigner-$D$ functions of the $X$ production and decay, respectively. The production angles, $\varphi^{J/\psi}_\gamma$ and $\theta^{J/\psi}_\gamma$, are the azimuthal and polar angles of the photon in the rest frame of the $J/\psi$ meson, where the quantization axis is defined by the flight directions of the beam particles. Similarly, $\varphi^X_p$ and $\theta^X_p$ represent the angles of the proton, defined in the helicity frame of the intermediate resonance $X$.

The transformation of the production helicity amplitudes in the radiative multipole amplitudes is given by

$$
\begin{aligned}
F^{J_{J/\psi}}_{\lambda_\gamma\lambda_X} = \sum_{J_\gamma} & \langle J_\gamma, -\lambda_\gamma; J_X, \lambda_X | J_{J/\psi}, \lambda_X - \lambda_\gamma \rangle \frac{1}{\sqrt{2}} [\delta_{\lambda_\gamma,1} + \delta_{\lambda_\gamma,-1} P_X (-1)^{J_\gamma - 1}] \\
& \cdot \alpha^{J/\psi}_{J_\gamma X},
\end{aligned} \tag{5.16}
$$

where $\alpha^{J/\psi}_{J_\gamma X}$ are the complex multipole amplitudes, which are usually left as free parameters in the fit, and $P_X$ represents the parity of $X$. The term $\langle J_\gamma, -\lambda_\gamma; J_X, \lambda_X | J_{J/\psi}, \lambda_X - \lambda_\gamma \rangle$ represent the Clebsch-Gordon coefficients, which can be found tabulated, for instance in reference [15]. The single terms in this expansion now correspond to magnetic or electric dipole (M1, E1), quadrupole (M2, E2) or octupole (M3, E3) transitions, depending on the spin and parity of the resonance $X$.

In the PWA fit, the decay helicity amplitudes will be expanded to the $LS$-scheme according to the relation

$$
V^{X\to p\bar{p}} F^{J_X}_{\lambda_p\lambda_{\bar{p}}} = \sum_{L,S} V^{X\to p\bar{p}}_L \langle J_{\bar{p}}, \lambda_{\bar{p}}; J_p, -\lambda_p | S_X, \lambda \rangle \langle L, 0; S_X, \lambda | J_X, \lambda \rangle \alpha^X_{LS}, \tag{5.17}
$$

with the two Clebsch-Gordon coefficients for the coupling of the (anti)proton spins to $S_X$, and for the coupling of $L$ and $S_X$ to $J_X$, where $\lambda$ is defined as $\lambda_{\bar{p}} - \lambda_p$. The terms $\alpha^X_{LS}$ are the complex $LS$-amplitudes. Just like the $\alpha^{J/\psi}_{J_\gamma X}$, the $\alpha^X_{LS}$ are in general left as free fit parameters. In this expansion, the dynamical part $V^{X\to p\bar{p}}$ is included, since this part contains an $L$-dependent component, as will be discussed in section 5.3.1.

As explained in section 5.2.2, not all helicity amplitudes are independent. Since parity is conserved in both the production and decay of $X$, the helicity amplitudes are connected via the relation in equation 5.10, resulting in

$$F^{J_{J/\psi}}_{\lambda_\gamma \lambda_X} = P_X(-1)^{J_X} F^1_{\lambda_\gamma \lambda_X} \quad \text{and} \quad F^{J_X}_{\lambda_p \lambda_{\bar{p}}} = P_X(-1)^{1-J_X} F^{J_X}_{\lambda_p \lambda_{\bar{p}}},$$

and, therefore, almost a factor two less fit parameters.

The functions $V^{a\to b+c}$ describe the $X$ production and decay dynamics, and can be parametrized in various ways, hereby introducing additional fit parameters. For the mass-independent study of $J/\psi \to \gamma p\bar{p}$, discussed in chapter 6, the dynamical part of the amplitude is omitted. However, for the PWA of the $\eta_c$ region, discussed in chapter 7, a dynamical description is included. In the following section, the different ingredients of this dynamical function will be discussed.

## 5.3 The dynamical part of the amplitude

For the mass-independent study of $J/\psi \to \gamma p\bar{p}$, the $p\bar{p}$ invariant mass range is divided in more than 200 bins, that are each studied separately. This procedure allows for the omission of the dynamical part of the amplitude. However, for the mass-dependent study of the $\eta_c$ meson, a description of the dynamical part of the amplitude is necessary. Where the angular distributions contain information on the spins of intermediate states, the invariant mass distributions of final-state particle combinations contain information about the energy dependency, and thus also about the mass and line shape of intermediate states. Resonances, like the $\eta_c$ meson, show up as peaks in the invariant mass distribution of the $p\bar{p}$ system. The line shape of a resonance depends on several factors, such as mechanisms related to its production and decay, threshold effects, interference effects with overlapping resonances, and the available phase space. First, a correction factor related to the threshold effects will be discussed, followed by a description of the $\eta_c$ line-shape parametrization.

### 5.3.1 Barrier factors

In general, fundamental particles are considered pointlike. However, bound states of quarks, like charmonium, must have finite, nonzero dimensions. This results in a potential well that limits the maximum angular momentum $L$ by $qR/\hbar c$, with $q$ the relative momentum of the daughter particles in the resonance's rest frame, and $R$ the interaction radius [135]. Slowly moving daughter particles cannot generate sufficient angular momentum to conserve the spin of the decaying resonance, therefore, these decays are suppressed. This suppression of higher angular momenta is called the centrifugal barrier effect and results in a distortion of the line shape close to the decay threshold. Similarly, the line shape of resonances at high invariant masses, close to the production threshold of the final state, is distorted, as they are produced with a small breakup momentum and an orbital angular momentum larger than zero. For $J/\psi \to \gamma p\bar{p}$, the measured invariant mass of the $p\bar{p}$ pair ranges approximately from two times the proton mass up to the $J/\psi$ mass, with small deviations due to experimental restrictions. Line shape distortions are thus expected near the $p\bar{p}$-threshold, around the $X(1835)$ mass peak, as well as close to the $J/\psi$ mass, around the $\eta_c$ mass peak.

The $L$-dependent distortion of the line shape can be described by the Blatt-Weisskopf barrier factors [136]. Aside from the angular momentum, these barrier factors depend on momentum $q$ of the daughter particles and scale parameter $q_R = \hbar c/R$, with $R$ being the interaction radius of the bound state under study. For decays of heavier bound states, like charmonium, this radius is commonly chosen to be $R = 0.3$ fm [128, 137, 138]. Although, the line shape distortion does not strongly depend on the value of parameter $R$ [138], the same radius $R = 0.3$ fm will be used in the study of $J/\psi \to \gamma p\bar{p}$. The first four barrier factors $b_L(z)$, with $z = (q/q_R)^2$, are listed in table 5.3. Note that the barrier factors introduce a strong $q^L$ dependence, and that the factors are equal to 1, for $q = q_R$, or $z = 1$.

The Blatt-Weisskopf barrier factors can be normalized to the breakup momentum $q_0$ of a resonance with mass $m_0$ to [128]

$$B_L(q, q_0) = \frac{b_L(q)}{b_L(q_0)}. \tag{5.18}$$

| $L$ | $b_L(z)$ |
|---|---|
| 0 | 1 |
| 1 | $\sqrt{\frac{2z}{z+1}}$ |
| 2 | $\sqrt{\frac{13z^2}{(z-3)^2+9z}}$ |
| 3 | $\sqrt{\frac{277z^3}{z(z-15)^2+9(2z-5)^2}}$ |
| 4 | $\sqrt{\frac{12746z^4}{(z^2-45z+105)^2+25z(2z-21)^2}}$ |

**Table 5.3:** Blatt-Weisskopf barrier factors for angular momenta $L \leqslant 4$ [139].

The normalized barrier factors for the decay and production of resonances are implemented into the PAWIAN software [140] and incorporated into the parametrization of the dynamical part of the amplitude.

### 5.3.2 Line-shape parametrization

A common parametrization of the line shape of a resonance is given by the relativistic Breit-Wigner function, which can be written as [128]

$$BW(m) = \frac{m_0 \Gamma B_L(q,q_0)}{m_0^2 - m^2 - i\frac{\rho(m)}{\rho(m_0)}\Gamma B_L^2(q,q_0)}, \tag{5.19}$$

where $m_0$ and $\Gamma$ denote the mass and width of the resonance, $B_L(q, q_0)$ the normalized barrier factors introduced in the previous section, and $\rho(m)$ represent the phase space factors defined by

$$\rho(m) = \sqrt{\left(1 - \left(\frac{m_b + m_c}{m}\right)\right)\left(1 - \left(\frac{m_b - m_c}{m}\right)\right)}, \tag{5.20}$$

with $m_b$ and $m_c$ the masses of the two daughter particles. Since $m_p = m_{\bar{p}}$, the phase factor reduces to $\rho(m) = \sqrt{1 - (2m_p/m)}$ for $X \to p\bar{p}$.

As discussed in section 1.4.2, the $\eta_c$ line shape cannot be described by a simple Breit-Wigner without taking into account the production part.

Therefore, the standard Breit-Wigner will be extended with two additional terms. Since $J/\psi \to \gamma\eta_c$ is described by a pure M1-transition [141], a term describing the energy dependence of the M1-transition matrix element will be added. For pure M1 transitions, the matrix element depends on the photon energy $E_\gamma$ as $\propto E_\gamma^3$ [79]. The additional $E_\gamma^3$ term improves the description of the $\eta_c$ line shape, however, it also introduces a diverging tail towards higher photon energies, corresponding to lower $p\bar{p}$ invariant masses. Hence a damping function, $f_d$, is added as second term to prevent the unphysical divergence. The resulting line shape description thus reads as

$$V_{\eta_c}(m) = E_\gamma^3 f_d BW(m). \tag{5.21}$$

The CLEO-c collaboration was the first to fit the energy spectrum of the radiative photon with a relativistic Breit-Wigner function, modified with the photon energy dependence and an empirical damping function [73]. In the CLEO-c analysis, the damping function was defined as $f_d^{CLEO}(E_\gamma) = e^{-E_\gamma^2/8\beta^2}$, with $\beta = 65$ MeV. Later, the KEDR collaboration proposed another empirical damping function [74] defined as

$$f_d^{KEDR}(E_\gamma) = \frac{E_0^2}{E_\gamma E_0 + (E_\gamma - E_0)}, \tag{5.22}$$

with $E_0$ the most probable transition energy,

$$E_0 = \frac{m_{J/\psi}^2 - m_{\eta_c}^2}{2m_{J/\psi}}. \tag{5.23}$$

Here, $m_{J/\psi}$ and $m_{\eta_c}$ refer to the nominal masses of $J/\psi$ and $\eta_c$, respectively. Note that $E_\gamma$ can be described in the same way to get

$$E_\gamma(m) = \frac{m_{J/\psi}^2 - m^2}{2m_{J/\psi}}, \tag{5.24}$$

giving the relation between $m$ and $E_\gamma$.

Figure 5.2 shows the differences between the general Breit-Wigner and the modified descriptions, and figure 5.3 demonstrates the effect of the barrier factors on the line shape. Although the KEDR and CLEO-c descriptions look quite similar, in the KEDR description the diverging tail is slightly less damped. As is customary in the BESIII collaboration, the KEDR descrip-tion will be adopted for the study of the $\eta_c$ meson, described in chapter 7. The CLEO-c description is used to provide an estimate of systematic un-certainties.

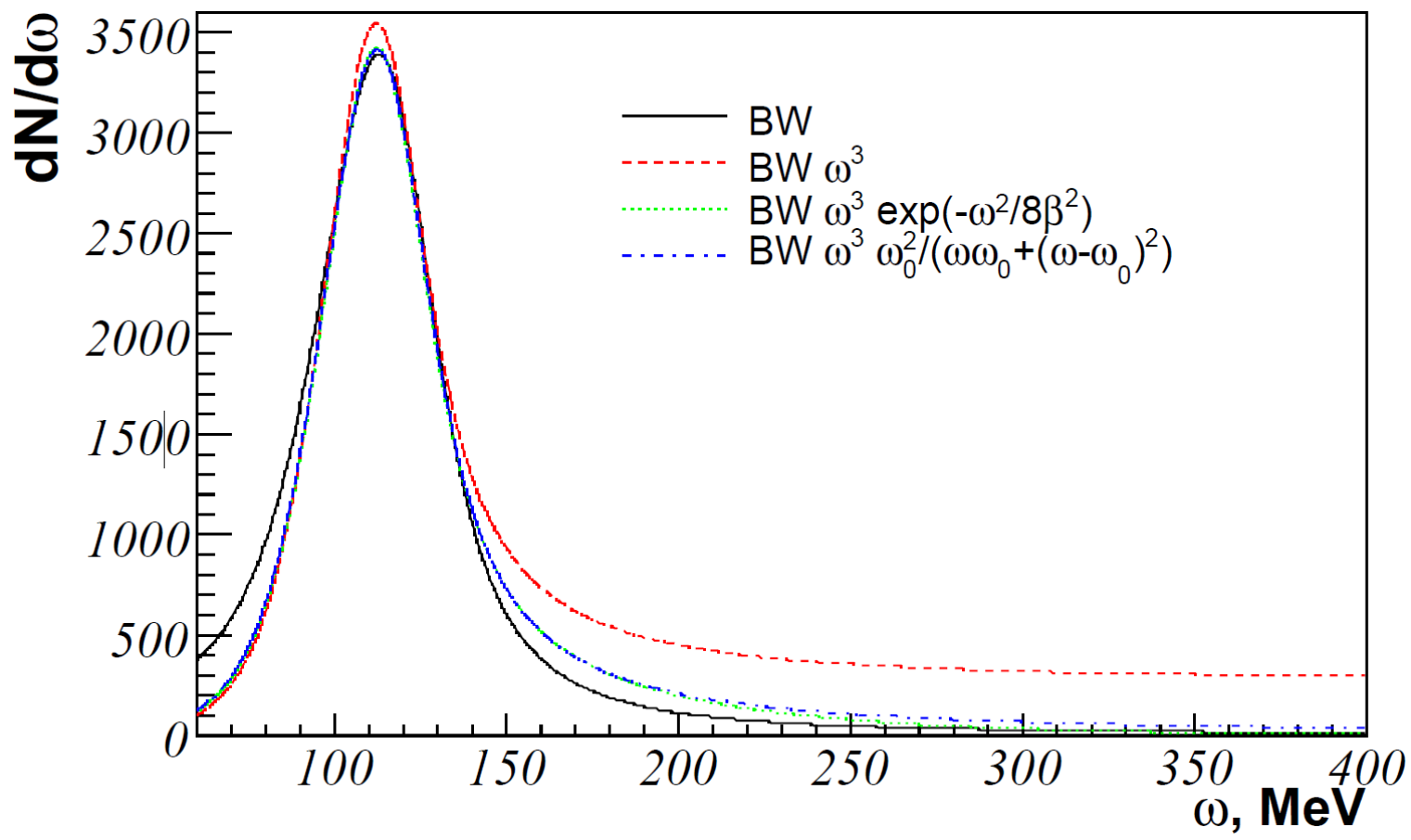

**Figure 5.2:** The resulting line shapes for the different descriptions, with the CLEO-c and KEDR descriptions in green and blue, respectively. In the figure, $\omega$ corresponds to the $E_\gamma$ in the text [74].

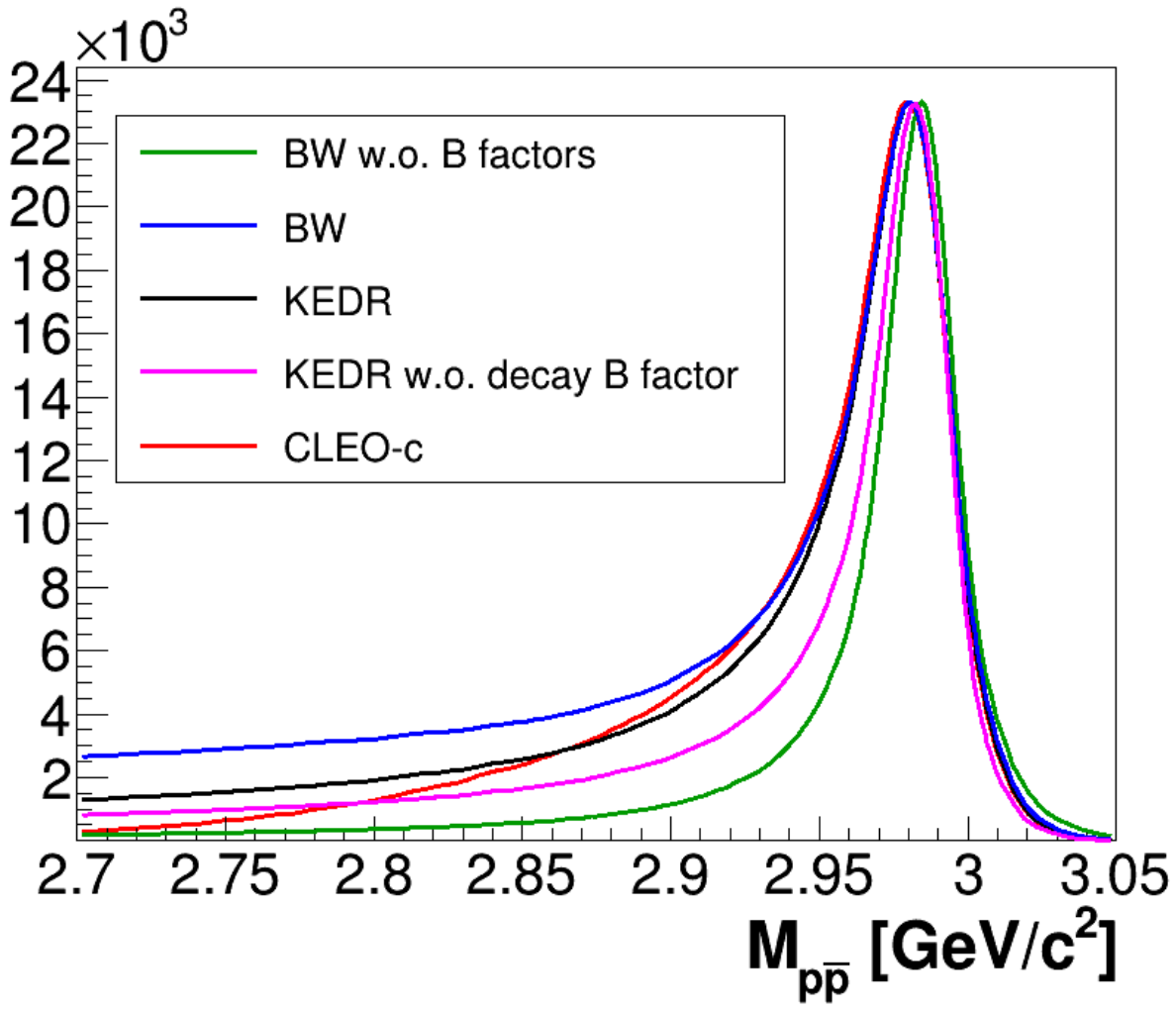

**Figure 5.3:** The resulting $\eta_c$ line shapes with both the decay and production barrier factors included, unless stated otherwise in the legend.

## 5.4 Fitting procedure

With both the description of the dynamical part and the angular part of the amplitudes defined, the total amplitude of the complete decay chain $J/\psi \to \gamma X$, $X \to p\bar{p}$ can be constructed. However, in our case only the five-fold differential cross-section $\frac{d\sigma}{d\tau}$ is experimentally accessible. The differential cross-section can be characterized as a normalized probability density function $w(\vec{x}, \theta) \propto \frac{d\sigma}{d\tau}$ that describes the transition probability from the initial to the final state, for each point in phase-space $\vec{x}$, and set of fit parameters $\theta$. The actual PWA consists of defining $w$ for different (number of) contributing $J^{PC}$-waves, so-called hypotheses. For each hypothesis, an unbinned-maximum-likelihood-fit of $w$ to the selected data will be performed. The detector acceptance and reconstruction efficiency are taken into account by using reconstructed phase-space-distributed MC events undergoing exactly the same selection criteria as applied for the data events. With this MC sample included in the likelihood function, the phase-space factor and other constant factors can be omitted in the description of $w$ [127]. The final step of the PWA consists of selecting the significantly best hypothesis based on the different fit results. In the following sections, these different steps will be discussed in more detail.

### 5.4.1 The function $w$ and its parameters

For the complete decay chain $J/\psi \to \gamma X$, $X \to p\bar{p}$, the probability density function $w(\vec{x}, \theta)$ can be described from the total amplitude via

$$\begin{aligned}
w(\vec{x},\theta) &= \sum_{\lambda_p,\lambda_{\bar{p}}=-\frac{1}{2},\frac{1}{2}} \sum_{\lambda_\gamma,M=-1,1} \left| A^{M,\lambda_\gamma}_{\lambda_p\lambda_{\bar{p}}}(J/\psi \to \gamma p\bar{p}) \right|^2 \\
&= \sum_{\lambda_p,\lambda_{\bar{p}}=-\frac{1}{2},\frac{1}{2}} \sum_{\lambda_\gamma,M=-1,1} \Bigg| \sum_{J_X,J_\gamma,\lambda_X} N_{J_\gamma} N_{J_X} D^{J_{J/\psi}*}_{M,\lambda_X-\lambda_\gamma}(\varphi^{J/\psi}_\gamma, \theta^{J/\psi}_\gamma, 0) \\
&\cdot D^{J_X*}_{\lambda_X,\lambda_p-\lambda_{\bar{p}}}(\varphi^X_p, \theta^X_p, 0) V^{J/\psi\to\gamma X} \\
&\cdot \left( \sum_{J_\gamma} \langle J_\gamma, -\lambda_\gamma; J_X, \lambda_X | J_{J/\psi}, -\lambda_\gamma \rangle \frac{1}{\sqrt{2}} [\delta_{\lambda_\gamma,1} + \delta_{\lambda_\gamma,-1} P_X (-1)^{J_\gamma-1}] \alpha^{J/\psi}_{J_\gamma X} \right) \\
&\cdot \left( \sum_{L,S} V^{X\to p\bar{p}}_L \langle J_{\bar{p}}, \lambda_{\bar{p}}; J_p, -\lambda_p | S_X, \lambda \rangle \langle L, 0; S_X, \lambda | J_X, \lambda \rangle \alpha^X_{LS} \right) \Bigg|^2,
\end{aligned} \tag{5.25}$$

with $\lambda_i$ the helicity of particle $i \in \{\gamma, p, \bar{p}\}$, and the amplitudes $A^{M,\lambda_\gamma}_{\lambda_p\lambda_{\bar{p}}}(J/\psi \to \gamma p\bar{p})$ as defined in equation 5.15, with the expansions from equations 5.16 and 5.17 implemented. For BESIII, only $J/\psi$ spin projections $M = \pm 1$ are possible, as the $J/\psi$ meson is produced in an electromagnetic process via $e^+e^-$ annihilation (virtual photon). For the mass-dependent study of the $\eta_c$ meson, the dynamical part of the amplitude, $V^{J/\psi\to\gamma X}$, contains the $\eta_c$ parametrization and Blatt-Weisskopf barrier factors discussed in the previous sections. These barrier factors depend on the orbital angular momentum $L$. Since $L$ is not defined in the radiative multipole production amplitudes, the minimum value of the orbital angular momentum for a given resonance $X$ will be used. The barrier factors are properly taken into account for the decay of $X$, because here the orbital angular momentum is directly accessible due to the chosen $LS$-amplitudes. For the mass-independent study, the dynamical descriptions are omitted.

In the function $w$, the different possible spins and helicities for intermediate resonances are summed coherently. The relative phase of the complex amplitudes in the coherent sum corresponds to the potential interference between different contributions. Additionally, all initial and final state helicities that are not fixed or determined by the experiment must be summed incoherently[1] [127, 132]. Note that this results in a relatively large number of coherent sums for $J/\psi \to \gamma p\bar{p}$. Since all three final-state particles have a nonzero spin, $2^3$ different combinations of helicities are possible for the final state. Together with the two possible spin-projections for the initial state, this results in $2^4 = 16$ coherent sums for the calculation of $w$.

For all hypotheses, the probability function $w$ is fitted to the selected data by varying its free parameters. These parameters are given by the production ($\alpha^{J/\psi}_{J_\gamma X}$) and decay ($\alpha^X_{LS}$) amplitudes, and, for the mass-dependent analysis, the parameters corresponding to the mass and width of the $\eta_c$ meson. Note that, in PAWIAN, the normalization factors, $N_{J_\gamma}$ and $N_{J_X}$, are incorporated in the fit parameters. The number of amplitudes in $w$ depends on what intermediate resonances are included in the fit hypothesis, and which multipole and $(L, S)$ combinations are allowed for the chosen resonances. Each one of these complex amplitudes can be expressed by a real magnitude and a phase, yielding two distinct fit parameters per amplitude. Due to dependencies, some of these parameters can be fixed, resulting in fewer free parameters.

In table 5.4, the number of total and free parameters are listed for the five single-wave hypotheses that will be considered. The fixing of parameters is based on a few principles. First, the product of two complex numbers is itself a complex number that can be fully described by two real parameters. Therefore, without any loss of generality, the magnitude and phase of either one production or decay amplitude can be fixed for each resonance. This fixing of the magnitude and phase is only for one of the possible multipole or $(L, S)$ combinations, since the relative phases and magnitudes between the different combinations are relevant. However, since $w$ consists of the absolute square of the sum of the amplitudes, the overall phase has no physical meaning. Consequently, one additional phase can be fixed without any loss of generality. Hence, $n - K = 3$ for the hypotheses in table 5.4. For hypotheses including multiple resonances, the total number of free parameters is often more than the sum of $K$ listed in the table. This is because, when more than one resonance contributes to the process, all their phases will come in individually, and only one overall phase can be fixed.

---

[1] For a coherent sum, the amplitudes are summed and then squared, while for an incoherent sum one first squares and then sums.

The relative phases now have a physical meaning and need to be free parameters. Additionally, the description of the mass dependence introduces further parameters related to the mass and width of the $\eta_c$ meson.

| $J^{PC}$ of $X$ | $X$-production | $X$-decay | $n$ | $K$ |
|---|---|---|---|---|
| $0^{-+}$ | M1 | $(0,0)$ | 4 | 1 |
| $0^{++}$ | E1 | $(1,1)$ | 4 | 1 |
| $1^{++}$ | E1, M2 | $(1,1)$ | 6 | 3 |
| $2^{-+}$ | M1, E2, M3 | $(2,0)$ | 8 | 5 |
| $2^{++}$ | E1, M2, E3 | $(1,1)$<br>$(3,1)$ | 10 | 7 |

**Table 5.4:** Possible $(L,S)$ and multipole combinations for different accessible $J^{PC}$ quantum numbers of resonance $X$ in $J/\psi \to \gamma X, X \to p\bar{p}$, and their corresponding number of parameters, $n$, and free parameters, $K$.

### 5.4.2 Likelihood optimization

For all hypotheses considered in the PWAs described in chapters 6 and 7, unbinned-maximum-likelihood fits are performed. The extended likelihood function is defined as [142, 143][2]

$$\mathcal{L}(\theta) \propto N_{data}! \cdot \exp\left(-\frac{(N_{data}-\bar{N})^2}{2N_{data}}\right) \prod_{i=1}^{N_{data}} \frac{w(\vec{x}_i,\theta)}{\int w(\vec{x},\theta)\epsilon(\vec{x})d\tau}. \tag{5.26}$$

Here, $N_{data}$ is the number of selected data events and $\theta$ represents the parameter set of the probability density function $w(\vec{x},\theta)$, as introduced in the previous section. The function $\epsilon(\vec{x})$ describes the detector acceptance, and the reconstruction and selection efficiencies at phase-space position $\vec{x}$. The symbol $d\tau$ denotes an infinitesimally small element of the phase-space.

[2] Note that listed references use the Q-factor method to handle specific backgrounds. As this method is not used in our study, the $Q_i$ weights are omitted.

Finally, $\bar{N}$ is defined as

$$\bar{N} = N_{data} \cdot \frac{\int w(\vec{x},\theta)\epsilon(\vec{x})d\tau}{\int \epsilon(\vec{x})d\tau}. \tag{5.27}$$

The function $\epsilon(\vec{x})$ and its phase-space integral can be approximated with use of a phase-space-distributed MC sample. For this approximation, the exclusive $J/\psi \to \gamma p\bar{p}$ MC sample, introduced in section 3.1, is propagated through the software package that simulates the BESIII detector, followed by the same reconstruction and selection criteria as used to analyze the data sample. The terms containing $\epsilon(\vec{x})$ can now be replaced via

$$\int w(\vec{x},\theta)\epsilon(\vec{x})d\tau = \sum_{j=1}^{N_{MC}} w(\vec{x}_j,\theta) \quad \text{and} \quad \int \epsilon(\vec{x})d\tau = N_{MC},$$

with $N_{MC}$ the total number of MC events. The best description of the data corresponds to a maximum in the likelihood function $\mathcal{L}(\theta)$. For practical purposes, $\mathcal{L}(\theta)$ will be transformed in a log-likelihood function $\ell(\theta)$. Since logarithms are strictly increasing functions, the maximum in $\mathcal{L}(\theta)$ is equivalent to the maximum in $\ell(\theta)$. However, the product in equation 5.26 is now transformed into a sum, reducing the computational costs. Additionally, a negative sign is added to the log-likelihood, converting the maximization problem into a minimization problem, so that the minimizer Minuit2 can be used. The negative log-likelihood (NLL) function that will be minimized is given by

$$-\ln \mathcal{L}(\theta) = -\ell(\theta) = -\sum_{i=1}^{N_{data}} \ln\left(w(\vec{x}_i)\right) + N_{data} \ln\left(\sum_{j=1}^{N_{MC}} w(\vec{x}_j,\theta)\right) + \frac{N_{data}}{2}\left(\frac{\sum_{j=1}^{N_{MC}} w(\vec{x}_j,\theta)}{N_{MC}} - 1\right)^2. \tag{5.28}$$

### 5.4.3 Hypothesis selection

After all hypotheses are fitted with the negative-likelihood function, one hypothesis has to be selected as the significant best. Especially for the mass-independent fit, presented in chapter 6, a standard selection crite-rion is necessary. In this analysis, 31 different hypotheses are fitted in more than 200 separate invariant-mass bins. The resulting NLL value from equation 5.28 is not a suitable quality description, since it correlates strongly with the number of free parameters, tending to prefer more complex models that might overfit the data.

So-called nested hypotheses can be compared via the likelihood-ratio test (LRT). Here, nested hypotheses means that the more simple hypothesis is a subset of the more complex model, so for instance the hypothesis $0^{-+}$ is a subset of the hypothesis $0^{-+}2^{++}$. In this test, the significance of the improvement between a more complex and a simple hypothesis can be calculated via [144]

$$\mathrm{LRT} = -2\ln\left(\frac{\mathcal{L}(\hat{\theta}_s)}{\mathcal{L}(\hat{\theta}_c)}\right) = -2\left(\ell(\hat{\theta}_s) - \ell(\hat{\theta}_c)\right), \tag{5.29}$$

where $\hat{\theta}_s$ and $\hat{\theta}_c$ are the parameter sets that maximize the (log-)likelihood for the simple and more complex hypothesis, respectively. The LRT is distributed according to a chi-squared distribution, with the degrees-of-freedom equal to the difference in the number of free parameters between the two models. However, this test is limited to nested hypotheses, and cannot be used to compare non-nested hypotheses, such as the $0^{-+}$ versus the $1^{++}2^{-+}$ scenario. Therefore, the standard selection criteria that will be used is based on two different information criteria from the field of model selection theory, that do not require hypotheses to be nested. Both selection criteria depend on the NLL value, with an additional penalty term for the number of free parameters in the hypothesis. The purpose of the penalty term is to prevent overfitting, and thus to balance the goodness-of-fit with the complexity of the hypothesis. The following descriptions are based on reference [145]. In this reference, the background of, and more details about, the different criteria for the model selection can be found.

The first criterion is the Bayesian-Information-Criterion (BIC), a minimization of the value based on the maximum value of the log-likelihood, $\ell(\hat{\theta})$, the number of free parameters $K$, and the number of data points $n$:

$$\mathrm{BIC} = -2\ell(\hat{\theta}) + K\ln(n). \tag{5.30}$$

Since the penalty term rises logarithmically with $n$, the BIC favors hypotheses with less parameters for large data samples. For a proper functioning of the BIC, it is required that $n$ is significantly larger than $K$. This condition can be met for all fits performed in the analyses described in both chapters 6 and 7.

The second criterion is a minimization of the Akaike-Information-Criterion (AIC), defined as

$$\mathrm{AIC} = -2\ell(\hat{\theta}) + 2K, \tag{5.31}$$

which is independent of $n$. Compared to the BIC, the penalty term is weaker, leading to a larger probability of overfitting the data. For small sample sizes of $n/K < 40$, the second-order Akaike-Information-Criterion (AICc) is preferred. This criterion is defined as

$$\mathrm{AICc} = \mathrm{AIC} + \frac{2K(K+1)}{n-K-1} = -2\ell(\hat{\theta}) + 2K + \frac{2K(K+1)}{n-K-1}. \tag{5.32}$$

The additional term accounts for the finite sample size, and increases the penalty term for small sample sizes, albeit not as strong as the BIC penalty term. Note that the last term in equation 5.32 decreases as $n$ increases. Thus, in the limit of $n \gg K$, the AICc converges to the AIC. Therefore, the AICc will be used besides the BIC.

Both criteria, the AIC(c) and the BIC, do not have a meaning as an absolute value, and cannot be used to compare hypotheses that are fitted to different datasets. However, if different hypotheses are fitted to the same dataset, the hypothesis that is closer to the true description will give a smaller value. Therefore, when different hypotheses are compared for a fixed dataset, the one with the smallest value is preferred. In general, if there is no systematic uncertainty present, a difference of $\Delta\mathrm{BIC} > 10$ or $\Delta\mathrm{AIC(c)} > 10$ is enough to eliminate the candidate hypothesis [145, 146].

In figure 5.4, the performance of the different selection criteria is demonstrated in terms of the number of free parameters. Due to the stronger penalty term, the BIC favors hypotheses with fewer parameters compared to AICc. The chance of overfitting is thus smaller when the BIC is used. However, theoretical considerations related to the accuracy and the overall performance lead to a preference towards the AICc [145]. As a compromise, the sum of the AICc and BIC will be used to select the best hypothesis, although, in general, the AICc and the BIC agree on the same best hypotheses [140, 143].

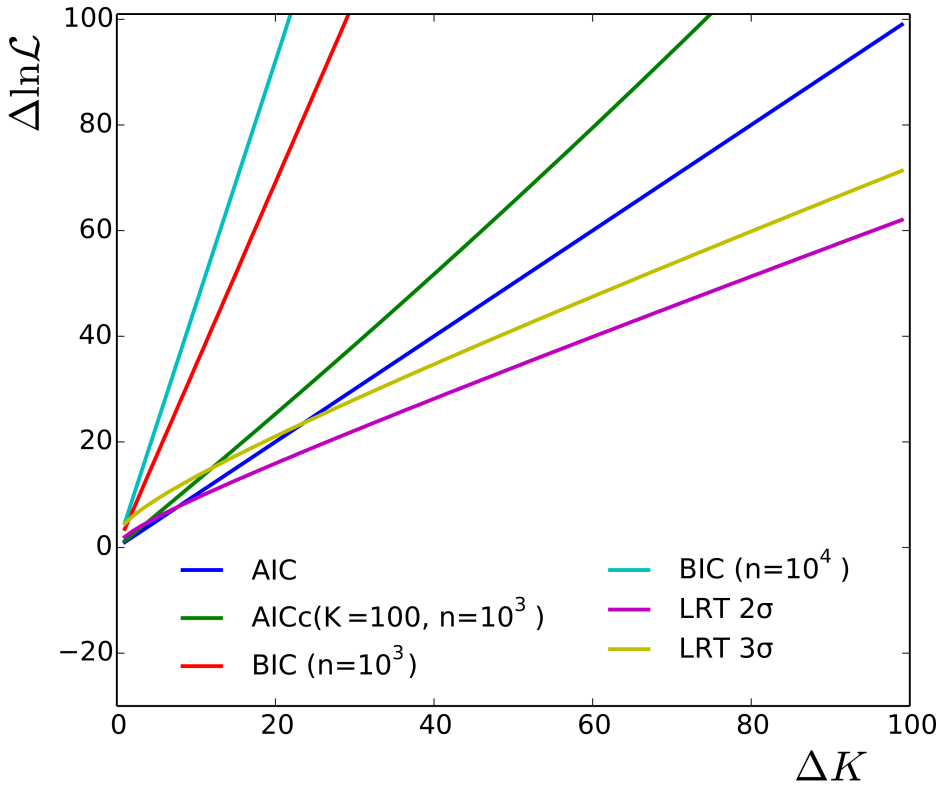

**Figure 5.4:** The improvement needed in $\ln \mathcal{L}$ for which the more complex hypothesis, with $\Delta K$ more parameters, is chosen by the different selection criteria [147]. The LRT is plotted for two different levels of significance $\sigma$.

## 5.5 The PAWIAN software package

The partial-wave analyses are performed with the software package PAWIAN (PArtial-Wave Interactive ANalysis). PAWIAN is developed at the Ruhr University Bochum and is a user-friendly and highly-modular PWA software package written in C++. It follows an object-oriented approach with a wide range of flexibilities. The user input is based on plain-text configuration files, where different spin-formalisms or line-shape parametrizations can be selected with the use of keywords. Amongst the included spin-formalisms are the canonical, Rarita-Schwinger, helicity and helicity-multipole formalisms. The dynamical parametrizations that can be selected contain the Breit-Wigner, Flatté, $K$-matrix and both the CLEO-c and the KEDR descriptions of the $\eta_c$ meson produced in radiative charmonium decays. Additionally, the code can easily be extended to include further decay models, complete amplitudes or other descriptions for the dynamics. PAWIAN is able to perform the fits in parallel via the server-client mode and supports coupled channel analyses. Aside from the $e^+e^-$ annihilations, proton-antiproton annihilations, pion-pion and pion-proton scattering processes, and two-photon fusion are supported by PAWIAN as well. The data are fitted with an event-based maximum-likelihood fit, where the minimization is carried out with the external Minuit2 package. Further information can be found in references [148], [140] and [130] or on the PAWIAN wiki page [149]. Published results from previous analyses based on PAWIAN can, for instance, be found in references [143], [140] and [150].

## 5.6 Concluding remarks

The described amplitudes, combined with PAWIAN software package, provide a strong set of analysis tools for the PWA of $J/\psi \to \gamma p\bar{p}$. However, the descriptions are not perfect and have some weaknesses. First of all, in the mass-dependent analysis, described in chapter 7, the dynamical part is parametrized by (modified) Breit-Wigner line shapes. It is known that a Breit-Wigner description is not reliable when there are overlapping resonances present. In the $\eta_c$ range, we expect only one isolated resonance to be present, so the formalism is expected to work properly. If the mass-dependent study would be extended to lower $p\bar{p}$ invariant masses, another description should be considered, since overlapping resonances can be expected in this range. A commonly used alternative is the $K$-matrix formalism [128,151]. This formalism can describe overlapping resonances, although it introduces another problem. For a correct description, all intermediate resonances, and all the final states that these resonances couple to, must be considered in the corresponding $K$-matrix parameterization. Here the problem arises, since the decay $J/\psi \to \gamma p\bar{p}$ contains poorly measured intermediate states. Therefore, the information needed for a correct $K$-matrix description is lacking. The $K$-matrix parameterization will thus not be used in this work.

In the mass-independent analysis, presented in the following chapter, the issue with mass description does not play a role, although, it should be noted that this analysis is still not fully model independent. Even this analysis is still based on the assumption that the decay $J/\psi \to \gamma p\bar{p}$ can be described properly by an isobar model.

# 6. Mass-independent Partial-Wave Analysis

In the previous chapter, the general concepts of a Partial-Wave Analysis (PWA) were introduced. To be able to perform a full PWA over the full available $p\bar{p}$ invariant mass range, one needs a proper model to describe all the relevant dynamics regarding possible intermediate resonances and final-state interactions (FSI). There are, for instance, multiple studies which show that the hadronic FSI in the $p\bar{p}$ system is significant [69,70,87–89]. In a 2012 BESIII study of $J/\psi \rightarrow \gamma p\bar{p}$, a mass-dependent PWA was performed on the invariant-mass range $M_{p\bar{p}} < 2.2$ GeV/$c^2$ [68]. However, this mass-dependent PWA was performed without accounting for FSI. This previous study was based on $225 \times 10^6$ $J/\psi$ events and focused solely on the near-threshold region. Nowadays, a total sample of $10^{10}$ $J/\psi$ events is available. The large statistics allows for a mass-independent PWA of the full $p\bar{p}$ invariant-mass range. Figure 6.1 shows the resulting $p\bar{p}$ invariant-mass spectrum for the present set of data, after reconstruction and selection, together with the two subtracted background models that were introduced in chapter 3.

The decay $J/\psi \rightarrow \gamma p\bar{p}$ is suitable for a mass-independent analysis, since intermediate resonances (and FSI) are expected to occur only in the $p\bar{p}$ system, and not in the $\gamma p$ or $\gamma\bar{p}$ system, as was concluded from figure 4.1 in chapter 4. For the present mass-independent approach, the data are divided into equally-sized bins of the invariant $p\bar{p}$ mass. Partial-wave fits for different hypotheses are performed for each mass bin individually. In these fits, the dynamical, or mass dependent, parts of the amplitudes are omitted from all hypotheses. The intensity of the contributions from different $J^{PC}$-partial waves is extracted for the best hypothesis in each bin. The resulting $p\bar{p}$ mass spectrum can give insights in intermediate resonances that are present. Note that this mass-independent approach is not a quantitative analysis. Hence, the underlying structures cannot be studied in detail. Resonance parameters, such as the mass and width, can thus not be extracted. However, a mass-independent PWA provides a qualitative analysis of the structures that show up in the $p\bar{p}$ mass spectrum.

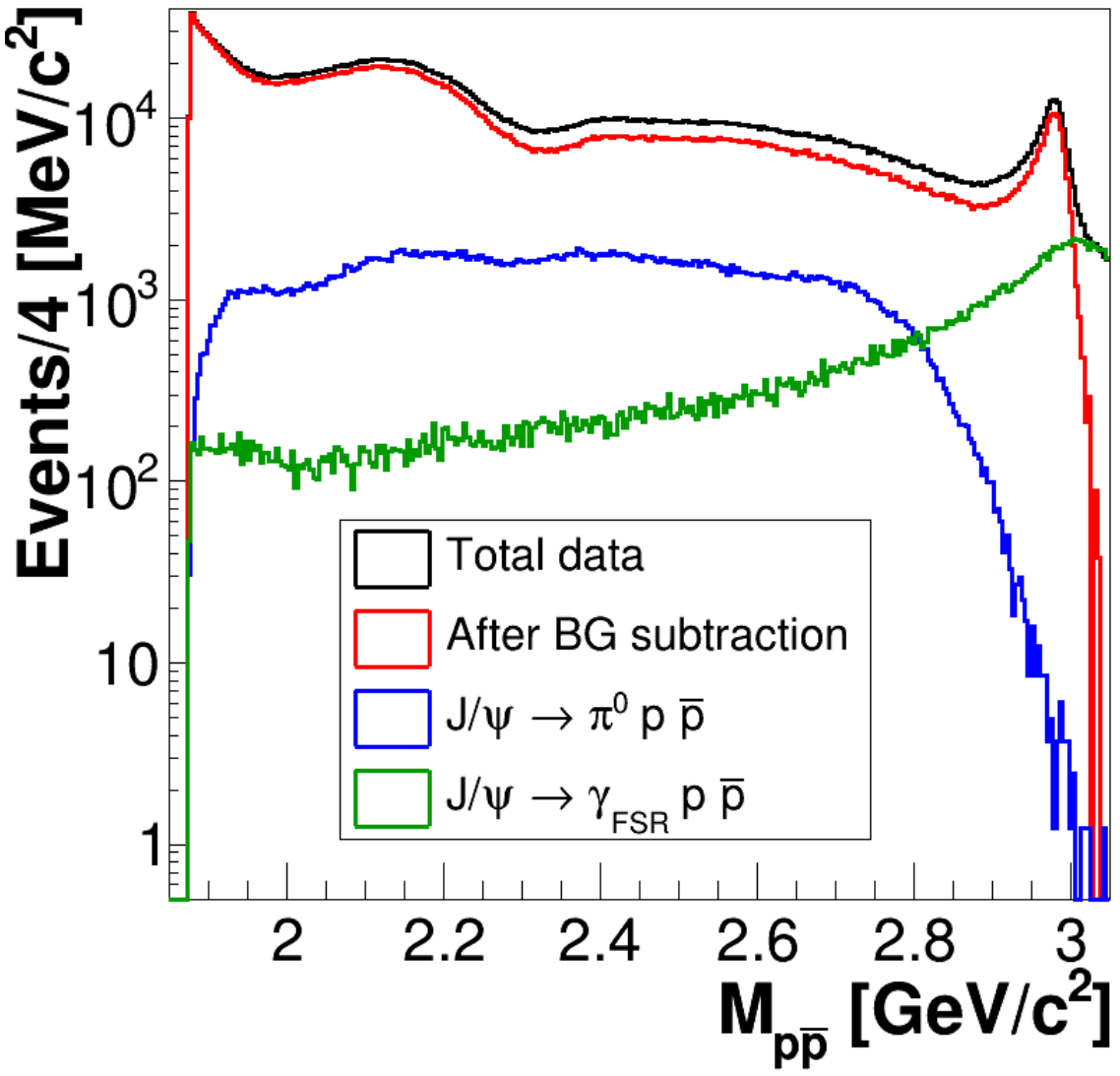

**Figure 6.1:** The resulting spectrum of the $p\bar{p}$ invariant-mass, extracted from BESIII data taken in 2009, 2012, 2017 and 2018, after reconstruction and selection. The black and red lines represent the data before and after background subtraction. The blue and green lines represent the two background models that are subtracted.

As explained in chapter 5, the detector acceptance and reconstruction efficiency are taken into account by using reconstructed phase-space-distributed MC events undergoing exactly the same selection criteria as applied for the data events. Generally, the number of MC events is chosen to be higher than the number of data events. However, the production-threshold region of the data contains strong enhancements, whereas the number of entries in this region are relatively low for a phase-space distributed MC sample. Therefore, additional MC events are generated in the production-threshold region, assuring more MC than data events in each bin along the full $p\bar{p}$ invariant-mass range.

## 6.1 The fitting procedure

The first step in the fitting procedure is determining the optimal size of the $p\bar{p}$ invariant-mass bins. To allow for the omission of the dynamical part of the amplitude, the bins should be chosen as fine as possible. On the other hand, it is meaningless to choose bins that are narrower than the detector resolution. Figure 3.7 in chapter 3, shows that the mass resolution is smaller than 3.8 MeV/$c^2$ over the full $p\bar{p}$ invariant-mass range. Another constraint on the minimum bin width is related to the number of data events, as each bin should contain enough events to be able to extract reasonable fit results. The minimum required number of events depends on the dimensionality of the problem and the number of free parameters in the fit. In our case, this can be of order $10^4$ events by following the arguments given in references [152–154]. With the amount of $J/\psi$ events available in our data sample, the statistics turned out not to be the limiting factor anymore. We decided, therefore, to take the width of the mass bins as 5 MeV/$c^2$, corresponding to 232 individual bins for the full mass range. This bin width is smaller than both the near $p\bar{p}$-production threshold and $\eta_c$ peaks, which are the narrowest structures visible in the $p\bar{p}$ invariant-mass spectrum. Hence, these structures can be studied with the mass-independent approach.

Secondly, one needs to decide about the different possible hypotheses to be included in the fitting procedure. In this analysis, it is decided to include all possible contributions with $L \leqslant 2$, which are listed in table 6.1. With these contributions, a total of 31 different hypotheses can be constructed, ranging from one, up to a maximum of five different contributions included in the hypothesis. The simplest hypothesis consists of just a $0^{-+}$ (or $0^{++}$) wave, leading to one single free fit parameter, as explained in section 5.4.1. For the most complex hypothesis, including all five different $J^{PC}$ contributions, the number of fit parameters increases to 21.

The analysis procedure based on the aforementioned decisions leads to a total of 7192 single fits to be performed. The hypotheses selection is based on the Akaike-Information-Criterion (AICc) and Bayesian-Information-Criterion (BIC), which were defined in chapter 5. For each mass bin, the hypothesis that gives the smallest sum of the AICc and BIC values is assigned as the best hypothesis. To prevent bias, each bin is analyzed individually, without imposing any condition concerning the continuity of the solution in the $p\bar{p}$ invariant-mass.

With a toy MC sample, corresponding to the $J/\psi \to \gamma\eta_c, \eta_c \to p\bar{p}$ channel and based upon the JPE model, the fitting method was tested. The JPE model is specifically constructed for vector decays into a photon and a pseudoscalar meson, compatible with the reaction of interest $J/\psi \to \gamma\eta_c$. The subsequent decay, $\eta_c \to p\bar{p}$, is described by a phase-space model. In this MC study, the assigned hypothesis in each mass bin coincided with the expected hypothesis consisting of a single $0^{-+}$ contribution. All other spin-parity assignments, except the $0^{++}$, could be statistically excluded. However, it was noted that the hypothesis consisting of a single $0^{++}$ contribution gives exactly the same AICc and BIC values in every fit. The ambiguity arises because the only difference between the two amplitudes is a minus sign. Since only the differential cross-sections, corresponding to the absolute square of the amplitudes, are experimentally accessible, the hypothesis consisting of a single $0^{-+}$ contribution is not distinguishable from the hypothesis containing just a $0^{++}$ contribution. As soon as any other contribution is added to the hypothesis, the sign difference results into a relative effect, due to interference, leading to distinct AICc and BIC values.

| $J^{PC}$ of $X$ | $X$-production | $X$-decay |
|---|---|---|
| $0^{-+}$ | M1 | $(0, 0)$ |
| $0^{++}$ | E1 | $(1, 1)$ |
| $1^{++}$ | E1, M2 | $(1, 1)$ |
| $2^{-+}$ | M1, E2, M3 | $(2, 0)$ |
| $2^{++}$ | E1, M2, E3 | $(1, 1)$<br>$(3, 1)$ |

**Table 6.1:** Possible $(L, S)$ and combinations of multipoles for different accessible $J^{PC}$ quantum numbers of resonance $X$ in $J/\psi \to \gamma X, X \to p\bar{p}$.

After the MC study, all PWA fits were performed on the data sample obtained from the BESIII data taken in 2009, 2012, 2017 and 2018. The data consist of preprocessed events after reconstruction, selection and background subtraction as described in chapter 3. In figures 6.2–6.4, the one-dimensional fit projections of the best hypotheses for an arbitrary bin are shown. These results are obtained for a bin within the invariant-mass interval of 2500 MeV/$c^2$ $< M_{p\bar{p}} <$ 2505 MeV/$c^2$, which contains 13792 events. All distributions show a good agreement between the data and the fit. These one-dimensional distributions are used to extract a general pa-rameter representing the goodness of fit, given by

$$\overline{\chi^2/n} = \frac{1}{N_{\text{hist}}} \sum_{i=1}^{N_{\text{hist}}} \frac{\chi_i^2}{n_{i,\text{bin}}}, \tag{6.1}$$

where $\chi_i^2$ is given by the general description

$$\chi_i^2 = \sum_{j=1}^{n_{i,\text{bin}}} \text{pull}_j^2. \tag{6.2}$$

Here, $n_{i,\text{bin}}$ is the number of nonzero bins in each one-dimensional histogram and $N_{\text{hist}}$ the number of histograms. The term 'pull$_j$' is defined as $\text{pull}_j = (n_{\text{dat}} - n_{\text{fit}})/\sqrt{\sigma_{\text{dat}}^2 + \sigma_{\text{fit}}^2}$, where $n_{\text{dat}}$ and $n_{\text{fit}}$ represent the bin contents of the data and fit histograms $j$, respectively, and $\sigma_{\text{dat}}$ and $\sigma_{\text{fit}}$ the corresponding bin (statistical) errors. To achieve a more reliable result, only histograms for which at least 10% of the bins contained entries were considered as input for the goodness-of-fit parameter. With this requirement, the two histograms corresponding to $\cos\theta_\gamma^{\gamma p}$ and $\cos\theta_\gamma^{\gamma\bar{p}}$ are excluded for the 11 lowest $p\bar{p}$ invariant-mass bins. For the other 221 mass bins, all ten histograms, such as those depicted in figures 6.2–6.4, are included. For the presented $p\bar{p}$ invariant-mass bin, the resulting value of the quality of the fit is $\overline{\chi^2/n} = 1.041$.

From the fit result, the intensity of the contributions of the different $J^{PC}$-partial waves can be extracted. For the exemplary $p\bar{p}$ invariant-mass bin, the best hypothesis contains a major $0^{-+}$ contribution, together with a small $1^{++}$ contribution. In a similar fashion, the intensity of all the invariant-mass bins, and thus the full $p\bar{p}$ invariant-mass range, can be extracted. The results in the presented bin are obtained by selecting the best hypothesis. Aside from solely selecting the best hypothesis, a more extensive study has been performed, as will be introduced in the following section.

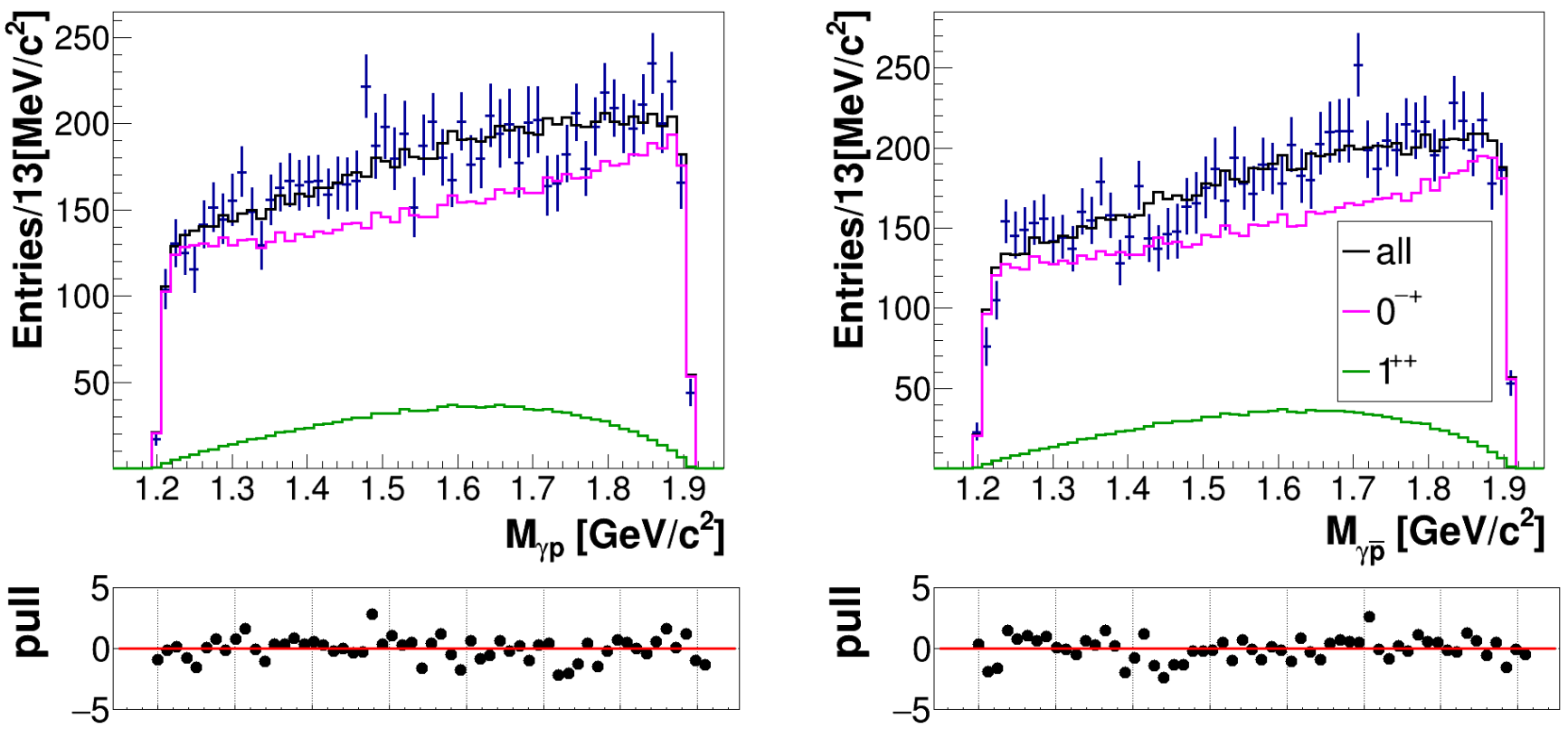

**Figure 6.2:** The one-dimensional invariant-mass projections for the selected hypothesis, and the $p\bar{p}$ invariant-mass bin 2500 − 2505 MeV/$c^2$. The total fit is shown in black, and the different contributions of the $0^{-+}$ and $1^{++}$ are shown by the magenta and green lines, respectively. The pull distributions represent pull $= (n_{\rm dat} - n_{\rm fit})/\sqrt{\sigma^2_{\rm dat} + \sigma^2_{\rm fit}}$, where $n_{\rm dat}$ and $n_{\rm fit}$ represent the bin contents of the data and fit histograms, respectively, and $\sigma_{\rm dat}$ and $\sigma_{\rm fit}$ the corresponding bin (statistical) errors. In general, $\sigma_{\rm fit}$ is much smaller than $\sigma_{\rm dat}$. Interference contributions are not explicitly shown.

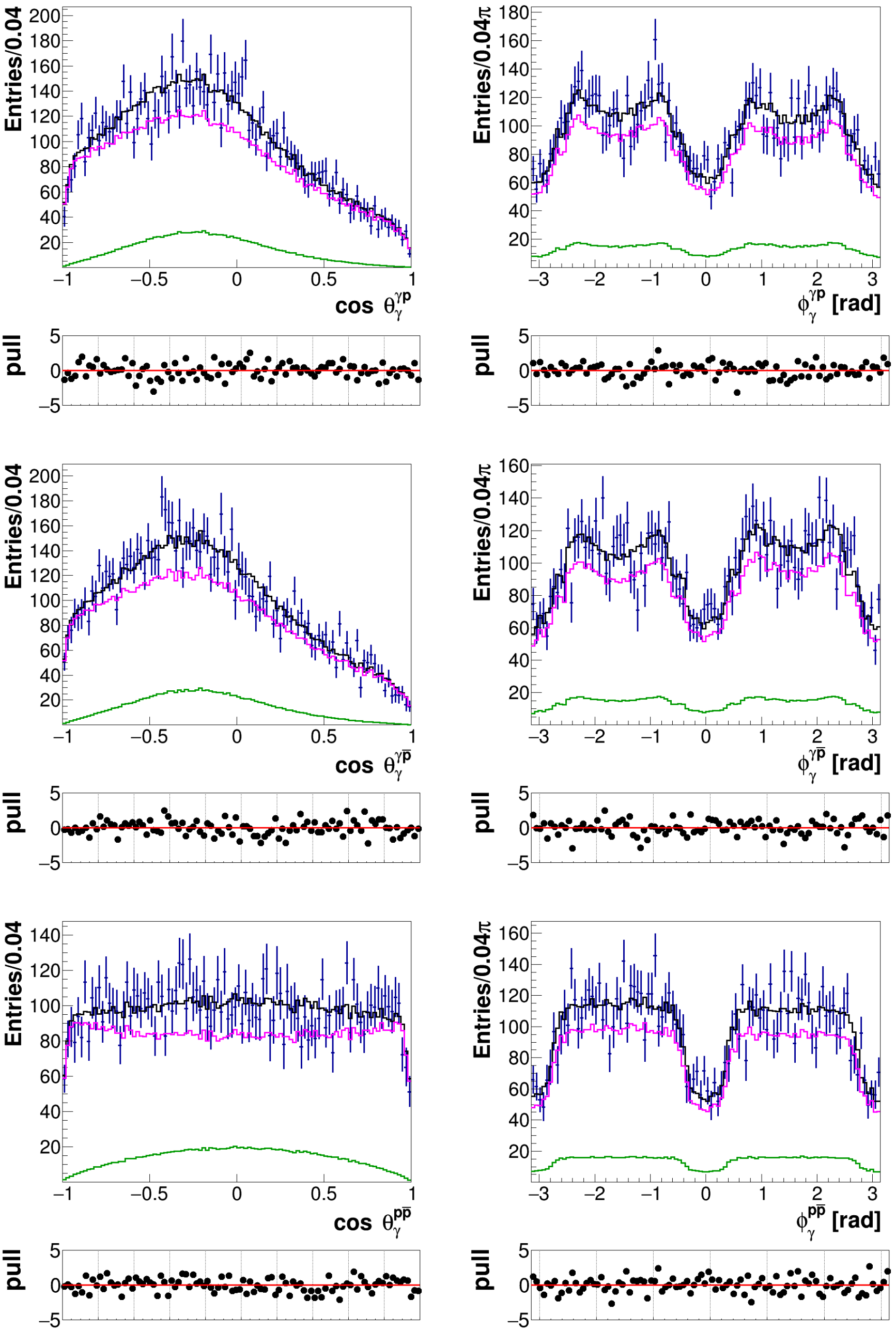

**Figure 6.3:** Same as figure 6.2, but for the angular distributions. Here, $\theta_b^a$ and $\phi_b^a$ represent the polar and azimuthal helicity angles, as defined in chapter 5, of particle $b$ in the rest frame of $a$.

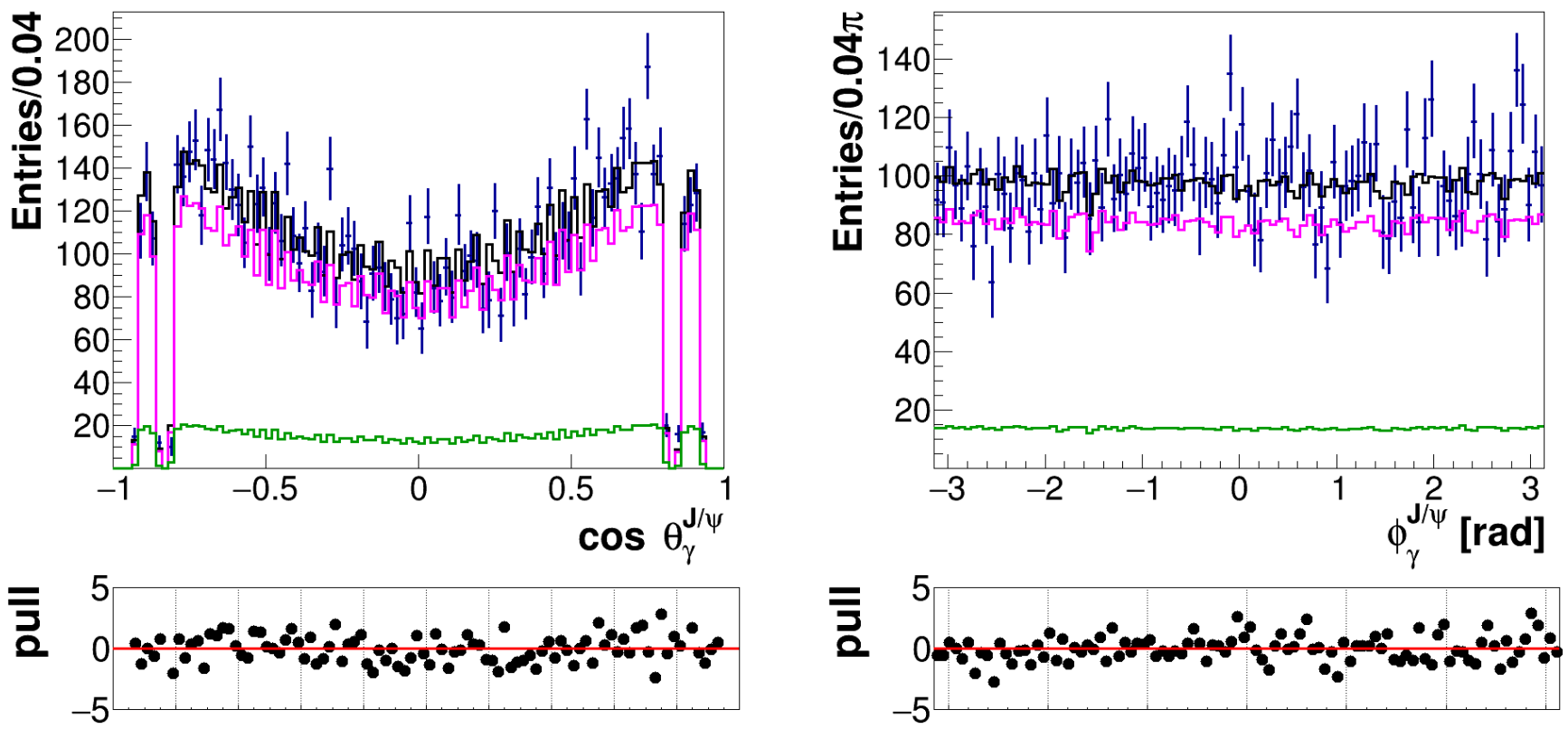

**Figure 6.4:** Follow up of figure 6.3.

## 6.2 Using weights and fixing higher-order multipoles

As explained in chapter 5, the value of the AICc and BIC itself has no meaning, and only the relative difference ($\Delta_i$) is an important measure. The best hypothesis will give the smallest AICc ($\text{AICc}_{min}$) or BIC value ($\text{BIC}_{min}$). Instead of selecting one single best hypothesis for each invariant-mass bin, the AICc and BIC values can be used to assign a weight $w_i$ for the best few hypotheses $i$ via [145, 146]

$$w_i = \frac{\exp(-\Delta_i/2)}{\sum_{i=1}^{H} \exp(-\Delta_i/2)}, \tag{6.3}$$

with $H$ the number of hypotheses to be considered. In this analysis, the sum of the AICc and BIC values are used. Therefore, $\Delta_i$ is defined as

$$\Delta_i = (\text{AICc}_i + \text{BIC}_i) - (\text{AICc}_{min} + \text{BIC}_{min}). \tag{6.4}$$

Aside from including multiple hypotheses with corresponding weights, we have looked at the effects of fixing additional fit parameters. For the five possible $J^{PC}$ combinations that are considered in this analysis, three combinations allow for more than one multipole transition; see table 6.1. Generally, when several multipole transitions can occur, the lowest one is dominant. Thus, the E1 and M1 transitions are expected to be the most important ones. Therefore, a study was done on how the results in each bin are affected when the magnitude of one or more of the 5 higher-order multipoles was fixed to zero. Fixing additional parameters results in a smaller number of free fit parameters. Since both the AICc and BIC criterion contain a penalty term on the number of free fit parameters, fixing parameters related to higher-order multipoles might lead to larger contributions of one or more of the $J^{PC}$ combinations $1^{++}$, $2^{-+}$ and $2^{++}$ in the final result.

All possible combinations of fixing one, up to five, of the magnitudes of the higher-order multipoles leads to another 62176 single fits for the full invariant-mass range. Note that, for this procedure, only hypotheses that contain the specific fixed parameter(s) have to be fitted. The results of these additional fits will be treated as extra available hypotheses in each of the mass bins. This results in 268 additional hypotheses, and thus, together with the aforementioned 31, a total of $268 + 31 = 299$ hypotheses in each bin.

The discussed options with the weights, the fixing of additional parameters, and a combination of the two, have been incorporated in this analysis.

## 6.3 Results and discussion

The resulting data sample was fitted for the 31 hypotheses in all available bins of the full $p\bar{p}$ invariant mass range, running from 1875 MeV/$c^2$ up to 3035 MeV/$c^2$. Subsequently, for each bin, the hypotheses are ranked based on the AICc+BIC criterion, and the intensity of the contributions are extracted for the two different scenarios:

**1a.** Solely selecting the best hypothesis in each bin;
**1b.** Use weights $w_i$ for the best three hypotheses in each bin.

First, for the best hypothesis in each bin, the goodness of the fits was studied with the parameter $\overline{\chi^2/n}$, as defined in equation 6.1. In figure 6.5, the resulting value for each bin is plotted. It is observed that for $p\bar{p}$ invariant-masses of about 2.2 GeV/$c^2$ and higher, the selected fits provide in general a good description of the data, whereas for lower masses, the extracted values show a systematic and significant deviation from the expected value of one. This deviation is probably not related to the $\pi^0$ background, as figure 6.1 demonstrates that the fraction of the $\pi^0$ background is smaller in this range than for the range of $p\bar{p}$ invariant masses between 2.2 GeV/$c^2$ and ∼2.8 GeV/$c^2$. The higher values of $\overline{\chi^2/n}$ in the lower-mass area might indicate that the assumed isobar model does not provide a proper description in this area. It was noted that the higher $\overline{\chi^2/n}$ values were often caused by relatively large deviations in the histograms related to the $\gamma p$ and $\gamma\bar{p}$ invariant masses. In figures 6.6 and 6.7, these histograms are shown for two different $p\bar{p}$ invariant-mass bins. In the assumed isobar model, resonances in the $\gamma p$ and $\gamma\bar{p}$ system are not taken into account. Discrepancies in the $\gamma p$ and $\gamma\bar{p}$ invariant masses might hint towards a presence of a $N^*$ resonance. Another reason for a deviation from the isobar model might, for instance, be related to $X \rightarrow p\bar{p}$ decays of resonances $X$ that couple to channels with other final states, such as $K\bar{K}$ and $\phi\phi$. In figure 6.5, the $\phi\phi$ production threshold is indicated by the vertical dashed line. Note that the $K\bar{K}$ production threshold lies well below the $p\bar{p}$ production threshold. A coupled-channel analysis could provide a better description of the low $p\bar{p}$ invariant-mass range.

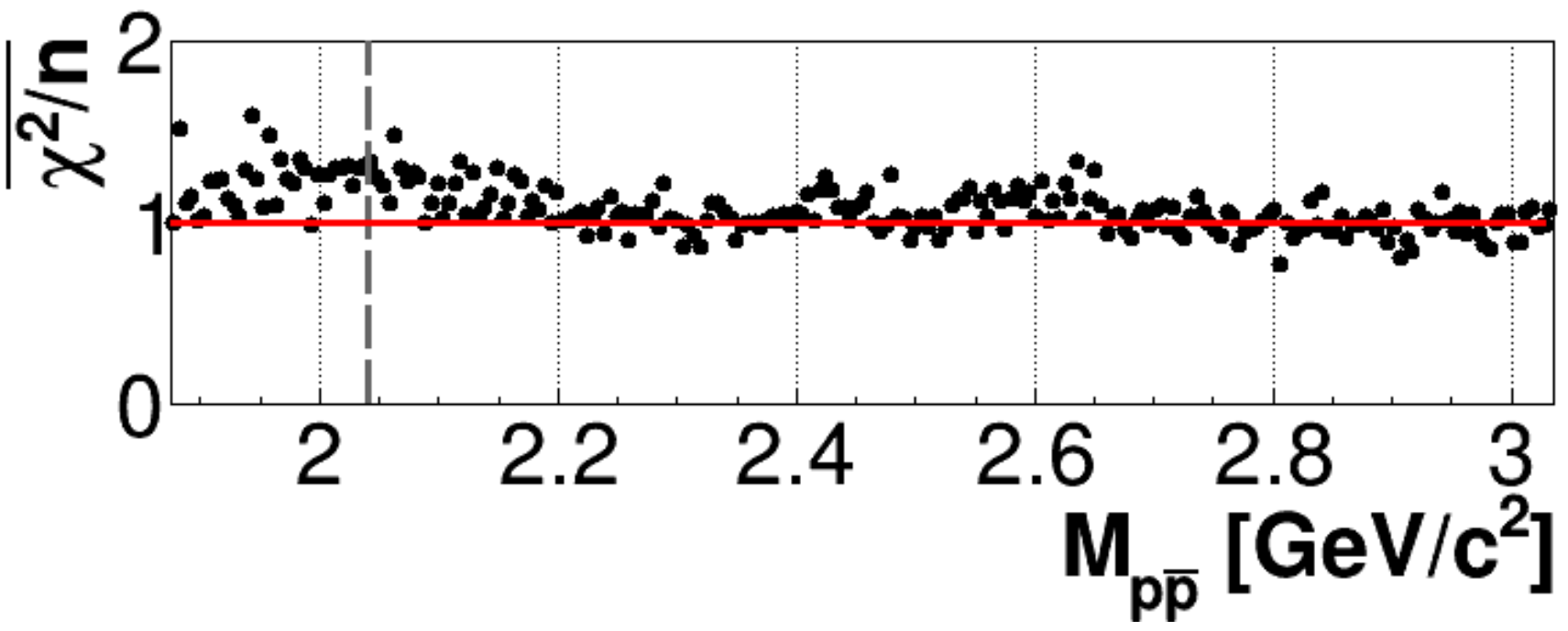

**Figure 6.5:** Resulting distribution of the goodness-of-fit parameter $\overline{\chi^2/n}$, defined in equation 6.1, obtained for each $p\bar{p}$ invariant-mass bin. The vertical gray dashed line indicates the $\phi\phi$ threshold.

Secondly, for the selected hypotheses, the intensity of the different contributions are extracted for scenarios 1a and 1b. In figures 6.8 and 6.9, the resulting contributions of the five different waves are shown for the full $p\bar{p}$ invariant-mass spectrum. In a similar fashion, the intensities are extracted for the two additional scenarios:

- **2a.** Fix higher-order multipoles to zero and select the best hypothesis;
- **2b.** Fix higher-order multipoles to zero and use weights $w_i$ for the best three hypotheses.

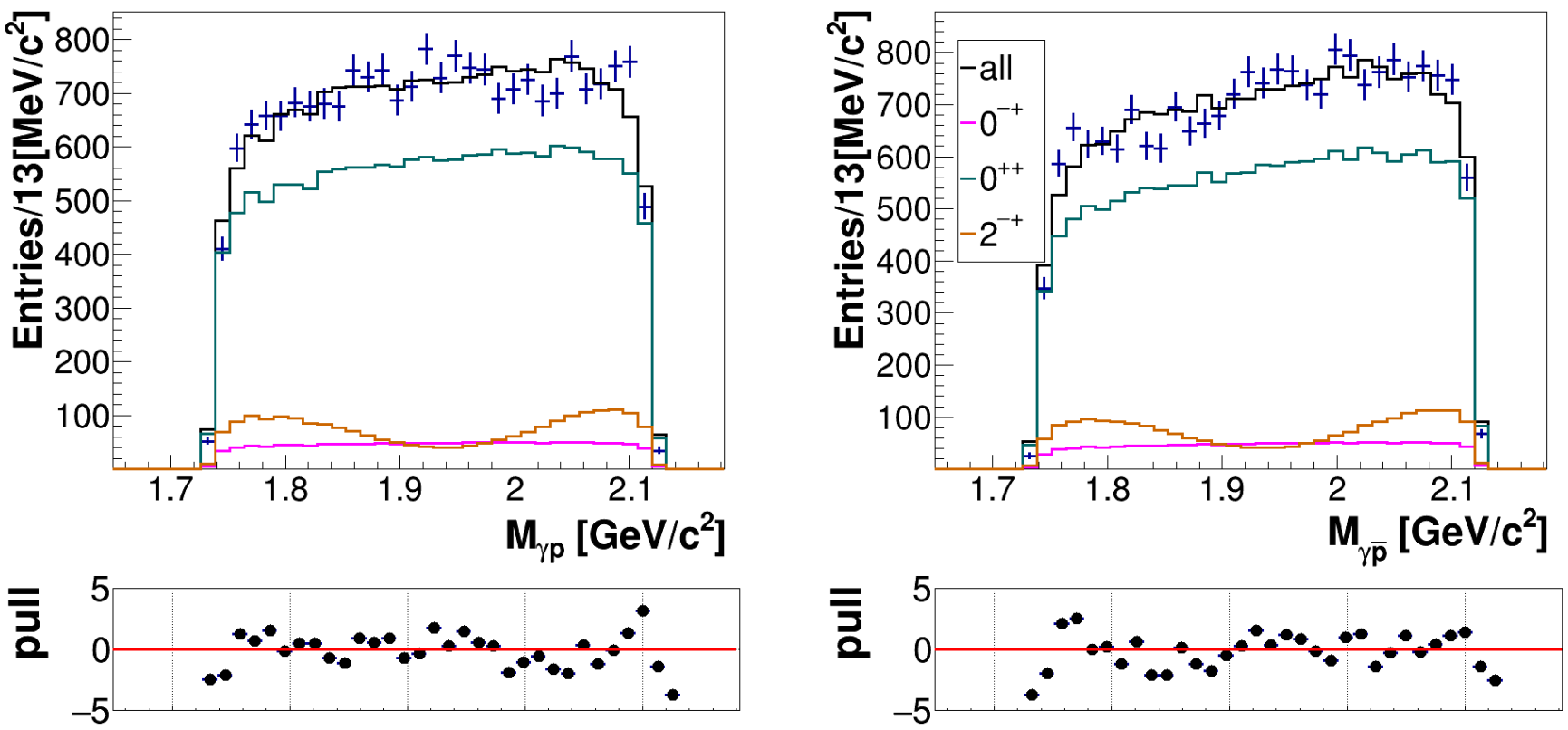

**Figure 6.6:** The one-dimensional invariant-mass projections for the selected hypothesis, and the $p\bar{p}$ invariant-mass bin 1955 – 1960 MeV/$c^2$. The total fit is shown in black, and the different contributions of the $0^{-+}$, $0^{++}$ and $2^{-+}$ are shown by the magenta, teal and brown lines, respectively. The pull distributions represent pull $= (n_{\text{dat}} - n_{\text{fit}})/\sqrt{\sigma^2_{\text{dat}} + \sigma^2_{\text{fit}}}$, where $n_{\text{dat}}$ and $n_{\text{fit}}$ represent the bin contents of the data and fit histograms, respectively, and $\sigma_{\text{dat}}$ and $\sigma_{\text{fit}}$ the corresponding bin (statistical) errors. In general, $\sigma_{\text{fit}}$ is much smaller than $\sigma_{\text{dat}}$. Interference contributions are not explicitly shown.

The resulting contributions of the two scenarios are shown in figures 6.10 and 6.11. Note that the two hypotheses consisting of a single $0^{-+}$ or $0^{++}$ wave give exactly the same AICc and BIC value. If these two hypotheses are assigned as the best hypothesis in the bin, the intensity in this bin is assigned to the $0^{-+}$ histogram. Additionally, to get a clearer overview, the hypothesis consisting of only the $0^{++}$ wave is not considered in the weighing procedure.

What immediately stands out in figures 6.8–6.11 are the bin-to-bin fluctuations. To check whether this might be related to statistics, the fitting procedure was repeated with bin widths of 10 MeV/$c^2$, and thus about twice as much events per bin. Additionally, to test the dependency on the start parameters of the fits, the fitting process was repeated with the final fit parameters of the neighboring bin used as start parameters for the fits in each bin. The resulting figures are shown in appendix A. Both the results from a change in statistics and the different start parameters still contain bin-to-bin fluctuations with similar patterns. Therefore, the fluctuations are probably related to ambiguities in selecting the hypotheses. Thus, the mass-independent approach does not provide enough information to distinguish between the different possible hypotheses. It is observed that the fluctuations are stronger in figures 6.10 and 6.11, where more hypotheses are available in each bin. This observation supports the assumption that the fluctuations are related to ambiguities. The ambiguities appear to be the strongest between hypotheses containing large fractions of either $0^{-+}$ or $0^{++}$. In appendix A, the one-dimensional fit projections of two neighboring bins are compared. One cause of ambiguities lies in the fact that the channel $J/\psi \to \gamma p\bar{p}$ has stable final-state hadrons with a nonzero spin. Information on the spin-helicities of the hadrons can thus not be experimentally extracted. This is contrary to, for instance, $J/\psi \to \gamma\omega\omega$, where the $\omega$'s will decay, providing information on their helicities [128].

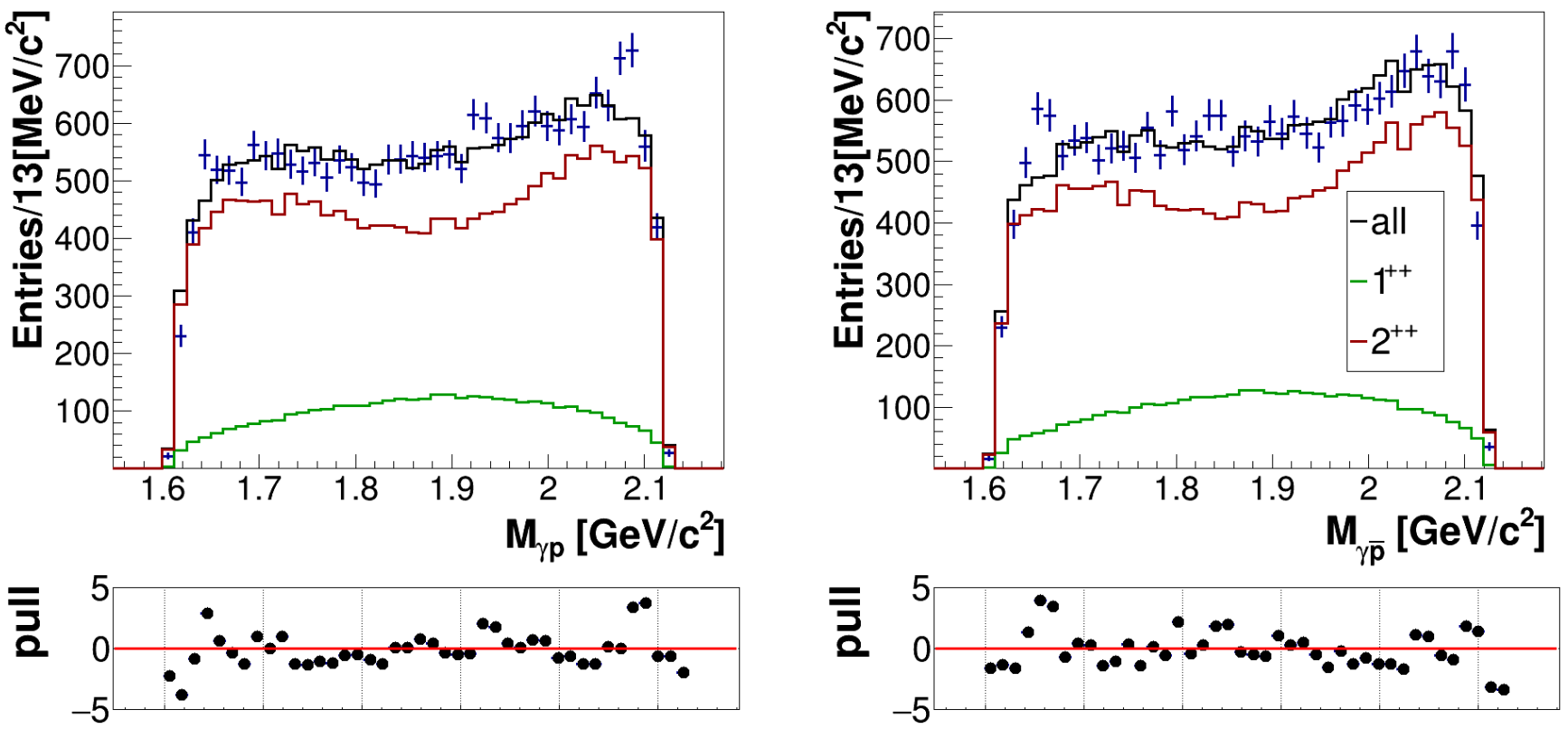

**Figure 6.7:** Same as 6.6, but for the $p\bar{p}$ invariant-mass bin 2060 − 2065 MeV/$c^2$, with the $1^{++}$ and $2^{++}$ contributions shown in green and dark-red, respectively.

In spite of the bin-to-bin fluctuations, the overall global structures observed in the resulting figures is always the same. Thus, fixing higher-order multipoles does not result in a significantly larger contribution of the $1^{++}$, $2^{-+}$ or $2^{++}$ partial waves. In all results, the strong enhancement at the $p\bar{p}$-production threshold is in accordance with quantum numbers $0^{-+}$, as extracted in previous experiments [15, 68]. Similarly, the peak around 2.98 GeV/$c^2$ is compatible with a $0^{-+}$ resonance, as one would expect for the $\eta_c$ meson [15]. Furthermore, the enhancement around 2.0 – 2.2 GeV/$c^2$ seems to be a complex mixture of various resonances. The mixture contains a significant $2^{++}$ contribution, which coincides with the resonance $f_2(2010)$. According to the Particle Data Group (PDG), the state $f_2(2010)$ is an established particle with a width of about 200 MeV [15]. This state has been observed to decay to $\phi\phi$ and $K\bar{K}$ [15, 155]. Thus far, a decay into $p\bar{p}$ has not been observed. The results of this analysis hint towards a significant $f_2(2010)$. Therefore, it would be interesting to study the sequential decay $J/\psi \to \gamma f_2(2010), f_2(2010) \to p\bar{p}$ and try to extract both branching fractions $\mathcal{B}(J/\psi \to \gamma f_2(2010))$ and $\mathcal{B}(f_2(2010) \to p\bar{p})$ for the first time. In the same range, there is a clear $0^{++}$ contribution present. Possible scalar states that can represent this contribution are $f_0(2020)$, $f_0(2100)$ and $f_0(2200)$, which all have widths of a few hundred MeV [15]. Additionally, a few bins around 2.1 GeV/$c^2$ reveal a large $1^{++}$ contribution. The PDG lists the poorly-established state $a_1(1930)$ with a width of about 150 MeV, which could explain the behavior of the $1^{++}$ contribution. However, since just a few bins show a large contribution, it is also possible that the intensity was wrongly attributed to this wave due to the difficulties in separating the three $J^{++}$ contributions. On the other hand, the $1^{++}$ peak appears to be more prominent in the result of the procedure using bin widths of 10 MeV/$c^2$.

Finally, some concluding remarks. This analysis is called mass-independent due to the omission of the dynamical part of the amplitude. However, the study it not fully model-independent as it is based on the assumption that the decay $J/\psi \to \gamma p\bar{p}$ can be described properly by an isobar model. Additionally, it is assumed that there are no resonances present in both the $M_{\gamma p}$ and $M_{\gamma\bar{p}}$ invariant-mass spectra and that the structures in $M_{p\bar{p}}$ can be described by a combination of partial waves with $L \leqslant 2$. In chapter 4, it is demonstrated that the data indeed do not reveal a resonance in either the $M_{\gamma p}$ or $M_{\gamma\bar{p}}$ spectrum. Nonetheless, in figure 6.5 it was observed that the range $M_{p\bar{p}} < 2.2$ GeV/$c^2$ give systematically worse values for the goodness-of-fit parameter. The complex mixture of resonances that is present in the range $2.0 < M_{p\bar{p}} < 2.2$ GeV/$c^2$ probably couples to other final states and can thus not be described by the isobar model. Additionally, it could be that in this mixture, there is an intermediate resonance present with $L > 2$. Furthermore, it was found that it sometimes is hard to distinguish between the different partial waves if one zooms in on a small $M_{p\bar{p}}$ range. This can be explained, at least partially, by the three final states with nonzero spin.

Together with the two possible spin-projections for the initial state, this results in $2^4 = 16$ coherent sums for the calculation of the probability density function $w(\vec{x}, \theta)$, see chapter 5. Lastly, the mass-independent approach can only reveal which partial waves are present in the different structures observed in a spectrum. To extract quantitative numbers, a mass-dependent PWA is necessary. Such a mass-dependent PWA of the range containing the $\eta_c$ meson will be presented in the next chapter.

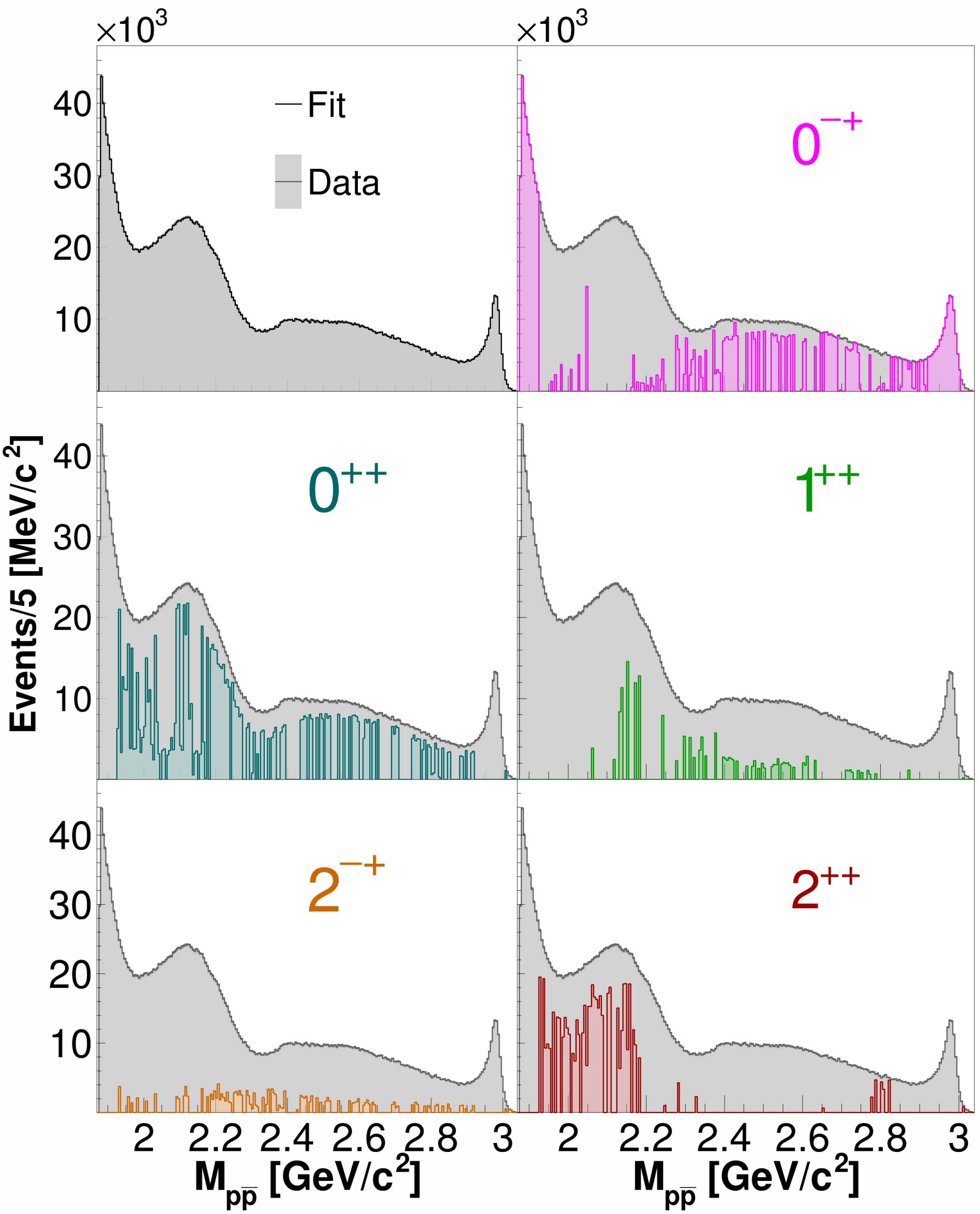

**Figure 6.8:** Extracted intensities of the different partial waves for the best hypothesis in each bin (scenario 1a).

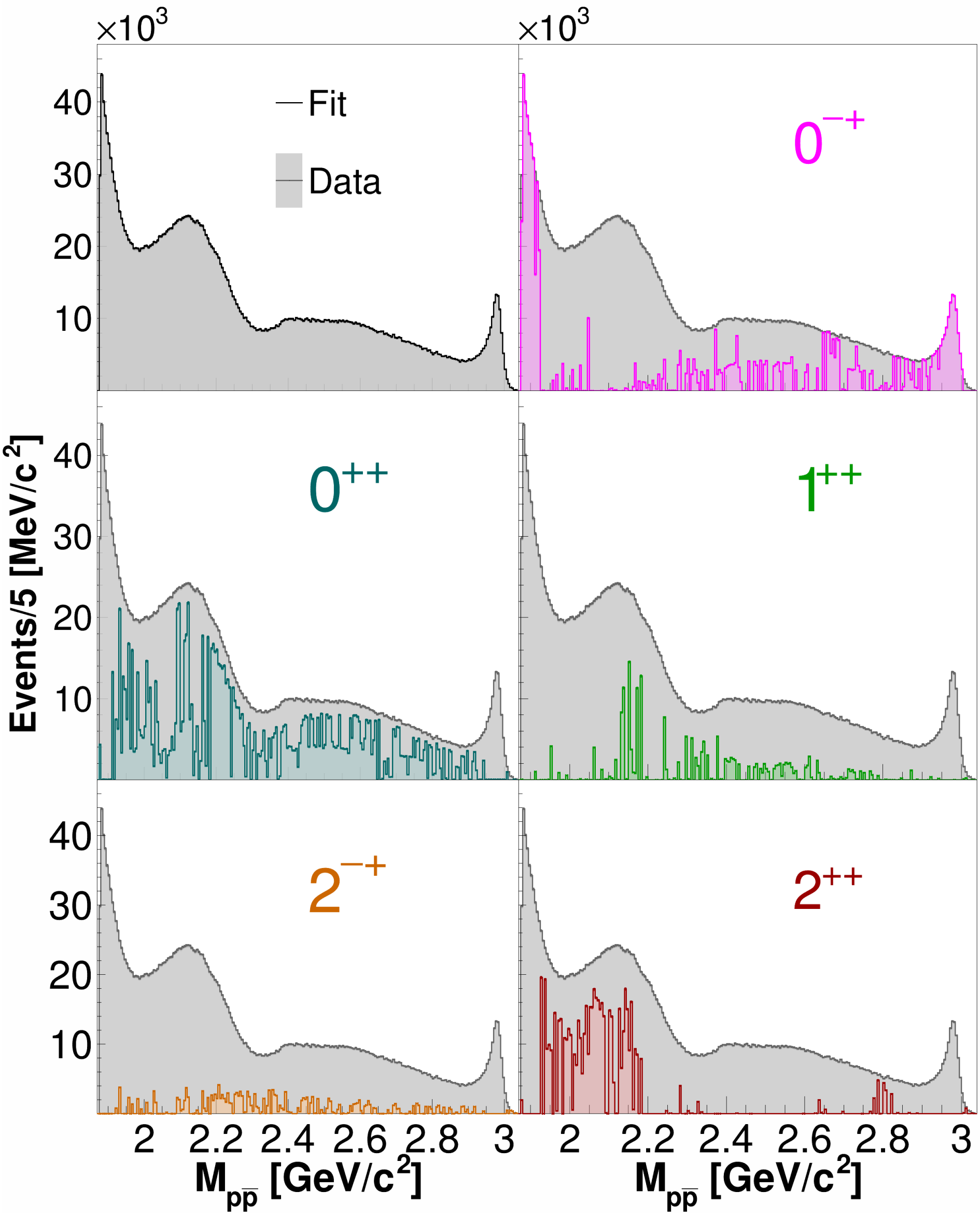

**Figure 6.9:** Extracted intensities of the different partial waves in each bin after weighing the best three hypotheses (scenario 1b).

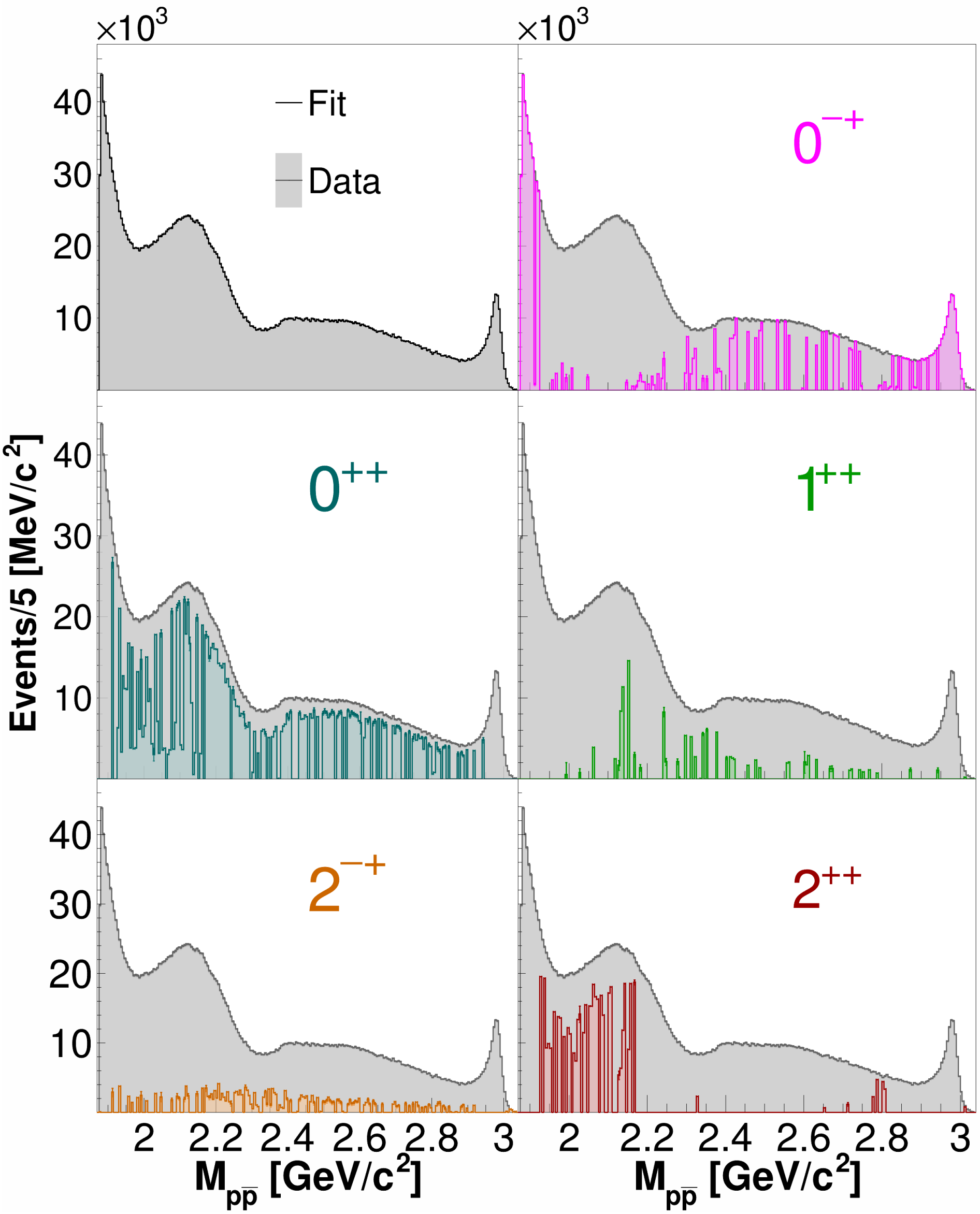

**Figure 6.10:** Extracted intensities of the different partial waves in each bin, after extending the available hypotheses with the fixing of higher-order multipoles to zero (scenario 2a).

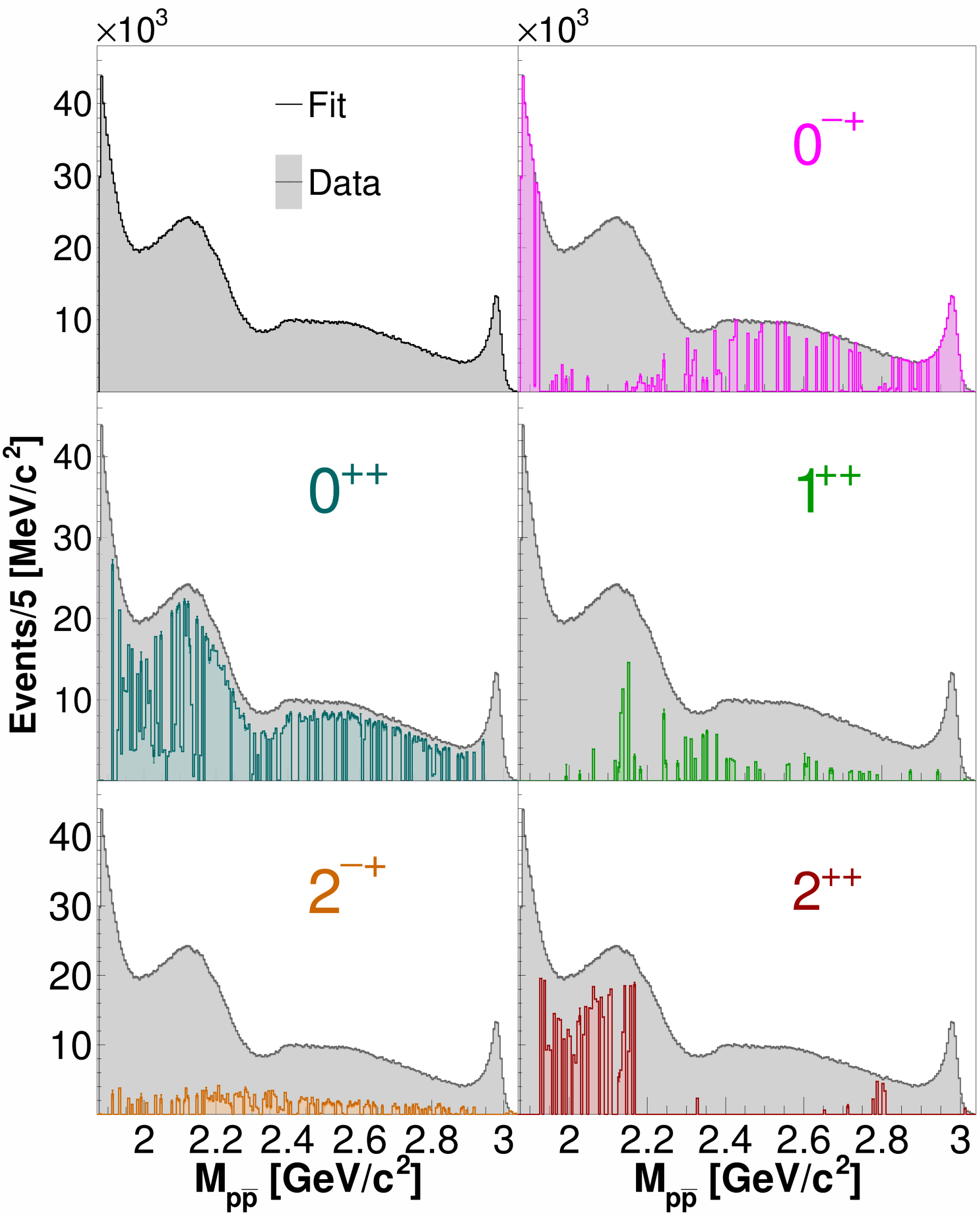

**Figure 6.11:** Extracted intensities of the different partial waves in each bin, after extending the available hypotheses with the fixing of higher-order multipoles to zero, followed by weighing the best three hypotheses (scenario 2b).

# 7. Mass-dependent Partial-Wave Analysis $\eta_c$ range

Previous studies of $J/\psi \to \gamma\eta_c$ found that interference effects play a significant role for the extraction of the basic resonance parameters of the $\eta_c$ and the radiative transition rate $J/\psi \to \gamma\eta_c$, as discussed in detail in section 1.4. In previous studies of $J/\psi \to \gamma\eta_c$, the mass, the width, and the branching fraction were obtained via a one-dimensional fit of the proton-antiproton invariant-mass spectrum, including a possible interference with a non-resonant background. Such a fit gives two possible phases, leading to a destructive and constructive interference effect, with the same fit quality. Particularly, this leads to an ambiguity in the extraction of the branching fraction. In this work, a Partial-Wave Analysis (PWA) of the decay $J/\psi \to \gamma p\bar{p}$ is performed with the aim to unambiguously extract the $J/\psi \to \gamma\eta_c$ branching fraction. The relevant concepts involved in this PWA were discussed in chapter 5. Here, for the study of the $\eta_c$ resonance, the PWA concepts will be applied to the $p\bar{p}$ invariant masses larger than 2.7 GeV/$c^2$. In the following sections, the PWA procedure and results will be discussed.

## 7.1 PWA procedure

The starting point of a mass-dependent PWA is defining a reasonable base hypothesis. The base hypothesis will be varied by adding or removing possible contributions. If an additional contribution results in an improvement of the fit, the contribution will be added to the hypothesis, otherwise it will be removed. The procedure will then be repeated with the extended hypothesis, until there are no further improvements of the fit.

To define a base hypothesis for $J/\psi \to \gamma p\bar{p}$, events that fall within the $p\bar{p}$ invariant mass range in the vicinity of the $\eta_c$ mass are used as input. To be precise, we selected events for which the $p\bar{p}$ invariant mass is larger than 2.7 GeV/$c^2$. This range is dominated by a clear resonance structure of the $\eta_c$ meson and a continuum non-resonance-like contribution, as can be seen in figure 7.1. Furthermore, aside from the $\eta_c$ meson, there are no intermediate resonances listed in the relevant mass range by the particle data group (PDG) [15]. The base hypothesis in this analysis will thus naturally include the $\eta_c$ contribution with quantum numbers $0^{-+}$, described by the KEDR lineshape as discussed in section 5.3.2. It is evident that a hypothesis containing just the $\eta_c$ contribution will provide a bad description of the data, as the data clearly shows contributions that go beyond the $\eta_c$ resonance. Since it is, a priori, not clear what quantum numbers the continuum part should have, the hypothesis is extended by each of the possible contributions with angular momentum $J \leqslant 2$, leading to five different base hypotheses. The results of this first analysis step are summarized in table 7.1, based on the two selection criteria, Akaike-Information-Criterion (AICc) and Bayesian-Information-Criterion (BIC), as defined in chapter 5, and their sum. It is worth repeating here that the value of the AICc and BIC itself has no meaning and that one should consider the relative differences. The best hypothesis will give the smallest AICc or BIC value, and the difference between hypotheses is considered statistically significant if the change in AICc or BIC is larger than 10, corresponding to a probability smaller than 0.01 [145,156]. Naively, one might question the importance of a difference of 10, when the absolute AICc and BIC values are large. However, large absolute AICc and BIC values contain large scaling constants, while the differences are free of such constants [145]. Hence, only these differences are interpretable as the strength of statistical evidence.

| **Hypothesis:** $\eta_c + \ldots$ | $0^{-+}$ | $0^{++}$ | $1^{++}$ | $2^{-+}$ | $2^{++}$ |
|---|---|---|---|---|---|
| $n_{dof}$ | 5 | 4 | 6 | 9 | 11 |
| AICc | −19950.9 | −19334.0 | −17795.6 | −16470.0 | −18999.7 |
| BIC | −19906.7 | −19298.6 | −17742.6 | −16390.5 | −18902.5 |
| AICc+BIC | −39857.6 | −38632.6 | −35538.2 | −32860.5 | −37902.2 |

**Table 7.1:** Fit results for the given hypotheses, with $n_{dof}$ the number of degrees-of-freedom of the fit.

Both the AICc and BIC agree on the best extension of the hypothesis by adding a $0^{-+}$ wave. All other extensions give a worse fit that can be seen visually by inspecting, for instance, the one-dimensional fit projections, as demonstrated in figure 7.2 for the $p\bar{p}$ invariant-mass projection of the second-best ($\eta_c + 0^{++}$, 7.2a) versus the best option ($\eta_c + 0^{-+}$, 7.2b). Here, the fit result with $\eta_c + 0^{++}$ clearly differs from the data towards lower $p\bar{p}$ invariant-masses. With just the hypothesis $\eta_c + 0^{-+}$, all one-dimensional fit projections are in good agreement with the data, as shown in detail in appendix B.

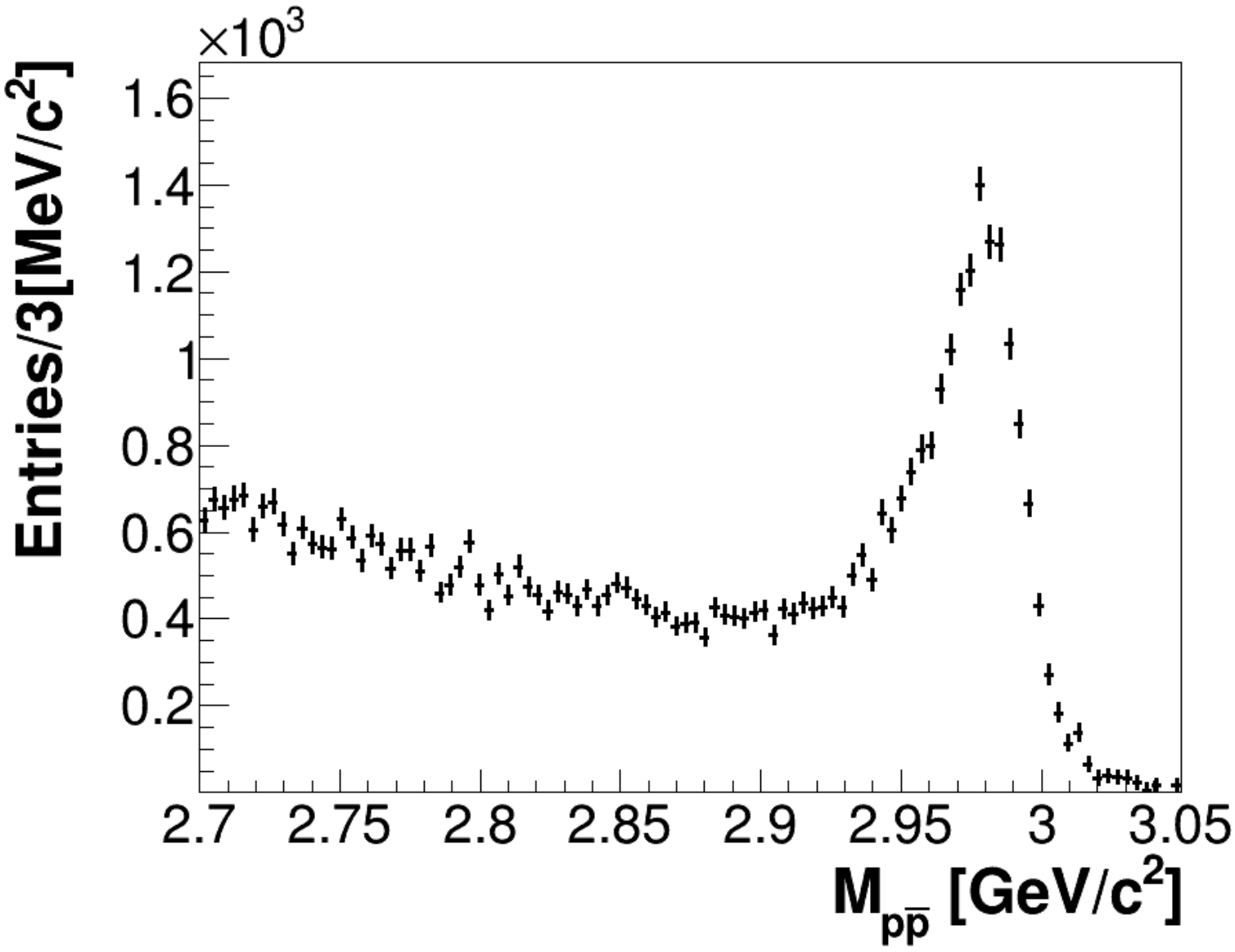

**Figure 7.1:** The resulting $p\bar{p}$ invariant-mass distribution obtained for BESIII data taken in 2009 and 2012, after reconstruction, selection and background subtraction.

| **Hypothesis:** $\eta_c + 0^{-+} + \ldots$ | $0^{++}$ | $1^{++}$ | $2^{-+}$ | $2^{++}$ |
|---|---|---|---|---|
| $n_{dof}$ | 6 | 10 | 11 | 13 |
| AICc | −20044.3 | −20121.9 | −20080.4 | −20188.7 |
| BIC | −19991.3 | −20033.5 | −19983.2 | −20073.9 |
| AICc+BIC | −40035.6 | −40155.4 | −40063.6 | −40262.6 |

**Table 7.2:** Fit results for the given hypotheses, with $n_{dof}$ the numbers of degrees-of-freedom of the fit.

In table 7.2, the results of including one more additional wave are summarized. From these extensions, the hypothesis $\eta_c + 0^{-+} + 2^{++}$ is the significant best based on the AICc and BIC values. Remarkably, none of the one-dimensional fit projections shows any visible improvement by the addition of the $2^{++}$ contribution. Similarly, the Dalitz spectrum of the pull distributions shows no significant improvement, as presented in figure 7.3. In figure 7.2, the $p\bar{p}$ invariant-mass projection is shown with (7.2c) and without (7.2b) the addition of the $2^{++}$ contribution. All one-dimensional projections are compared in appendix B. It is likely that the difference between the hypotheses lies in correlations of the different observables that are not covered by a Dalitz spectrum. By plotting one-dimensional fit projections, all correlations get lost. Effectively, an $n$-dimensional problem is reduced to $n$ one-dimensional problems by solely comparing the fit projections. On the other hand, the AICc and BIC are designed to select the best hypothesis based purely on statistics. The implied assumption is that there are no systematic effects present. Hence, assuming that the background subtractions and the interactions with the detector are very well understood, with uncertainties much smaller than statistical effects, which is not true. Systematic effects can, for instance, be caused by over- or under-estimations and corrections for subtracting the background, or systematic discrepancies related to the description of the interactions of low-energy photons with the detector. Further studies with other purely-statistical selection criteria, such as the likelihood-ratio test, will just assign the same best hypothesis as the AICc and BIC, and will give no further insights in possible systematic effects.

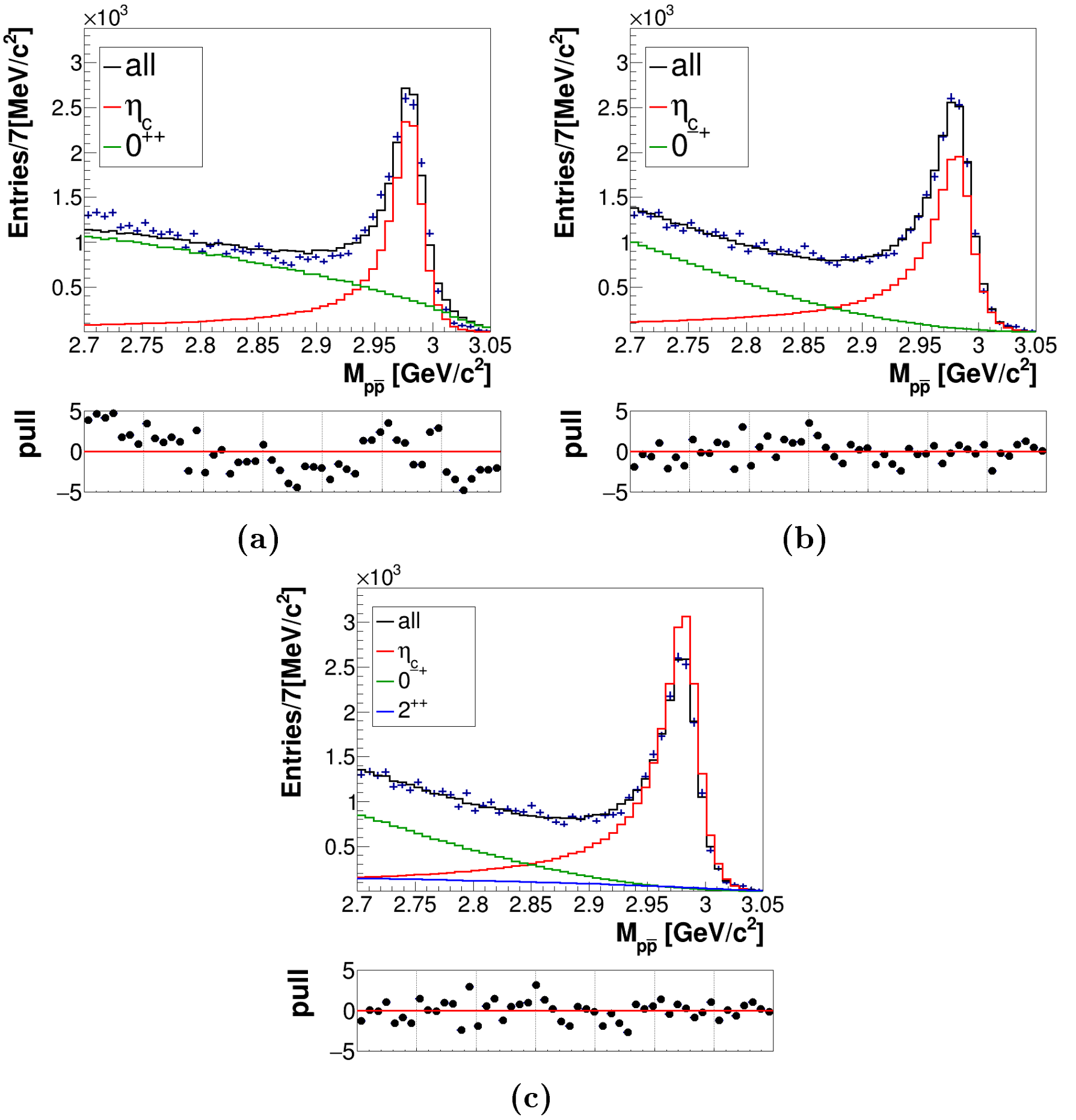

**Figure 7.2:** The $p\bar{p}$ invariant-mass projection for different hypotheses. The total fit is shown in black, and the different contributions of the $\eta_c$, $0^{\pm+}$ and $2^{++}$ are shown by the red, green and blue lines, respectively. The pull distributions represent pull $= (n_{\text{dat}} - n_{\text{fit}})/\sqrt{\sigma^2_{\text{dat}} + \sigma^2_{\text{fit}}}$, where $n_{\text{dat}}$ and $n_{\text{fit}}$ represent the bin contents of the data and fit histogram, respectively, and $\sigma_{\text{dat}}$ and $\sigma_{\text{fit}}$ the corresponding bin (statistical) errors. In general, $\sigma_{\text{fit}}$ is much smaller than $\sigma_{\text{dat}}$. Interference contributions are not explicitly shown.

The systematic uncertainties are considered to be strongly influenced by low-energy photons, corresponding to large $p\bar{p}$-invariant masses. Here, uncertainties arise due to photon clusters related to beam-background or other non-physical effects. Selection and subtraction steps are implemented with the purpose of removing the non-relevant photon clusters from the data. However, these procedures are not perfect, resulting in relatively large uncertainties for low photon energies. In chapter 4, it was, for instance, demonstrated that the number of remaining events having a photon energy smaller than about 80 MeV is amplified by the efficiency correction. Hence, it was studied how the fits were affected when events were excluded with $p\bar{p}$-invariant-masses larger than 3 GeV/$c^2$. Most striking are the variations in the $\eta_c$ yield while varying the mass range, as demonstrated in table 7.3. For the hypothesis $\eta_c + 0^{-+}$, the $\eta_c$ yield becomes a bit smaller for a slightly smaller fit range, as one would expect, since the acceptance reduces with a tighter range. However, for the hypothesis $\eta_c + 0^{-+} + 2^{++}$, the $\eta_c$ yield fluctuates strongly, and becomes about $\sim$1.5 times smaller or larger if the fit range is extended by just 10 MeV$/c^2$. The hypothesis $\eta_c + 0^{-+} + 2^{++}$ thus gives unreliable results for the $\eta_c$ yield, and thus for the calculation of the branching fraction $J/\psi \to \gamma\eta_c$. Additionally, the mass-independent analysis presented in chapter 6 does not show a contribution of $2^{++}$ in the $p\bar{p}$ invariant mass range of 2.7 GeV$/c^2$ and higher. These results, together with the fact that $\eta_c$+$0^{-+}$ is the simplest hypothesis, led us decide to choose the fit with hypothesis $\eta_c + 0^{-+}$, and the fit range $2.70 - 3.00$ GeV$/c^2$, as the nominal description.

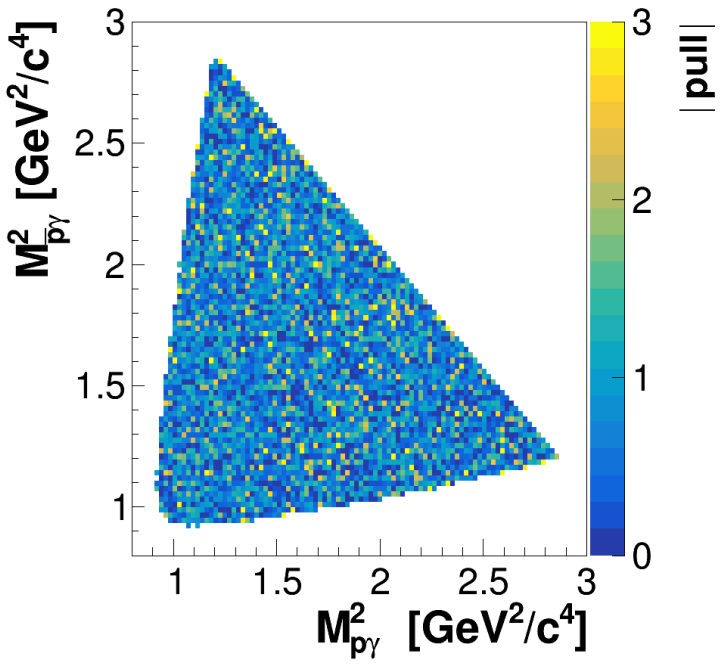

(a) $\eta_c + 0^{-+}$

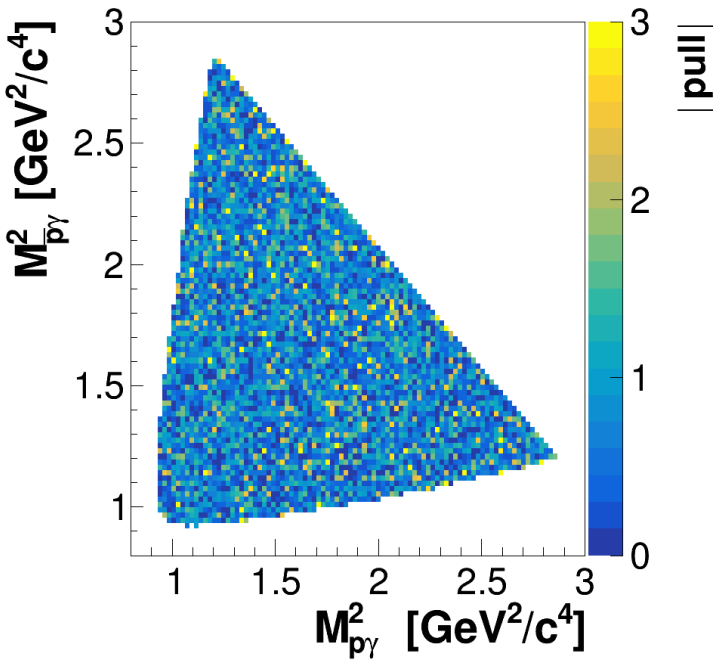

(b) $\eta_c + 0^{-+} + 2^{++}$

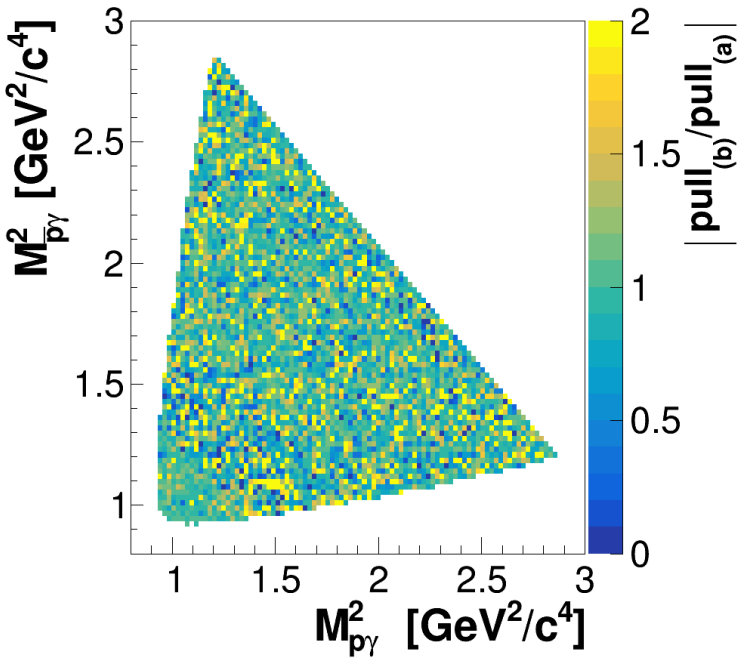

(c)

**Figure 7.3:** The two upper panels show the Dalitz spectra of the difference between the data and fit for two different hypotheses. The distributions are filled via pull $= (n_{\text{dat}} - n_{\text{fit}})/\sqrt{\sigma^2_{\text{dat}} + \sigma^2_{\text{fit}}}$, where $n_{\text{dat}}$ and $n_{\text{fit}}$ represent the bin contents of the data and fit Dalitz histogram, respectively, and $\sigma_{\text{dat}}$ and $\sigma_{\text{fit}}$ the corresponding bin (statistical) errors. The bottom panel shows the result of dividing the two Dalitz spectra.

| Hypothesis $\eta_c + 0^{-+} + \ldots$ | Mass: 2.7−... [GeV/$c^2$] | 3.10 | 3.02 | 3.01 | 3.00 |
|---|---|---|---|---|---|
| - | $\eta_c$ yield | 20166 | 19637 | 19358 | 18925 |
| | AICc | −19950.9 | −19392.2 | −19418.8 | −19200.9 |
| | BIC | −19906.7 | −19348.1 | −19374.6 | −19156.8 |
| $0^{++}$ | $\eta_c$ yield | 19099 | 19174 | 19107 | 18425 |
| | AICc | −20044.3 | −19435.5 | −19417.0 | −19331.7 |
| | BIC | −19991.3 | −19488.5 | −19364.3 | −19278.8 |
| $1^{++}$ | $\eta_c$ yield | 19311 | 19238 | 28202 | 18673 |
| | AICc | −20121.9 | −19557.9 | −19480.5 | −19387.6 |
| | BIC | −20033.5 | −19469.6 | −19392.3 | −19299.4 |
| $2^{-+}$ | $\eta_c$ yield | 20155 | 19735 | 19387 | 18928 |
| | AICc | −20080.4 | −19567.0 | −19470.4 | −19373.0 |
| | BIC | −19983.2 | −19469.9 | −19373.4 | −19276.0 |
| $2^{++}$ | $\eta_c$ yield | 29279 | 18681 | 27688 | 18431 |
| | AICc | −20188.7 | −19596.8 | −19532.6 | −19323.3 |
| | BIC | −20073.9 | −19482.0 | −19417.8 | −19208.8 |

**Table 7.3:** The $\eta_c$ yields obtained from the fits of different $p\bar{p}$ invariant-mass ranges for the listed hypotheses. Note that the BIC and AICc values cannot be compared for the different mass ranges, since the values are related to the number of events in the data sample.

Note that only the effects of the extension with a $2^{++}$ contribution are discussed here, since this extension gives the best AICc and BIC values. The three other hypotheses from table 7.2 also give better AICc and BIC values than the hypothesis $\eta_c + 0^{-+}$. As can be seen in table 7.3, similar to the $2^{++}$ extension, the addition of a $1^{++}$ contribution gives an unstable fit. Only the extension consisting of a $2^{-+}$ contribution shows the expected behavior of a smaller yield for a tighter range. Therefore, the two strongly unstable hypotheses, $\eta_c + 0^{-+} + 1^{++}$ and $\eta_c + 0^{-+} + 2^{++}$, are rejected for further studies. Since the $\eta_c$ yield fluctuations for the addition of a $0^{++}$ contribution are relatively small, this hypothesis, together with the stable hypothesis $\eta_c + 0^{-+} + 2^{-+}$, will be used to estimate part of the systematic uncertainty. As the systematic effects are not taken into account when calculating AICc or BIC, these criteria are less useful in the choice of the best hypothesis. Hence, in this analysis, the AICc and BIC do not have a decisive role in the selection of the partial waves.

## 7.2 Extracting the efficiency

Besides the $\eta_c$ yield, the reconstruction and selection efficiency is needed to calculate the $J/\psi \to \gamma\eta_c$ branching fraction. The efficiency can be obtained directly from the partial-wave-analysis software, resulting in the most sophisticated description of the efficiency in all dimensions of the phase space. This procedure makes use of two MC samples. The first consists of phase-space distributed events for which the detector response is simulated and the same reconstruction and selection algorithms are applied as for the data. The second MC sample contains all the generated phase-space distributed events, without any form of reconstruction or selection, the so-called MC-truth. For the extraction of efficiency $\epsilon$, both samples are weighted by the amplitude model obtained from the best fit:

$$\epsilon = \frac{\sum_{i=1}^{N_{rec}} w(\vec{x}_i, \hat{\theta})}{\sum_{i=1}^{N_{gen}} w(\vec{x}_i, \hat{\theta})}. \tag{7.1}$$

Here, $N_{rec}$ and $N_{gen}$ are the number of events in the reconstructed and in the total generated MC samples. The amplitude model is described by the weight function $w(\vec{x}, \hat{\theta})$, as introduced in section 5.4.1, with $\hat{\theta}$ the set of amplitudes determined by the best fit. The result of the best fit provides the correct weight for each point in phase-space $\vec{x}_i$.

For the nominal result, the extracted overall efficiency is equal to $\epsilon = 55.9\%$. Several one-dimensional efficiency corrected distributions are shown in figure 7.4, all one-dimensional corrected distributions can be found in appendix C. A beneficial effect of the extraction of the efficiency based on the multi-dimensional fit result is that even regions with acceptance gaps can be covered. The characteristics of a detector can produce gaps in one-dimensional distributions as a results of zero acceptance. For BESIII, there are, for instance, gaps visible in the distributions of the photon production angle, due to space between the barrel and endcaps of the EMC detector (see chapter 2). The discussed efficiency extraction allows for an extrapolation to these gap regions, as demonstrated in figure 7.4b for the production angle. Similarly, the efficiency corrected distribution goes beyond the applied cut of 3.0 GeV/$c^2$ on the $p\bar{p}$ invariant mass, as visible in figure 7.4a. Additionally, the total value of the different fit parameters can be calculated after efficiency correction. This could, for example, be utilized to find the contribution of one type of multipole transition when several multipole transitions are allowed. The efficiency dips in 7.4c are due to the requirement that the angle between a charged track and a cluster in the EMC should be larger than 10°.

For the determination of the total systematic uncertainty, different fits are performed, as will be described in the next section. For all fits, the efficiency was individually determined using the method described above.

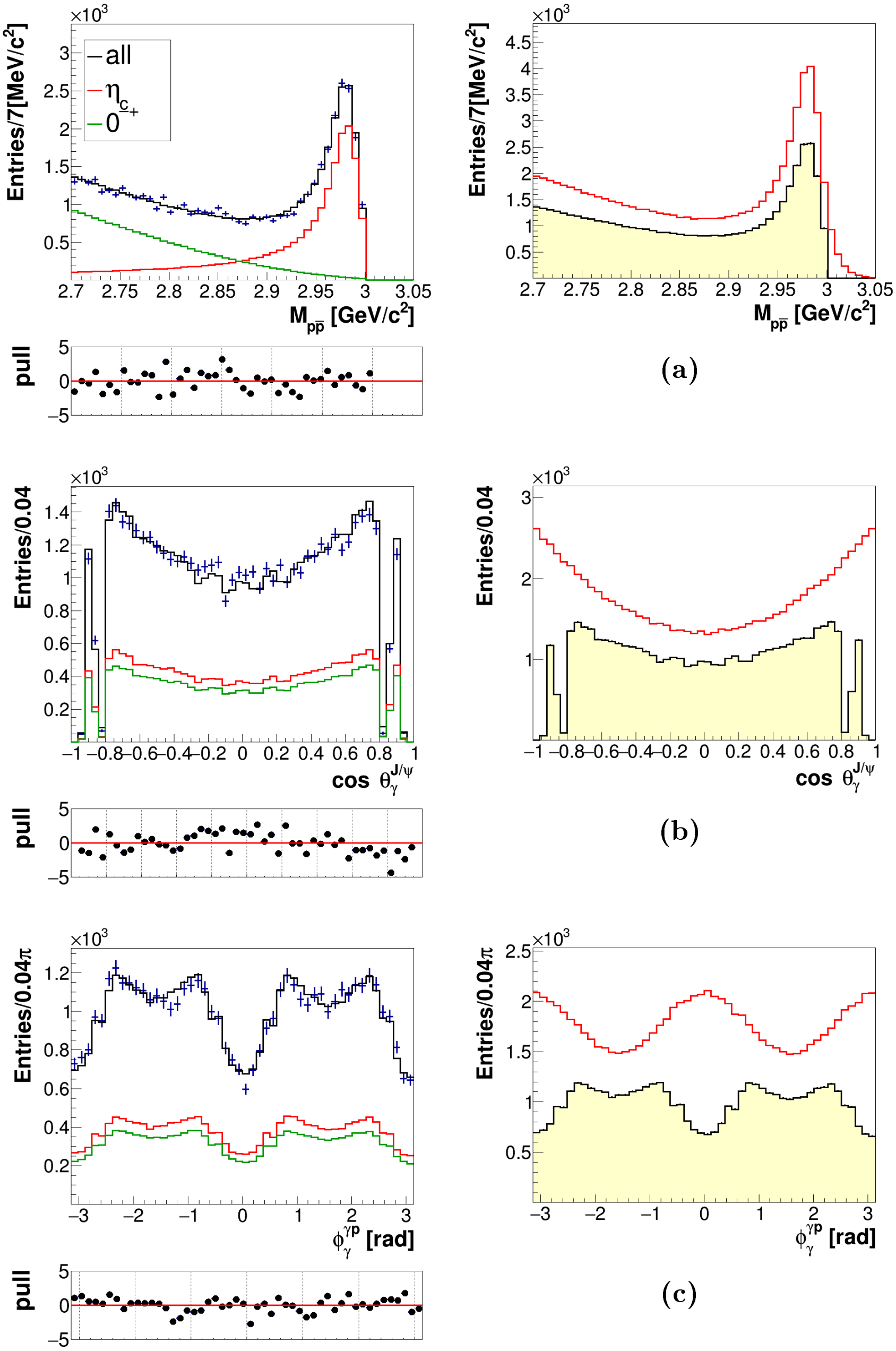

**Figure 7.4:** The nominal fit results (left), with the lines and pull distributions as explained in figure 7.2, and $\theta_b^a$ and $\phi_b^a$ (the polar and azimuthal helicity angles) defined for particle $b$ in the rest frame of $a$. The right panels show the efficiency-corrected distributions (red) together with the fit results (black).

## 7.3 Systematic error analysis

For the determination of the branching fraction, the extracted number of reconstructed $\eta_c$ events, $N_{\eta_c}$, has to be corrected with the previously determined efficiency $\epsilon$. All intermediate resonances and their branching fractions must also be considered. Using the number of $J/\psi$ events present in the data sample, external branching fractions, the efficiency calculated using the amplitude model and the yield of the fitted $\eta_c$ signal, the branching fraction can be derived via

$$\mathcal{B}(J/\psi \to \eta_c) = \frac{N_{\eta_c}}{\epsilon N_{J/\psi}\mathcal{B}(\eta_c \to p\bar{p})}, \tag{7.2}$$

with the nominal values given as $N_{\eta_c} = 18925$, $\epsilon = 55.9\%$, $N_{J/\psi} = 1310.6 \times 10^6$, and $\mathcal{B}(\eta_c \to p\bar{p}) = 1.44 \times 10^{-3}$ [15]. All the parameters in the extraction of the branching fraction introduce systematic uncertainties. Some of these uncertainties are relevant for the extraction of the mass and width of the $\eta_c$ as well. The systematic uncertainties arising due to the experimental characteristics of BESIII have been evaluated extensively by the collaboration and will be summarized briefly. Additional systematic uncertainties arise due to the specific selection criteria and model choices applied in the present analysis. In the following, the determination of the different types of systematic uncertainties will be discussed. The effect of these uncertainties on the extraction of the $\eta_c$ mass, width and branching fraction are presented in table 7.4.

### 7.3.1 Photon and charged-track detection

The general procedure in finding the photon and charged-track detection efficiencies starts with choosing an appropriate control sample. The relevant efficiencies are then determined for both the data of the control sample, and data from MC simulations. The differences in the efficiencies between the recorded data and the MC events are then considered as the uncertainties of the efficiencies.

The photon detection efficiency of BESIII was studied with a control sample of the initial state radiation process $e^+e^- \to \gamma\mu^+\mu^-$ at center-of-mass energies corresponding to the $J/\psi$ and $\psi(3770)$ resonances [113]. The difference in the photon detection efficiency of the data and the MC sample is determined to be 1%. Since there is just one photon to be detected in this analysis, that results in a total uncertainty of 1%.

The BESIII uncertainties of tracking and PID efficiencies for a proton or antiproton were investigated using the almost background-free control sample $J/\psi \to p\bar{p}\pi^+\pi^-$ [112,122]. For both the tracking and PID efficiency, an uncertainty of 1% per track was estimated. Since a proton and an antiproton need to be detected, an uncertainty of 2% is used for both tracking and PID.

### 7.3.2 Selection criteria

In this analysis, three additional channel-specific selection criteria were imposed: a maximum value for the $\chi^2_{4C}$, $M^2_{miss}$ and $\theta_{p\bar{p}}$ above which events are discarded; see section 3.2.2. The values used were found by optimizing the significance $\frac{S}{\sqrt{S+B}}$. To estimate the systematic error, the stepsizes used in the optimization process are applied to alter the maximum limits of the cuts. The implemented stepsizes are $\chi^2_{4C} = 0.5$, $M^2_{miss} = 0.001$ GeV$^2/c^4$ and $\theta_{p\bar{p}} = 0.1°$. For each of the three cuts, the fit is repeated twice, with the maximum limit altered to the maximum plus, or minus, the stepsize. The largest deviation is used as an estimate of the systematic error.

### 7.3.3 Background subtraction

To estimate the errors from the background subtraction of the $J/\psi \to p\bar{p}$ sample, the fit is repeated twice whereby we alter the weight factor by plus or minus the uncertainty of the $J/\psi \to p\bar{p}$ branching fraction, the largest deviation is included as an uncertainty. Additionally, the model-dependent effect is estimated by repeating the procedure with a MC sample generated with a generic phase-space model of $J/\psi \to p\bar{p}$, instead of the J2BB1 model. For the background subtraction of the data-driven $\pi^0$ model, the model dependence is estimated by repeating the procedure with bins that are a factor 2 narrower, so ~0.006 GeV$^2/c^4$ in both directions, instead of 0.012 GeV$^2/c^4$ bins. To estimate the statistical fluctuations, the data-driven $\pi^0$ model was constructed from a fraction of the 2018 $J/\psi$ data, with roughly the same events as the data sample used for the nominal model. Furthermore, the weight factor is changed by plus or minus 1%, to include the uncertainties related to the detection of the extra photon from

$J/\psi \to \pi^0 p\bar{p}, \pi^0 \to \gamma\gamma$.

### 7.3.4 Choice of hypothesis

The results obtained with the hypotheses containing an additional $0^{++}$ or $2^{-+}$ contribution are used to estimate the uncertainties related to the choice of hypothesis. The results are extracted for the nominal-fit range $2.70 - 3.00$ GeV/$c^2$. For all three extracted parameters, the $0^{++}$ addition gives the largest deviations of the two additions. It is worth noting that the other possible extensions ($1^{++}$ and $2^{++}$) give lower or comparable deviations.

### 7.3.5 Mass resolution

The mass resolution is taken into account by using the results of the nominal fit to generate a MC sample. This sample is exposed to the detector response simulation, and the same reconstruction and selection criteria as the data. Finally, the PWA fit is repeated. The results from this fit give the uncertainties in the $\eta_c$ mass, width and yield.

### 7.3.6 $\eta_c$ line shape

As discussed in section 5.3.2, both the $\eta_c$ line shape model from KEDR and CLEO-c are empirical models. Since there is no theoretical basis to choose one over the other, a PWA with the CLEO-c model was performed to estimate the uncertainties related to the line shape.

### 7.3.7 Fit range

The effects of the fit range are estimated by changing the fit range. For the high-mass limit, the fit is repeated for the full range of $2.70 - 3.10$ GeV/$c^2$. The lower-mass limit is changed by $\pm 10$ MeV/$c^2$, the largest deviation is included as the systematic error.

### 7.3.8 Angle between the photon and (anti)proton in the EMC

In chapter 3, the general BESIII selection criteria are discussed. One such criterion is a requirement that the angle between an EMC cluster and any charged-particle track should be larger than 10°. The purpose is to eliminate showers related to the charged track. However, in figure 7.4c it is observed that this requirement strongly affects the variation of the efficiency in one of the decay angles. To account for possible model dependencies induced by this criterion, the full analysis was repeated with the 10° requirement replaced by a 20° requirement. Here, 20° is chosen, since this number is used in several other BESIII studies where antiprotons are present. The differences between the acquired results are incorporated in the systematic error estimation.

### 7.3.9 Total number of $J/\psi$ and external branching fractions

A study [86] determined the total collected $J/\psi$ events in 2009 and 2012 to be $(1310.6 \pm 7.0) \times 10^6$. The quoted uncertainty of 0.53% will be taken into account in the present analysis. For the external branching fraction of $\eta_c \to p\bar{p}$, the value and its uncertainty as listed by the PDG are used [15]. This branching fraction is $\mathcal{B}(\eta_c \to p\bar{p}) = (1.44 \pm 0.14) \times 10^{-3}$, and hence adds up to an uncertainty of 9.72%.

### 7.3.10 Efficiency

For the calculation of all systematic errors, the efficiency was recalculated based on the variation of the selection criteria in question. For some systematic-error fits, one would expect the efficiency to change. For instance, if the fit range is changed, or the selection criteria are altered. For other fits used to estimate the systematic uncertainty, such as the choice of hypothesis, or the mass-resolution influence, the same efficiency would be expected as for the nominal fit. This has been confirmed by comparing the resulting efficiencies for these latter fits, where the largest discrepancy was only 0.12%. This maximum value is used to estimate the systematic error induced by extracting the efficiency using a multi-dimensional analysis.

All the systematic errors discussed above are summarized in table 7.4.

| **Source** | **Mass [$\times 10^{-4}$]** | **Width [%]** | **BF [%]** |
|---|---|---|---|
| (anti)proton PID | - | - | 2.0 |
| (anti)proton track | - | - | 2.0 |
| photon detection | - | - | 1.0 |
| CLEO-c line shape | 0.40 | 3.26 | 11.1 |
| Hypothesis | 4.86 | 0.74 | 2.61 |
| Cuts: $\chi^2_{4C}$ | 1.00 | 0.75 | 0.08 |
| Cuts: $M^2_{miss}$ | 0.94 | 0.97 | 0.58 |
| Cuts: $\theta_{p\bar{p}}$ | 1.23 | 1.19 | 0.78 |
| $p\bar{p}$ background: $\pm$ PDG sys. err. | 0.57 | 0.17 | 2.69 |
| $p\bar{p}$ background: PHSP model | 0.77 | 1.19 | 1.04 |
| $\pi^0$ background: $\pm 1\%$ | 0.90 | 0.77 | 0.07 |
| $\pi^0$ background: binning | 0.74 | 1.58 | 0.34 |
| $\pi^0$ background: stat. fluctuations | 0.67 | 1.73 | 0.38 |
| Fit range: 2.7-3.1 GeV/$c^2$ | 1.74 | 1.92 | 4.41 |
| Fit range: lower-mass limit | 1.07 | 0.94 | 4.10 |
| Mass resolution | 6.46 | 3.08 | 3.60 |
| photon-(anti)proton angle EMC | 0.30 | 0.55 | 0.58 |
| $N_{J/\psi}$ | - | - | 0.55 |
| $\mathcal{B}(\eta_c \to p\bar{p})$ | - | - | 9.72 |
| Efficiency | - | - | 0.12 |
| **Total** | 8.71 | 5.98 | 17.1 |
| **Total w.o. $\mathcal{B}(\eta_c \to p\bar{p})$** | 8.71 | 5.98 | 14.0 |
| **Nominal value** | 2986.26 MeV/$c^2$ | 33.56 MeV | $1.79 \times 10^{-2}$ |

**Table 7.4:** Summary of all the systematic errors, presented as relative errors with respect to the nominal values. The total error is obtained by the quadratic sum of the individual systematic uncertainties, assuming that the uncertainties are uncorrelated.

## 7.4 Results and discussion

All contributions to the systematic error are listed in table 7.4. This table also shows the quadratic sum of all systematic uncertainties, as well as the quadratic sum excluding the uncertainty of the branching fraction of the decay $\eta_c \to p\bar{p}$. The total systematic uncertainty of the branching fraction is dominated by two values: the external branching fraction $\eta_c \to p\bar{p}$ and the uncertainty introduced by the $\eta_c$ line shape. Therefore, the external uncertainty is denoted as a separate uncertainty in the final value of the branching fraction. The obtained results for the branching fraction, mass and width of the $\eta_c$ read as:

$$\boxed{\begin{aligned} \mathcal{B}(J/\psi \to \gamma\eta_c) &= (1.79 \pm 0.01 \pm 0.25 \pm 0.17) \times 10^{-2} \\ m_{\eta_c} &= 2986.26 \pm 0.003 \pm 2.60 \text{ MeV}/c^2 \\ \Gamma_{\eta_c} &= 33.56 \pm 0.007 \pm 2.01 \text{ MeV} \end{aligned}} \quad (7.3)$$

The first uncertainties are the statistical uncertainties, the second the systematic uncertainties, and the third error on $\mathcal{B}(J/\psi \to \gamma\eta_c)$ is from the external branching fraction $\mathcal{B}(\eta_c \to p\bar{p})$. All three results coincide with the values listed by the Particle Data Group (PDG) [15], see figures 7.5, 7.6 and 7.7 for an overview of all the values. This is the first time that the branching fraction $\mathcal{B}(J/\psi \to \gamma\eta_c)$ is extracted from a PWA, ensuring a correct description of the interference in all dimensions. Including all uncertainties, the branching fraction has a 13% smaller error than the value listed by the PDG. With just the 2009 and 2012 data, the statistical uncertainties are negligibly small compared to the systematic uncertainties. Therefore, it is irrelevant to perform a PWA with the full available dataset. It should be noted that the statistical uncertainties listed here are the numerically calculated uncertainties based on the covariance-error-matrix provided by Minuit2. In a recent study [140], it was shown that the uncertainties obtained in this fashion might be systematically too small, and that the actual statistical uncertainties could be roughly a factor 1-5 higher. For the branching fraction, the obtained statistical uncertainty approximates $1/\sqrt{N_{\eta_c}}$, which is compatible with the result calculated by Minuit2. Since the statistical uncertainties for the branching fraction are in agreement, it is assumed that the other statistical errors are calculated properly as well.

For the branching fraction, the dominant contribution of the internal systematic uncertainty comes from uncertainties related to the radiative-line-shape description of the $\eta_c$. Both models, from the CLEO-c and KEDR collaborations, are empirical models. The difference between the two descriptions is mainly due to a discrepancy in the description of the tail towards lower $p\bar{p}$-invariant masses. Since the used KEDR description has a less damped tail, the errors related to the fit range are larger than they would be with the CLEO-c description. Furthermore, the $\eta_c$ width has a relatively large error related to this description as well. To gain a better accuracy, especially in the branching fraction, it is essential that more clarity in the correct description becomes available, from either theoretical or experimental inputs. Theoretical input might, for instance, come from advanced lattice or non-relativistic QCD calculations [80–83, 157]. Further experimental input might be challenging, especially since both the $\eta_c$ and the non-resonant contribution have the same quantum numbers. However, one might think of experimental results from other M1 transitions with well-known resonance states. Another possibility could be to perform a combined fit for data samples of $J/\psi \to \gamma\eta_c$ and $\psi' \to \gamma\eta_c'$.

In figure 7.6, it is seen that the present analysis has a relatively large error for the $\eta_c$ mass compared to previous BESIII studies. The dominant contribution in the systematic uncertainty of the mass comes from the presently used procedure to estimate the effects of the mass resolution of the detector. This procedure, described in section 7.3.5, is needed since the fit lacks to incorporate the resolution effect of the detector, and has no well-defined procedure to correct for that. Currently, a new procedure to incorporate the mass resolution with PAWIAN is being developed. For the PWA minimization process in PAWIAN, the detector resolution cannot be taken into account directly. This would require the knowledge of a proper $n$-dimensional convolution function, which is outside the scope of this thesis, and it will take a significant amount of time to incorporate this. Therefore, in the new procedure, the correction of the one-dimensional mass-shapes will be performed in a second procedure, using the outcome of the PWA-fit result. In this procedure, the obtained mass shapes will be convoluted with a Gaussian function representing the detector resolution. If the aim of this analysis were to extract the $\eta_c$ mass with a maximum precision, it could be relevant to incorporate this new procedure to minimize the systematic error.

For the present analysis, it was decided to not blindly trust the purely-statistical criteria AICc and BIC. It would be interesting to understand how systematic effects influence these values, and what numerical difference belongs to a significant better value when systematics are taken into account. From the present analysis, and discussions with colleagues, it appears that a difference of $\sim$1000 for the AICc or BIC gives a visible improvement, whereas a difference of $\sim$500 does not affect the one-dimensional fit projections. To better understand how systematical issues affect the AICc and BIC values, and what difference is large enough to choose the significant best fit, one could perform studies with toy MC data. Additionally, it would be favorable if the negative weights for the background subtraction could be free parameters in PAWIAN, even though the errors from the current fixed values do not play a dominant role in the systematic uncertainty. A more refined background subtraction might solve some of the issues in the tail for $p\bar{p}$ invariant masses larger than 3.0 GeV/$c^2$.

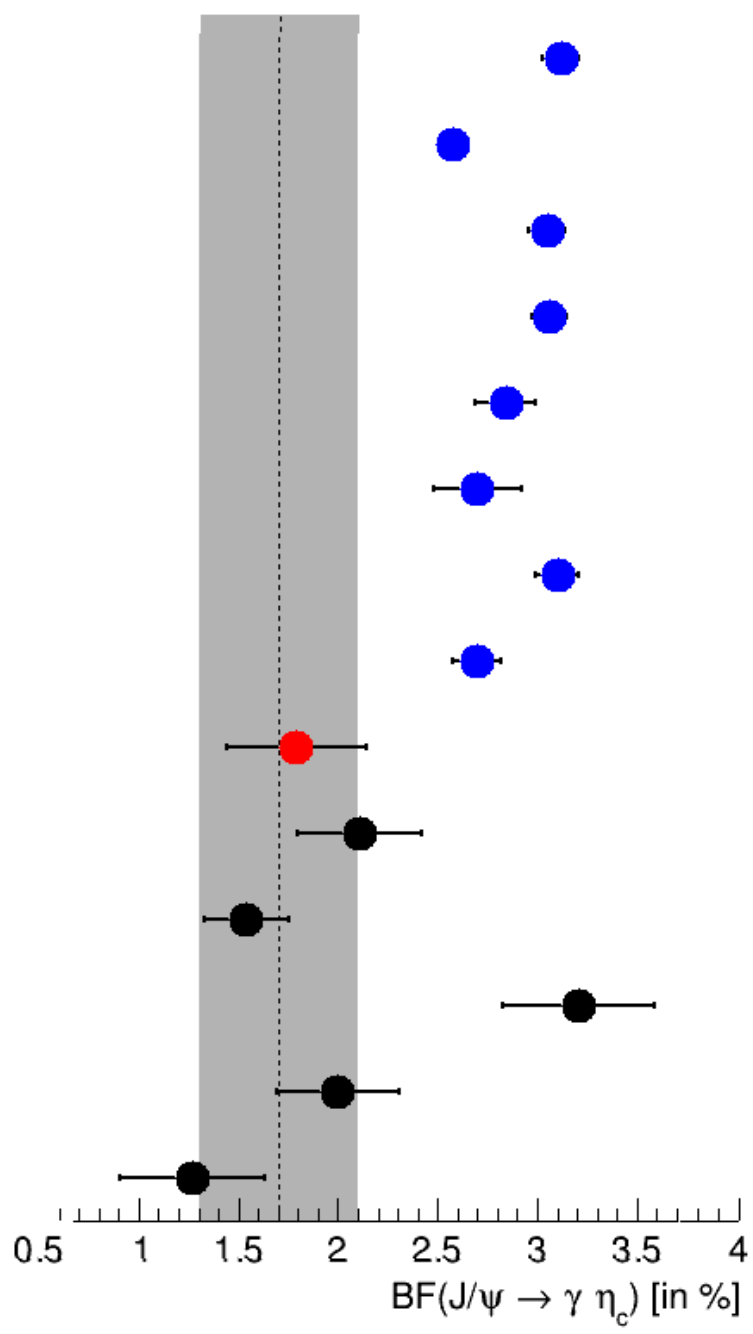

| **Source** | **Year** | **Ref.** |
|---|---|---|
| Non-relativistic potential model | 2008 | [77] |
| Godfrey-Isgur model | 2008 | [77] |
| Non-relativistic EFT | 2008 | [78] |
| Non-relativistic EFT | 2006 | [79] |
| Lattice QCD ($N_f = 2$) | 2013 | [80] |
| Lattice QCD ($N_f = 2 + 1$) | 2012 | [81] |
| Lattice QCD ($N_f = 2$) | 2011 | [82] |
| Lattice QCD ($N_f = 0$) | 2009 | [83] |
| BESIII: $J/\psi \to \gamma p\bar{p}$ | 2022 | This work |
| BESIII: $J/\psi \to \gamma K^0_S K^\pm \pi^\mp$ (D) | 2017 | [32] |
| BESIII: $J/\psi \to \gamma K^0_S K^\pm \pi^\mp$ (C) | 2017 | [32] |
| KEDR: $J/\psi \to \gamma \eta_c$ | 2014 | [84] |
| CLEO-c: $e^+e^- \to \gamma X$ | 2009 | [73] |
| Crystal Ball: $J/\psi \to \gamma X$ | 1986 | [85] |

**Figure 7.5:** Final result from the present work (red) for the branching fraction compared to theoretical predictions (blue) and experimental values (black). The gray band represents the PDG value [15]. $N_f$ stands for the number of quark flavors that are considered in the LQCD calculations. The two terms (D) and (C) refer to two equally good fits with either a destructive of constructive interference, details can be found in reference [32].

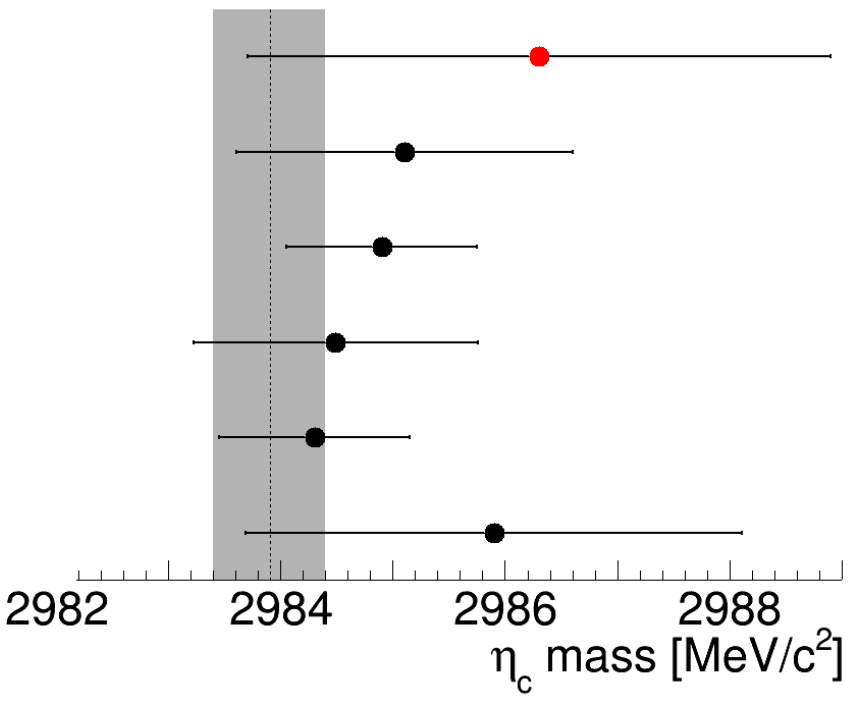

| Source | Year | Ref. |
|---|---|---|
| $J/\psi \to \gamma p\bar{p}$ | 2022 | This work |
| $J/\psi \to \gamma K^0_S K^{\pm}\pi^{\mp}$ (D) | 2017 | [32] |
| $J/\psi \to \gamma K^0_S K^{\pm}\pi^{\mp}$ (C) | 2017 | [32] |
| $\psi(2S) \to \gamma\pi^0$ hadrons | 2012 | [158] |
| $\psi(2S) \to \gamma\eta_c$ | 2012 | [159] |
| $J/\psi \to \gamma\omega\omega$ | 2019 | [160] |

**Figure 7.6:** Final result from the present work (red) for the $\eta_c$ mass, compared to other BESIII results (black). The gray band represents the PDG value [15].

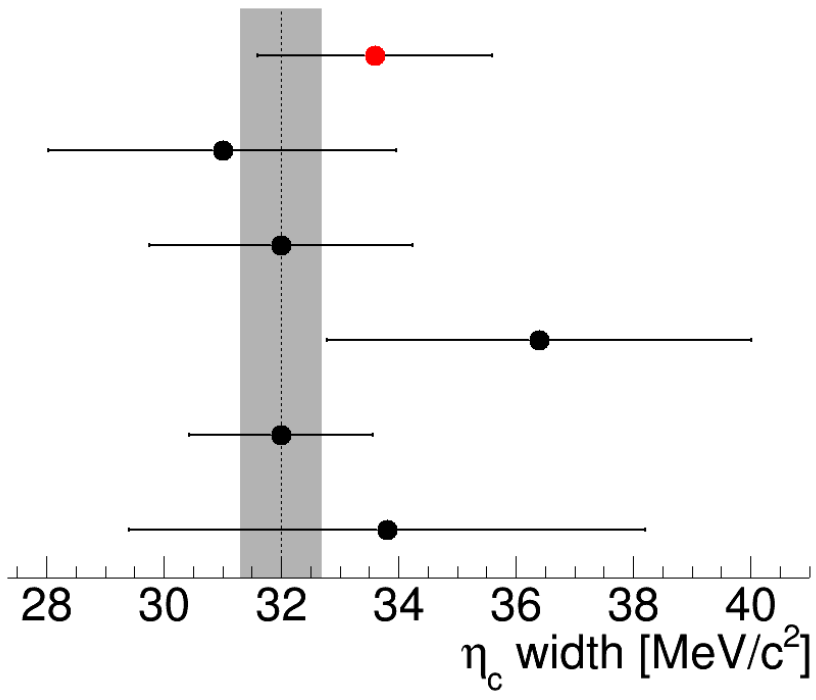

| Source | Year | Ref. |
|---|---|---|
| $J/\psi \to \gamma p\bar{p}$ | 2022 | This work |
| $J/\psi \to \gamma K^0_S K^{\pm}\pi^{\mp}$ (D) | 2017 | [32] |
| $J/\psi \to \gamma K^0_S K^{\pm}\pi^{\mp}$ (C) | 2017 | [32] |
| $\psi(2S) \to \gamma\pi^0$ hadrons | 2012 | [158] |
| $\psi(2S) \to \gamma\eta_c$ | 2012 | [159] |
| $J/\psi \to \gamma\omega\omega$ | 2019 | [160] |

**Figure 7.7:** Same as figure 7.6 for the $\eta_c$ width.

# 8. Summary, conclusions and outlook

The aim of this PhD was to explore BESIII data to study the radiative decay $J/\psi \to \gamma p\bar{p}$, including the intermediate resonance $\eta_c$. In the following, the contents of the different chapters will be summarized briefly. Thereafter, the general conclusions will be drawn, together with an outlook on future prospects.

## 8.1 Summary

The Standard Model (SM) is a widely accepted theory that describes the properties of the fundamental particles, called quarks and leptons, as well as describing how they interact according to three of the four fundamental forces, namely the electromagnetic, strong, and weak forces. The SM has gradually grown into a very successful theory for particle physics over the last decades of the $20^{\text{th}}$ century. One of its most compelling successes is the correct prediction of the existence of the Higgs boson, which was confirmed at the European Organization for Nuclear Research (CERN) in 2012 [3, 4]. But for all its power, there still remain unanswered questions on the origin of the mass of hadrons, the composite particles that make up all visible matter in the Universe. The Higgs mechanism explains only 1% of the mass of the well-known hadron named proton [5]. The remaining 99% arises from internal dynamics that are not yet fully understood. Furthermore, certain states that are possible according to the SM currently have not been observed unambiguously. These non-mesonic and non-baryonic states are called exotic matter, states mainly consisting of gluons (so-called glueballs) being one type of them. To be able to correctly identify possible exotic matter, a full understanding of the spectrum of hadrons is essential. Therefore, it is crucial to study the production of hadrons in different processes, and to cover as many decay modes as possible. The subfield of particle physics that studies the basic properties, such as the masses and decays, of the possible hadrons is called hadron spectroscopy, and the part of the SM that aims to describe the strong interaction is called Quantum ChromoDynamics (QCD). Charmonium states, such as $J/\psi$ and the ground-state $\eta_c$, are important states for hadron spectroscopy, since they provide good probes to study QCD, as the properties of charmonium are determined by the strong interaction. Moreover, the heavy-quark constituents of charmonium result in a relatively easily interpretable spectrum with narrow states that do not overlap. Particularly, the $J/\psi$ meson is a well-understood resonance, with known quantum numbers, that is produced abundantly in electron-positron colliders.

In this thesis, the radiative decay $J/\psi \to \gamma p\bar{p}$ is studied. The aim of this study is to get a better understanding of the full spectrum of the $p\bar{p}$ invariant mass, and, especially, to get more insight in poorly-understood properties of the intermediate $\eta_c$ resonance. Chapter 1 gives a general introduction in the SM, including QCD, and the different hadrons that can be constructed with these models. Additionally, the motivation for the study of the decay $J/\psi \to \gamma p\bar{p}$ is described in more detail there.

The Beijing Spectrometer (BES) III at the Beijing Electron Positron Collider (BEPC) II is an outstanding setup for experiments which aim to study conventional hadrons, like charmonium, and to search for exotic states, like glueballs. BESIII has collected the world's largest sample of $J/\psi$ data, consisting of $10^{10}$ $J/\psi$ events. In chapter 2, the relevant details of the cylindrical BESIII detector are discussed.

For the study of $J/\psi \to \gamma p\bar{p}$, the data events collected by BESIII are reconstructed and the selection criteria are optimized. In chapter 3, the datasets, both from the experiment and from Monte Carlo (MC) simulations, are briefly presented and the event-selection criteria are described and motivated. Additionally, the treatment of the remaining background events is discussed. The main background contributions are related to the channel $J/\psi \to \pi^0 p\bar{p}$ and final-state radiation of the decay $J/\psi \to p\bar{p}$.

The total $J/\psi$ data sample of BESIII, after reconstruction, selection and background subtraction, is used to extract the total branching fraction $\mathcal{B}(J/\psi \to \gamma p\bar{p})$. This analysis and the results are presented in chapter 4. The obtained value is $\mathcal{B}(J/\psi \to \gamma p\bar{p}) = (5.30 \pm 0.18) \times 10^{-4}$, where the error is dominated by systematics.

The high statistics provided by BESIII enables a Partial-Wave Analysis (PWA) of the reaction channel. In a PWA, the correlation between the momenta of final-state particles is analyzed to determine the spin-parities of the contributing waves, and masses, widths and branching ratios of intermediate resonances. In the BESIII study of $J/\psi \to \gamma p\bar{p}$, the full event topology is obtained by reconstructing the four-momenta of the photon, proton, and antiproton. This allows us to perform a PWA of the complete reaction chain, from the initial $e^+e^-$ state (or $J/\psi$) down to the three final particles of the decay. The general concepts of a PWA, and the utilized software package PArtial-Wave Interactive ANalysis (PAWIAN), are introduced in chapter 5.

To be able to perform a full PWA over the full available $p\bar{p}$ invariant mass range, one needs a proper model to describe all the relevant dynamics regarding possible intermediate resonances and Final-State Interactions (FSI). The large statistics allow for a mass-independent PWA of the full $p\bar{p}$ invariant-mass range, omitting the need to include FSI. For this mass-independent PWA, the data are divided into equally-sized bins of the invariant $p\bar{p}$ mass. Partial wave fits for different hypotheses are performed for each mass bin individually. Such a mass-independent approach is not a quantitative analysis. Hence, the underlying structures cannot be studied in detail. However, a mass-independent PWA provides a qualitative analysis on the global structures that show up in the $p\bar{p}$ mass spectrum. The results of the mass-independent PWA are presented in chapter 6.

Finally, in chapter 7, the PWA concepts are applied for the study of the $\eta_c$ resonance. This analysis consists of a mass-dependent PWA for $p\bar{p}$ invariant masses larger than 2.7 GeV$/c^2$. Contrarily to the two other analysis parts, the $\eta_c$ study is based on part of the total data sample corresponding to $(1310.6 \pm 7.0) \times 10^6$ $J/\psi$ events that were recorded in 2009 and 2012. The obtained results for the branching fraction, mass and width of the $\eta_c$ read as $\mathcal{B}(J/\psi \to \gamma\eta_c) = (1.79 \pm 0.01 \pm 0.25 \pm 0.17) \times 10^{-2}$, $m_{\eta_c} = 2986.26 \pm 0.003 \pm 2.60$ MeV$/c^2$ and $\Gamma_{\eta_c} = 33.56 \pm 0.007 \pm 2.01$ MeV. In these results, the first uncertainties are the statistical uncertainties, the second the systematic uncertainties, and the third error on $\mathcal{B}(J/\psi \to \gamma\eta_c)$ is from the external branching fraction $\mathcal{B}(\eta_c \to p\bar{p})$.

## 8.2 Conclusions and outlook

In this thesis, it was demonstrated that the experimental setup and high statistics of BESIII provided an adept combination to study the radiative channel $J/\psi \to \gamma p\bar{p}$. The full branching fraction, $\mathcal{B}(J/\psi \to \gamma p\bar{p})$, was only studied once before in 1984. The previous value of $\mathcal{B}(J/\psi \to \gamma p\bar{p}) = (3.8 \pm 1.0) \times 10^{-4}$ was obtained from a data sample of $1.32 \times 10^6$ $J/\psi$ events collected with the MARKII detector [90], where the contributions from the statistical and systematic uncertainties were of similar size. The value from this study, $\mathcal{B}(J/\psi \to \gamma p\bar{p}) = (5.30 \pm 0.18) \times 10^{-4}$, results in about one order of magnitude improvement in the relative error. In our result, the statistical error is negligibly small due to roughly $10^4$ times more $J/\psi$ data collected with BESIII compared to that of the experiment with MARKII. The systematical error has been reduced significantly as a result of the more sophisticated BESIII detector. The most relevant improvements compared to MARKII are a better signal-to-background ratio due to a photon-energy resolution that is improved by a factor three, and an acceptance of 93% of $4\pi$ for BESIII versus 65% for MARKII [90, 123].

Our branching fraction falls within two standard deviations of the MARKII result. We note, however, that the error of the MARKII measurement is very large compared to our obtained value. The fact that MARKII has a significantly smaller acceptance and signal-to-background ratio lends more confidence to our data and result, as the high resolution and close to $4\pi$ coverage of BESIII provide a better handle on the systematic effects. Despite these improvements, the systematic error of our result remains dominated by uncertainties related to the characteristics of the BESIII detector. Thus, for an even more accurate determination, further improvements in the understanding of the detector performances are necessary. For instance, the determination of the standard BESIII errors related to the tracking and particle-identification could be revisited. Currently, the relevant efficiencies were estimated by comparing control data samples of almost background-free channels with MC simulations [112, 113, 122]. It could be that the given values overestimate the uncertainty, and that the uncertainties can be reduced by another approach. One could, for instance, use a next generation MC sample as the reference channel. Alternatively, instead of using a MC sample, one could compare the results of data samples for two different, independent benchmark channels.

It is not expected that data from other experiments in the near future will improve the accuracy and precision on the branching fraction $\mathcal{B}(J/\psi \to \gamma p\bar{p})$ significantly. For instance, the forthcoming $\overline{\text{P}}$ANDA (antiProton ANnihilations at DArmstadt) detector [161, 162] will result in a substantially dirtier spectrum, as all quantum numbers are directly accessible and the hadronic background is expected to be significantly higher than for BESIII. Thereby, a multitude of background channels will open with signatures that could compete with the topology of the decay $J/\psi \to \gamma p\bar{p}$. It will thus be harder to retrieve an accurate value for the branching fraction $\mathcal{B}(J/\psi \rightarrow \gamma p\bar{p})$ with $\overline{\text{P}}$ANDA. On the other hand, $\overline{\text{P}}$ANDA can improve the accuracies on the mass and width of the $\eta_c$ meson significantly. As the $\eta_c$ can be produced directly, a resonance scan can be performed around its mass, hence omitting the uncertainties related to a radiative transition. Following the detailed balance procedure given in reference [31], it can be calculated that, with a luminosity of $L = 2 \cdot 10^{32}\text{cm}^{-2}\text{s}^{-1}$, about 3550 $\eta_c$ events per day are to be expected in the $\overline{\text{P}}$ANDA dataset, collected for the benchmark channel $\eta_c \rightarrow 2(\pi^+\pi^-\pi^0)$, with an assumed detection and reconstruction efficiency of 10%. In general, it is assumed that the width and mass determination of hadronic resonances measured with $\overline{\text{P}}$ANDA will have an accuracy 10 to 100 times better than in any electron-positron collider experiment, such as BESIII [162, 163]. Note that the expected higher accuracy is a combination of the direct formation of all conventional quantum numbers, and the high resolution expected for the antiproton beam. With electron-positron colliders, $1^{--}$ states can be scanned rather precisely, resulting, for instance, in the accurately determined $J/\psi$ mass. In our analysis, the results on the mass and width of the $\eta_c$ meson have relative large uncertainties compared to previous measurements. As the $J/\psi$ mass is known with a high accuracy, a precision measurement of the $\eta_c$ mass will result in an accurate value for the hyperfine splitting of ground-state charmonium. In our study of the $\eta_c$ meson, the main goal was to extract the $J/\psi \rightarrow \gamma\eta_c$ branching fraction unambiguously, which, as mentioned before, is likely out of reach for $\overline{\text{P}}$ANDA due to the higher background.

Previous results on the $J/\psi \rightarrow \gamma\eta_c$ branching fraction were obtained via a one-dimensional fit of the proton-antiproton invariant-mass spectrum, including a possible interference with a non-resonant background [32, 75]. The interference contribution was necessary to describe the non-symmetric line shape that was observed in the invariant-mass spectrum. In these studies, it was found that such a fit gives two possible solutions, corresponding to a destructive and constructive interference effect. Particularly, this leads to an ambiguity in the extraction of the branching fraction. In this work, the branching fraction $\mathcal{B}(J/\psi \rightarrow \gamma\eta_c)$ was extracted for the first time via a PWA, ensuring a correct description of the interference in all dimensions. All three results, on the relevant branching fraction, and the mass and width of the $\eta_c$ meson, coincide with the values listed by the Particle Data Group (PDG) [15]. The obtained branching fraction, $\mathcal{B}(J/\psi \rightarrow \gamma\eta_c) = (1.79 \pm 0.35) \times 10^{-2}$, has a 13% smaller error than the value listed by PDG. The error is dominated by systematics, where the largest uncertainty comes from uncertainties related to the radiative-line-shape description of the $\eta_c$. The model used in the nominal value originates from the KEDR collaboration, whereby the CLEO-c model is used to estimate the corresponding systematic uncertainty. Both models included in this analysis are empirical. To further improve the accuracy, it is essential that there will come more clarity in the correct description, from either theoretical or experimental inputs. Theoretical input might, for instance, come from advanced lattice calculations [80–83]. Further experimental input might be challenging, especially since both the $\eta_c$ and the non-resonant contribution have the same quantum numbers. However, one might think of combining experimental data from various M1 transitions with well-known resonance states. Additionally, one could use results from the stronger E1 transitions, such as $h_c \to \gamma\eta_c$. Another possibility could be to perform a combined fit for data samples of $J/\psi \to \gamma\eta_c$ and $\psi' \to \gamma\eta_c'$, where a challenge lies in the identification of the $\eta_c'$ resonance.

In the mass-dependent PWA, it was decided to not blindly trust the two purely-statistical second-order Akaike-Information-Criterion (AICc) and Bayesian-Information-Criterion (BIC), as it is unclear how the values are influenced by systematic effects. From the presented $J/\psi \to \gamma\eta_c$ analysis, and after discussions with colleagues, it appears that a difference of ∼1000 for the AICc or BIC corresponds to a visible improvement in the pull distributions of the one-dimensional fit projections, whereas a difference of ∼500 does not affect the one-dimensional fit projections significantly. To gain a better understanding in how systematical effects influence the AICc and BIC values, and what difference is large enough to choose the significant best fit, one could perform studies with MC data with and without the inclusion of systematic effects. Additionally, it would be favorable if the negative weights for the background subtraction could be free parameters in PAWIAN, even though the errors from the current fixed values do not play a dominant role in the systematic uncertainty. A more refined background subtraction might solve some of the issues in the tail for $p\bar{p}$ invariant masses larger than 3.0 GeV/$c^2$.

Contrary to the mass-dependent PWA, the AICc and BIC values are used for the hypothesis selection in the mass-independent PWA. The large number of individual fits included in the mass-independent approach make it difficult to compare all results in detail. More importantly, due to the narrow $p\bar{p}$ invariant-mass bins of 5 MeV/$c^2$, it is assumed that the systematic effects entering the AICc and BIC values are negligibly small. In the mass-independent PWA over the full $p\bar{p}$ invariant-mass range, it was found that both the near-threshold peak, and the $\eta_c$ peak coincided with the expected $0^{-+}$ quantum numbers. As one would expect, the $p\bar{p}$ invariant-mass ($M_{p\bar{p}}$) provided a clean spectrum around the narrow charmonium ground-state $\eta_c$. Below the open-charm thresholds, due to the limited number of available decay channels, the charmonium states are in general long-lived, narrow states, such as the $J/\psi$ and $\eta_c$. Above these thresholds, charmonium can decay into a pair of mesons containing both one (anti)charm quark. Therefore, a large number of possible decay channels becomes available, resulting in broader, overlapping charmonium states. A similar effect is known towards lower invariant-masses, where numerous light-meson states become available. This effect was observed in the mass-independent PWA, as the broad enhancement around $M_{p\bar{p}} \sim 2.1$ GeV/$c^2$ appears to be a complex mixture of resonances with various quantum numbers. Furthermore, towards higher $p\bar{p}$ invariant masses the assumed isobar model seems to describe the data properly, whereas for $M_{p\bar{p}} < 2.2$ GeV/$c^2$ it appears that the implied isobar model does not provide a proper description. In this lower-mass range, a large $2^{++}$ contribution is found, which hints towards a significant contribution of the established resonance $f_2(2010)$ [121]. This state has been observed to decay to $\phi\phi$ and $K\bar{K}$ [15, 155], but, thus far, a decay into $p\bar{p}$ has not been observed. Therefore, it would be interesting to study both the decays $J/\psi \to \gamma f_2(2010)$, and $f_2(2010) \to p\bar{p}$, and try to extract both branching fractions for the first time. Additionally, it would be interesting to perform a coupled-channel analysis. It might well be that the $f_2(2010) \to p\bar{p}$ decay couples to other decays, such as $f_2(2010) \to K\bar{K}$ and $f_2(2010) \to \phi\phi$, explaining the inadequate description by the isobar model. Furthermore, as mentioned earlier, with $\overline{\text{P}}$ANDA resonance scans of hadrons can be performed. These scans could provide more insight in the $p\bar{p}$ invariant-mass range of 2.0−2.2 GeV/$c^2$, and specifically in the supposed $f_2(2010)$ state. As the minimum $p\bar{p}$ collision-energy at $\overline{\text{P}}$ANDA will be 2 GeV [164], the near-threshold peak cannot be studied with this future experiment.

For a quantitative analysis of the near-threshold peak, one could perform a mass-dependent PWA with the inclusion of a well-known $N\bar{N}$-potential representing the Final-State Interaction (FSI) between the proton and antiproton. One could, for instance, incorporate the Nijmegen [88, 89] or Paris [165] $N\bar{N}$-potential in a PWA fit to the BESIII data of $J/\psi \to \gamma p\bar{p}$. Moreover, it would be especially interesting to combine this analysis with similar studies of the near-threshold regions of the decays $J/\psi \to \gamma n\bar{n}$ and $J/\psi \to \gamma\Delta\bar{\Delta}$. A PWA of $J/\psi \to \gamma p\bar{p}$, with a proper description of the FSI, can provide the necessary information to better understand the spectacularly appearance of the strong and narrow structure at the $p\bar{p}$ threshold. A possible theoretical interpretation of the structure is a $p\bar{p}$ (quasi)bound state [69, 70, 165], which requires a mass of $\sim$1.85 GeV/$c^2$ [68]. Another interpretation of the observed structure is a glueball, or at least that it contains a large glueball component [71].

In the coming years, the BESIII data taking is moving to higher intensities for higher center-of-mass energies up to about 6 GeV/$c^2$. This higher energy range is beneficial to gain a better understanding of the $XYZ$-states in the charmonium spectrum above the open-charm threshold. It is, however, less relevant for the main topics discussed in this thesis. Since the analyses presented in this thesis have relatively small statistical errors, it is not that beneficial to collect more data with similar characteristics. An advantage of increasing statistics is that a larger dataset becomes more sensitive to channels with branching fractions that are too small to be registered with existing data. To make a significant step forward in understanding the dynamics that give rise to the observed $p\bar{p}$ invariant-mass spectrum of the radiative $J/\psi$, it would be interesting to develop a detector that is capable to measure the polarizations of the final-state particles in addition to their momenta. Currently, a PWA of the decay $J/\psi \to \gamma p\bar{p}$ is relatively complex due to the redundancies introduced by the three final-states with nonzero spin. Together with the two possible spin-projections for the initial state, this results in $2^4 = 16$ coherent sums for the calculation of the probability density function to be optimized in the PWA. Therefore, it would be valuable if a future experiment would be able to give additional information on the polarizations of the proton and antiproton, and, ideally, on the photon as well.

# 9. Nederlandse samenvatting

Het Standaard Model (SM) is een algemeen aanvaarde theorie die de eigenschappen beschrijft van de fundamentele deeltjes: kwarks en leptonen. Daarnaast beschrijft het SM hoe deze deeltjes interageren volgens drie van de vier fundamentele krachten, namelijk de elektromagnetische, sterke en zwakke krachten. Het SM is in de laatste decennia van de 20e eeuw geleidelijk uitgegroeid tot een zeer succesvolle theorie voor deeltjesfysica. Een van de meest overtuigende successen is de juiste voorspelling van het bestaan van het Higgs-deeltje, dat in 2012 werd bevestigd door de Europese Organisatie voor Nucleair Onderzoek (CERN) [3, 4]. Maar ondanks al de kracht van het SM blijven er nog steeds onbeantwoorde vragen over de oorsprong van de massa van hadronen, de samengestelde deeltjes waaruit alle zichtbare materie in het heelal bestaat. Het Higgs-mechanisme verklaart slechts 1% van de massa van het bekende hadron genaamd proton [5]. De resterende 99% komt voort uit interne dynamica die nog niet volledig wordt begrepen. Daarnaast zijn bepaalde toestanden die volgens het SM mogelijk zijn op dit moment niet eenduidig waargenomen. Deze niet-mesonische en niet-baryonische toestanden worden exotische materie genoemd. Hieronder vallen bijvoorbeeld toestanden die hoofdzakelijk uit gluonen bestaan (zogenaamde gluonballen). Om mogelijk exotische materie correct te kunnen identificeren is een volledig begrip van het spectrum aan hadronen essentieel. Daarom is het van cruciaal belang om de productie van hadronen in verschillende processen te bestuderen en zoveel mogelijk vervalmodi te bestrijken. Het deelgebied van de deeltjesfysica dat de basiseigenschappen, zoals de massa's en vertakkingsfracties, van hadronen bestudeert wordt hadronspectroscopie genoemd, en het onderdeel van het SM dat de sterke interactie poogt te beschrijven wordt kwantumchromodynamica (QCD) genoemd. Charmoniumtoestanden, zoals $J/\psi$ en de grondtoestand $\eta_c$, zijn belangrijke toestanden voor hadronspectroscopie, omdat ze goede kanalen bieden om QCD te bestuderen, aangezien de eigenschappen van charmonium worden bepaald door de sterke interactie. Bovendien resulteren de zware-kwark bestanddelen van charmonium in een relatief gemakkelijk te interpreteren spectrum van smalle pieken die elkaar niet overlappen. In het bijzonder is het meson $J/\psi$ een goed begrepen resonantie, met bekende kwantumgetallen, die overvloedig wordt geproduceerd in elektron-positron-botsers.

In dit proefschrift wordt het stralingsverval van $J/\psi \to \gamma p\bar{p}$ bestudeerd. Het doel van deze studie is om een beter begrip te krijgen van het volledige spectrum van de $p\bar{p}$ invariante massa, en vooral om meer inzicht te krijgen in de slecht begrepen eigenschappen van de tussentijdse resonantie $\eta_c$. Hoofdstuk 1 geeft een algemene introductie tot het SM, inclusief QCD, en de verschillende hadronen die met dit model geconstrueerd kunnen worden. Bovendien wordt daar de motivatie voor het bestuderen van het verval $J/\psi \to \gamma p\bar{p}$ in meer detail beschreven.

De Beijing Spectrometer (BES) III aan de Beijing Electron Positron Collider (BEPC) II is een uitstekende opstelling voor experimenten die tot doel hebben conventionele hadronen, zoals charmonium, te bestuderen en om naar exotische toestanden, zoals gluonballen, te zoeken. BESIII heeft 's werelds grootste dataset van $J/\psi$ gebeurtenissen verzameld, bestaande uit $10^{10}$ $J/\psi$ gebeurtenissen. In hoofdstuk 2 worden de relevante details van de cilindrische BESIII detector besproken.

Voor de studie van het verval $J/\psi \to \gamma p\bar{p}$ worden de door BESIII verzamelde gegevens gereconstrueerd en de selectiecriteria geoptimaliseerd. In hoofdstuk 3 worden de datasets, zowel uit het experiment als uit Monte Carlo (MC) simulaties, kort gepresenteerd en worden de selectiecriteria beschreven en onderbouwd. Daarnaast wordt de behandeling van de overige achtergrondgebeurtenissen besproken. De belangrijkste achtergrond bijdragen komen van het kanaal $J/\psi \to \pi^0 p\bar{p}$ en de eindtoestand-straling van het verval $J/\psi \to p\bar{p}$.

De totale $J/\psi$ dataset van BESIII - na reconstructie, selectie en achtergrondaftrekking - wordt gebruikt om de totale vertakkingsfractie $\mathcal{B}(J/\psi \to \gamma p\bar{p})$ te bepalen. Deze analyse en de resultaten worden gepresenteerd in hoofdstuk 4. De verkregen waarde is $\mathcal{B}(J/\psi \to \gamma p\bar{p}) = (5{,}30 \pm 0{,}18) \times 10^{-4}$, waarbij de fout wordt gedomineerd door systematiek. De vertakkingsfractie is slechts één keer eerder geëxtraheerd in 1984. De relatieve fout van de huidige studie is ongeveer een factor tien kleiner dan die van het vorige resultaat.

De hoge statistiek die door BESIII wordt geleverd maakt een partiële golfanalyse (PWA) van het reactiekanaal mogelijk. In een PWA wordt de correlatie tussen de vierimpulsen van deeltjes in de eindtoestand geanalyseerd om zo de spin-pariteiten van bijdragende golven, en de massa's, breedtes en vertakkingsfracties van tussentijdse resonanties te bepalen. In de BESIII studie van $J/\psi \to \gamma p\bar{p}$ wordt de volledige vervaltopologie verkregen door de vierimpuls van het foton, proton en antiproton te reconstrueren. Dit stelt ons in staat om een PWA uit te voeren van de volledige reactieketen, van de initiële $e^+e^-$ (of $J/\psi$) toestand tot aan de drie uiteindelijke deeltjes van het verval. De algemene concepten van een PWA, en het gebruikte softwarepakket PArtial-Wave Interactive ANalysis (PAWIAN), worden geïntroduceerd in hoofdstuk 5.

Om een complete PWA uit te voeren over het volledige beschikbare $p\bar{p}$ invariante-massabereik heeft men een geschikt model nodig om alle relevante dynamica met betrekking tot mogelijke tussentijdse resonanties en eindtoestandsinteracties (FSI) te beschrijven. De beschikbare grote statistiek opent de mogelijkheid tot een massa-onafhankelijke PWA van het volledige $p\bar{p}$ invariante-massabereik. Voor deze massa-onafhankelijke PWA worden de data verdeeld in gelijkgrotige klassenbreedtes in de invariante $p\bar{p}$-massa. Voor elke klassenbreedte wordt de PWA op afzonderlijke wijze uitgevoerd voor de verschillende hypothesen. Een dergelijke massa-onafhankelijke benadering is geen kwantitatieve analyse. Daarom kunnen de onderliggende structuren niet in detail worden bestudeerd. Een massa-onafhankelijke PWA biedt echter een kwalitatieve analyse van de globale structuren die voorkomen in het $p\bar{p}$ massaspectrum. De resultaten van de massa-onafhankelijke PWA worden gepresenteerd in hoofdstuk 6.

Ten slotte worden in hoofdstuk 7 de PWA-concepten toegepast voor de studie van de $\eta_c$ resonantie. Deze analyse bestaat uit een massa-afhankelijke PWA voor $p\bar{p}$ invariante massa's groter dan 2,7 GeV/$c^2$. In tegenstelling tot de twee andere analysedelen is het $\eta_c$-onderzoek gebaseerd op slechts een deel van de totale data die overeenkomt met $(1310{,}6 \pm 7{,}0) \times 10^6$ $J/\psi$ gebeurtenissen die zijn geregistreerd in 2009 en 2012. De verkregen resultaten voor de vertakkingsfractie, massa en breedte van de $\eta_c$ zijn $\mathcal{B}(J/\psi \to \gamma\eta_c) = (1{,}79 \pm 0{,}01 \pm 0{,}25 \pm 0{,}17) \times 10^{-2}$, $m_{\eta_c} = 2986{,}26 \pm 0{,}003 \pm 2{,}60$ MeV/$c^2$ en $\Gamma_{\eta_c} = 33{,}56 \pm 0{,}007 \pm 2{,}01$ MeV. In deze resultaten zijn de eerste onzekerheden de statistische onzekerheden, de tweede zijn de systematische onzekerheden, en de derde onzekerheid op $\mathcal{B}(J/\psi \to \gamma\eta_c)$ komt van de externe vertakkingsfractie $\mathcal{B}(\eta_c \to p\bar{p})$. Dit is de eerste keer dat de vertakkingsfractie is geëxtraheerd door middel van een PWA, waardoor de interferenties correct worden beschreven. De verkregen waardes voor de vertakkingsfractie, massa en breedte komen overeen met de gemiddelde experimentele waardes genoteerd door de Particle Data Group (PDG) [15].

# Appendices

# A. Supplementary figures of the mass-independent PWA

This appendix contains additional figures related to the mass-independent Partial-Wave Analysis (PWA) of the $p\bar{p}$ invariant-mass range, as discussed in chapter 6. More details about the analysis can be found in that chapter.

For this mass-independent PWA, the data are divided into equally-sized bins of the invariant $p\bar{p}$ mass. Partial wave fits for different hypotheses are performed for each mass bin individually. In these fits, the dynamical, or mass dependent, parts of the amplitudes are omitted from all hypotheses. The intensity of the contributions from different $J^{PC}$-partial waves is finally extracted for the best hypothesis in each bin. Figures A.1, A.2 and A.3 show the resulting $p\bar{p}$ mass spectrum for certain specific conditions, as mentioned in the captions.

In figures A.4–A.8, the one-dimensional fit projections of two neighboring bins are compared. The shown fit results belong to the best hypotheses, selected based on the AICc and BIC values, for the nominal procedure with bin widths of 5 MeV/$c^2$. Note that by plotting the one-dimensional fit projections all correlations are lost, whereas the correlations are taken into account into the PWA fitting procedure.

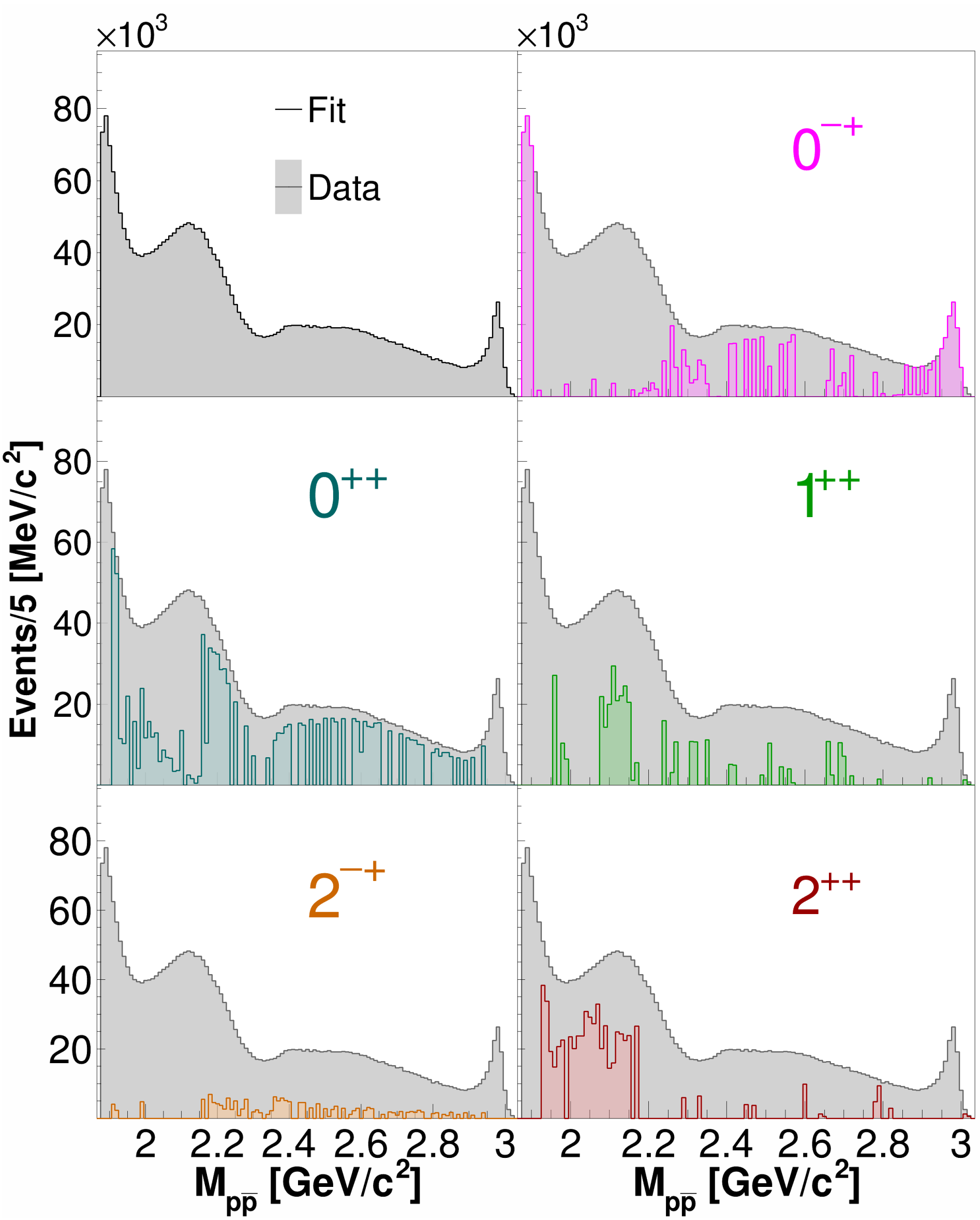

**Figure A.1:** Extracted intensities of the different partial waves in each bin, for bin widths of 10 MeV/$c^2$.

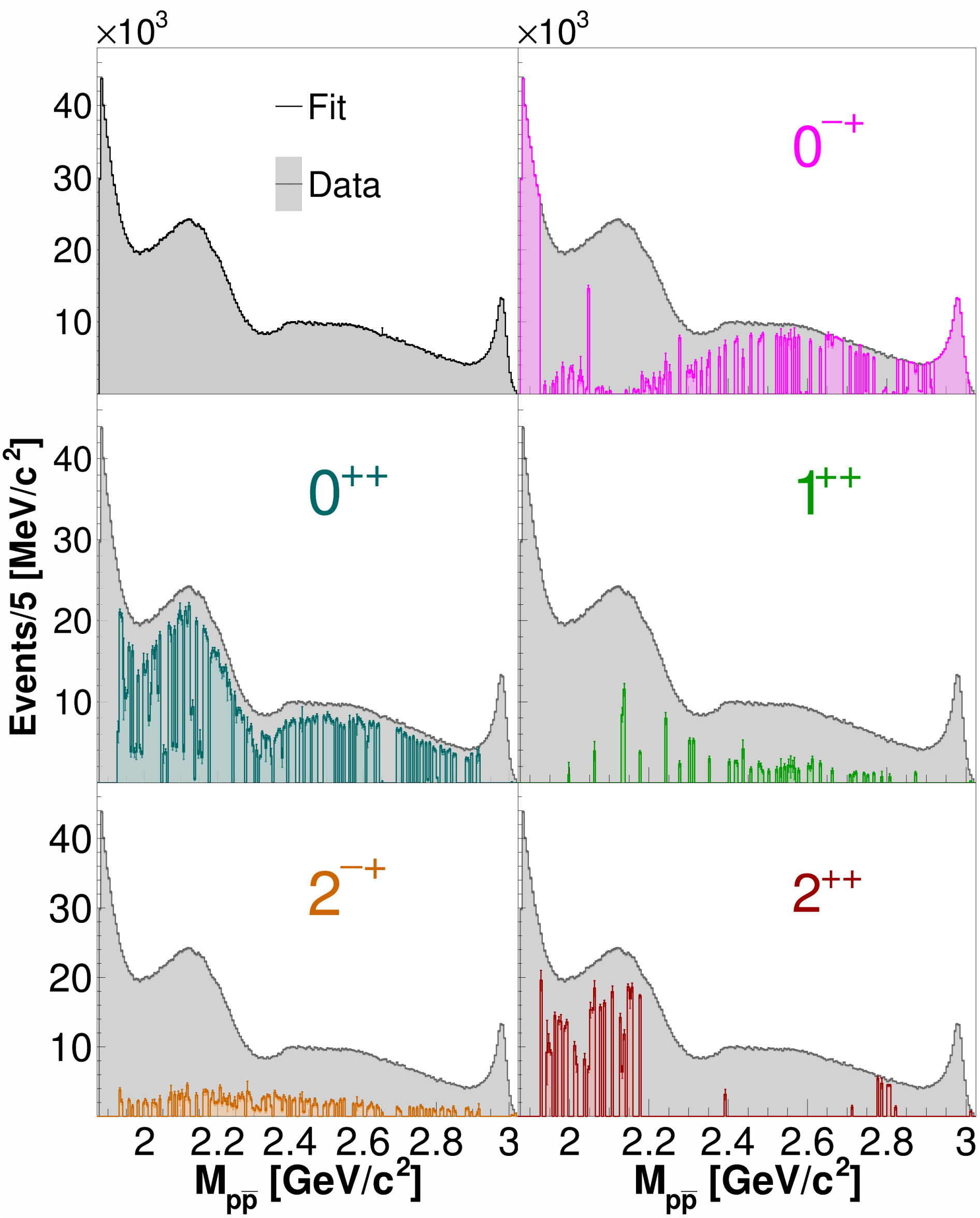

**Figure A.2:** Extracted intensities of the different partial waves in each bin, with the final fit parameters of neighboring bins used as start parameters for each 5 MeV/$c^2$ bin.

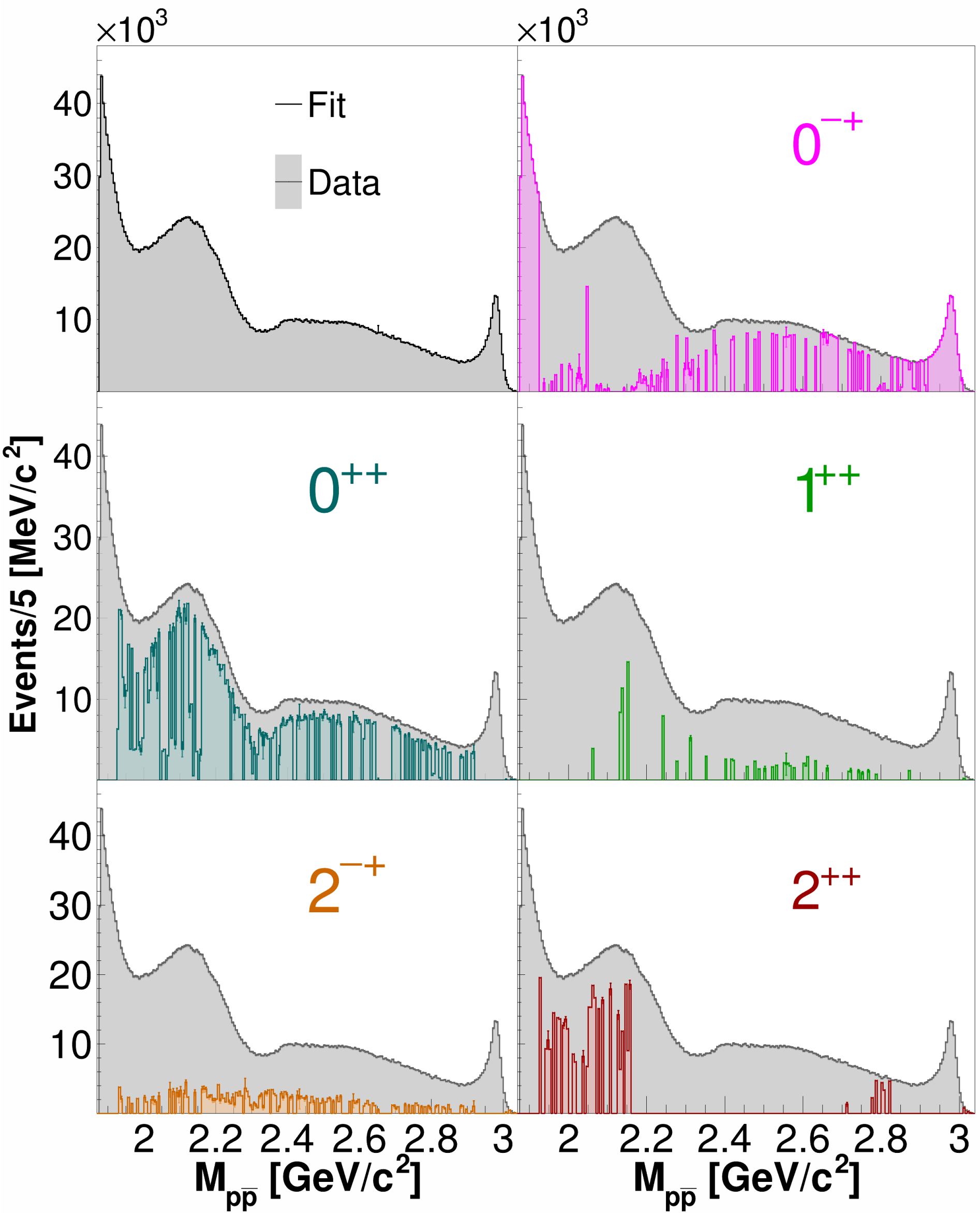

**Figure A.3:** Same as figure A.2, but here the results from the original fits, and from the fits with the neighboring final parameters used as input are both taken into account when selecting the best hypothesis in each bin.

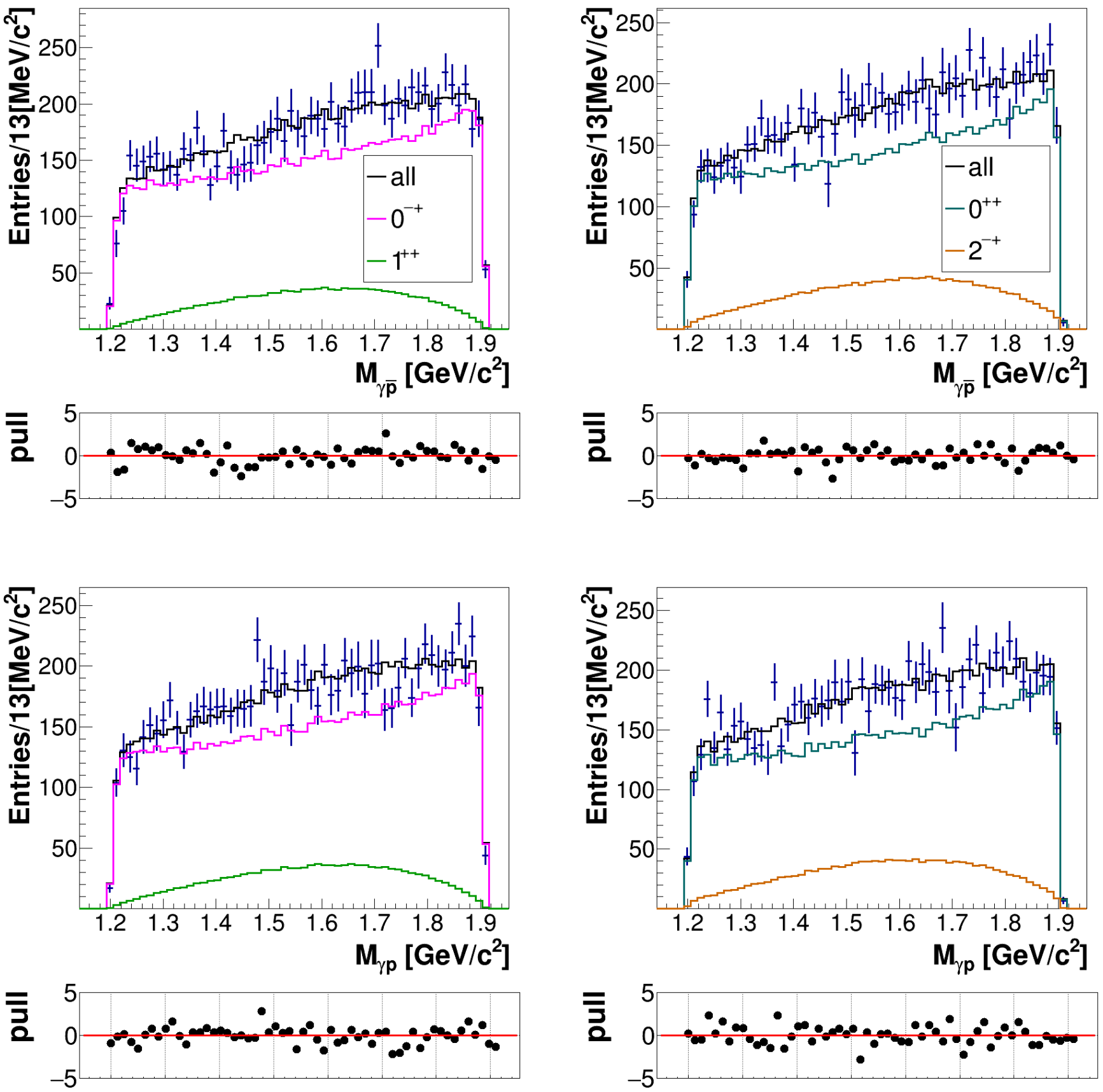

**Figure A.4:** The one-dimensional invariant-mass projections for the selected hypothesis, and the $p\bar{p}$ invariant-mass bins $2500-2505$ MeV/$c^2$ (left) and $2505-2510$ MeV/$c^2$ (right). The total fit is shown in black, and the different contributions of the $0^{-+}$, $1^{++}$, $0^{++}$ and $2^{-+}$ are shown by the magenta, green, teal and brown lines, respectively. The pull distributions represent pull $=(n_{\mathrm{dat}}-n_{\mathrm{fit}})/\sqrt{\sigma_{\mathrm{dat}}^2+\sigma_{\mathrm{fit}}^2}$, where $n_{\mathrm{dat}}$ and $n_{\mathrm{fit}}$ represent the bin contents of the data and fit histograms, respectively, and $\sigma_{\mathrm{dat}}$ and $\sigma_{\mathrm{fit}}$ the corresponding bin (statistical) errors. In general, $\sigma_{\mathrm{fit}}$ is much smaller than $\sigma_{\mathrm{dat}}$. Interference contributions are not explicitly shown.

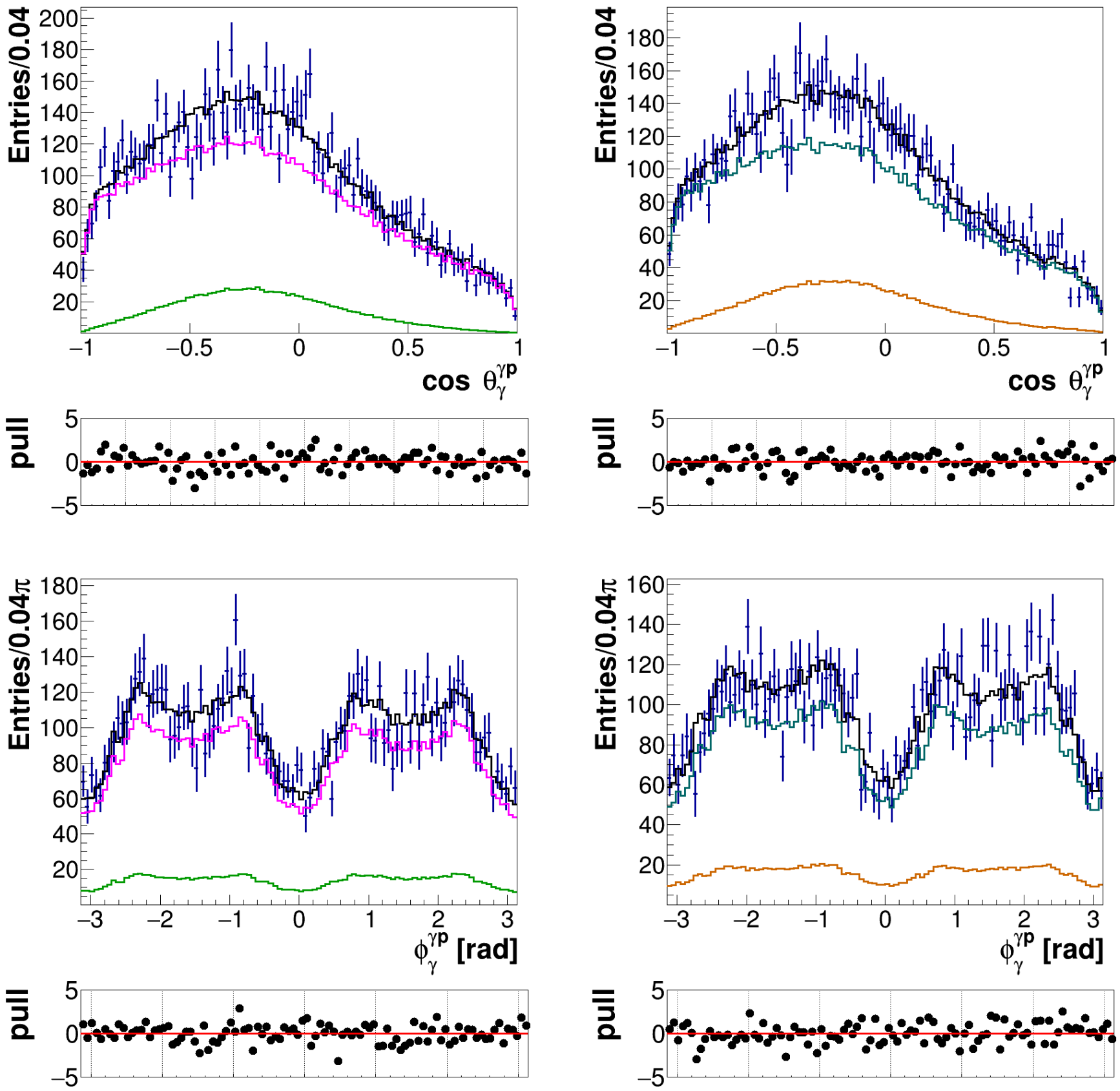

**Figure A.5:** Same as figure A.4, but for the angular distributions. Here, $\theta_b^a$ and $\phi_b^a$ represent the polar and azimuthal helicity angles, as defined in chapter 5, of particle $b$ in the rest frame of $a$.

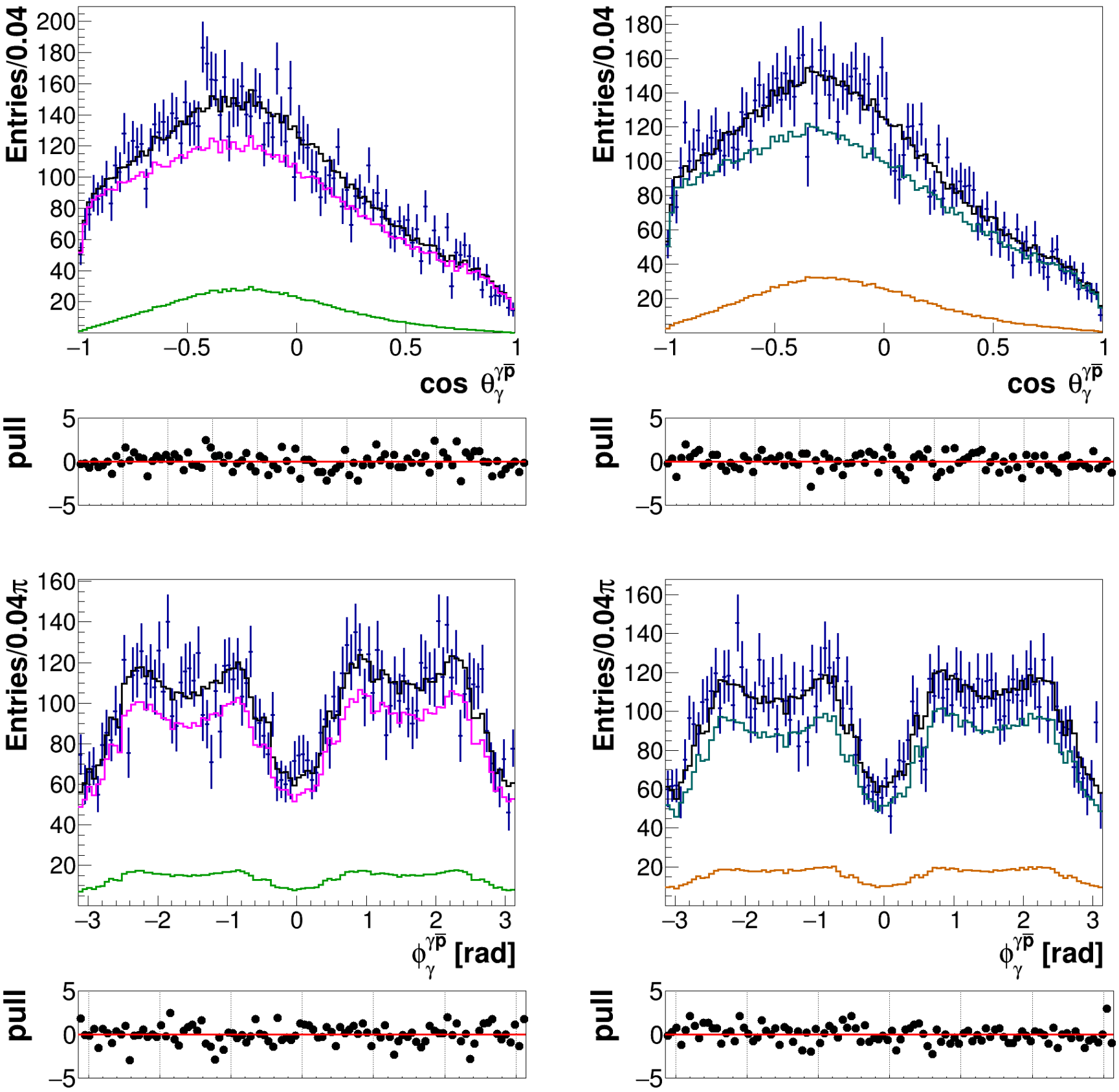

**Figure A.6:** Follow up of figure A.5.

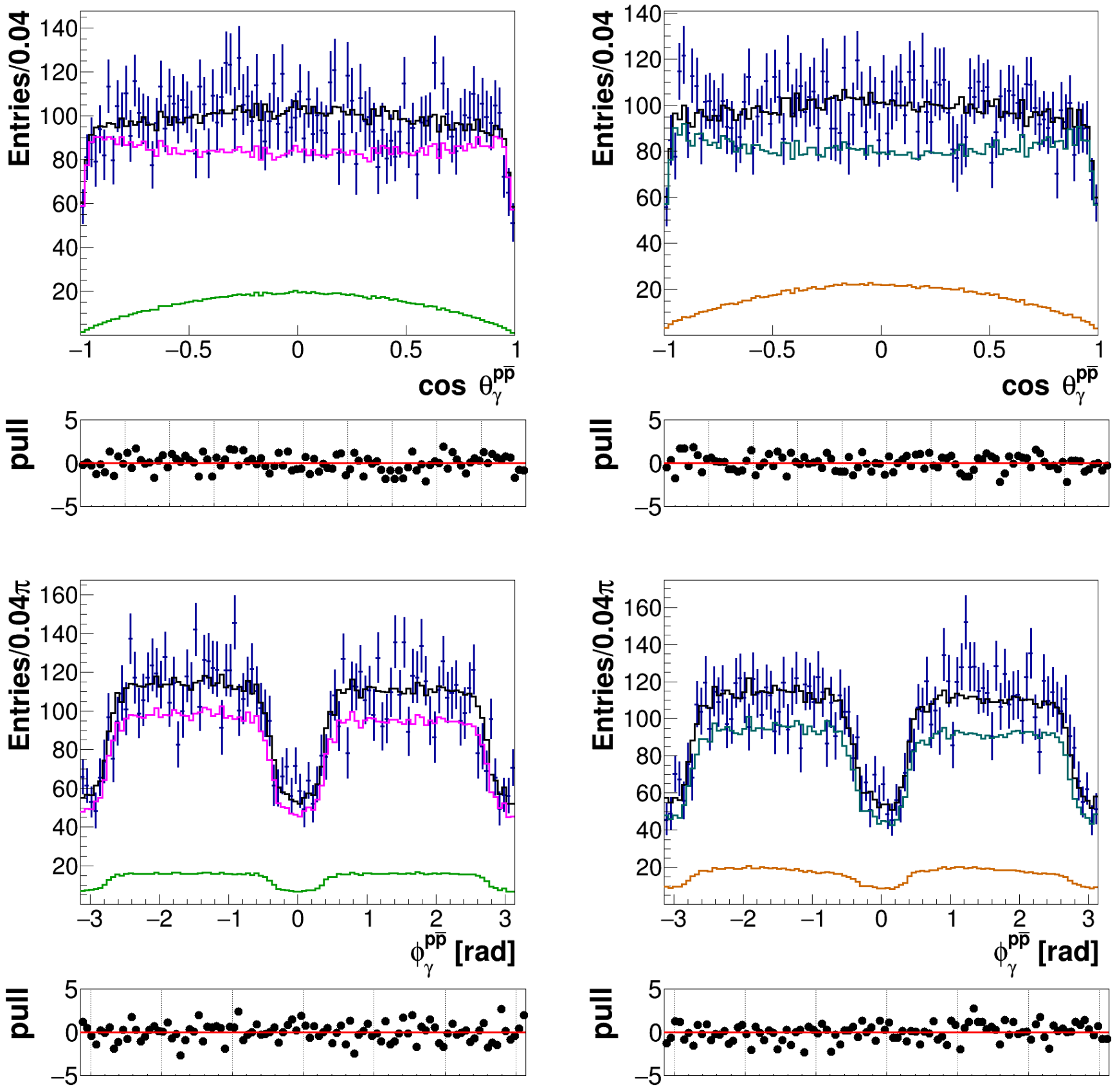

**Figure A.7:** Follow up of figure A.5.

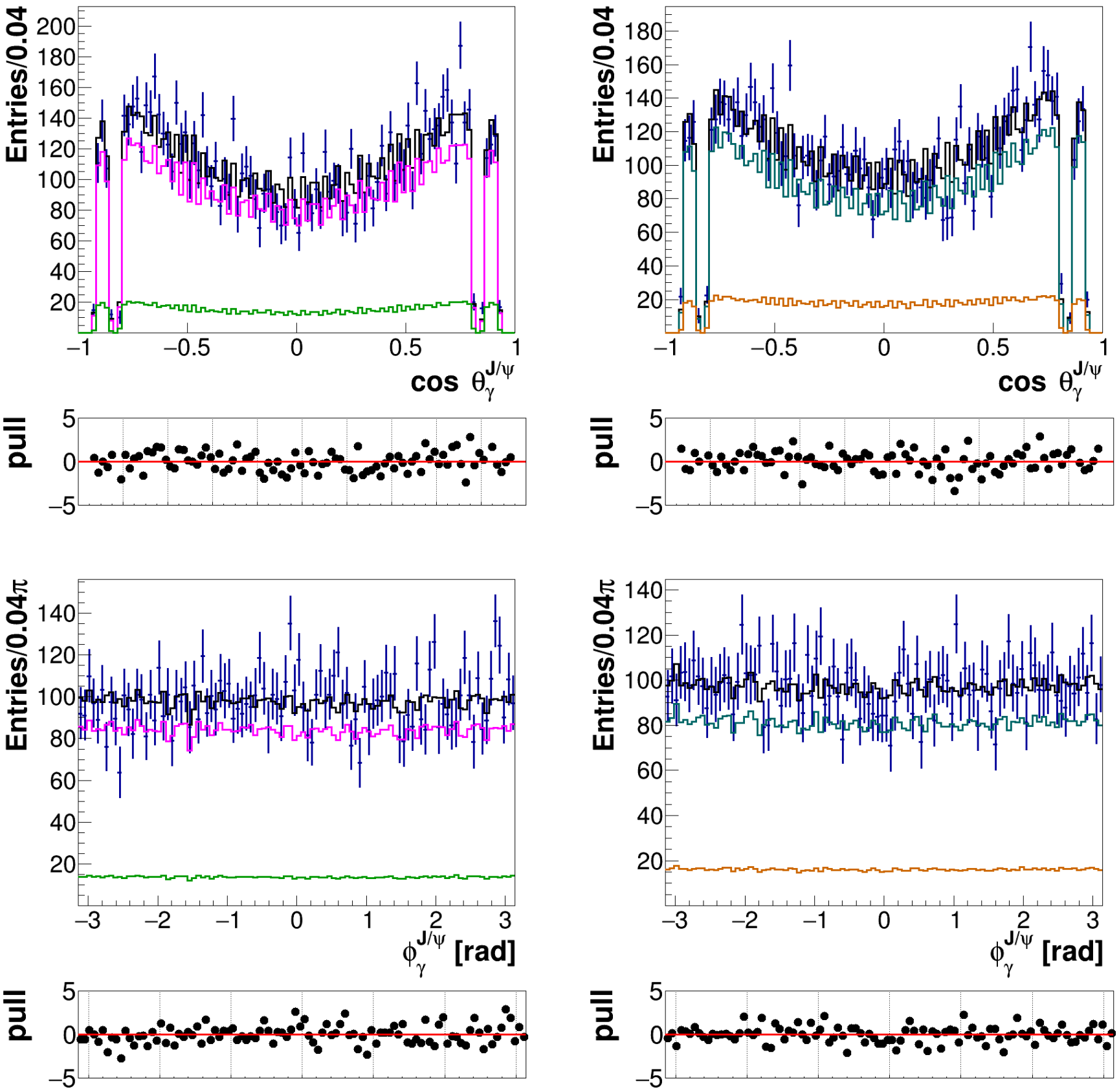

**Figure A.8:** Follow up of figure A.5.

# B. Comparison of the different hypotheses of the mass-dependent PWA

This appendix contains additional figures related to the outcome of the Partial-Wave Analysis (PWA) within the $p\bar{p}$ mass range in the vicinity of $\eta_c$ resonance, discussed in chapter 7. More details about the analysis can be found in that chapter.

In the following figures, the results for the two hypotheses $\eta_c + 0^{-+}$ and $\eta_c + 0^{-+} + 2^{++}$ are shown. All results are obtained from a Partial-Wave Analysis (PWA) in the range $2.7 - 3.1$ GeV/$c^2$, with the KEDR description for the $\eta_c$ line shape. All the figures show the hypothesis $\eta_c + 0^{-+}$ on the left-hand side, and $\eta_c + 0^{-+} + 2^{++}$ on the right-hand side. The total fit is shown in black, and the different contribution from $\eta_c$, $0^{-+}$ and $2^{++}$ are shown by the red, green, and blue lines, respectively. The pull distributions are filled via pull $= (n_{\mathrm{dat}} - n_{\mathrm{fit}})/\sqrt{\sigma^2_{\mathrm{dat}} + \sigma^2_{\mathrm{fit}}}$, where $n_{\mathrm{dat}}$ and $n_{\mathrm{fit}}$ represent the bin contents of the data and fit histogram, respectively, and $\sigma_{\mathrm{dat}}$ and $\sigma_{\mathrm{fit}}$ the corresponding bin (statistical) errors. In general, $\sigma_{\mathrm{fit}}$ is much smaller than $\sigma_{\mathrm{dat}}$. Interference contributions are not explicitly shown. Figure B.1 shows the different invariant mass combinations, and figures B.2–B.5 show the angular distributions. Here, $\theta^a_b$ and $\phi^a_b$ represent the polar and azimuthal helicity angles, as defined in chapter 5, of particle $b$ in the rest frame of $a$.

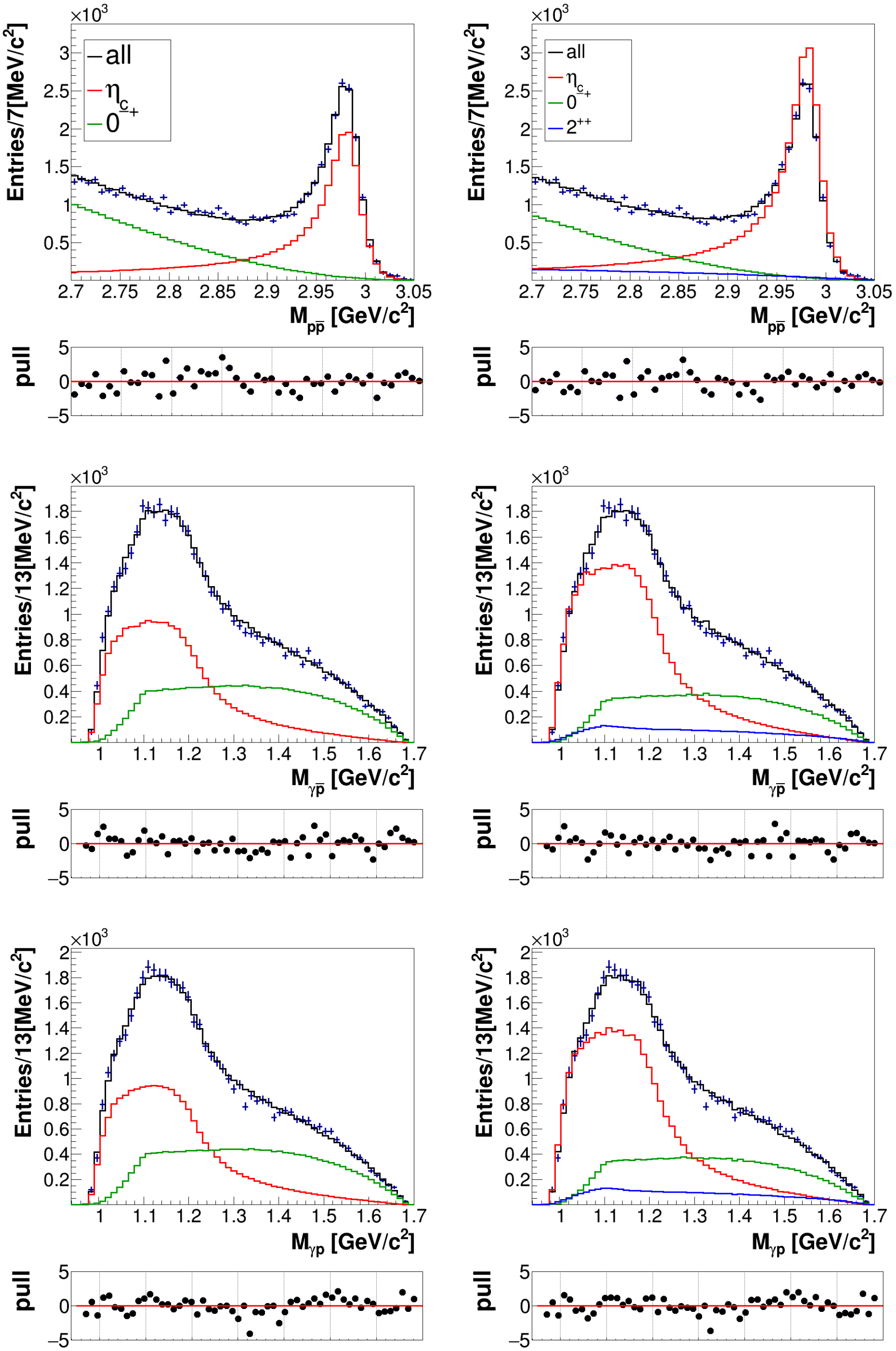

Figure B.1

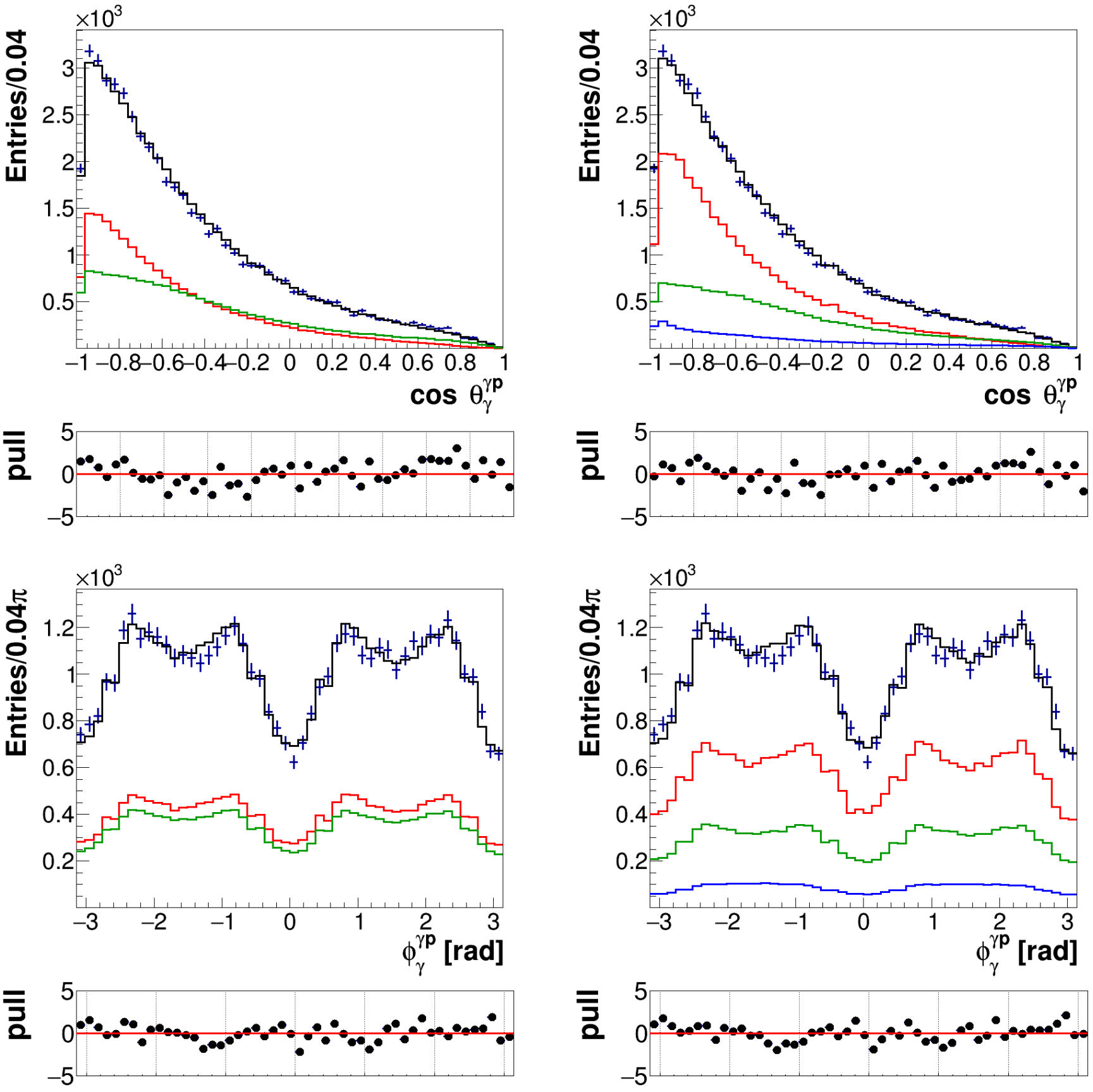

Entries/0.04
cos $\theta_\gamma^{\gamma p}$
pull
Entries/0.04π
$\phi_\gamma^{\gamma p}$ [rad]

**Figure B.2**

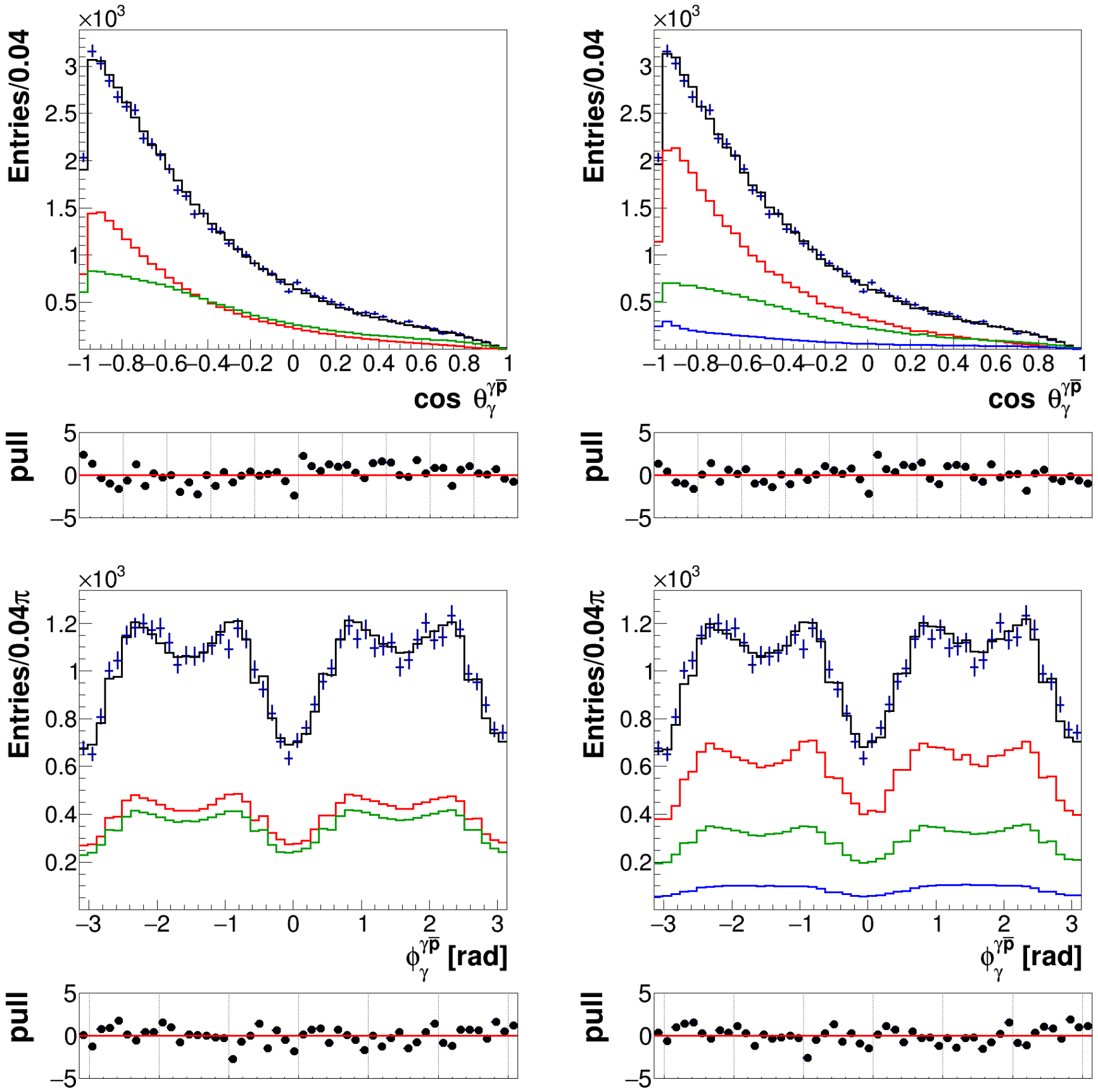

Figure B.3

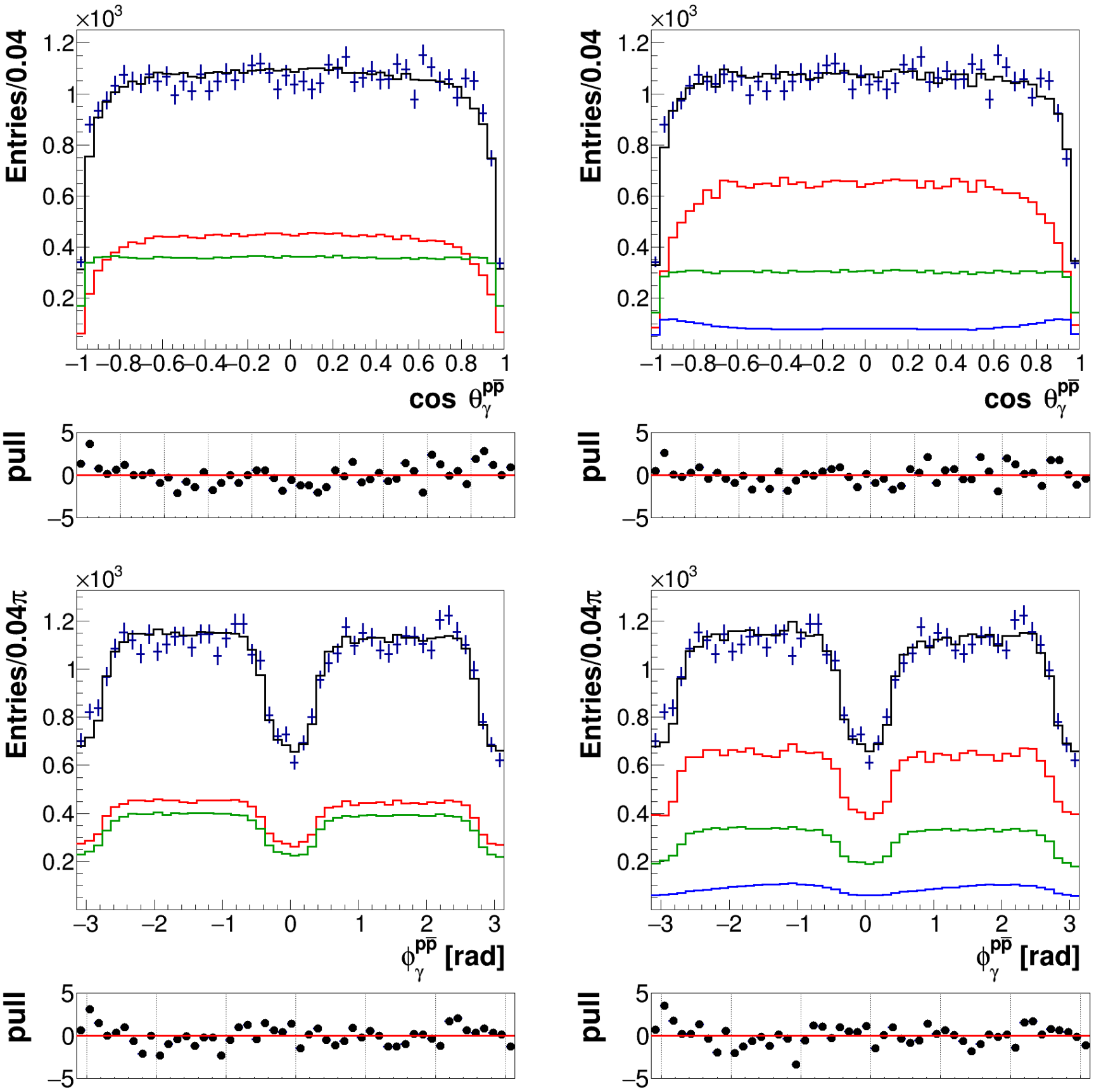

Figure B.4

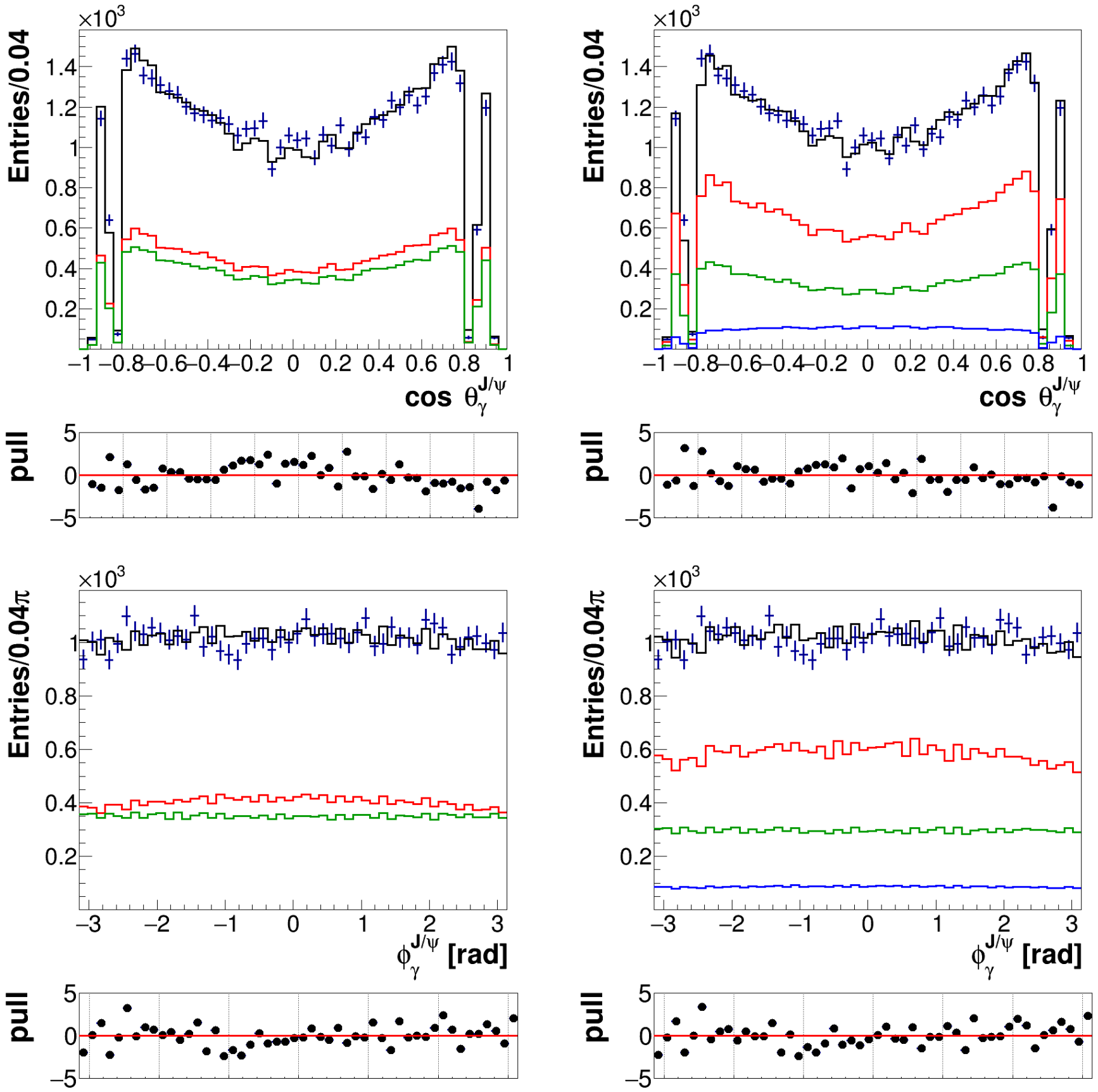

Figure B.5

# C. Plots of the final result and efficiency correction of the mass-dependent PWA

This appendix contains additional figures related to the outcome of the Partial-Wave Analysis (PWA) within the $p\bar{p}$ mass range in the vicinity of $\eta_c$ resonance, discussed in chapter 7. More details about the analysis can be found in that chapter. In appendix B, the results for two different hypotheses in the mass range $2.7 - 3.1$ GeV/$c^2$ are shown, whereas in this appendix figures displaying the final results for the hypothesis $\eta_c + 0^{-+}$, and the range $2.7 - 3.0$ GeV/$c^2$, are listed.

The PWA is performed with the KEDR description for the $\eta_c$ line shape. All the figures show the efficiency corrected histograms on the left-hand side, and the final PWA-fit result on the right-hand side. On both sides, the black line represent the total fit. The figures on the left additionally show the efficiency corrected histogram in red. The other figures include the different contributions from $\eta_c$ and $0^{-+}$ by the red and green lines, respectively. The pull distributions are filled via pull $= (n_{\text{dat}} - n_{\text{fit}})/\sqrt{\sigma_{\text{dat}}^2 + \sigma_{\text{fit}}^2}$, where $n_{\text{dat}}$ and $n_{\text{fit}}$ represent the bin contents of the data and fit histogram, respectively, and $\sigma_{\text{dat}}$ and $\sigma_{\text{fit}}$ the corresponding bin (statistical) errors. In general, $\sigma_{\text{fit}}$ is much smaller than $\sigma_{\text{dat}}$. Interference contributions are not explicitly shown. Figure C.1 shows the different invariant mass combinations, and figures C.2–C.5 show the angular distributions. Here, $\theta_b^a$ and $\phi_b^a$ represent the polar and azimuthal helicity angles, as defined in chapter 5, of particle $b$ in the rest frame of $a$.

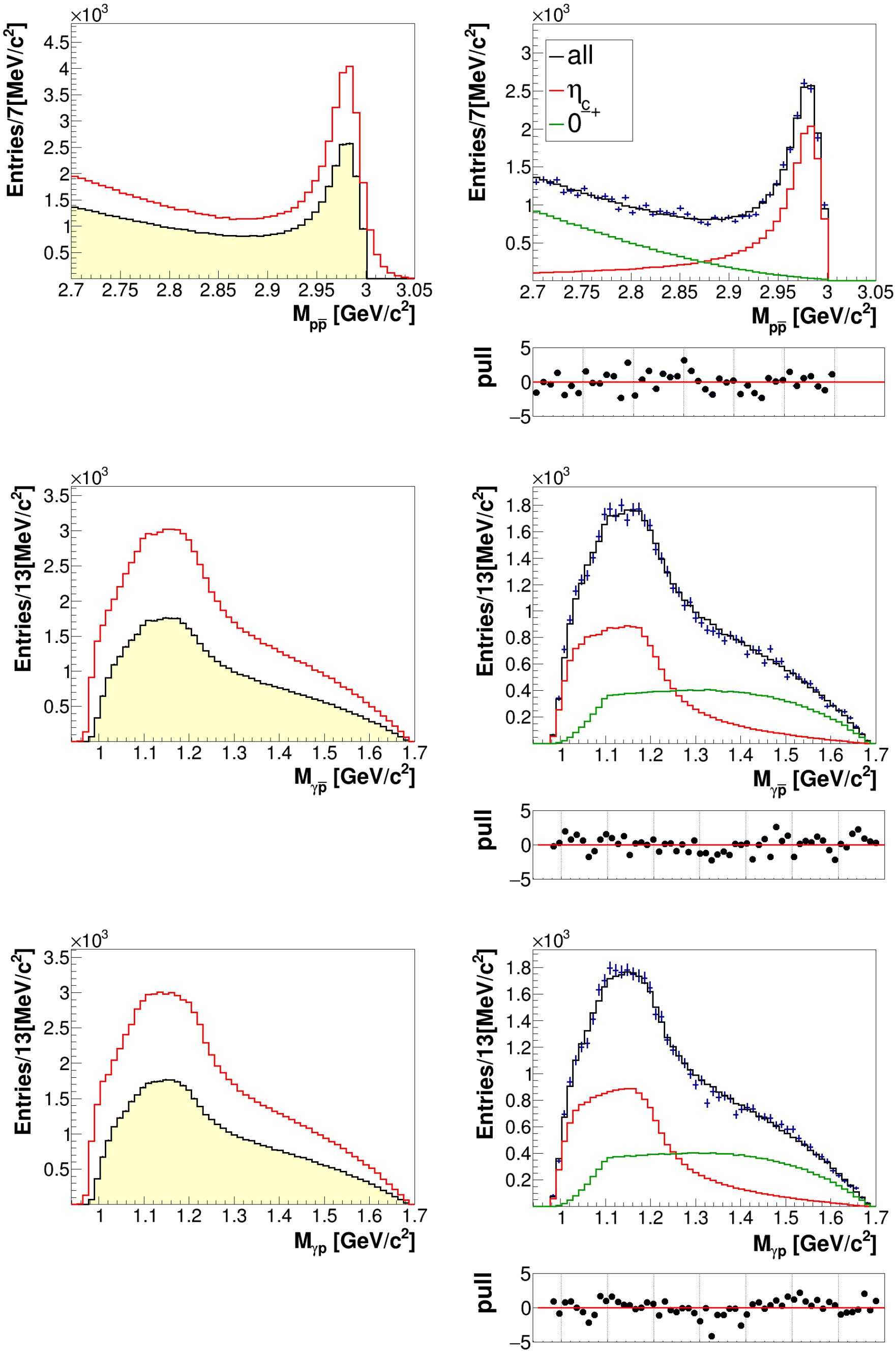

Figure C.1

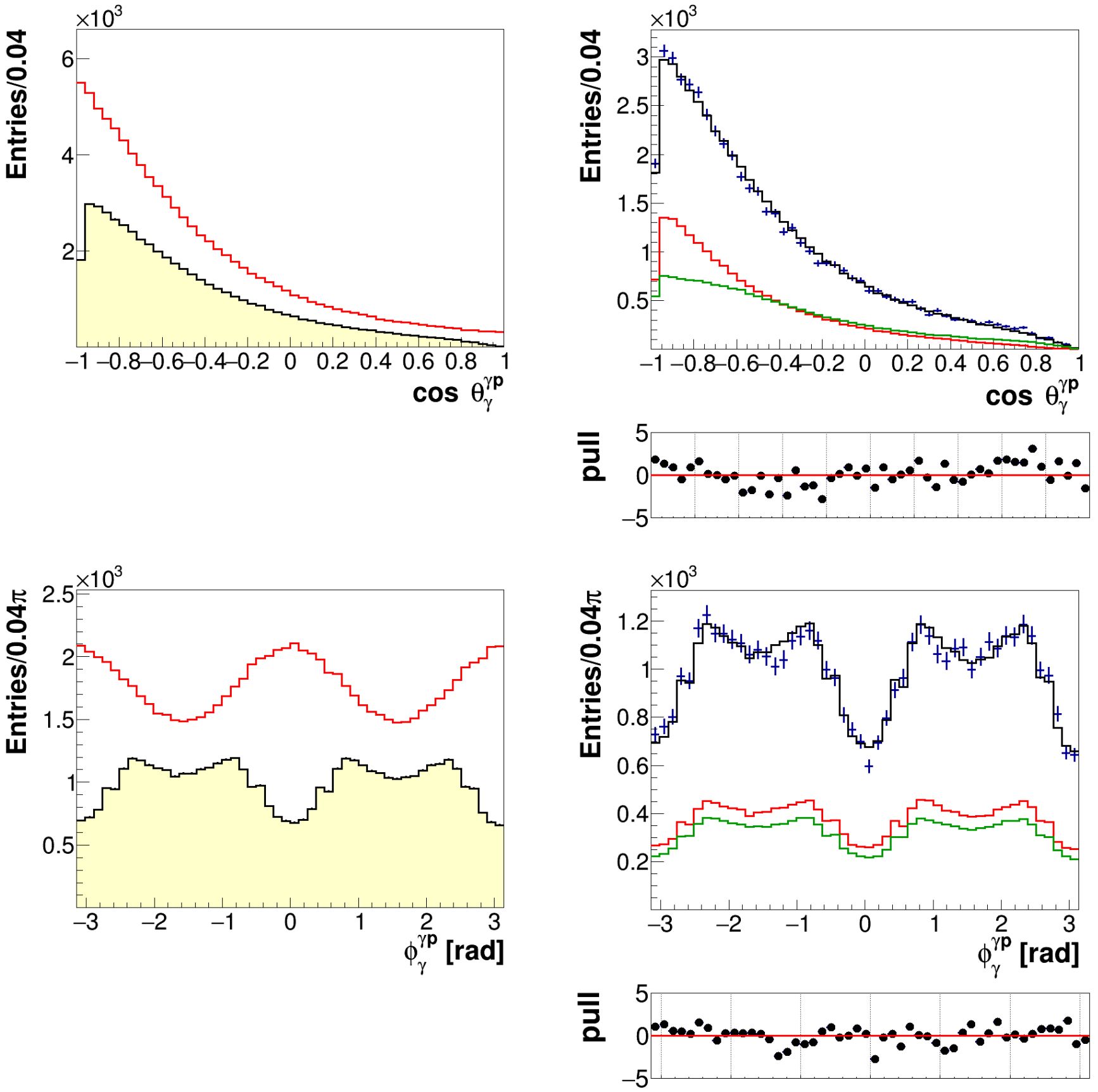

**Figure C.2**

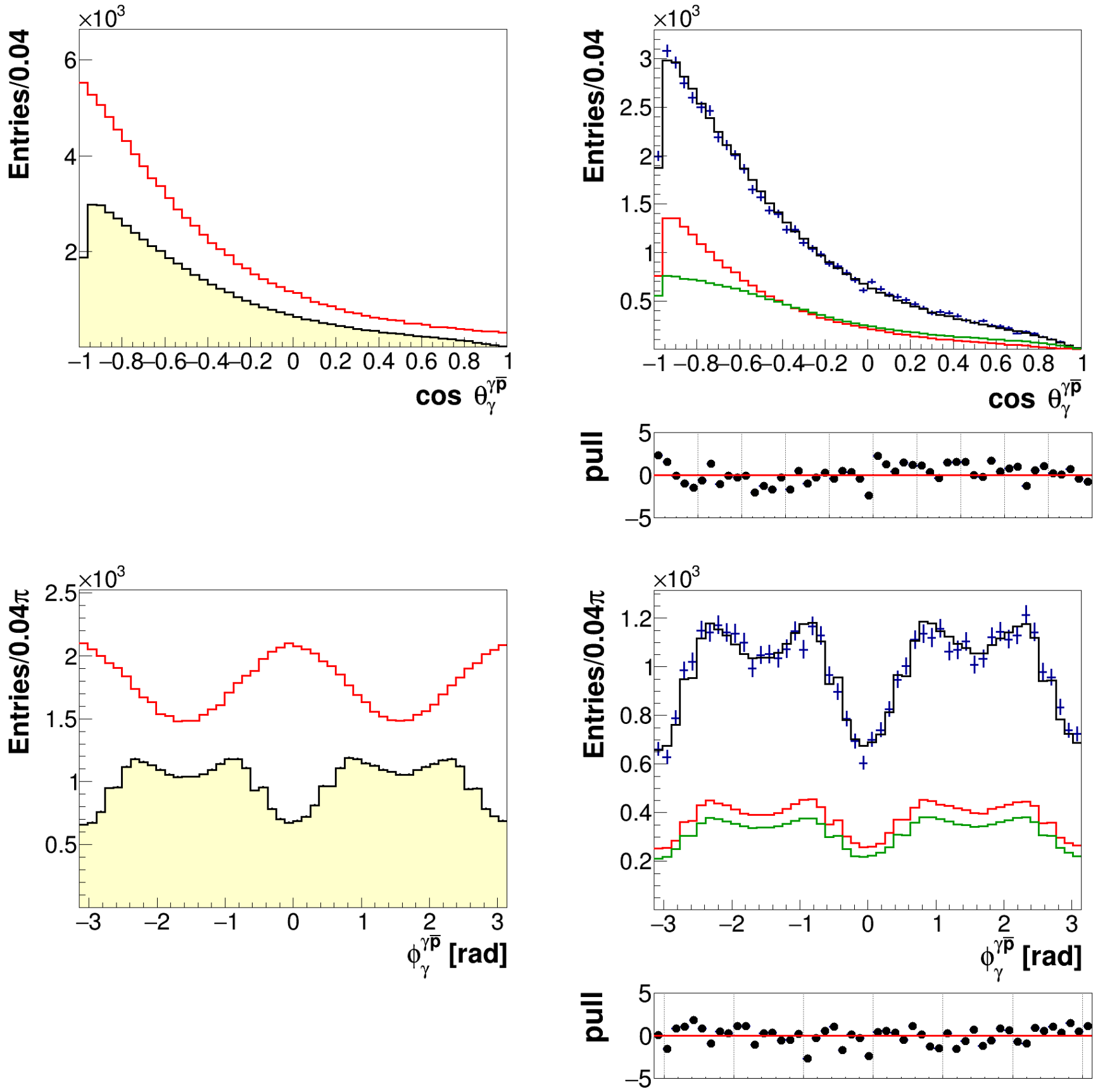

×10^3
Entries/0.04
6
4
2
−1 −0.8−0.6−0.4−0.2 0 0.2 0.4 0.6 0.8 1
cos θ_γ^γp̄
×10^3
Entries/0.04
3
2.5
2
1.5
1
0.5
pull
5
0
−5
×10^3
Entries/0.04π
2.5
2
1.5
1
0.5
−3 −2 −1 0 1 2 3
φ_γ^γp̄ [rad]
1.2
0.8
0.6
0.4
0.2

Figure C.3

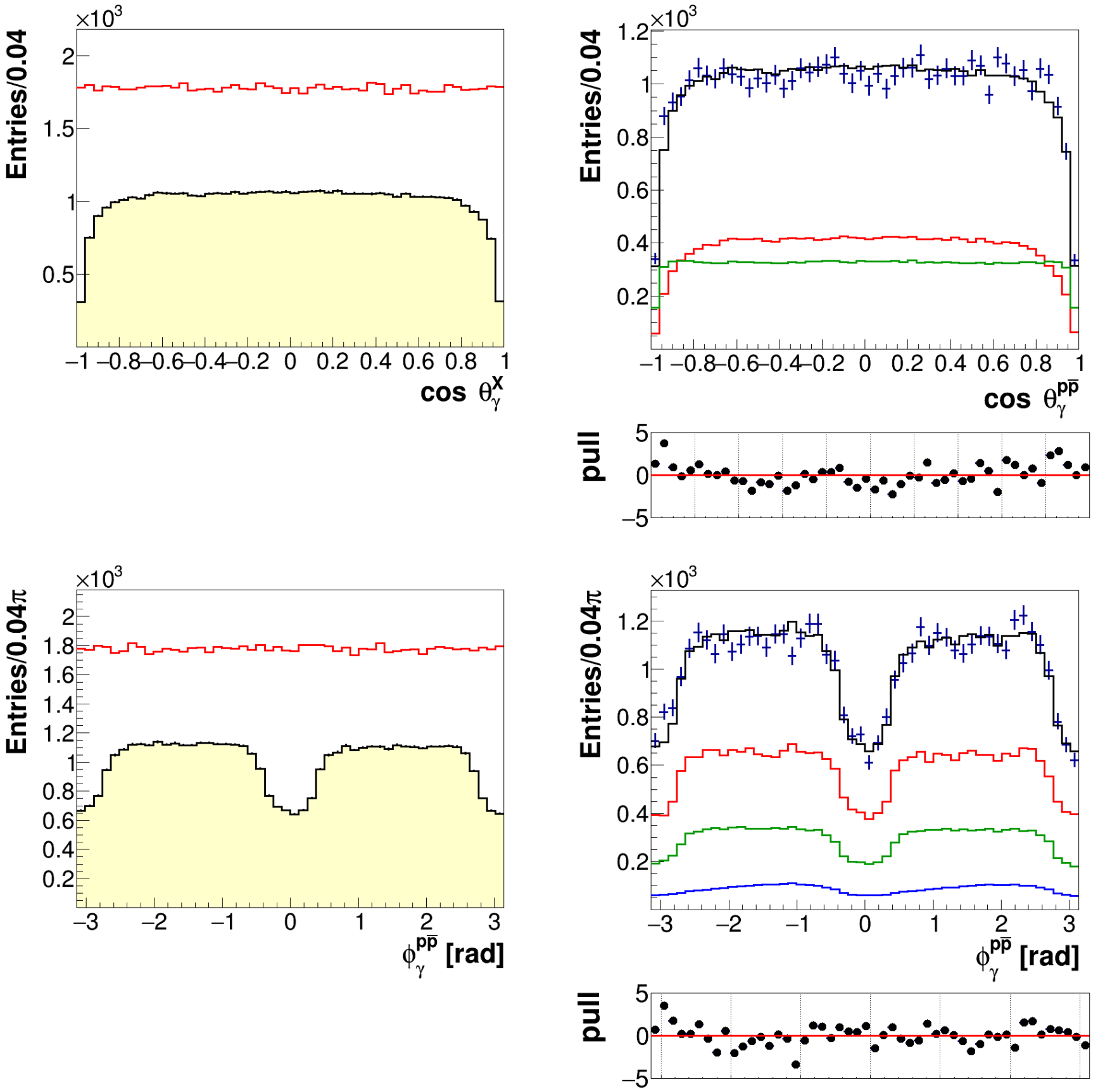

**Figure C.4**

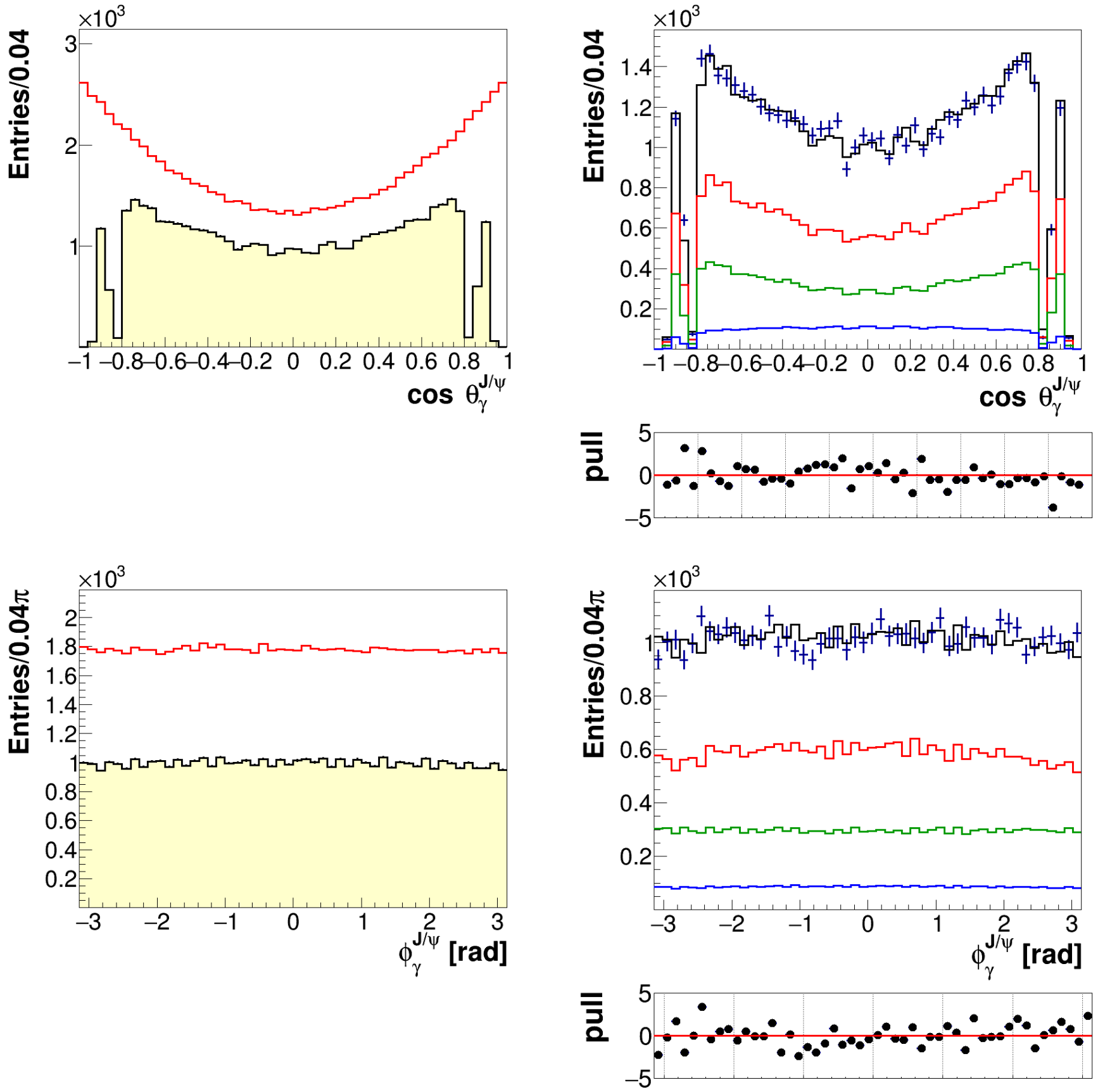

Figure C.5

# Acknowledgments

*I get by with a little help from my friends.*

— The Beatles

First of all, Johan, thank you for everything. I liked how your door was always open to come in and have a chat. Of course, that became a bit more challenging after you moved to Germany, but our Zoom sessions were long enough to discuss all the relevant developments on both sides. Aside from physics, I could not have imagined our views on the world would be so similar. It was great to keep each other up-to-date with (satirical) news articles. I hope we stay in touch.

Nasser, I am grateful to have been part of your group. I admire your temperamental personality. Your stories about customer services were entertaining and I enjoyed our many discussions about different cultures, religions and languages. I hope the new research direction of your group will turn out to be fruitful.

Thanks to the members of the BESIII collaboration, and especially the Bochum-delegation. It was a pleasure working with you and spending time together during the workshops and meetings in China and Europe. Bertram, thank you so much for all your quick responses and always being ready to help.

I would like to thank the members of my assessment committee for their thorough reading, followed by sincere questions and in-depth feedback.

Thanks to all my colleagues at the kvi-building. It has not always been easy during the reorganization and the pandemic, but it was always nice to have a chat with each of you in the hallways or at the coffee machine. For my more direct colleagues, I loved our traditional french-press coffee breaks in the afternoon. The watergun fight was memorable, as was the ice cream on every first hot day of the year. Maisam and Viktor, I am thankful to have you guys as my paranymphs. Viktor, you have been a wonderful office mate. Speciaal voor jou nog een paar woorden in het Nederlands. Bedankt voor alles, ik wens je een mooie toekomst in Nederland toe.

Last but not least, I would like to thank Steven, my family, his family and my friends. I believe you all know how important you are to me. A special thanks to my father Lambert for reading almost every version of my thesis and giving his feedback.

# List of Acronyms

**AIC** Akaike-Information-Criterion.
**AICc** second-order Akaike-Information-Criterion.

**BEPC** Beijing Electron Positron Collider.
**BES** Beijing Spectrometer.
**BIC** Bayesian-Information-Criterion.
**BOSS** BES Offline Software System.

**EMC** Electromagnetic Calorimeter.

**FSI** Final-State Interactions.
**FSR** Final-State Radiation.

**IHEP** Institute of High Energy Physics.

**JPE** $J/\psi$ to Photon Eta model.

**LQCD** Lattice Quantum ChromoDynamics.

**MC** Monte Carlo.
**MDC** Multilayer Drift Chamber.

**NRQCD** Non-Relativistic Quantum ChromoDynamics.

**PAWIAN** PArtial Wave Interactive ANalysis.
**PDG** Particle Data Group.
**PHSP** PHase-SPace model.
**PWA** Partial-Wave Analysis.

**QCD** Quantum ChromoDynamics.
**QED** Quantum ElectroDynamics.

**SM** Standard Model.

**TOF** Time-of-Flight.

# List of Figures

# List of Tables